\documentclass[11pt]{article}
\usepackage{graphicx,amsmath,amssymb}
\usepackage{indentfirst}
\usepackage[usenames]{color}
\usepackage[colorlinks=true, urlcolor=navyblue, linkcolor=navyblue, citecolor=navyblue]{hyperref}
\usepackage{epstopdf}
\usepackage{enumerate}
\usepackage{appendix}
\usepackage{lineno}
\usepackage{setspace}
\usepackage{wrapfig}
\usepackage{lipsum}
\usepackage{placeins}
\usepackage{ulem}
\definecolor{navyblue}{rgb}{0,0.08,0.45}

\begin{document}


{
\begin{flushright}
{\small JLAB-PHY-18-2760 \\ \vspace{3pt}
}
\end{flushright}
\begin{flushright}
{\small SLAC--PUB--17279 \\ \vspace{3pt}
}
\end{flushright}

\vspace{40pt}

\centerline{\huge \bf The Spin Structure of the Nucleon}

\vspace{80pt}

\centerline{Alexandre~Deur}

\vspace{5pt}

\centerline {\it Thomas Jefferson National Accelerator Facility, Newport News, VA 23606, USA}

\vspace{15pt}

\centerline{Stanley J. Brodsky}

\vspace{5pt}

\centerline {\it SLAC National Accelerator Laboratory, Stanford University, Stanford, CA 94309, USA}

\vspace{15pt}

\centerline{Guy F. de T\'eramond}

\vspace{5pt}

\centerline {\it Laboratorio de F\'isica Te\'orica y Computacional, Universidad de Costa Rica, San Jos\'e, Costa Rica}
\vspace{20pt}

{\small \centerline{\today}}

\vspace{60pt}

{\small
\centerline{\href{mailto:deurpam@jlab.org}{\tt deurpam@jlab.org},}

\vspace{1pt}

{\small
\centerline{\href{mailto:sjbth@slac.stanford.edu}{\tt sjbth@slac.stanford.edu}, \,~\href{mailto:gdt@asterix.crnet.cr}{\tt gdt@asterix.crnet.cr},}
}}

 \vspace{40pt}

\newpage

{\centerline{ \bf \large Abstract}

We review the present understanding of the spin structure of protons and neutrons, 
the fundamental building blocks of nuclei collectively known as nucleons. The field 
of nucleon spin provides a critical window for testing Quantum Chromodynamics (QCD), 
the gauge theory of the strong interactions, since it involves fundamental aspects of 
hadron structure which can be probed in detail in experiments, particularly deep 
inelastic lepton scattering on polarized targets.

QCD was initially probed in high energy deep inelastic lepton scattering with 
unpolarized beams and targets. With time, interest shifted from testing perturbative 
QCD to illuminating the nucleon structure itself. In fact, the spin degrees of freedom 
of hadrons provide an essential and detailed verification of both perturbative and 
nonperturbative QCD dynamics.

Nucleon spin was initially thought of coming mostly from the spin of its quark constituents, 
based on intuition from the parton model. However, the first experiments showed that this 
expectation was incorrect. It is now clear that nucleon physics is much more complex, 
involving quark orbital angular momenta as well as gluonic and sea quark contributions. 
Thus, the nucleon spin structure remains a most active aspect of QCD research, involving 
important advances such as the developments of generalized parton distributions (GPD) 
and transverse momentum distributions (TMD). 

Elastic and inelastic lepton-proton scattering, as well as photoabsorption experiments 
provide various ways to investigate non-perturbative QCD. Fundamental sum rules -- 
such as the Bjorken sum rule for polarized photoabsorption on polarized nucleons -- 
are also in the non-perturbative domain. This realization triggered a vig{orous program 
to link the low energy effective hadronic description of the strong interactions to fundamental 
quarks and gluon degrees of freedom of QCD. This has also led to advances 
in lattice gauge theory simulations of QCD and to the development of 
holographic QCD ideas based on the AdS/CFT or gauge/gravity correspondence, a novel 
approach providing a well-founded semiclassical approximation to QCD. Any QCD-based 
model of the nucleon's spin and dynamics must also successfully account for the observed 
spectroscopy of hadrons. Analytic calculations of the hadron spectrum, a long sought 
goal of QCD research, have now being realized using light-front holography and superconformal 
quantum mechanics, a formalism consistent with the results from nucleon spin studies.

We begin this review with a phenomenological description of nucleon structure in general 
and of its spin structure in particular, aimed to engage non-specialist readers. Next, we 
discuss the nucleon spin structure at high energy, including topics such as  Dirac's front 
form and light-front quantization which provide a frame-independent, relativistic description 
of hadron structure and dynamics, the derivation of spin sum rules, and a direct connection 
to the QCD Lagrangian. We then discuss experimental and theoretical advances in the 
nonperturbative domain -- in particular the development of light-front holographic QCD and 
superconformal quantum mechanics, their predictions for the spin content of nucleons, the 
computation of PDFs and of hadron masses.

}

\tableofcontents

\section{Preamble}
The study of the individual contributions to the nucleon spin provides a critical window
for testing detailed predictions of QCD for the internal quark and gluon structure of hadrons. 
Fundamental spin predictions can be tested experimentally  to high precision,
particularly in  measurements of deep inelastic scattering (DIS) of polarized 
leptons on polarized proton and nuclear targets.

The spin of the nucleons was initially thought to originate simply from the spin
of the constituent quarks, based on intuition from the parton model.
However, experiments have shown that this expectation was incorrect.
It is now clear that nucleon spin physics is much more complex, involving
quark and gluon orbital angular momenta (OAM) as well as gluon spin and sea-quark contributions.  
Contributions to the nucleon spin, in fact, originate from the 
nonperturbative dynamics associated with color confinement  as well as  from perturbative QCD (pQCD)  evolution.
Thus, nucleon spin structure  has become an active aspect of QCD
research, incorporating important  theoretical advances such as the development of
GPD and TMD.

Fundamental  sum rules, such as the Bjorken sum rule for polarized DIS or the Drell-Hearn-Gerasimov sum rule for polarized photoabsorption cross-sections, constrain critically the 
spin structure.  In addition, elastic lepton-nucleon scattering and other 
exclusive processes, e.g. Deeply Virtual Compton Scattering (DVCS), 
also determine important aspects of nucleon spin dynamics. 
This has led to a vigorous theoretical and experimental program to obtain an 
effective hadronic description of the strong force in terms of the
basic quark and gluon fields of QCD.  
Furthermore, the theoretical program for determining the spin structure of hadrons has 
benefited from advances in lattice gauge theory simulations of QCD
and the recent development of light-front holographic QCD ideas based on the AdS/CFT 
correspondence, an approach to hadron structure based on the
 holographic embedding of light-front dynamics in a higher dimensional gravity theory, together with
 the constraints imposed by the underlying superconformal algebraic structure.
This  novel approach to nonperturbative QCD and color confinement has provided
a well-founded semiclassical approximation to QCD. 
QCD-based models
of the nucleon spin and dynamics must also successfully account
for the observed spectroscopy of hadrons.  Analytic calculations of
the hadron spectrum, a long-sought goal, 
are now being carried out using  Lorentz frame-independent light-front holographic 
methods.

We begin this review by discussing why nucleon
spin structure has become a central topic of hadron physics (Section~\ref{overview}).   
The goal of this introduction is to engage  the non-specialist reader by providing a phenomenological
description of nucleon structure in general and  its spin structure
in particular.

We then discuss the scattering reactions (Section~\ref{Scattering spectrum})
which constrain nucleon spin structure, and the 
theoretical methods (Section~\ref{Computation methods}) used 
for perturbative or nonperturbative QCD calculations. 
A fundamental tool is Dirac's front form (\emph{light-front quantization})
which, while keeping a direct connection to the QCD Lagrangian, 
provides a frame-independent,  relativistic description of hadron
structure and dynamics, as well as a rigorous physical formalism 
that can be used to derive spin sum rules (Section ~\ref{sum rules}). 

Next, in Section~\ref{sec:data}, we discuss the existing spin structure
data, focusing on the inclusive lepton-nucleon scattering results, as well as 
other types of data, such as semi-inclusive deep
inelastic scattering (SIDIS) and proton-proton scattering.  
Section~\ref{sec:perspectives} provides an example of the
knowledge gained from nucleon spin studies which illuminates fundamental  features of hadron dynamics and structure.
Finally, we summarize in Section~\ref{cha:Futur-results} our present understanding of 
the nucleon spin structure and its impact on testing nonperturbative aspects of QCD.

A lexicon of terms specific to the nucleon spin 
structure and related topics is provided at the end of this review  to assist non-specialists.
Words from this list are italicized throughout the review. 
Also included is a list of acronyms used  in  this review.

Studying the spin of the nucleon is a complex subject because light 
quarks move relativistically within hadrons;   one needs special care 
in defining angular momenta beyond conventional  nonrelativistic treatments~\cite{Chiu:2017ycx}. 
Furthermore, the concept of gluon spin is gauge dependent; there is no gauge-invariant definition of the spin of gluons 
-- or gauge particles in general~\cite{Ji:2014lra, Kinoshita:1990nb};  the definition
of the  spin  content of the nucleon is thus dependent on the choice of gauge. 
 In the light-front form one usually takes the light-cone gauge~\cite{Chiu:2017ycx}  where the spin is well defined: there are no ghosts or
negative metric states in this transverse gauge (See Sec. \ref{LC dominance and LF quantization}).
Since nucleon structure is nonperturbative,  calculations based solely on first principles of QCD are difficult. 
These features make the nucleon spin structure an active and challenging field of study. 

There are several excellent previous reviews  which discuss the high-energy 
aspects of proton spin dynamics~\cite{Bass:2004xa, Chen:2005tda, Burkardt:2008jw, Kuhn:2008sy, Chen:2010qc, Aidala:2012mv, Blumlein:2012bf}.
This review will also cover  less conventional topics, such as how studies of 
spin structure illuminate aspects of the strong force in its nonperturbative domain, 
the consequences of color confinement, the origin of the QCD mass scale,
and the emergence of hadronic degrees of freedom from its partonic ones.

It is clearly important to know how the quark and gluon
spins combine with their OAM to form the total nucleon spin. 
A larger purpose is to use empirical information on the spin structure of hadrons   
to illuminate features of the strong force -- arguably the least understood fundamental force 
in the experimentally accessible domain. For example, the 
parton distribution functions (PDFs) are themselves nonperturbative quantities.  
Quark  and gluon OAMs -- which significantly 
contribute to the nucleon spin -- are directly connected to color confinement.

We will only briefly discuss 
some high-energy topics such as GPDs, TMDs, and the nucleon spin observables 
sensitive to final-state interactions such as the Sivers effect. These topics
are well covered in the reviews mentioned above. A recent review on the transverse 
spin in the nucleon is given in Ref.~\cite{Perdekamp:2015vwa}. 
These topics are needed to understand the details of nucleon spin structure at 
high energy, but they only provide qualitative information on our main topic, the
nucleon spin~\cite{Jaffe:1991kp}.   
For example, the large transverse spin asymmetries measured in 
singly-polarized lepton-proton and proton-proton
collisions hint at significant transverse-spin--orbit coupling in
the nucleon. This provides an important motivation for the TMD and GPD studies which 
constrain OAM contributions to nucleon spin. 


\section{Overview of QCD and the nucleon structure  \label{overview}}

The description of phenomena given by the Standard Model is
based on a small number of basic elements: the fundamental particles (the six
quarks and six leptons, divided into three families), the four fundamental interactions
(the electromagnetic, gravitational, strong and weak nuclear forces) through 
which these particles interact, and
the Higgs field which is at the origin of the masses of the fundamental particles.
Among the four interactions, the strong force is the least 
understood in the presently accessible experimental domains.  
QCD, its gauge theory, describes the interaction of quarks 
via the exchange of vector gluons, the gauge bosons
associated with the color fields.  Each quark carries a ``color" charge 
labeled blue, green or red,  and they interact by the exchange of colored gluons 
belonging to a color octet.

QCD is best understood and well
tested at small distances thanks to the property of \emph{asymptotic freedom}~\cite{Gross:1973id}: 
the strength of the interaction between color charges effectively
decreases as they get closer.  The formalism of pQCD can therefore
be applied at small distances; {\it i.e.}, at high momentum transfer, and it has met with remarkable success.
This important feature allows one to validate QCD as the correct fundamental theory
of the strong force.  However, most natural phenomena involving hadrons, including color confinement, 
are governed by nonperturbative aspects of QCD.

\emph{Asymptotic freedom} also implies that the binding of quarks becomes stronger
as their mutual separation increases.  Accordingly, the quarks
confined in a hadron react increasingly coherently as the
characteristic distance scale at which the hadron is probed becomes
larger: The nonperturbative distributions of all quarks and gluons within
the nucleon can participate in the reaction.  
In fact, even in the perturbative domain, the nonperturbative
dynamics which underlies  hadronic bound-state structure is nearly always 
involved and is incorporated in distribution amplitudes, structure functions, and quark and gluon jet fragmentation
functions.  This is why, as a general rule,
pQCD cannot predict the analytic form and magnitude of such distributions, but only their evolution
with a change of scale, such as the momentum transfer of the probe.  
For a complete understanding of the strong force
and of the hadronic and nuclear matter surrounding us (of which $\approx95\%$ of the 
mass comes from the strong force), it is essential
to understand QCD in its nonperturbative domain. The key
example of a nonperturbative mechanism 
which is still not clearly understood is color confinement.

At large distances, where the internal structure cannot be resolved, effective degrees of freedom emerge;
thus the fundamental degrees of freedom of QCD, quarks and gluons, are effectively replaced by baryons and mesons. 
The emergence of relevant degrees of freedom associated with
an effective theory is a standard occurence in physics;  {\it e.g.}, Fermi's theory for
the weak interaction at large distances, molecular physics with its
effective Van der Waals force acting on effective degrees of
freedom (atoms), or geometrical optics whose essential degree of freedom
is the light ray.  Even outside of the field of physics, a science based
on natural processes often leads to an effective theory in which the complexity
of the basic phenomena is simplified by the introduction of effective
degrees of freedom, sublimating the underlying effects that become
irrelevant at the larger scale.  For example, biology takes root from
chemistry, itself based on atomic and molecular physics which in part are based on effective
degrees of freedom such as nuclei.  Thus the importance of understanding
the connections between the fundamental theories and effective theories
 to satisfactorily  unify  knowledge on a single
theoretical foundation.  An important avenue of research in  
QCD belongs to this context: to understand the connection
between the fundamental description of nuclear matter in terms
of quarks and gluons and its effective description in terms of the
baryons and mesons.  A part of this review will discuss how spin
helps with this endeavor.

QCD is most easily studied with the nucleon, since it is
stable and its structure is determined by the strong force.
As a first step, one studies its structure without accounting for the
spin degrees of freedom. This simplifies both theoretical and
experimental aspects. Accounting for spin then tests QCD in detail. 
This has been made possible due to continual technological 
advances such as polarized beams and polarized targets.

A primary way to study the nucleon is to scatter beams of particles 
-- leptons or hadrons -- on a fixed target.  The interaction
between the beam and  target typically occurs by the exchange of 
a photon or a $W$ or $Z$ vector boson. 
The momentum of the exchanged quantum
controls the time and distance scales of the probe.

Alternatively, one can  collide  two beams.  Hadrons  either
constitute one or both beams (lepton-hadron or hadron-hadron colliders)
or are generated during the collision ($e^+$--$e^-$ colliders). 
The main facilities where nucleon spin structure has been studied are 
SLAC in California, USA (tens of GeV electrons impinging on fixed proton or nuclear targets),  
CERN in France/Switzerland (hundreds of GeV muons colliding with fixed targets),
DESY in Germany (tens of GeV electrons in a ring scattering off an internal gas target),
Jefferson Laboratory (JLab) in Virginia, USA (electrons with energy up to 11 GeV with fixed targets), 
the Relativistic Heavy Ion Collider (RHIC)  at Brookhaven Laboratory in New York, 
USA (colliding beams of protons or nuclei  with energies about 10 GeV per nucleon),  and
MAMI  (electrons of  up to 1.6 GeV on fixed targets) in Germany.
%
%
We will now survey the formalism describing the
various reactions just described.

\subsection{Charged lepton-nucleon scattering}

\begin{wrapfigure}{r}{0.45\textwidth} 
\vspace{-1.5cm}
\hspace{0.6cm}
\includegraphics[width=6.0cm]{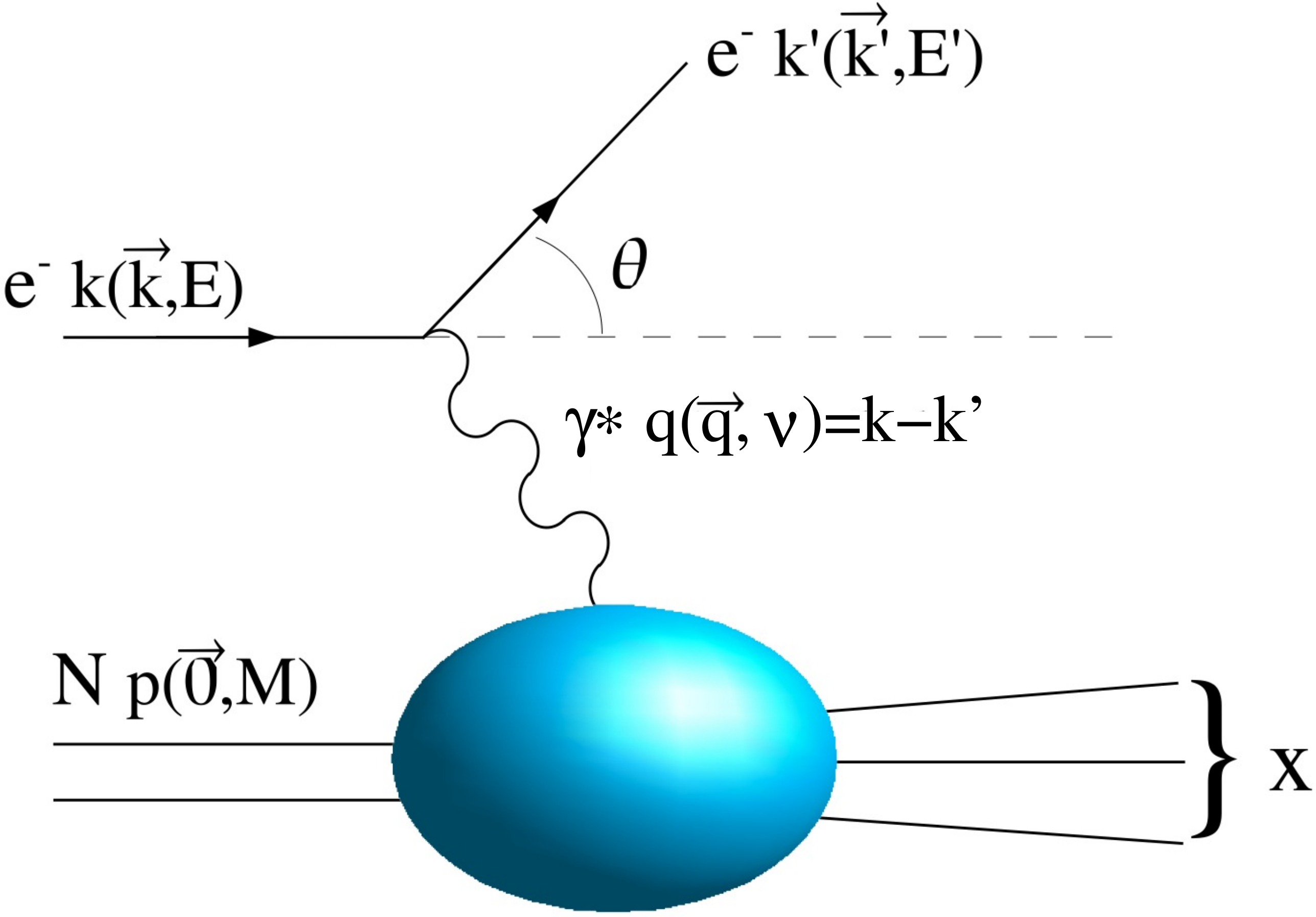}
\vspace{-0.5cm}
\caption{\label{Flo:electron scattering}\small Inclusive electron scattering off
a nucleon, in the first Born approximation. 
The blob represents the nonperturbative response of the target
to the photon.}
\vspace{-0.5cm}
\end{wrapfigure}

We start our discussion with experiments where charged leptons  scatter
off a fixed target. We focus on the ``inclusive" case
where only the scattered lepton is detected. The interactions involved
in the reaction are the electromagnetic force (controlling the scattering of the lepton)
and the strong force (governing the nuclear or nucleon structures).
Neutrino scattering,
although it is another important probe of nucleon structure, will not 
be discussed in detail here because  the small weak interaction cross-sections, and the unavailability of large 
polarized targets, have so far prohibited its use for detailed spin structure studies.  
Nonetheless, as we shall discuss, neutrino scattering off unpolarized  targets and parity-violating electron
scattering  yield constraints  on nucleon spin~\cite{Alberico:2001sd}. 
The formalism for inelastic lepton scattering, including the weak interaction,
can be found {\it e.g.} in Ref.~\cite{Anselmino:1994gn}.

\subsubsection{The first Born approximation {\label{sub:born1}}}

The electromagnetic interaction of a lepton with a hadronic or nuclear target proceeds
by the exchange of a virtual photon.  The first-order amplitude, known as the first
Born approximation,  corresponds to a single photon exchange, see Fig.~\ref{Flo:electron scattering}. 
In the case of electron scattering, where the lepton mass is small, 
higher orders in perturbative quantum electrodynamics (QED)
are needed to account for bremsstrahlung (real photons emitted by the incident or the scattered
electron), vertex corrections (virtual photons emitted by the incident
electron and re-absorbed by the scattered electron) and ``vacuum polarization" diagrams
(the exchanged photon temporarily turning into  pairs of charged
particles). In some cases, such as high-$Z$ nuclear targets, it  is also necessary to account for the cases
where the interaction between the electron and the target is transmitted
by the exchange of multiple photons (see {\it e.g.}~\cite{Guichon:2003qm}).
This correction will be negligible for the reactions and kinematics discussed here. Perturbative techniques 
can be applied to the electromagnetic probe, since  the QED coupling $\alpha \approx 1/137$, but not to the target structure 
whose reaction to the absorption of the photon is governed by the strong force at large distances where 
the QCD coupling $\alpha_s$ can be large.

\subsubsection{Kinematics }
\begin{wrapfigure}{r}{0.5\textwidth} 
\vspace{-1.95cm}
  \includegraphics[scale=0.65]{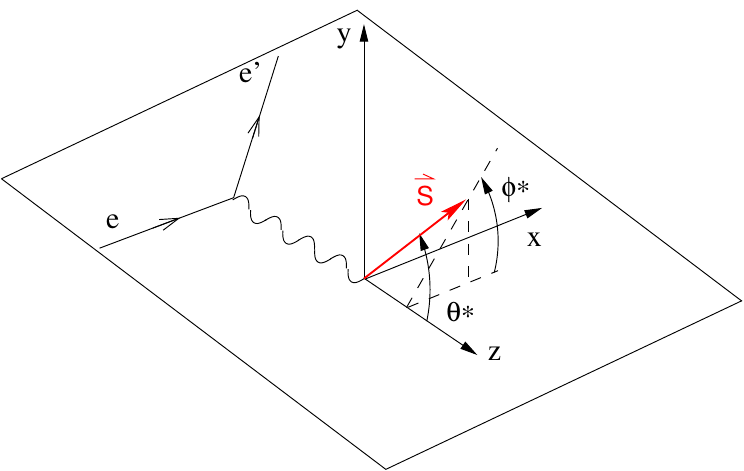}
 \vspace{-0.5cm}\caption{\label{fig:spinangles}\small Definitions of the polar angle $\theta^{*}$ and azimuthal angle $\phi^{*}$
of the target spin $\vec{S}$. The scattering plane is defined by $x\otimes z$.}
\vspace{-0.5cm}
\end{wrapfigure}
In inclusive reactions the final state system
$X$ is not detected. In the case of an ``elastic" reaction, the target particle emerges without structure modification.
Alternatively, the target nucleon or nucleus can emerge as excited states which promptly decay by emitting new particles (the resonance region), 
or the target can fragment, with additional particles produced in the final state as in DIS.

We first consider measurements in the laboratory frame where the nucleon or nuclear  target 
is at rest (Figs.~\ref{Flo:electron scattering} and \ref{fig:spinangles}).   
The laboratory energy of the virtual photon is $\nu\equiv E-E'$.  
The direction of the momentum  $\overrightarrow{q}\equiv\overrightarrow{k}-\overrightarrow{k'}$ of the virtual photon 
defines  the $\overrightarrow{z}$ axis,
while $\overrightarrow{x}$  is in the $(\overrightarrow{k},\overrightarrow{k'})$
plane.  $\overrightarrow{S}$ is the target
spin, with $\theta^{*}$ and $\phi^{*}$ its polar and  azimuthal angles, 
respectively. In inclusive reactions, two variables
suffice to characterize the kinematics; in the elastic case, they are related, and one variable is enough. 

During an experiment, the transferred energy $\nu$ and  the scattering angle $\theta$ are typically varied. Two of the following
relativistic invariants are used to characterize the kinematics:

$\bullet$ The exchanged 4-momentum squared $Q^2 \equiv -(k-k')^2 =4EE'\sin^2\frac{\theta}{2}$ 
for ultra-relativistic leptons. For a real photon, $Q^2=0$.

$\bullet$  The invariant mass squared $W^2\equiv(p+q)^2=M_t^2+2M_t\nu-Q^2$,
where $M_t$ is the mass of the target nucleus. $W$
is the mass of the system formed after the lepton-nucleus collision; {\it e.g.}, a nuclear excited state.

$\bullet$ The Bjorken variable $x_{Bj} \equiv \frac{Q^2}{2p.q} = \frac{Q^2}{2M_t\nu}$. 
This variable was introduced by Bjorken in the context of scale invariance in 
DIS;  see Section~\ref{DISscaling}.
One has $0<x_{Bj}<M_t/M$, where $M$ the nucleon mass, since $W \geq M_t$, $Q^2>0$ and $\nu>0$.

$\bullet$ The laboratory energy transfer relative to the incoming lepton energy $y=\nu/E$.

\noindent Depending on the values of $Q^2$ and $\nu$, the target
can emerge in different excited states.  It is advantageous to study
the excitation spectrum in terms of $W$ since each excited state
corresponds to specific a value of $W$ rather than $\nu$, see Fig.~\ref{fig:gross}.

\subsubsection{General expression of the reaction cross-section \label{sub:general XS}}
 In what follow, ``hadron'' can refer to either a nucleon or a nucleus.
The reaction cross-section is obtained from the scattering amplitude $T_{fi}$ for an initial state $i$ 
and final state $f$. $T_{fi}$ is computed from the photon propagator and the leptonic current contracted with the
electromagnetic current of the hadron for the exclusive reaction $\ell H \to \ell^\prime H^\prime$, 
or a tensor in the case of an incompletely known final state. 
These quantities are conserved at the leptonic and hadronic vertices (gauge invariance). 
\begin{wrapfigure}{r}{0.4\textwidth}
\vspace{-0.8cm}
 \includegraphics[scale=0.55]{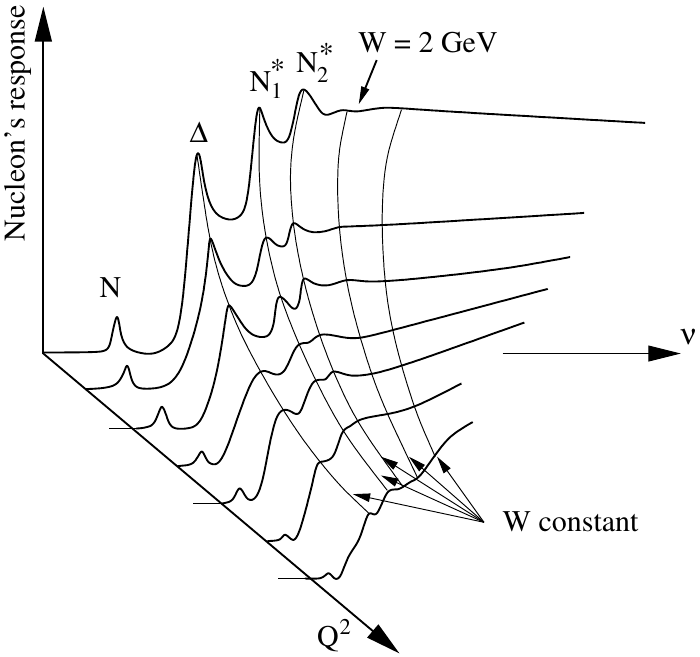} 
\vspace{-0.5cm}
\caption{{\label{fig:gross}\small Response of the nucleon to the electromagnetic probe
as a function of $Q^2$ and $\nu$. The $\nu$-positions of the
peaks (N, $\Delta$,...) change as $Q^2$ varies.
(After~\cite{Gross:1985dn}.)} }
\end{wrapfigure}

In the first Born approximation:
\vspace{-0.3cm}
\begin{equation}
\vspace{-0.3cm}
T_{fi}=\left\langle k'\right|j^{\mu}(0)\left|k\right\rangle \frac{1}{Q^2}\left\langle P_X\right|J_{\mu}(0)\left|P\right\rangle, 
\label{eq:amplitran}
\end{equation}
where the leptonic current is $j^{\mu}=e\overline{\psi_l}\gamma^{\mu}\psi_l$
with $\psi_l$ the lepton spinor, $e$ its electric charge and $J^\mu$  the quark current. The exact expression of the hadron's current matrix element 
$\left\langle P_X\right|J_{\mu}(0)\left|P\right\rangle$ is unknown because of our
ignorance of the nonperturbative hadronic structure and, for non-exclusive experiments, that of the
final state. However, symmetries  (parity, time reversal, hermiticity, and current conservation) constrain  the matrix elements of $J^\mu$
to a generic form written in terms of the vectors and tensors pertinent to the reaction. Our ignorance of 
the hadronic structure is thus parameterized by functions which can be either measured, 
computed numerically, or modeled.  These are called either 
``form factors" (elastic scattering, see Section~\ref{elastic scatt}),
``response functions" (quasi-elastic reaction, see Section~\ref{qel}) 
or ``structure functions" (DIS case, see Section~\ref{DIS}). 
A significant advance of the late 1990s and early 2000s is the unification of
form factors and structure functions under the concept of GPDs. 
%
The differential cross-section $d\sigma$ is obtained from the absolute square of the amplitude 
(\ref{eq:amplitran}) times the lepton flux and a phase space factor, given {\it e.g.}, in Ref.~\cite{Olive:2016xmw}.

\subsubsection{Leptonic and hadronic tensors, and cross-section parameterization\label{tensors}}

The leptonic tensor $\eta^{\mu\nu}$ and the hadronic tensor $W^{\mu\nu}$ are defined such that 
$d\sigma\propto\left|T_{fi}\right|^2=\eta^{\mu\nu}\frac{1}{Q^{4}}W_{\mu\nu}$. That is, 
$\eta^{\mu\nu}\equiv\frac{1}{2}\sum  j^{\mu *} j^{\nu}$, where all the possible final states of the lepton
have been summed over ({\it e.g.}, all of the lepton final spin states for the unpolarized experiments), and the  tensor
\vspace{-0.3cm}
\begin{equation}
\vspace{-0.3cm}
W^{\mu\nu}=\frac{1}{4\pi} \int  d^4 \xi \, e^{iq_\alpha \xi^\alpha} \left\langle P \right|\left[J^{\mu \dagger} (\xi),J^\nu (0)\right] \left| P \right\rangle,
\label{hadronic tensor}
\end{equation}
follows from the optical theorem by computing the forward matrix element of a product of currents in the proton state.
The  contribution to $W^{\mu\nu}$ which is symmetric in ${\mu , \nu}$ -- thus constructed from  the hadronic vector current --
contributes to the unpolarized cross-section, whereas its antisymmetric part 
-- constructed from the pseudo-vector (axial) current -- yields the spin-dependent contribution.

In the unpolarized case;  {\it i.e.}, with summation over all spin states, 
the cross-section can be parameterized with six photoabsorption terms. 
Three terms originate from the three possible polarization states of the virtual 
photon. (The photon spin is a 4-vector but for a virtual photon, 
only three components are independent because of the constraint from gauge invariance.
The unphysical fourth component is called a \emph{ghost} photon.)
The other three terms stem from the multiplication of the two
tensors. They depend in particular on the azimuthal scattering angle, which is integrated over 
for inclusive experiments. Thus,  these three terms disappear and
\vspace{-0.35cm}
\begin{equation}
\vspace{-0.35cm}
\left|T_{fi}\right|^2=\frac{e^2}{Q^2(1-\epsilon)}\left[(w_{RR}+w_{LL})+2\epsilon w_{ll}\right],
\label{eq:sep}
\end{equation}
where $R$, $L$ and $l$ label the photon helicity state (they are not Lorentz indices) and  \\
$\epsilon\equiv1/\left[1+2\left(\nu^2/Q^2+1\right)\tan^2
(\theta/2)\right]$ is the virtual photon degree of polarization in the $m_e=0$ approximation. 
The right and left helicity terms are $w_{RR}$ and $w_{LL}$, respectively. 
The longitudinal term $w_{ll}$ is non-zero only for virtual photons. 
It can be isolated by varying $\epsilon$~\cite{Rosenbluth:1950yq}, but $w_{RR}$ and $w_{LL}$
cannot be separated. Thus, writing $w_T=w_{RR}+w_{LL}$ and $w_L=w_{ll}$, 
the cross-section takes the form: 
\vspace{-0.3cm}
\begin{equation}
\vspace{-0.3cm}
d\sigma\propto\left|T_{fi}\right|^2=\frac{e^2}{Q^2(1-\epsilon)}\left[w_T+2\epsilon w_{L}\right].
\label{eq:tensor}
\end{equation}
The total unpolarized inclusive cross-section is expressed
in terms of two photoabsorption partial cross-sections, $\sigma_{L}$ and $\sigma_T$.
The parameterization in term of virtual photoabsorption quantities 
is convenient because the leptons create the virtual photon flux probing the target.
For doubly-polarized  inclusive inelastic scattering, where both the beam and target are polarized, 
two additional parameters are required:  $\sigma_{TT}$ and $\sigma_{LT}'$. (The reason for the prime $'$  
is explained below). The $\sigma_{TT}$  term stems from the interference of the amplitude involving one of 
the two possible transverse photon helicities with the amplitude involving the other
transverse photon helicity. Likewise, $\sigma_{LT}'$ originates from the
imaginary part of the longitudinal-transverse interference amplitude. The real part, which produces 
$\sigma_{LT}$, disappears in inclusive experiments because all angles defined by variables describing the hadrons produced during
the reaction are averaged over.  This term, however, appears in exclusive or semi-exclusive reactions, 
see {\it e.g.}, the review~\cite{Burkert:2004sk}.

\subsubsection{Asymmetries} 
The basic observable for studying nucleon spin structure in doubly polarized lepton scattering 
is the cross-section asymmetry with respect to the lepton and nucleon spin directions.
Asymmetries can be absolute: $A=\sigma^{\downarrow\Uparrow}-\sigma^{\uparrow\Uparrow}$, or relative: 
$A=(\sigma^{\downarrow\Uparrow}-\sigma^{\uparrow\Uparrow})/(\sigma^{\downarrow\Uparrow}+\sigma^{\uparrow\Uparrow})$. 
The $\downarrow$ and $\uparrow$ represent the leptonic beam helicity in the laboratory frame
whereas $\Downarrow$ and $\Uparrow$  define the direction of the target polarization (here, along the beam direction).
Relative asymmetries convey less information, the absolute magnitude of the process 
being lost in the ratio, but are easier to measure than absolute asymmetries or cross-sections
since the absolute normalization ({\it e.g.}, detector acceptance, target density, or inefficiencies) cancels in the ratio. 
Measurements of absolute asymmetries can also be advantageous, since the contribution
from any unpolarized material present in the target cancels out. The optimal choice between relative 
and absolute asymmetries thus depends on the experimental conditions;
see Section~\ref{Structure functions extraction}.

One can readily understand  why the asymmetries appear physically, 
and why they are related to the spin distributions of the quarks in the nucleon.  
Helicity is defined as the projection of spin in the direction of motion. 
In the Breit frame where the massless lepton and the quark flip their spins 
after the interaction, the polarization of the incident relativistic  leptons sets the polarization of the probing 
photons because of angular momentum conservation; {\it i.e.}, 
these photons must be transversally polarized and have helicities $\pm 1$.  
Helicity conservation requires that a photon of a given helicity couples only to quarks of 
opposite helicities, thereby probing the quark helicity (spin) distributions in the nucleon. 
Thus the difference of scattering probability between leptons of $\pm 1$ helicities (asymmetry) 
is  proportional to the difference of the population of quarks of different helicity. This is the basic physics 
of quark longitudinal polarization as characterized by the target hadron's longitudinal spin structure function.
Note also that virtual photons can also be longitudinally polarized, {\it i.e.}, with 
helicity 0, which will also contribute to the lepton asymmetry at finite $Q^2$.

\subsection{Nucleon-Nucleon scattering \label{n-n scattering}}
Polarized proton--(anti)proton scattering, as done at RHIC (Brookhaven, USA), 
is another way to access the nucleon spin structure. 
Since hadron structure is independent of the measurement, the PDFs measured in lepton-nucleon and
nucleon-nucleon scattering should be the same. 
This postulate of pQCD factorization underlies the ansatz that PDFs are universal.
Several processes in nucleon-nucleon scattering are available to access PDFs, see Fig. \ref{Flo:Drell-Yan}. 
Since different 
PDFs contribute differently in different processes, investigating all of these reactions 
will allow us to disentangle the contributing PDFs. 
The analytic effects of evolution generated by pQCD is known at least to 
next-to-leading order (NLO) in $\alpha_s$ for these processes, which permits
the extraction of the PDFs to high precision. The most studied processes which access  nucleon spin structure are:

\begin{figure}[ht]
\centering
\vspace{-0.cm}
\includegraphics[width=9.0cm]{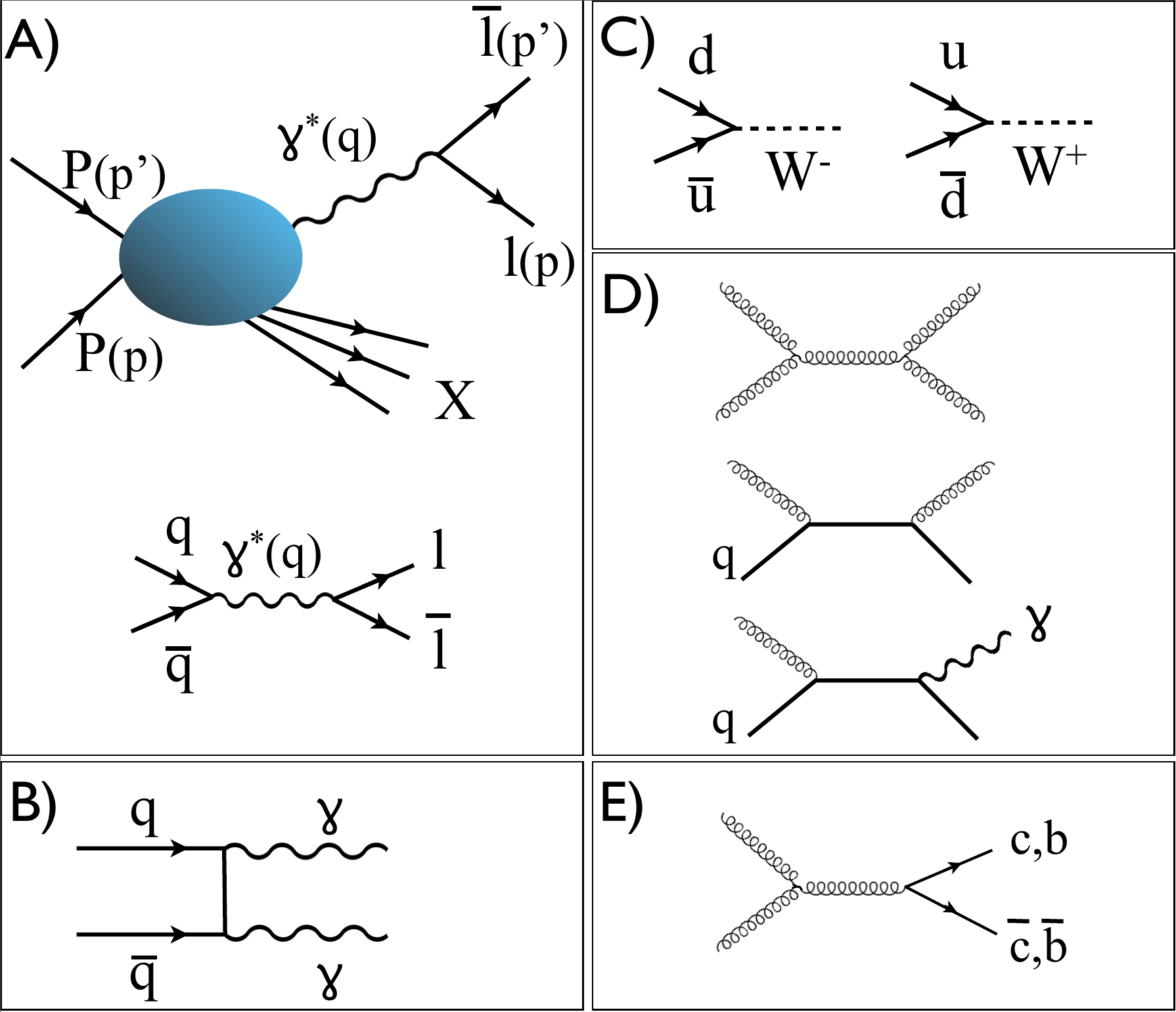}
\vspace{-0.3cm}
\caption{\label{Flo:Drell-Yan}\small Various $\protect\overrightarrow{p} \protect\overrightarrow{p}$ reactions
probing the proton spin structure. 
Panel A: Drell-Yan process and its underlying LO diagram. 
Panel B: Direct diphoton production at LO.
Panel C: $W^{+ / -}$  production at LO.
Panel D: LO process dominating photon, pion and/or Jet production 
in $\protect\overrightarrow{p} \protect\overrightarrow{p}$ scattering.
Panel E: heavy-flavor meson production at LO.}
\end{figure}

\noindent  \textbf{A) The Drell-Yan process} 
A lepton pair detected in the final state corresponds to 
the Drell-Yan process, see Fig.~\ref{Flo:Drell-Yan}, panel A. 
In the high-energy limit, this process is described as the annihilation of a 
quark from a proton with an antiquark from the other (anti)proton, the
resulting timelike photon then converts into a lepton-antilepton pair. 
Hence, the process is sensitive to the convolution of the quark 
and antiquark polarized PDFs $\Delta q(x_{Bj})$ and  $\Delta \overline{q}(x_{Bj})$. (They will be
properly defined by Eq.~(\ref{eq:pdf def.}).)  
Another process that leads to the same final state is lepton-antilepton pair creation from a virtual photon 
emitted by a single quark. However, this process
requires large virtuality to produce a high energy lepton--anti-lepton pair, and it is thus 
kinematically suppressed compared to the panel A case. 

An important complication is that the Drell-Yan process is sensitive to double initial-state 
corrections, where both the quark and antiquark before annihilation interact  
with the spectator quarks of the other projectile. Such corrections are 
``leading \emph{twist}"; {\it i.e.}, they are not power-suppressed at high lepton pair virtuality. 
They induce strong modifications of 
the lepton-pair angular distribution and  violate the Lam-Tung relation~\cite{Boer:2002ju}. 

A fundamental QCD prediction is that a naively time-reversal-odd 
distribution function, measured \emph{via} Drell-Yan should change sign compared to a
SIDIS measurement~\cite{Brodsky:2002cx,Collins:2002kn,Brodsky:2002rv,Brodsky:2013oya}.
  An example is the Sivers function~\cite{Sivers:1989cc}, a transverse-momentum dependent 
distribution function sensitive to spin-orbit effects inside the polarized proton.

\noindent \textbf{B) Direct diphoton production}
Inclusive diphoton production $\overrightarrow{p}\overrightarrow{p} \rightarrow \gamma \gamma+X$ is
another process sensitive to $\Delta q(x_{Bj})$ and $\Delta \overline{q}(x_{Bj})$.
The underlying leading order (LO) diagram is shown on panel B of Fig.~\ref{Flo:Drell-Yan}.

\noindent \textbf{C) $W^{+ / -}$ production} 
The structure functions probed in lepton scattering involve the quark charge squared 
(see Eqs.~(\ref{eq:eqf1parton}) and (\ref{eq:eqg1parton})):
They are thus only sensitive to $\Delta q + \Delta \overline{q}$. 
$W^{+ / -}$ production  is sensitive to $\Delta q(x_{Bj})$ and $\Delta \overline{q}(x_{Bj})$ separately.
Panel C  in Fig.~\ref{Flo:Drell-Yan}
shows how $W^{+ / -}$ production allows the measurement of both mixed $\Delta u \Delta \overline{d}$ and 
$\Delta d \Delta \overline{u}$ combinations; 
thus combining $W^{+ / -}$ production data and data providing 
$\Delta q + \Delta \overline{q}$ ({\it e.g.}, from lepton scattering) permits
individual quark and antiquark contributions to be separated. The produced $W$
is  typically identified  \emph{via} its leptonic decay to $\nu l$, with the $\nu$ escaping detection.

\noindent \textbf{D) Photon, Pion and/or Jet production}
These processes are 
$\overrightarrow{p}\overrightarrow{p} \rightarrow \gamma+X$,
$\overrightarrow{p}\overrightarrow{p} \rightarrow \pi+X$,
$\overrightarrow{p}\overrightarrow{p} \rightarrow jet(s)+X$ and
$\overrightarrow{p}\overrightarrow{p} \rightarrow \gamma+jet+X$.
At high momenta, such reactions are dominated by either
gluon fusion or gluon-quark Compton scattering with a gluon or photon in the final state; 
See  panel D  in Fig.~\ref{Flo:Drell-Yan}.  These processes are sensitive to the polarized gluon distribution 
$\Delta g(x,Q^2)$. 

\noindent \textbf{E) Heavy-flavor meson production} 
Another process which is sensitive to $\Delta g(x,Q^2)$ is $D$ or $B$ heavy meson production
\emph{via} gluon fusion $\overrightarrow{p}\overrightarrow{p} \rightarrow D+X$ or 
$\overrightarrow{p}\overrightarrow{p} \rightarrow B+X$. See  panel E  in Fig.~\ref{Flo:Drell-Yan}. The 
heavy mesons subsequently decay into charged leptons which are detected.

\subsection{$e^+ ~ e^-$ annihilation \label{e^+e^- annihilation}}
\begin{wrapfigure}{r}{0.45\textwidth} 
\vspace{-1.35cm}
\centering
\includegraphics[width=5.0cm]{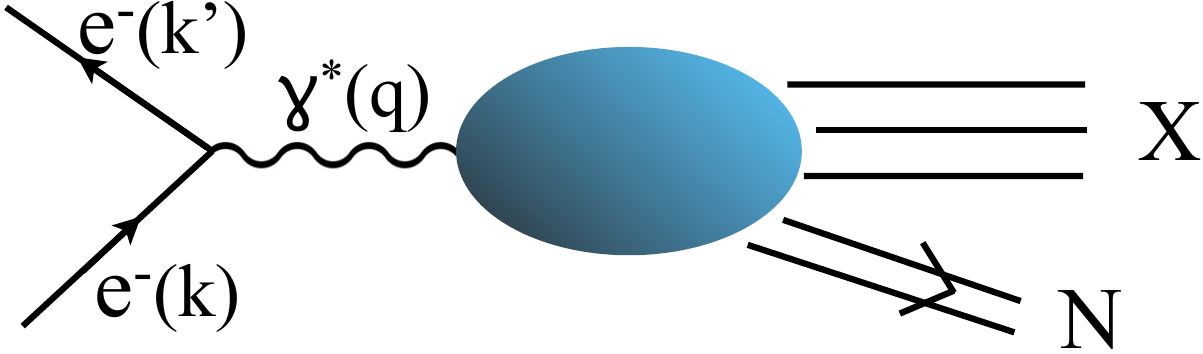}
\vspace{-0.45cm}
\caption{\label{Flo:ep_annihilation}\small Annihilation of $e^+ ~ e^-$ with only one detected hadron from the final state.}
\vspace{-0.5cm}
\end{wrapfigure}
The $e^+ ~ e^-$ annihilation process  where only one  hadron is detected in the final state (Fig.~\ref{Flo:ep_annihilation})
is the timelike version of DIS if the final state hadron is a nucleon.  The nucleon structure is parameterized
by \emph{fragmentation functions}, whose  analytic form is limited -- as for the spacelike case -- by fundamental symmetries.

\section{Constraints on spin dynamics from scattering processes \label{Scattering spectrum}}

We now discuss the set of inclusive scattering processes which are sensitive to 
the polarized parton distributions and provide the cross-sections for each type of reaction.
We start with DIS where the nucleon structure is best understood. 
DIS was also historically the first \emph{hard}-scattering reaction which provided an understanding of fundamental hadron dynamics. 
Thus, DIS is the prototype -- and it remains the archetype -- of tests of QCD. 
We will then survey other inclusive reactions and explore their connection to 
exclusive reactions such as  elastic lepton-nucleon scattering.

\subsection{Deep inelastic scattering \label{DIS}}

\subsubsection{Mechanism\label{sub:MecaDIS}}

The kinematic domain of DIS  where leading-twist Bjorken scaling is valid requires 
$W \gtrsim 2$ GeV
and $Q^2  \gtrsim 1$~GeV$^2$.
Due to   \emph{asymptotic freedom}, QCD 
can be treated perturbatively in this domain, and standard gauge theory calculations are possible. 
In the Bjorken limit where 
$\nu\to\infty$ and $Q^2\to\infty$, with $x_{Bj}=Q^2/(2M\nu)$ fixed, 
DIS can be  represented in the first approximation by a  lepton scattering elastically off 
a fundamental quark or antiquark constituent of the target nucleon, 
as in Feynman's parton model.  The momentum distributions of the quarks (and gluons) 
in the nucleon, which determine the DIS cross-section, reflect its nonperturbative bound-state structure. 
The ability to separate, at high lepton momentum transfer,  perturbative photon-quark interactions from
the nonperturbative nucleon structure is known as the \emph{factorization 
theorem}~\cite{factorization theorem}  -- a direct consequence of  \emph{asymptotic freedom}.  It is
an important ingredient in establishing the validity of QCD as a description of the strong interactions.

The momentum distributions of quarks and gluons are parameterized by the structure functions:
These distributions 
are universal; 
{\it i.e.}, they are properties of the hadrons themselves,  and thus should be independent of the
particular high-energy reaction used to probe the nucleon.
In fact, all of the  interactions within the nucleon which occur before the
lepton-quark interaction, including the dynamics, are contained in the frame-independent 
light-front (LF) wave functions (LFWF) of the nucleon -- the eigenstates of the QCD LF Hamiltonian. They thus reflect
the nonperturbative underlying confinement dynamics of QCD;  we discuss 
how this is assessed in models and confining theories such as Light Front Holographic QCD (LFHQCD) in Section~\ref{sec:LFHQCD}.
Final-state interactions -- processes happening after the lepton interacts 
with the struck quark -- also exist. They lead to novel phenomena such as diffractive DIS (DDIS), 
$\ell p \to \ell^\prime  p^\prime X$,  
or the pseudo-T-odd Sivers single-spin asymmetry $\vec S_p \cdot \vec q \times \vec p_q$  which is observed in 
polarized SIDIS. These processes also contribute at ``leading twist"; {\it i.e.}, they contribute to the
Bjorken-scaling DIS cross-section.

\subsubsection{Bjorken scaling\label{DISscaling}}

DIS is effectively represented by the elastic scattering of leptons on the pointlike 
quark constituents of the nucleon in the Bjorken limit.
Bjorken predicted that the  hadron structure functions would depend only on the dimensionless  
ratio $x_{Bj}$, 
and that the structure functions reflect \emph{conformal} invariance; {\it i.e.}, they will be  $Q^2$-invariant.   
This is in fact the prediction of ``conformal" theory -- a quantum field theory of pointlike quarks with no
fundamental mass scale. Bjorken's expectation was verified by the first measurements at 
SLAC~\cite{Breidenbach:1969kd}  in the domain $x_{Bj} \sim 0.25$.
However,  in a  gauge theory such as QCD, Bjorken scaling is 
broken by logarithmic corrections from pQCD processes, such as gluon radiation --  see 
Section~\ref{sub:pQCD}. One also predicts  deviations from Bjorken scaling  due to 
power-suppressed $M^2/Q^2$ corrections called 
\emph{higher-twist} processes. They reflect finite mass corrections and \emph{hard} scattering  
involving two or more quarks. The effects become  particularly evident at low $Q^2$ 
($\lesssim1$~GeV$^2$), see Section~\ref{OPE}.    
The underlying \emph{conformal} features of chiral QCD (the massless quark limit) also has important consequence for color confinement and hadron dynamics at low $Q^2$. This 
perspective will be discussed in Section~\ref{sec:LFHQCD}.

\subsubsection{DIS: QCD on the light-front \label{LC dominance and LF quantization}}

An essential point of DIS is that the lepton interacts  via the exchange of a virtual photon 
with the quarks of the proton --  not at the same \emph{instant time} $t$ (the ``\emph{instant form}" as defined by Dirac), 
but at  the time along the LF,  in analogy to a flash photograph.  
In effect DIS provides a measurement of hadron structure at  fixed LF time $\tau = x^+ = t + z/c$.

The LF coordinate system in position space is based on the LF variables
$x^{\pm} = (t \pm z) $.  The choice of the $\hat z = {\hat x}^3$ direction is arbitrary. The two other orthogonal vectors
defining the LF coordinate system are written as $x_{\bot}=(x,y)$. 
They are perpendicular to the $(x^+,x^-)$ plane. Thus $x^2 = x^+ x^- -  x_\perp^2$.
Similar definitions are applicable to momentum space: $p^{\pm} = (p^0 \pm p^3)$, 
$p_{\bot}=(p_1,p_2)$. 
The product of two vectors $a^{\mu}$ and $b^{\mu}$ in LF coordinates is 
\vspace{-0.15cm}
\begin{equation}
\vspace{-0.15cm}
a^{\mu} b_{\mu} = \frac{1}{2} (a^+b^- + a^-b^+ ) - a_{\bot} b_{\bot}.
\label{lc vector product}
\end{equation}
The relation between covariant and contravariant vectors is $a^+=a_-$,  $a^-=a_+$ and the relevant metric is:
\vspace{-0.1cm}
\footnotesize
\begin{equation}
\left(
\begin{array}{cccc}
0 & 1 & 0 & 0\\
1 & 0 & 0 & 0\\
0 & 0 & -1 & 0\\
0 & 0 & 0 & -1
\end{array}
\right).
\nonumber
\vspace{-0.1cm}
\end{equation}
\normalsize
Dirac matrices $\gamma^\mu$ adapted to the LF coordinates can also be defined~\cite{Kogut:1969xa}.

The LF coordinates 
provide the natural coordinate system for DIS and other \emph{hard} reactions.
The LF formalism, called the ``Front Form" by Dirac, is Poincar\'{e} invariant 
(independent of the observer's Lorentz frame) and 
``causal"  (correlated information is only possible as allowed by the finite speed of light).   
The momentum and spin distributions of the quarks which are probed in DIS experiments 
are in fact determined by the LFWFs  of the target hadron -- the eigenstates 
of the QCD LF Hamiltonian $H_{LF}$ with the Hamiltonian defined at fixed $\tau$.  
$H_{LF}$ can be computed directly from the QCD Lagrangian.
This explains why quantum field theory quantized at fixed $\tau $ (\emph{LF quantization}) is the  
natural formalism underlying DIS experiments. The LFWFs being independent of 
the proton momentum, one obtains the same predictions for DIS at an electron-proton collider
as for a fixed target experiment where the struck proton is at rest.

Since  important nucleon spin structure information is derived from DIS experiments, 
it is relevant to outline the basic elements of the LF formalism here.
The evolution operator in LF time is $P^- = P^0 - P^3$, while $P^+ = P^0 + P^3$ and $P_\perp$  are kinematical. 
This leads to the definition of the Lorentz invariant LF Hamiltonian 
$H_{LF} = P^\mu P_\mu = P^- P^+ - P^2_\perp$.   The LF Heisenberg equation derived from the QCD LF Hamiltonian is
\vspace{-0.3cm}
\begin{equation}
\vspace{-0.3cm}
 H_{LF} |\Psi_H \rangle   = M^2_H |\Psi_H \rangle,
 \end{equation}
 where the eigenvalues $M_H^2$ are the squares of the  masses of the hadronic eigenstates.
 The eigensolutions $|\Psi_H \rangle$ projected on the free parton eigenstates $|n\rangle$
 (the Fock expansion) are the boost-invariant hadronic  LFWFs, 
 $ \langle n | \Psi_H \rangle = \Psi_n(x_i, \vec k_{\perp i},  \lambda_i)$, which underly 
 the DIS structure functions.  Here $x_i= k^+_i/P^+$, with
 $\sum_i x_i = 1$, are the LF momentum fractions of the quark and gluon constituents 
 of the hadron eigenstate in the $n$-particle Fock state, the $\vec k_{\perp i}$ are the 
 transverse momenta of the $n$ constituents where $\sum_i\vec k_{\perp i} = 0_\perp $; 
 the variable $\lambda_i$ is the spin projection of constituent $i$ in the $\hat z$ direction.

A critical point is that \emph{LF quantization} provides the LFWFs
describing relativistic bound systems, independent of the observer's Lorentz frame; 
{\it i.e.}, they are boost invariant. In fact, the LF provides an exact and rigorous framework to
study nucleon structure in both the perturbative and nonperturbative domains of QCD~\cite{Brodsky:1997de}.  

Just as the energy $P^0$ is the conjugate of the standard time $x^0$ in the \emph{instant form}, 
the conjugate to the LF time $x^+$ is the operator $P^- = i \frac{d}{d x^+} $. It represents the LF time evolution operator
\vspace{-0.1cm}
\begin{equation}
\vspace{-0.1cm}
P^- \Psi =\frac{(M^2+P_{\bot}^2)}{2P^+} \Psi,
\label{LF Hamiltonian}
\end{equation}
and generates the translations normal to the LF.  

The structure functions measured in DIS are computed from integrals of the square of the LFWFs, 
while the hadron form factors measured in elastic lepton-hadron scattering are given by the 
overlap of LFWFs. The power-law fall-off of the form factors at high-$Q^2$ are predicted from first principles by simple 
counting rules which reflect the composition of the hadron~\cite{Brodsky:1973kr, Matveev:1973ra}.  
One also can predict observables such as the DIS spin asymmetries for polarized targets~\cite{Brodsky:1994kg}. 

\emph{LF quantization} differs from the traditional equal-time quantization at fixed $t$~\cite{Weinberg:1966jm} 
in that eigensolutions of the Hamiltonian defined at a fixed time $t$ depend on the hadron's momentum $\vec P$.  
The boost of the \emph{instant form} wave function is then a complicated dynamical problem; 
even the Fock state structure depends on $P^\mu$.  Also, interactions
of the lepton with quark pairs (connected time-ordered diagrams) created from the \emph{instant form} vacuum
must be accounted for.  
Such complications are absent in the LF formalism.  
The LF vacuum is defined as the state with zero $P^-$; {\it i.e.},
invariant mass zero and thus $P^\mu=0$.  Vacuum loops do not 
appear in the LF vacuum since $P^+$ is conserved at 
every vertex; one thus cannot create particles with   $k^+ \ge 0$ from the LF vacuum.

It is sometimes useful to simulate \emph{LF quantization} by using \emph{instant time} 
in a Lorentz frame where the observer has ``infinite momentum" $ P^z \to - \infty.$  
However, it should be stressed that the LF formalism is frame-independent; it  is 
valid in any frame,  including the hadron rest frame. It reduces to standard nonrelativistic 
Schr\"odinger theory if one takes $c \to \infty$.
The \emph{LF quantization} is thus the natural, physical, formalism for QCD.

As we shall discuss below, the study of dynamics 
with the LF holograpic  approach which incorporates the exact   \emph{conformal} symmetry of the classical  QCD Lagrangian in the chiral limit,
provides a successful description of color confinement and  nucleon structure at low $Q^2$~\cite{Brodsky:2014yha}.
An example is given in Section~\ref{unpo cross-section} where 
nucleon form factors emerge naturally from the LF framework
and are computed in  LFHQCD.

\noindent \textbf{Light-cone gauge} 

The gauge condition often chosen in the LF framework is the ``light-cone"  (LC) gauge 
defined as $A^+ = A^0 + A^3= 0$; it is an
axial gauge condition in the LF frame. The LC gauge is analogous to the usual 
Coulomb or radiation gauge since there are no longitudinally polarized nor 
\emph{ghosts} (negative-metric) gluon. Thus, Fadeev--Popov \emph{ghosts}~\cite{Faddeev:1967fc} are
also not required. In LC gauge one can show that $A^-$ is a function of $A_\bot$. 
Therefore, this physical gauge simplifies the study of hadron structure since the 
transverse degrees of freedom of the gluon field $A_\bot$ are the only independent
dynamical variables.
The LC gauge also insures that at LO, \emph{twist}-2 expressions do not explicitly involve the gluon 
field, although the results retain color-gauge 
invariance~\cite{Jaffe:1996zw}.   Instead a LF-instantaneous interaction proportional to 
$1\over {k^+}^2$  appears in the LF Hamiltonian, analogous to the \emph{instant time} instantaneous 
${1}\over {\vec k}^2$ interaction which appears in  Coulomb (radiation) gauge in QED.

\noindent \textbf{Light-cone dominance} 

\begin{wrapfigure}{r}{0.4\textwidth} 
\vspace{-1.2cm}
\centering
\includegraphics[width=6.0cm]{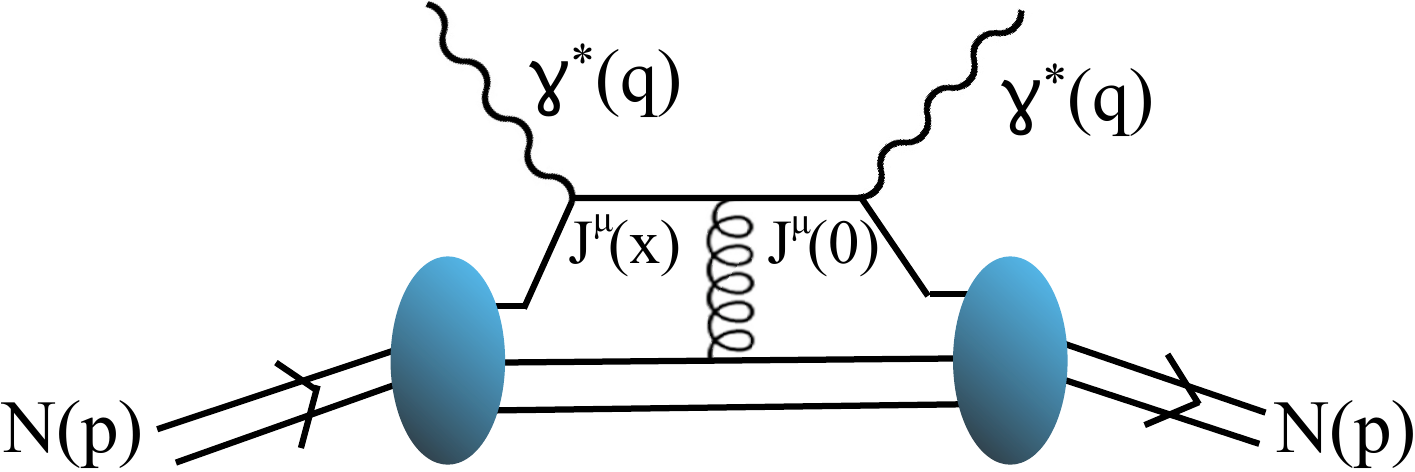}
\vspace{-0.4cm}
\caption{\label{Handbag+HT}\small Forward virtual Compton scattering with a Wilson line.}
\vspace{-0.3cm}
\end{wrapfigure}
Using unitarity, the hadronic tensor $W^{\mu \nu}$, Eq.~(\ref{hadronic tensor}), can be computed 
from the imaginary part of the forward virtual Compton scattering amplitude
$\gamma^*(q) N(p) \to \gamma^* (q) N(p)$, see Fig.~\ref{Handbag+HT}.  At large $Q^2$, the quark propagator 
which connects the two currents  in the DVCS amplitude goes far-off shell; as a result, the invariant 
spatial separation $x^2 = x_\mu x^\mu$ between the currents 
$J^\mu(x)$  and $ J^\nu (0) $ acting on the quark line vanishes as 
$x^2 \propto {1\over Q^2}$.  Since $x^2 = x^+ x^- - x^2_\perp \to 0$, 
this  domain is referred to as ``light-cone dominance".  
The interactions of gluons with this quark propagator  are referred to as the \emph{Wilson line}. 
It represents the final-state interactions between the struck quark and the target spectators
(``final-state", since the imaginary part of the amplitude in Fig.~\ref{Handbag+HT}
is related by the \emph{Optical Theorem} to the DIS cross-section
with the \emph{Wilson line} connecting the outgoing quark to the nucleon remnants).
Those can contribute to leading-\emph{twist} -- e.g. the Sivers effect~\cite{Sivers:1989cc} or DDIS, or can
generate \emph{higher-twists}.
In QED such final-state interactions are related to the ``Coulomb phase".

More explicitly, one can  choose coordinates such  that $ q^+ = -M x_{Bj} $ and 
$q^-  = (2\nu + M x_{Bj}) $ with $q_{\bot}=0$.  Then 
$q^\mu \xi_\mu = \big[ (2\nu+M x_{Bj})\xi^+ -M x_{Bj} \xi^- \big] $, with 
$\xi$ the integration variable in Eq.~(\ref{hadronic tensor}).  In the Bjorken limit, 
$\nu \to \infty$ and $x_{Bj}$ is finite.  One verifies then that the cross-section
is dominated by $\xi^+   \to  0$, $\xi^- \propto 1/(M x_{Bj})$ in the Bjorken limit, that is $\xi^+ \xi^- \approx 0$,
and the reaction happens on the LC specified by $\xi^+ \xi^- = \xi^2 = 0$. Excursions out of the LC generate
$M^2/Q^2$  \emph{twist}-4 and higher corrections ($M^{2n}/Q^{2n}$ power corrections), see Section~\ref{OPE}.  

It can be shown that LC kinematics also
dominates Drell-Yan lepton-pair reactions (Section~\ref{n-n scattering}) 
and inclusive hadron production in $e^+ ~ e^-$ annihilation 
(Section~\ref{e^+e^- annihilation}).

\noindent \textbf{Light-front quantization }

The two currents appearing in DVCS  (Fig.~\ref{Handbag+HT}) effectively couple to the nucleon 
as a local operator at a single LF time in the Bjorken limit.   
The nucleon is thus described,
in the Bjorken limit, as distributions of  partons along $x^-$ at a fixed LF time $x^+$ 
with  $x_\bot =0$.  At finite $Q^2$ and $\nu$  one becomes sensitive to 
distributions with nonzero $x_\bot$.   It is often convenient to expand the operator 
product appearing in DVCS as a sum of ``good"  operators,
such as $\gamma^+ = \gamma^0 + \gamma^-$, which have simple interactions 
with the quark field.    In contrast, ``bad" operators such as $\gamma^-$  have a complicated 
physical interpretation since they can connect the electromagnetic current to more than one 
quark in the hadron Fock state via LF instantaneous interactions.

The equal LF time condition, $x^+ = $ constant,  defines a plane, rather than a cone, tangent to the LC, 
thus the name ``Light-Front".
In high-energy scattering, the leptons and partons being ultrarelativistic,
it is often useful for purposes of intuition to interpret the DIS kinematics 
in  the Breit frame, or to use the \emph{instant form} in the infinite momentum frame (IMF).  
However, since a change of frames requires Lorentz boosts in the \emph{instant form},
it mixes the dynamics
and kinematics of the bound system, complicating the study of the hadron dynamics and structure.
In contrast, the LF description of the nucleon structure is  frame independent.
The LF momentum carried by a quark $i$ is $x_i = k^+_i/P^+$ and identifies with 
the scaling variable, $x_i=x_{Bj}$, and $P^+=\sum_i k_i^+$.
Likewise, the hadron LFWF is the sum 
of individual Fock state wave functions \emph{viz} the states corresponding to a specific number of partons in 
the hadron. 

One can use the QCD LF equations to reduce the 4-component Dirac spinors  
appearing in LF quark wave functions to a description based 
on two-component Pauli spinors by using the LC gauge.
The upper two components  of the quark field are  the dynamical quark field proper; it yields the 
\emph{leading-twist} description, understood on the LF as the quark 
probability density in the hadron eigenstate.
This procedure allows an interpretation in terms of a transverse confinement 
force~\cite{Burkardt:2008ps, Abdallah:2016xfk};  it is thus of prime 
interest for this review.   
The lower two components of the quark spinor link to a field depending on 
both the upper components and the gluon independent dynamical fields $A_\bot$;  
it thus interpreted as a correlation of both quark and gluons \emph{higher-twists}:
They are further discussed in Sections~\ref{OPE} and \ref{sub:HT Extraction}.  
Thus, LF formalism allows for a frame-independent description of the nucleon structure
with clear interpretation of the parton wave functions, of the Bjorken scaling variable
and of the meaning of \emph{twists}.
\noindent There are other advantages for studying QCD on the LF:

\noindent $\bullet$ As we have noted, the vacuum eigenstate in the LF formalism is the eigenstate 
of the LF Hamiltonian with $P^\mu =0$;  it thus has zero invariant mass $M^2 = P^\mu P_\mu = 0.$
Since $P^+ = 0$ for the LF vacuum, and  $P^+ $ is conserved at every vertex, all disconnected 
diagrams vanish. 
The LF vacuum structure is thus simple, without the complication of vacuum loops of
particle-antiparticle pairs.  The dynamical effects normally associated with the \emph{instant form} 
vacuum,  including quark and gluon condensates, are replaced by the nonperturbative dynamics 
internal to the hadronic eigenstates in the front form.

\noindent  $\bullet$ The LFWFs are universal objects which describe hadron structure at all  scales.
In analogy to parton model structure functions, LFWFs have a probabilistic interpretation: 
their projection on an $n$-particle Fock state is  the probability amplitude
that the hadron has  that number of partons at a fixed LF time $x^+$  -- the probability to be in a 
specific Fock state. 
This probabilistic interpretation remains valid
regardless of the level of analysis performed on the data; this contrasts with standard 
analyses of PDFs which can only be interpreted
as parton densities at lowest pQCD order  ({\it i.e.}, LO in $\alpha_s$), see Section~\ref{parton model}.
The probabilistic interpretation implies that PDFs, \emph{viz} structure functions, are  
thus identified with the sums of the LFWFs squared.
In principle it allows for an exact nonperturbative 
treatment of confined constituents.  One thus can approach the challenging problems 
of understanding the role of color confinement in hadron structure
and the transition between physics at short and long distances.
Elastic form factors also emerge naturally from LF QCD: they are
overlaps of the LFWFs based on matrix elements of the local operator $J^+ = \bar \psi \gamma^+ \psi $. %
In practice, approximations and additional constraints are required to carry out
calculations in 3+1 dimensions, such as the \emph{conformal} symmetry
of the chiral QCD Lagrangian. This will be discussed in Section~\ref{sec:LFHQCD}. 
Phenomenological LFWFs can also be constructed using quark models;  see {\it e.g.}, 
Refs.~\cite{Ma:1997gy}-\cite{Chabysheva:2012fe}.
Such models can provide predictions for polarized PDFs due to contributions  to
nucleon spin from the \emph{valence quarks}. While higher Fock states are typically not present in these models,
some do account for gluons or $q\bar{q}$ pairs~\cite{Braun:2011aw,Pasquini:2007iz}.
Knowledge of the effective LFWFs 
is relevant for the computation of form factors,
PDFs, GPDs, TMDs and parton distribution amplitudes~\cite{Chabysheva:2012fe}, 
for both unpolarized and polarized parton  distributions~\cite{Maji:2017ill}-\cite{Nikkhoo:2017won}. 
LFWFs also allow the study of the GPDs skewness 
dependence~\cite{Traini:2016jko}, and to 
compute other parton distributions, {\it e.g.}, the Wigner distribution 
functions~\cite{Gutsche:2016gcd, Chakrabarti:2016yuw}, which encode 
the correlations between the nucleon spin and the spins or OAM of its quarks~\cite{She:2009jq, Lorce:2011ni, Lorce:2011kd}.
Phenomenological models of parton distribution functions  based on the LFHQCD 
framework~\cite{Gutsche:2013zia, Maji:2015vsa, Abidin:2008sb}
 use as a starting point the convenient analytic form of GPDs  found in Refs.~\cite{Brodsky:2007hb}.

\noindent $\bullet$  A third benefit of QCD on the LF is its rigorous formalism to implement the DIS parton model,
alleviating the need to choose a specific frame, such as the IMF.   QCD evolution
equations (\emph{DGLAP}~\cite{Gribov:1972ri}, \emph{BFKL}~\cite{BFKL} and \emph{ERBL}~\cite{ERBL} 
(see Sec.~\ref{sub:pQCD}) can be derived using the LF framework. 

\noindent $\bullet$  A fourth advantage of LF QCD is that in  the LC gauge,  
gluon quanta only have  transverse polarization. The difficulty to define physically 
meaningful gluon spin and angular momenta~\cite{Ji:2012gc, Ji:2013fga, Hatta:2013gta} is thus circumvented; 
furthermore, negative metric degrees of freedom \emph{ghosts} and 
Fadeev--Popov \emph{ghosts}~\cite{Faddeev:1967fc} are unnecessary.

\noindent $\bullet$  A fifth advantage of LF QCD is that 
the LC gauge  allows one to identify the sum of gluon 
spins with 
$\Delta G$~\cite{Anselmino:1994gn}
in the longitudinal spin sum rule, Eq.~(\ref{eq:spin SR}). It will be discussed more in Section~\ref{SSR components}.

The LFWFs fulfill conservation of total angular momentum:
$J^z=\sum_{i=1}^n s^z_i + \sum_{j=1}^{n-1} l^z_j , $ Fock state by Fock state.  
Here  $s^z_i$ labels each constituent spin, and the $l^z_j$ are the $n-1$ independent 
OAM of each $n$-particle Fock state projection.
Since $[H_{LF}, J^z] =0$, each Fock component of 
the LFWF eigensolution  has fixed angular momentum $ J^z $ for any choice of the 3-direction $\hat z$. 
$J^z$ is also conserved at every vertex in LF time-ordered perturbation theory. 
The OAM can only change by zero or one unit at any vertex in a 
renormalizable theory.  This provides a useful constraint on the spin structure of amplitudes in pQCD~\cite{Chiu:2017ycx}.

While the definition of spin is unambiguous for non-relativistic objects,
several definitions exist for relativistic spin~\cite{Chiu:2017ycx}.
In the case of the front form,  LF ``helicity"  is the spin projected on the same $\overrightarrow z$  
direction used to define LF time. Thus, by definition,  LF
helicity is the projection $S^z$ of the particle spin which
contributes to the sum rule for $J^z$ conservation.  
This is in contrast to the usual ``Jacob-Wick"  helicity  defined as the projection of each 
particle's spin vector along the particle's 3-momentum; The Jacob-Wick helicity is thus not conserved.
In that definition,  after a Lorentz boost from
the particle's rest frame -- in which the spin is defined --
to the frame of interest, the particle momentum does not in general coincide with the $z$-direction. 
Although helicity is a Lorentz invariant quantity regardless of its definition, the spin $z$-projection is not Lorentz invariant
unless it is defined on the LF~\cite{Chiu:2017ycx}. 

In the LF analysis the OAM $L^z_i $ of each particle  in  a
composite state~\cite{Chiu:2017ycx,Brodsky:2000ii} is also defined as the projection on 
the $\hat z$ direction;  thus the total $J^z$ is conserved and is the same for each Fock projection of the eigenstate.
Furthermore, the LF spin of each fermion is conserved at each vertex in QCD if $m_q=0.$ 
One does not need to  choose a specific frame, such as the Breit frame, nor require high
momentum transfer (other than $Q \gg m_q$). Furthermore, the LF definition  preserves
the LF gauge $A^+=0$.

We conclude by an important prediction of LFQCD for nucleon spin structure: a non-zero anomalous 
magnetic moment for a hadron requires a non-zero quark transverse OAM
$\mbox{L}_\bot$ of its components~\cite{Brodsky:1980zm, Burkardt:2005km}. 
Thus the discovery of the proton anomalous magnetic moment in the 1930s by Stern and 
Frisch~\cite{Stern:1933} actually gave the 
first evidence for the proton's composite structure, although this was not recognized at that time.

\subsubsection{Formalism and structure functions \label{Formalism}}

Two structure functions are measured in unpolarized DIS:
$F_1(Q^2,\nu)$ and $F_2(Q^2,\nu)$\footnote{Not to be confused with the
Pauli and Dirac form factors for elastic scattering, see Section~\ref{unpo cross-section}}, 
where $F_1$ is proportional to the
photoabsorption cross-section of a transversely polarized virtual photon, {\it i.e.}, $F_1 \propto \sigma_T$. 
Alternatively, instead of $F_1$ or $F_2$, one can define $F_L = F_2 /(2x_{Bj}) - F_1$,  a structure function proportional
to the photabsorption of a  purely longitudinal virtual photon. 
Each of these structure functions can be related to the imaginary part of the corresponding forward double 
virtual Compton scattering amplitude $\gamma^* p \to \gamma^* p$ through the \emph{Optical Theorem}.

The inclusive DIS cross-section for the scattering of polarized leptons off  of a polarized nucleon
 requires four structure functions (see Section~\ref{tensors}).
The additional two polarized structure functions are denoted by $g_1(Q^2,\nu)$ and $g_2(Q^2,\nu)$:
The function $g_1$ is proportional to the transverse photon scattering asymmetry.  Its
first moment  in the Bjorken scaling limit is related to  the nucleon axial-vector current 
$\langle P | \overline{\psi}(0) \gamma^{\nu}\gamma_5 \psi(\xi) | P \rangle$,  which provides 
a direct probe of the nucleon's spin content (see Eq.~(\ref{hadronic tensor}) and below). 
The second function, $g_2$, has no simple interpretation, but $g_t=g_1+g_2$ is proportional 
to the scattering amplitude of a virtual photon which has transverse polarization in its initial state 
and  longitudinal polarization in its final state~\cite{Jaffe:1996zw}. 
If one considers all possible Lorentz invariant combinations formed with the available vectors and tensors, 
three spin structure functions emerge after applying the usual
symmetries (see Section \ref{sub:general XS}).
One ($g_1$, \emph{twist}-2) is associated with the $P^+$ LC vector.
Another one ($g_3$, \emph{twist}-4, see Eq.~(\ref{eq:g3})) is associated with the $P^-$ direction. The third one,
($g_t$, \emph{twist}-3) is associated with the transverse direction;  {\it i.e.}, it represents effects arising
from the nucleon spin polarized transversally to the LC.  
Only $g_1$ and $g_2$ are typically considered in polarized DIS analyses  because $g_t$ and $g_3$ are suppressed as 
$1/Q$ and $1/Q^2$, respectively.

The DIS cross-section involves the contraction of the hadronic and leptonic tensors. If the 
target is polarized in the beam direction one has~\cite{Roberts}:
\vspace{-0.3cm}
\begin{eqnarray}
\vspace{-0.3cm}
\left(\frac{d^2\sigma}{d\Omega dE'}\right)_{\parallel} = \sigma_{Mott} 
\bigg\{ \frac{F_1(Q^2,\nu)}{E'} \tan^2\frac{\theta}{2} + \frac{{2E'F}_2(Q^2,\nu)}{M\nu}  \nonumber \\
\pm \frac{4}{M}\tan^2\frac{\theta}{2} 
\bigg[ \frac{E+E'\cos\theta}{\nu}g_1(Q^2,\nu) -  \gamma^2 g_2(Q^2,\nu)\bigg]\bigg\} ,
\label{eq:sigmapar} 
\end{eqnarray} 
where $\pm$ indicates that the initial lepton is polarized 
parallel {\it vs.} antiparallel to the beam direction.  Here $\gamma^2\equiv Q^2/\nu^2$.  At fixed
$ x_{Bj} = Q^2/(2M\nu)$, the contribution from $g_2$ is suppressed as $\approx1/E$ in the target rest frame.

It is useful to define 
$\sigma_{Mott}$, the photoabsorption cross-section for a point-like, infinitely heavy, target in its rest frame:
\vspace{-0.3cm}
\begin{equation}
\vspace{-0.3cm}
\sigma_{Mott} \equiv \frac{\alpha^2\cos^2(\theta/2)}{4E^2\sin^{4}(\theta/2)}.
\label{eq:mott}
\end{equation}
The $\sigma_{Mott}$ factorization thus isolates  the effects of the hadron structure. 

If the target polarization is perpendicular to both the beam direction and the 
lepton scattering plane, then:
\vspace{-0.3cm}
\begin{eqnarray}
\vspace{-0.3cm}
\left(\frac{d^2\sigma}{d\Omega dE'}\right)_{\perp} = \sigma_{Mott} 
\bigg\{\frac{F_1(Q^2,\nu)}{E'}\tan^2\frac{\theta}{2}+\frac{{2E'F}_2(Q^2,\nu)}{M\nu}  \nonumber \\
\pm\frac{4}{M}\tan^2\frac{\theta}{2}E'\sin\theta \bigg[\frac{1}{\nu}g_1(Q^2,\nu)+\frac{2E}{\nu^2}g_2(Q^2,\nu) \bigg] \bigg\},
\label{eq:sigmaperp}
\end{eqnarray}
In this case $g_2$ is not suppressed compared to $g_1$, since typically $\nu \approx E$ in DIS in the nucleon target rest frame.
The unpolarized contribution is evidently identical in Eqs.~(\ref{eq:sigmapar}) and~(\ref{eq:sigmaperp}). 
Combining them provides the cross-section for any target polarization direction within the plane of the lepton scattering.
The general formula for any polarization direction,  including nucleon spin normal to the lepton plane, is given in Ref.~\cite{Jaffe:1989xx}.

From Eqs.~(\ref{eq:sigmapar}) and~(\ref{eq:sigmaperp}), the cross-section relative asymmetries are:
\vspace{-0.3cm}
\begin{equation}
\vspace{-0.3cm}
A_{\Vert}\equiv\frac{\sigma^{\downarrow\Uparrow}-\sigma^{\uparrow\Uparrow}}{\sigma^{\downarrow\Uparrow}+\sigma^{\uparrow\Uparrow}}=\frac{4\tan^2\frac{\theta}{2}\left[\frac{E+E'\cos\theta}{\nu}g_1(Q^2,\nu)- \gamma^2 g_2(Q^2,\nu)\right]}{M\left[\frac{F_1(Q^2,\nu)}{E'}\tan^2\frac{\theta}{2}+\frac{{2E'F}_2(Q^2,\nu)}{M\nu}\right]},
\label{Apar}
\end{equation}
\begin{equation}
\vspace{-0.2cm}
A_{\bot}\equiv\frac{\sigma^{\downarrow\Rightarrow}-\sigma^{\uparrow\Rightarrow}}{\sigma^{\downarrow\Rightarrow}+\sigma^{\uparrow\Rightarrow}}=\frac{4\tan^2\frac{\theta}{2}E'\sin\theta\big[\frac{1}{\nu}g_1(Q^2,\nu)+\frac{2E}{\nu^2}g_2(Q^2,\nu)\big]}{M\left[\frac{F_1(Q^2,\nu)}{E'}\tan^2\frac{\theta}{2}+\frac{{2E'F}_2(Q^2,\nu)}{M\nu}\right]}.
\label{Aperp}
\end{equation}
%

\subsubsection{Single-spin asymmetries}
The beam and target must  both be polarized to produce non-zero  asymmetries in an inclusive cross-section.
The derivation of these asymmetries typically assumes the ``first Born approximation", 
a purely electromagnetic interaction, and the standard symmetries -- in particular C, P and T invariances.
In contrast, single-spin asymmetries (SSA) arise when one of these assumptions is invalidated;  {\it e.g.}, in 
SIDIS by the selection
of a particular direction corresponding to the 3-momentum of a produced hadron.
Note that T-invariance  should be distinguished from ``pseudo T-odd" asymmetries.   
For example, the final-state interaction in single-spin SIDIS  $ \ell p_\updownarrow \to  \ell' H X$ 
with a polarized proton target produces correlations such as 
$i \vec S_p \cdot \vec q \times \vec p_H$.
Here $\vec S_p$ is the proton spin vector
and $\vec p_H$ is the 3-vector of the tagged final-state hadron.
This triple product changes sign under time reversal  $T \to -T$; however, the factor $i$, 
which arises from the struck quark FSI on-shell cut diagram,
provides a signal which retains time-reversal invariance.

The single-spin asymmetry  measured in SIDIS thus  can access effects beyond the naive 
parton model described in Section~\ref{parton model}~\cite{DAlesio:2007bjf} such as rescattering 
or ``lensing" corrections~\cite{Brodsky:2002cx}.  
Measurements of SSA  have in fact become a vigorous research area  of QCD called ``Transversity".

The observation of parity violating (PV) SSA in DIS can test  fundamental symmetries 
of the Standard Model~\cite{Wang:2014bba}.  When one allows for $Z^0$ exchange, the
PV effects are enhanced by the interference between the $Z^0$ and virtual photon interactions.
Parity-violating interactions in the  elastic and resonance region of DIS can also reveal novel aspects 
of  nucleon structure~\cite{Armstrong:2012bi}.

Other SSA phenomena; {\it e.g.}, correlations arising \emph{via} two-photon exchange,
have been investigated both theoretically~\cite{DeRujula:1973pr} and 
experimentally~\cite{Zhang:2015kna}. 
In the inclusive quasi-elastic experiment reported in Ref.~\cite{Zhang:2015kna}, for which the target
was polarized vertically ({\it i.e.}, perpendicular to the  scattering plane), the SSA is
sensitive to departures from the single photon time-reversal conserving contribution.  

\subsubsection{Photo-absorption asymmetries\label{opt. theo. and photo-abs}}
In electromagnetic photo-absorption reactions, the probe is the photon. Thus, instead of
lepton asymmetries, $A_{\Vert}$ and $A_{\bot}$, one can also consider the physics of 
photoabsorption with polarized photons. The effect of polarized photons can be deduced from combining  
$A_{\Vert}$ and $A_{\bot}$ (Eq.~(\ref{eq:asyp}) below).
The photo-absorption cross-section is related to the 
imaginary part of the forward virtual Compton scattering amplitude
by the  \emph{Optical Theorem}.
Of the ten angular momentum-conserving Compton amplitudes, only
four are independent because of parity and time-reversal symmetries. 
The following ``partial cross-sections" are typically used~\cite{Roberts}:
\vspace{-0.3cm}
\begin{equation}
\vspace{-0.3cm}
\sigma_{T,3/2} = \frac{4\pi^2\alpha} {M \kappa_{\gamma^*}} \left[F_1(Q^2,\nu) - g_1(Q^2, \nu) + \gamma^2 g_2(Q^2,\nu) \right] ,
\label{eq:sigma3/2}
\end{equation}
\begin{equation}
\vspace{-0.1cm}
\sigma_{T,1/2}=\frac{4\pi^2\alpha}{M\kappa_{\gamma^*}}\left[F_1(Q^2,\nu)+g_1(Q^2,\nu)-\gamma^2g_2(Q^2,\nu)\right],
\label{eq:sigma1/2}
\end{equation}
\begin{equation}
\vspace{-0.1cm}
\sigma_{L,1/2}=\frac{4\pi^2\alpha}{M\kappa_{\gamma^*}}\left[-F_1(Q^2,\nu)+\frac{M}{\nu}(1+\frac{1}{\gamma^2})F_2(Q^2,\nu)\right],
\label{eq:sigma1/2L}
\end{equation}
\begin{equation}
\vspace{-0.1cm}
{\sigma}_{LT,3/2}'=\frac{4\pi^2\alpha}{\kappa_{\gamma^*}}\frac{\gamma}{\nu}\left[g_1(Q^2,\nu)+g_2(Q^2,\nu)\right],
\label{eq:sigma3/2TL}
\end{equation}
where {\scriptsize \emph{T,1/2} } and  {\scriptsize \emph{T,3/2}} refer to the absorption 
of a photon with its spin antiparallel or parallel, 
respectively, to that of the spin of the  longitudinally polarized  target.  As  a result,  
{\scriptsize1/2} and {\scriptsize 3/2} are the total spins in the 
direction of the photon momentum.  The notation {\scriptsize \emph{L}} refers to longitudinal virtual photon absorption and
{\scriptsize \emph{LT}} defines the contribution from the transverse-longitudinal interference. 
The effective cross-sections can be negative and depend on the convention chosen for 
flux factor of the virtual photon, which is proportional to
the ``equivalent energy of the virtual photon'' $\kappa_{\gamma^*}$. 
(Thus, the nomenclature of ``cross-section" can be misleading.)
The expression for $\kappa_{\gamma^*}$ is arbitrary 
but must match the real photon energy  $\kappa_{\gamma}=\nu$ 
when $Q^2\to0$. In the Gilman convention, $\kappa_{\gamma^*}=\sqrt{\nu^2+Q^2}$~\cite{Gilman:1967sn}. 
The Hand convention~\cite{Hand:1963bb}  $\kappa_{\gamma^*}=\nu-Q^2/(2M)$ has also been
widely used. Partial cross-sections 
must be normalized by  $\kappa_{\gamma^*}$ since the total cross-section, which is
proportional to the virtual photon flux times a sum of partial cross-sections is an observable and thus
convention-independent. We define:
\vspace{-0.3cm}
\begin{equation}
\vspace{-0.3cm}
\sigma_T\equiv \frac{\sigma_{T,1/2}+\sigma_{T,3/2}}{2}=\frac{8\pi^2\alpha}{M\kappa_{\gamma^*}}F_1 , \\\
 ~~~~~\sigma_{L}\equiv\sigma_{L,1/2},
\nonumber
\end{equation}
\begin{equation}
\vspace{-0.1cm}
 \sigma_{TT} \equiv \frac{\sigma_{T,1/2} - \sigma_{T,3/2}}{2} \equiv -\sigma_{TT}' = \frac{4\pi^2\alpha}{M\kappa_{\gamma^*}}(g_1-\gamma^2g_2), \label{sigmaTT} \\\
~~~~ \sigma_{LT}'\equiv\sigma_{LT,3/2}' ,
\end{equation}
\begin{equation}
\vspace{-0.1cm}
 R\equiv\frac{\sigma^{L}}{\sigma^T}=\frac{1+\gamma^2}{2x}\frac{F_2}{F_1}-1,
\label{R(Q2)}
\end{equation}
as well as the two asymmetries 
%
$
A_1\equiv \sigma^{TT} / \sigma^T, 
A_2\equiv \sigma^{LT} / \left(\sqrt{2}\sigma^T\right), 
$
%
with $\left|A_2\right|\leq R$, since
$\left|\sigma^{LT}\right|<\sqrt{\sigma^T\sigma^{L}}$.
A tighter constraint can also be derived: the ``Soffer bound"~\cite{Soffer:1994ww}  which is also based on 
\emph{positivity constraints}. These constraints can be used to improve PDF determinations~\cite{Leader:2005kw}.
Positivity also constrains the other structure functions and their moments, e.g. $\left|g_1\right|\leq F_1$. This is readily 
understood when structure functions are interpreted in terms of PDFs, as discussed in the next section. 
The $A_1$ and $A_2$  asymmetries are related to those defined by:
\vspace{-0.2cm}
\begin{equation}
\vspace{-0.2cm}
A_{\Vert}=D(A_1+\eta A_2), \label{eq:asyp} \\\
~~~~~A_{\bot}=d(A_2-\zeta A_1),
\end{equation}
where $D\equiv\frac{1-\epsilon E'/E}{1+\epsilon R}$ ,  $d\equiv D\sqrt{\frac{2\epsilon}{1+\epsilon}}$,  
$\eta\equiv\frac{\epsilon\sqrt{Q^2}}{E-\epsilon E'}$, $\zeta \equiv\eta\frac{1+\epsilon}{2\epsilon}$, and
$\epsilon$ is given below Eq.~(\ref{eq:sep}).
\subsubsection{Structure function extraction}\label{Structure functions extraction}
One can use the relative asymmetries  $A_1$ and $A_2$, 
or the cross-section differences $\Delta \sigma_{\parallel}$ and $\Delta \sigma_{\perp}$ 
in order to extract $g_1$ and $g_2$,
The SLAC, CERN and DESY experiments used the asymmetry method, whereas the 
JLab experiments have used both techniques.

\noindent \textbf{Extraction using relative asymmetries\label{par:Extraction asy}}
This is the simplest method: only relative measurements are necessary and 
normalization factors (detector acceptance and inefficiencies,
incident lepton flux, target density, and data acquisition inefficiency)  cancel out with high accuracy.
Systematic uncertainties are therefore minimized. 
However, measurements of the unpolarized structure functions  $F_1$ and $F_2$ (or
equivalently $F_1$ and their ratio $R$, Eq.~(\ref{R(Q2)})) must be used as input. In addition,
the measurements must be corrected for any unpolarized materials present in and around the target.
These two contributions increase the total systematic uncertainty.
Eqs.~(\ref{Apar}), (\ref{Aperp}) and (\ref{eq:asyp}) yield
\vspace{-0.3cm}
\begin{equation}
\vspace{-0.3cm}
A_1=\frac{g_1-\gamma^2g_2}{F_1}, \label{eq:A1} \\\
~~~~~A_2=\frac{\gamma\left(g_1+g_2\right)}{F_1},
\end{equation}
and thus
\begin{equation}
\vspace{-0.1cm}
g_1=\frac{F_1}{1+\gamma^2}\left[A_1+\gamma A_2\right] = 
\frac{y(1+\epsilon R)F_1}{(1-\epsilon)(2-y)}\left[A_{\parallel}+\tan(\theta/2) A_{\perp}\right],
\nonumber
\end{equation}
\begin{equation}
g_2=\frac{F_1}{\gamma(1+\gamma^{2})}\left[A_2-\gamma A_1\right]=\frac{y^2(1+\epsilon R)F_1}{2(1-\epsilon)(2-y)}\left[\frac{E+E'\cos\theta}{E'\sin\theta}A_{\perp}-A_{\parallel}\right].
\nonumber
\end{equation}

\noindent \textbf{Extraction from cross-section differences \label{par:Extraction-par-Xs}}
The advantage of this method is that it eliminates all unpolarized material contributions. 
In addition, measurements of $F_1$ and $F_2$ are not needed. However, measuring absolute quantities is
usually more involved, which may lead
to a larger systematic error. According to Eqs.~(\ref{eq:sigmapar})
and (\ref{eq:sigmaperp}),
\vspace{-0.3cm}
\begin{equation}
\vspace{-0.3cm}
\Delta\sigma_{\parallel}\equiv\frac{d^2\sigma^{\downarrow\Uparrow}}{dE'd\Omega}-\frac{d^2\sigma^{\uparrow\Uparrow}}{dE'd\Omega}=\frac{4\alpha^2}{MQ^2}\frac{E'}{E\nu}\left[g_1(E+E'\cos\theta)-Q^2\frac{g_2}{\nu}\right],
\nonumber
\end{equation}
\vspace{-0.3cm}
\begin{equation}
\vspace{-0.1cm}
\Delta\sigma_{\perp}\equiv\frac{d^2\sigma^{\downarrow\Rightarrow}}{dE'd\Omega}-\frac{d^2\sigma^{\uparrow\Rightarrow}}{dE'd\Omega}=\frac{4\alpha^2}{MQ^2}\frac{E'^2}{E\nu}\sin\theta\left[g_1+2E\frac{g_2}{\nu}\right],
\nonumber
\end{equation}
which yields
\vspace{-0.3cm}
\small
\begin{equation} 
\vspace{-0.3cm}
g_1=\frac{2ME\nu Q^2}{8\alpha^2E'(E+E')}\left[\Delta \sigma_{\parallel}+\tan(\theta/2)\Delta \sigma_{\perp}\right], 
\\
~
g_2=\frac{M\nu^2Q^2}{8\alpha^2E'(E+E')}\left[\frac{E+E'\cos\theta}{E'\sin\theta}\Delta \sigma_{\perp}-\Delta \sigma_{\parallel}\right].
\nonumber
\end{equation}
\normalsize

\subsubsection{The Parton Model \label{parton model}}
\noindent \textbf{DIS in the Bjorken limit}

The moving nucleon in the Bjorken limit is effectively described as bound states of  
nearly collinear partons. The underlying dynamics manifests itself by the fact that partons have
both  position and momentum distributions. The  partons are assumed to be loosely bound, and the lepton scatters incoherently 
only on the point-like quark or antiquark constituents  since gluons are electrically neutral. In this simplified description
the hadronic tensor takes a form similar to  that of the leptonic tensor.
This simplified model, the ``Parton Model",  was introduced by Feynman~\cite{Feynman:1969wa} 
and applied to DIS by Bjorken and Paschos~\cite{Bjorken:1969ja}. 
Color confinement, quark and nucleon masses, transverse momenta
and transverse quark spins are neglected  and Bjorken scaling is satisfied. Thus, in this approximation, 
studying the spin structure of the nucleon is reduced to studying its helicity structure. It
is a valid description only in the IMF~\cite{Weinberg:1966jm}, or equivalently, the frame-independent  
Fock state picture of the LF. After integration over the quark momenta
and the summation over quark flavors,  the measured hadronic tensor can be matched to the hadronic tensor parameterized by the structure functions to obtain: 
%
\vspace{-0.3cm}
\begin{equation}
\vspace{-0.3cm}
F_1(Q^2,\nu) \to F_1(x)=\sum_i\frac{e_i^2}{2}\left[q_i^\uparrow(x)+q_i^\downarrow(x)+\overline{q}_{i}^\uparrow(x)+\overline{q}_i^\downarrow(x)\right],
\label{eq:eqf1parton}
\end{equation}
\vspace{-0.3cm}
\begin{equation}
F_{_2}(Q^2,\nu) \to F_2(x)=2xF_1(x),
\label{eq:Callan-gross}
\end{equation}
\vspace{-0.3cm}
\begin{equation}
g_1(Q^2,\nu) \to g_1(x)=\sum_i\frac{e_i^2}{2}\left[q_i^\uparrow(x)-q_i^\downarrow(x)+\overline{q}_i^\uparrow(x)-\overline{q}_i^\downarrow(x)\right],
\label{eq:eqg1parton}
\end{equation}
\vspace{-0.3cm}
\begin{equation}
\vspace{-0.3cm}  
g_2(Q^2,\nu) \to g_2(x)=0,
\end{equation}
where $i$ is the quark flavor, $e_i$ its charge and $q^\uparrow(x)$
($q^\downarrow(x)$) the probability that 
its spin  is aligned (antialigned) with the nucleon spin at a given
$x$.
Electric charges are squared in Eqs.~(\ref{eq:eqf1parton}) and~(\ref{eq:eqg1parton}), thus the inclusive DIS 
cross-section in the parton model is unable to distinguish antiquarks from quarks.

The unpolarized and polarized PDFs are  respectively
\vspace{-0.3cm}
\begin{equation}
\vspace{-0.3cm}
q_i(x)\equiv q_i^\uparrow(x)+q_i^\downarrow(x), \\\
~~~~~\Delta q_i(x)\equiv q_i^\uparrow(x)-q_i^\downarrow(x).
\label{eq:pdf def.}
\end{equation}
\begin{figure}
\center
\includegraphics[scale=0.49]{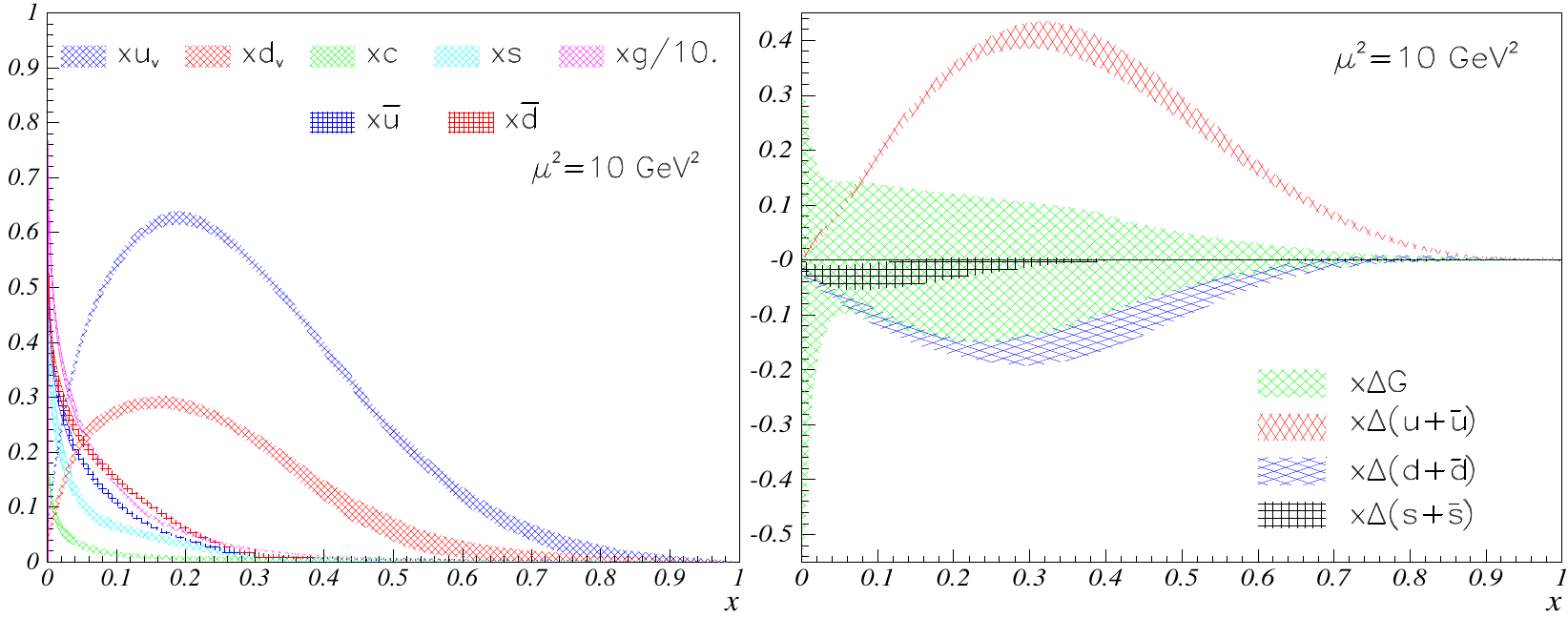}
\vspace{-0.3cm}
\caption{\label{fig:pdf_dist}\small{Left: Unpolarized PDFs as function
of $x$ for the proton from NNPDF~\cite{Ball:2013lla,Ball:2017nwa}.   The
\emph{valence quarks} are denoted $u_{v}$ and
$d_{v}$, with $q_v(x) = q(x) - \bar q(x)$ normalized to the valence content of the proton: $\int_0^1 u_v(x)= 2$ and $\int_0^1 dx d_v(x) = 1$.
The gluon distribution $g$ is divided by 10 on the figure.
Right: Polarized PDFs for the proton. The  $\mu^2$ values refer to scale at which the
PDFs are calculated.}}
\vspace{-0.6cm}
\end{figure}
%
These distributions can be extracted from inclusive DIS (see e.g. Fig.~\ref{fig:pdf_dist}). 
The gluon distribution, also shown in Fig.~\ref{fig:pdf_dist},
can be inferred from sum rules and global fits of the DIS data.  However, 
the identification of the specific contribution of quark and gluon OAM to the nucleon spin
(Fig.~\ref{fig:Lq_dist}) is beyond the parton model analysis.
Note that Eq.~(\ref{eq:pdf def.}) imposes  the constraint $\left|\Delta q_i(x)\right| \leq q_i(x)$,  which together with 
Eqs.~(\ref{eq:eqf1parton}) and~(\ref{eq:eqg1parton}) yields the 
\emph{positivity constraint}  $\left|g_1\right|\leq F_1$.

Eqs.~(\ref{eq:eqf1parton}) and~(\ref{eq:eqg1parton}) are derived assuming that there is no  interference of amplitudes 
for the lepton scattering at high momentum transfer on one type of quark or another; the  final states in the parton model
are distinguishable  and depend on which quark participates in the scattering and is  
ejected from the  nucleon target; likewise, the derivation of  Eqs.~(\ref{eq:eqf1parton}) and~(\ref{eq:eqg1parton})
assumes that quantum-mechanical coherence is not possible for different quark scattering amplitudes 
since the quarks are assumed to be quasi-free. Such interference and coherence effects can arise
at lower momentum transfer where quarks can coalesce into specific hadrons and thus participate
together in  the scattering amplitude. In such a  case, the specific quark which scatters cannot  be identified
as the struck quark. This resonance regime is discussed in Sections~\ref{resonance region} 
and~\ref{elastic scatt}.

The parton model naturally predicts
1)  Bjorken scaling: the structure functions depend only on $x = x_{Bj}$; 
2) the Callan-Gross relation~\cite{Callan:1969uq}, $F_2=2xF_1$, reflecting the spin-$1/2$ nature of quarks;
{\it i.e.}, $F_L=0$ (no absorption of longitudinal photons in DIS due to helicity conservation); 
3) the interpretation of $x_{Bj}$ as the momentum fraction carried by the struck quark 
in the IMF~\cite{Weinberg:1966jm},  or equivalently, the quarks' LF momentum fraction $x= k^+ / P^+$;  and 
4) a probabilistic interpretation of the structure functions: they are 
the square of the parton wave functions and can be constructed
from individual quark distributions and polarizations in momentum space.
The parton model interpretations of $x_{Bj}$ and of structure functions is only valid in the DIS limit
and at LO in  $\alpha_s$. For example, unpolarized PDFs extracted at NLO may be 
negative~\cite{Ball:2013lla, Leader:2005ci}, see also~\cite{Brodsky:2002ue}. 

In the parton model, only two structure functions are needed to  describe the nucleon.
The vanishing of $g_2$ in the parton model does not mean it is zero
in pQCD. In fact, pQCD predicts a non-zero value for $g_2$, see Eq.~(\ref{eq:g2ww}).
The structure function $g_2$ appears when $Q^2$ is finite  due to 
1)  quark interactions, and 
2)  transverse momenta and spins (see {\it e.g.},~\cite{Anselmino:1994gn}).
It also should be noted that the parton model cannot  account for DDIS  
events $\ell p \to \ell' p X$, where the proton remains intact in the final state.  
Such events contribute to roughly 10\% of the total DIS rate.

DIS experiments are typically performed at  beam energies for which at most the three or four 
lightest quark flavors can appear in the final state.
Thus, for the proton and the neutron, with three active quark flavors:
\vspace{-0.3cm}
\begin{eqnarray} 
\vspace{-0.3cm}
F_1^p(x) & = & \frac{1}{2} \bigg(\frac{4}{9} \big(u(x)+ \overline{u} (x)\big)+
\frac{1}{9}\big(d(x)+\overline{d}(x)\big)+\frac{1}{9}\big(s(x)+\overline{s}(x)\big)\bigg),
\nonumber
\\
\vspace{-0.2cm}
g_1^p(x) & = &\frac{1}{2}\bigg(\frac{4}{9}\big(\Delta u(x)+\Delta\overline{u}(x)\big)+
\frac{1}{9}\big(\Delta d(x)+\Delta\overline{d}(x)\big)+\frac{1}{9}\big(\Delta s(x)+\Delta\overline{s}(x)\big)\bigg),
\nonumber
\\
\vspace{-0.1cm}
F_1^n(x) & = & \frac{1}{2}\bigg(\frac{1}{9}\big(u(x)+\overline{u}(x)\big)+
\frac{4}{9}\big(d(x)+\overline{d}(x)\big)+\frac{1}{9}\big(s(x)+\overline{s}(x)\big)\bigg),
\nonumber
\\
\vspace{-0.0cm}
g_1^n(x) & = & \frac{1}{2}\bigg(\frac{1}{9}\big(\Delta u(x)+\Delta\overline{u}(x)\big)+
\frac{4}{9}\big(\Delta d(x)+\Delta\overline{d}(x)\big)+\frac{1}{9}\big(\Delta s(x)+\Delta\overline{s}(x)\big)\bigg),
\nonumber
\end{eqnarray}
where the PDFs $q(x), ~\overline q(x), ~ \Delta q(x)$, and $  \Delta \overline q(x)$ correspond to the longitudinal light-front momentum fraction distributions of the quarks inside the  nucleon.
This analysis assumes SU(2)$_f$ charge symmetry, which typically is believed to hold at the 1\% level~\cite{Miller:2006tv, Cloet:2012db}.

In the Bjorken limit, this description provides spin information in terms of $x$ (or $x$ and $Q^2$ at lower energies, as discussed below). 
The spatial spin distribution is also accessible, \emph{via} the nucleon axial form factors. This is analogous to the fact that 
the nucleon's electric charge and current distributions are accessible through the electromagnetic form factors 
measured in elastic lepton-nucleon scattering (see Sec. \ref{elastic scatt}).
Form factors and  particle distributions functions are linked by  GPDs and Wigner Functions, which correlate both the spatial and 
 longitudinal momentum 
information~\cite{Burkardt:2000za}, including that of OAM~\cite{Lorce:2017wkb}.

\subsubsection{Perturbative QCD at finite $Q^2$ \label{sub:pQCD}}

In pQCD,  the struck quarks in DIS can radiate gluons; 
the simplicity of Bjorken scaling is then broken by computable logarithmic corrections.
The lowest-order $\alpha_s$ corrections arise from
1) vertex correction, where a gluon links the incoming and outgoing quark lines; 
2) gluon bremsstrahlung on either the incoming and outgoing quark lines;
3) $q$-$\overline{q}$ pair creation or annihilation. This latter leads to the axial anomaly and makes gluons
to contribute to the nucleon spin (see Sec.~\ref{DIS SR}). 
These corrections introduce a power of $\alpha_s$ at each order, which   
leads to logarithmic dependence in $Q^2$,
corresponding to the  behavior of the strong coupling $\alpha_s(Q^2)$ at high $Q^2$~\cite{Deur:2016tte}.

Amplitude calculations,  including gluon radiation, exist up to
next-to-next-to leading order (NNLO) in $\alpha_s$
\cite{Gorishnii:1990vf}. In some particular cases, calculations or assessments exist up to fourth order 
{\it e.g.}, for the Bjorken sum rule, see Section~\ref{DIS SR}. 
These gluonic corrections are similar to the effects derived from photon emissions (radiative corrections)
in QED;   they are  therefore called pQCD radiative corrections. 
As in QED, canceling  infrared and ultraviolet divergences appear and calculations must
be regularized and then renormalized. Dimensional regularization
is often used for pQCD (minimal subtraction scheme,
$\overline{MS}$)~\cite{Bardeen:1978yd}, although several other schemes are also commonly used.  
The pQCD radiative corrections are described to first approximation  by the 
\emph{DGLAP evolution equations}~\cite{Gribov:1972ri}.
This formalism  correctly  predicts the $Q^2$-dependence of structure functions 
in  DIS.
The pQCD radiative corrections are renormalization scheme-independent  at any order if one applies the 
\emph{BLM/PMC}~\cite{the:BLM, the:PMC} scale-setting procedure.  

The small-$x_{Bj}$ power-law Regge behavior of structure functions can be 
related to the exchange of the Pomeron trajectory using
the \emph{BFKL equations}~\cite{BFKL}. 
Similarly  the $t$-channel exchange of the isospin $I=1$ Reggeon trajectory with 
$\alpha_R = 1/2$ in DVCS can explain the observed behavior 
$F_{2p}(x_{Bj},Q^2) - F_{2n}(x_{Bj},Q^2) \propto \sqrt {x_{Bj}}$,  
as shown by Kuti and Weisskopf~\cite{Kuti:1971ph}. 
This small-$x$ Regge behavior is incorporated in the LFHQCD structure for the $t$-vector meson exchange~\cite{deTeramond:2018ecg}.
A general discussion of the application of Regge dynamics to DIS structure functions is given in 
Ref.~\cite{Landshoff:1970ff}.
The evolution of $g_1(x_{Bj},Q^2$) at  low-$x_{Bj}$ has been  investigated by 
Kirschner and Lipatov, and Blumlein and Vogt~\cite{Kirschner:1983di}, 
by Bartels, Ermolaev and Ryskin~\cite{Bartels:1995iu};
and more recently by Kovchegov, Pitonyak and Sievert~\cite{Kovchegov:2015pbl};
See~\cite{Blumlein:2012bf} for a summary of small-$x_{Bj}$ behavior of the PDFs.
The distribution and evolution at low-$x_{Bj}$ of the gluon spin contributions 
$\Delta g(x_{Bj})$ and $\mbox{L}_g(x_{Bj})$ is discussed 
in~\cite{Hatta:2016aoc}, with the suggestion that in this domain, $\mbox{L}_g(x_{Bj}) \approx -\Delta g(x_{Bj})$.
In addition to structure functions, the evolution of the  \emph{distribution amplitudes} in $\ln (Q^2)$ 
defined from the valence LF Fock state is also known and given
by the \emph{ERBL} equations~\cite{ERBL}.

Although the evolution of the $g_1$ structure function  is
known to NNLO~\cite{Moch:2014sna},  we will focus here  on  the leading order (LO) analysis 
in order to demonstrate the general formalism.  At \emph{leading-twist}  one finds
\vspace{-0.3cm}
\begin{equation}
\vspace{-0.3cm}
g_1(x_{Bj},Q^2)=\frac{1}{2} \sum_i e_i^2 \Delta q_i(x_{Bj},Q^2),
\label{g_1 LT evol}
\end{equation}
where the polarized quark distribution functions $\Delta q$ obey the evolution equation
 \vspace{-0.3cm}
 \begin{equation}
 \vspace{-0.3cm}
\frac{\partial \Delta q_i(x,t)}{\partial t}= \frac{\alpha_s(t)}{2\pi} \int_{x}^1 \frac{dy}{y}
\left[ \Delta q_i(y,t)  P_{qq}\left(\frac{x}{y}\right) +
\Delta g(y,t)  P_{qg}\left(\frac{x}{y}\right) \right]   , 
\label{quark LO evol}
\end{equation}
with $t=\ln(Q^2/\mu^2)$. Likewise, the evolution equation for the polarized gluon distribution function $\Delta g$ is
 \vspace{-0.3cm}
 \begin{equation}
 \vspace{-0.3cm}
\frac{\partial\Delta g(x,t)}{\partial t}= \frac{\alpha_s(t)}{2\pi} \int_{x}^1 \frac{dy}{y}
\bigg[\sum_{i=1}^{2f} \Delta q_i(y,t)  P_{gq}\left(\frac{x}{y}\right) +
\Delta g(y,t) P_{gg}\left(\frac{x}{y}\right) \bigg]  . 
\label{gluon LO evol}
\end{equation}

At LO the splitting functions  $P_{\alpha \beta}$ appearing in Eqs. (\ref{quark LO evol}) and (\ref{gluon LO evol}) are given by
\vspace{-0.3cm}
\begin{eqnarray}
\vspace{-0.3cm}
P_{qq}(z) &=& \frac{4}{3}\frac{1+z^2}{1-z}+2\delta(z-1), \nonumber \\
 P_{qg}(z) &=& \frac{1}{2}\big(z^2-(1-z)^2 \big), \nonumber \\
 P_{gq}(z) &=& \frac{4}{3}\frac{1-(1-z)^2}{z}, \nonumber \\
P_{gg}(z) &=& 3\bigg[\big(1+z^4\big)\bigg(\frac{1}{z}+\frac{1}{1+z} \bigg) - 
\frac{(1-z)^3}{z}\bigg]+\bigg[\frac{11}{2}-\frac{f}{3} \bigg]\delta(z-1). \nonumber
\end{eqnarray}
\vspace{-0.6cm}

These  functions are related to Wilson coefficients defined in the operator product expansion (OPE), see Section~\ref{OPE}. 
They can be  interpreted as the probability that:

\noindent$P_{qq}$: a quark emits a gluon and retains $z=x_{Bj}/y$ of its initial momentum;

\noindent$P_{qg}$: a gluon splits into $q$-$\overline{q}$, with the quark having a fraction $z$ of the gluon momentum;

\noindent$P_{gq}$: a quark emits a gluon with a fraction $z$ of the initial quark momentum;

\noindent$P_{gg}$: a gluon splits in two gluons, with one having the fraction $z$ of the initial momentum.

The presence of $P_{qg}$ allows inclusive polarized DIS to access the polarized gluon 
distribution $\Delta g(x_{Bj},Q^2)$, and thus its moment $\Delta G \equiv \int_0^1 \Delta g~dx$, 
albeit with limited accuracy. The evolution of $g_2$ at LO in $\alpha_s$  is obtained from the 
above equations applied to the Wandzura-Wilczek relation, Eq.~(\ref{eq:g2ww}).

In general, pQCD can predict $Q^2$-dependence, but not the $x_{Bj}$- dependence of the parton 
 distributions which is derived from nonperturbative dynamics (see Section~\ref{DIS}). 
 The high-$x_{Bj}$ domain is an exception (see Section~\ref{sec: high-x}). 
The intuitive \emph{DGLAP} results are recovered more formally using the OPE,
see Section~\ref{OPE}.

\subsubsection{The nucleon spin sum rule and the ``spin crisis" \label{spin crisis}}

The success of modeling the nucleon with quasi-free \emph{valence quarks} and  with
\emph{constituent quark} models (see Section~\ref{CQM}) 
suggests that only quarks contribute to the nucleon spin:
\vspace{-0.3cm}
\begin{equation}
\vspace{-0.2cm}
J= \frac{1}{2}\Delta\Sigma+\mbox{L}_q  =\frac{1}{2},
\end{equation}
where $ \Delta \Sigma $ is the quark spin contribution to the nucleon spin $J$;
\vspace{-0.3cm}
\begin{equation}
\vspace{-0.3cm}
\Delta \Sigma = \sum_q \int_0^1 dx \, \Delta q(x),
\end{equation}
\begin{wrapfigure}{r}{0.5\textwidth} 
\vspace{-0.7cm}
\includegraphics[scale=0.37]{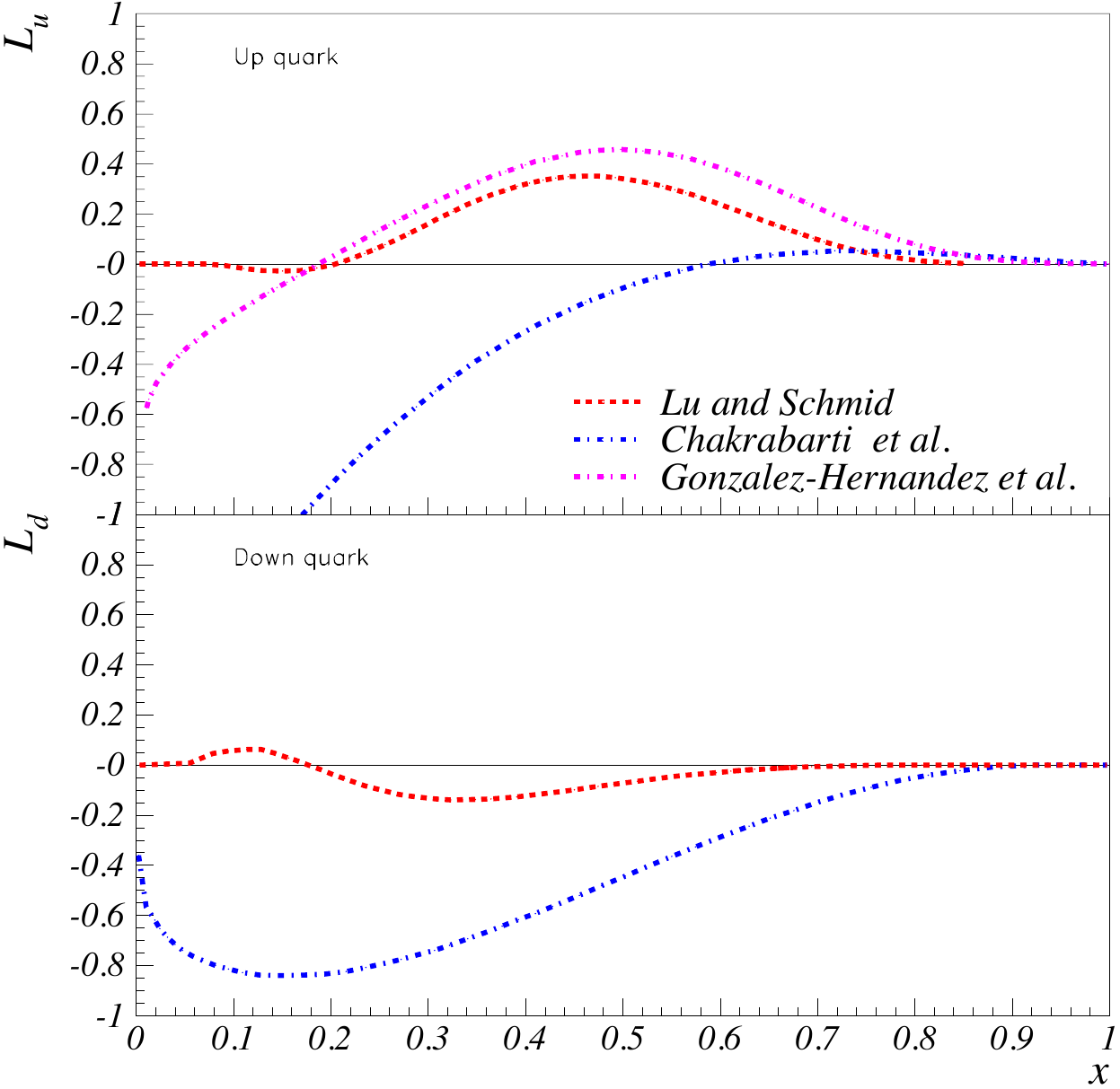}
\vspace{-0.5cm}
\caption{\label{fig:Lq_dist}\small{Models predictions 
for the quark kinematical OAM $L_z$, from~Refs.~\cite{GonzalezHernandez:2012jv} (dot-dashed line),
\cite{Chakrabarti:2016yuw} (dots), and
\cite{Lu:2010dt} (dashes).
}}
\vspace{-0.5cm}
\end{wrapfigure}
%
and $ \mbox{L}_q$ is the quark OAM contribution. Extracted 
polarized PDFs and modeled quark OAM distributions are shown in Figs.~\ref{fig:pdf_dist} and \ref{fig:Lq_dist}.
It should be emphasized that the existence of the proton's anomalous magnetic moment requires nonzero 
quark OAM~\cite{Brodsky:1980zm}. For instance, in the Skyrme model, chiral symmetry implies a 
dominant nonperturbative contribution to the proton spin from quark OAM~\cite{Brodsky:1988ip}. 
It is interesting to quote the conclusion from Ref.~ \cite{Sehgal:1974rz}: ``Nearly 40\% of the angular 
momentum of a polarized proton arises from the orbital motion of its constituents. 
In the geometrical picture of hadron structure, this implies
 that  \emph{a polarized proton possesses a significant amount of rotation contribution 
 to $S_z$ and} $\mbox{L}_z$ \emph{comes from the valence quarks}.'' (emphasis by the author).
 QCD radiative effects introduce corrections to the spin dynamics from gluon emission and 
 absorption which evolve in $\ln Q^2$. 
It was generally expected that the radiated gluons would contribute to the nucleon spin, but only as a small correction
(beside their effect of introducing a $Q^2$-dependence to the different contributions to the nucleon spin).
The speculation that  polarized gluons contribute significantly to nucleon spin, 
whereas their sources -- the quarks --  do not, is unintuitive, although it is a scenario that
was (and still is by some) considered (see e.g. the bottom left panel of Fig.~\ref{spin history} 
on page~\pageref{spin history}).
A small contribution to the nucleon spin from gluons would also imply a small role of the 
\emph{sea quarks}, so that $ \Delta \Sigma $ and the quark OAM
would then  be understood as coming mostly from \emph{valence quarks}. 
In this framework, it was determined that   the quark OAM contributes to about 20\%~\cite{Sehgal:1974rz, Ellis:1973kp} 
based on the values for $F$ and $D$, the
weak hyperon decay constants (see Section~\ref{DIS SR}), SU(3)$_f$  flavor symmetry and 
$\Delta s=0$~\cite{Isgur:1978xj, Jaffe:1989jz, Brodsky:1994fz}.
This prediction was made in 1974 and predates the first spin structure measurements by SLAC E80~\cite{Alguard:1976bm}, 
E130~\cite{Alguard:1978gf} and CERN EMC~\cite{Ashman:1987hv}.

The origin of the quark OAM was later understood as due to 
relativistic kinematics~\cite{Jaffe:1989jz, Brodsky:1994fz}, whereas $ \Delta \Sigma $ comes from the quark axial currents
(see discussion below Eq.~(\ref{hadronic tensor})). 
For a nonrelativistic quark, the lower component of the Dirac spinor is 
negligible; only the upper component contributes to the axial current. 
In hadrons, however, quarks are confined in a small volume and are thus relativistic. The 
lower component, which is in  a $p$-wave, with its spin anti-aligned to that of the nucleon, 
contributes and reduces $ \Delta \Sigma $. 
At that time, it seemed reasonable to neglect gluons, thus predicting 
a nonzero  contribution to $J$ from the quark OAM.   
The result was the initial expectation $\Delta\Sigma\approx$ 0.65 and thus  the quark OAM was about 18\%.
Since this review is also concerned with 
spin composition of the nucleon at low energy, it is interesting to remark that a large  quark OAM  contribution would essentially be a confinement effect.

The first high-energy measurements of  $ g_1(x_{Bj},Q^2) $ was performed at  
SLAC in the E80~\cite{Alguard:1976bm} and E130~\cite{Alguard:1978gf} experiments. 
The data covered a limited $x_{Bj}$ range and agreed with the naive model described above. 
However, the later EMC experiment at CERN~\cite{Ashman:1987hv} measured $g_1(x_{Bj},Q^2)$  
over a  range of $x_{Bj}$ sufficiently large to evaluate moments. It showed the conclusions 
based on the SLAC measurements to be incorrect.
The EMC  measurement suggests instead  that $ \Delta \Sigma \approx 0$, with large 
uncertainty. This contradiction with the naive model became known as the ``spin crisis".

Although more recent measurements  at COMPASS, HERMES and Jlab are consistent 
with a value of $ \Delta \Sigma \approx 0.3$, the EMC indication still stands that
gluons and/or gluon and quark OAM are more important than had been foreseen;  
see {\it e.g.}, Ref.~\cite{Myhrer:2007cf}.   
Since gluons are possibly important, 
 $J$ must obey the total angular momentum  conservation  law
known as the ``nucleon spin sum rule" 
\vspace{-0.3cm}
\begin{equation}
\vspace{-0.3cm}
J = \frac{1}{2}\Delta\Sigma(Q^2)+\mbox{L}_q(Q^2) + \Delta G(Q^2) + \mbox{L}_g(Q^2) = \frac{1}{2},
\label{eq:spin SR}
\end{equation}
at any scale $Q$. The gluon spin  $\Delta G$ represents with $\mbox{L}_g$ a single term, $ \Delta G+\mbox{L}_g$, 
since the individual $\Delta G$ and $\mbox{L}_g$ 
contributions are not separately gauge-invariant.  (This is discussed in more detail  
in the next section.) 
The terms in Eq.~(\ref{eq:spin SR}) are obtained by utilizing \emph{LF-quantization} 
or the IMF and the LC gauge,  writing 
the hadronic angular momentum tensor in terms of the quark and gluon fields~\cite{Jaffe:1989jz}. 
In the gauge and frame-dependent partonic formulation, in which $\Delta G$ and $\mbox{L}_g$ can be separated,
Eq.~(\ref{eq:spin SR}) is referred to as the Jaffe-Manohar decomposition.  An alternative formulation is given by Ji's 
decomposition. It is  gauge/frame independent, but its partonic interpretation is not as direct as for
the Jaffe-Manohar decomposition~\cite{Ji:2012vj}. 

The quantities in Eq.~(\ref{eq:spin SR}) 
are  integrated over $x_{Bj}$.  They
have been determined at a moderate value of  $Q^2$, typically 3 or 5 GeV$^2$. 
Eq.~(\ref{eq:spin SR}) does not  separate \emph{sea}
and \emph{valence quark} contributions. Although  DIS observables do not 
distinguish them, separating them is an important task.
In fact, recent data and theoretical developments 
indicate that the \emph{valence quarks} are dominant contributors to $ \Delta \Sigma$.  
We also note that the strange and anti-strange sea quarks can contribute differently 
to the nucleon spin~\cite{Brodsky:1996hc}.
Finally, a separate analysis of spin-parallel and antiparallel PDFs
is clearly valuable since they have different nonperturbative inputs.

A transverse spin sum rule similar to Eq.~(\ref{eq:spin SR}) has also been derived~\cite{Bakker:2004ib, Leader:2011za}. 
Likewise, transverse versions of the Ji sum rule (see next section) 
exist~\cite{Leader:2011cr, Ji:2012ba}, together with debates on which version is correct. 
Transverse spin not being the focus of this review, we will not discuss this issue further.

The $Q^2$-evolution of quark and gluon spins discussed in Section~\ref{sub:pQCD} provides
the $Q^2$-evolution of $\Delta\Sigma$ and $\Delta G$. The evolution equations are known to at least NNLO and are
discussed in Section~\ref{DIS SR}. The evolution of the quark and gluon OAM is known to 
NLO~\cite{Ji:1995cu, Gluck:1994uf, Gluck:1995yr, Wakamatsu:2007ar, Altenbuchinger:2010sz}. 
The evolution of the nucleon spin sum rule components at LO is given in Ref.~\cite{Ji:1995cu}:
\vspace{-0.3cm}
\begin{equation}
\Delta\Sigma(Q^2)  = \mbox{constant}, \nonumber 
\vspace{-0.1cm}
\end{equation}
\begin{equation}
\mbox{L}_q(Q^2) =  \frac{-\Delta \Sigma(Q^2)}{2} +   \frac{3n_f}{32+6n_f} + 
 \bigg(\mbox{L}_q(Q_0^2) + \frac{-\Delta \Sigma(Q^2_0)}{2} - \frac{3n_f}{32+6n_f}\bigg)(t/t_0)^{-\frac{32+6n_f}{9\beta_0}},  \nonumber 
\end{equation}
\begin{equation}
\Delta G(Q^2)  = \frac{-4\Delta \Sigma(Q^2)}{\beta_0} + \bigg(\Delta G(Q_0^2)+ \frac{4\Delta \Sigma(Q^2_0)}{\beta_0}\bigg)\frac{t}{t_0}, \nonumber 
\end{equation}
\begin{equation}
\vspace{-0.1cm}
\mbox{L}_g(Q^2)  =  -\Delta G(Q^2) +  \frac{8}{16+3n_f} +
 \bigg(\mbox{L}_q(Q_0^2) + \Delta G(Q_0^2) - \frac{8}{16+3n_f} \bigg)(t/t_0)^{-\frac{32+6n_f}{9\beta_0}}  
\label{eq:LO evol. of spin SR}
\end{equation}
with $t=\ln(Q^2/\Lambda_s^2)$ and $Q_0^2$ the starting scale of the evolution. The \emph{QCD $\beta$-series}
 is defined here such that $\beta_0=11-\frac{2}{3}n_f$. The NLO equations can be found in Ref.~\cite{Altenbuchinger:2010sz}.

\subsubsection{Definitions of the spin sum rule components \label{SSR components}}

Values for the components of Eq.~(\ref{eq:spin SR}) obtained from experiments, Lattice Gauge 
Theory or models are given in Section~\ref{nucleon spin structure at high energy} and in the Appendix.
It is important to recall that these values are convention-dependent for several reasons. 
One is that the axial anomaly shifts contributions between $\Delta \Sigma$ 
and $\Delta G$, depending on the choice of renormalization scheme, 
\emph{even at arbitrary high $Q^2$} (see Section~\ref{DIS SR}). 
This effect was suggested as a cause for the smallness of $\Delta \Sigma$ compared to the naive quark model
expectation: a large value $\Delta G \approx 2.5$ would increase the measured $\Delta \Sigma$ to about 0.65. Such large value
of $\Delta G$ is nowadays excluded. Furthermore, it is unintuitive to use a specific renormalization scheme in
which the axial anomaly contributes, to match quark models that do not need renormalization.
Another reason is that the definitions of  $\Delta G$, $\mbox{L}_q$, $\mbox{L}_g$ 
are also conventional. 
This was known before the spin crisis~\cite{Jaffe:1989jz} but 
the discussion on what the best operators are
has been renewed by the derivation of the Ji sum rule~\cite{Ji:1996ek}:
\vspace{-0.3cm}
\begin{equation}
\vspace{-0.3cm}
J_{q,g}= \frac{1}{2} \int_{-1}^{1} x \big[E_{q,g}(x,0,0)+H_{q,g}(x,0,0)\big] dx,
\label{Eq. Ji SR}
\end{equation}
with $\sum_q J^q + J^g = \frac{1}{2}$ being frame and gauge invariant  and
$J_{q,g}$ and the GPDs $ E_{q,g}$ and $H_{q,g}$ stand either for quarks or gluons. 
For quarks, $J_q \equiv \Delta \Sigma/2 + \mbox{L}_q$. For gluons, $J_g$ cannot be separated
into spin and OAM parts in a frame or gauge invariant way. (However, 
it can be separated in the IMF, with an additional  ``potential" angular momentum term~\cite{Ji:2012sj}.)

Importantly, the Ji sum rule provides a model-independent access to $ \mbox{L}_q$, whose measurability 
had been until then uncertain. Except for Lattice Gauge Theory (see Section~\ref{Ji's LGT method}) the theoretical
assessments of  the quark OAM are model-dependent. We  mentioned the
relativistic quark model that predicted  about 20\%} even before the occurrence of the spin crisis. 
More recently, investigation within an unquenched quark model suggested  that the unpolarized 
\emph{sea} asymmetry $\overline{u} -  \overline{d}$ is proportional to the
nucleon OAM:
\vspace{-0.3cm}
\begin{equation}
\vspace{-0.3cm}
\mbox{L}(Q^2) \equiv \mbox{L}_q(Q^2)+\mbox{L}_g(Q^2) \propto \big(\overline{u}(Q^2) -  \overline{d}(Q^2)\big),
\label{Eq. L propto sea}
\end{equation}
where $\overline{q}(Q^2)=\int_0^1 \overline{q}(x,Q^2)dx$. 
The non-zero  $\overline{u} -  \overline{d}$ distribution is well measured~\cite{Baldit:1994jk}
and causes the violation of the Gottfried  sum rule~\cite{Gottfried:1967kk, Amaudruz:1991at}.
The initial derivation of Eq.~(\ref{Eq. L propto sea}) by Garvey~\cite{Garvey:2010fi} indicates a strict equality,
$L = (\overline{u} -  \overline{d}) = 0.147 \pm 0.027$, while a derivation in a chiral quark model~\cite{Bijker:2014ila}
suggests $L= 1.5(\overline{u} -  \overline{d}) = 0.221 \pm 0.041$. 
The lack of precise polarized PDFs at low-$x_{Bj}$ does not allow yet
to verify this remarkable prediction~\cite{Nocera:2016zyg}. 
Another  quark OAM prediction is from LFHQCD: $\mbox{L}_q(Q^2 \leq Q_0^2)=1$ in the strong regime of QCD, evolving to 
 $\mbox{L}_q=0.35 \pm 0.05$ at $Q^2=5$~GeV$^2$, see Section~\ref{sec:LFHQCD}. 

Beside Eq.~(\ref{Eq. Ji SR}) and possibly Eq.~(\ref{Eq. L propto sea}),  the quark OAM can also be accessed from
the two-parton \emph{twist}-3 GPD $G_2$~\cite{Penttinen:2000dg}:
\vspace{-0.2cm}
\begin{equation}
\vspace{-0.2cm}
 \mbox{L}_q=-\int\  G^q_2(x,0,0) dx,
\label{Eq. quark OAM from twist-3}
\end{equation} }
 or generalized TMD (GTMD)~\cite{Lorce:2011ni, Lorce:2011kd, Hatta:2011ku}.
TMD allow to infer $\mbox{L}_q$ model-dependently~\cite{Gutsche:2016gcd}.

Jaffe and Manohar set the original convention to define the angular momenta~\cite{Jaffe:1989jz}. They expressed 
Eq.~(\ref{eq:spin SR}) using the canonical angular momentum and momentum tensors. 
This choice is natural since it follows from Noether's theorem~\cite{Leader:2011za}. 
For angular momenta, the relevant symmetry is the 
rotational invariance of QCD's Lagrangian. The ensuing conserved quantity ({\it i.e.}, that commutes with the Hamiltonian) 
is the generator of the rotations. This definition provides the four angular momenta of the longitudinal spin sum rule,
Eq.~(\ref{eq:spin SR}). A similar transverse spin sum rule was also derived~\cite{Bakker:2004ib, Leader:2011za}.
A caveat of the canonical definition is that in Eq.~(\ref{eq:spin SR}), 
only  $J$ and $\Delta\Sigma$ are gauge invariant, {\it i.e.}, are measurable. In the 
light-cone gauge, however, the gluon spin term coincides with the measured observable $\Delta G$.
(This is true also in the $A^0=0$ gauge~\cite{Anselmino:1994gn}.)
The fundamental reason for the gauge dependence of the other  components of Eq.~(\ref{eq:spin SR}) is their
derivation in the IMF.
 
What triggered the re-inspection of the Jaffe-Manohar decomposition and subsequent discussions 
was that Ji proposed another decomposition using the Belinfante-Rosenfeld energy-momentum 
tensor~\cite{Belinfante tensor}, which lead to the Ji sum rule~\cite{Ji:1996ek}, Eq.~(\ref{Eq. Ji SR}).  
The Belinfante-Rosenfeld tensor originates from General Relativity in which the canonical momentum 
tensor is modified so that it becomes symmetric and conserved (commuting with the Hamiltonian): 
in a world without angular momentum, the canonical momentum tensor would be symmetric.  
However, adding spins breaks its symmetry.  An appropriate combination of canonical momentum 
tensor and spin tensor yields the Belinfante-Rosenfeld tensor, which is symmetric and thus natural for General Relativity
where it identifies to its field source ({\it i.e.} the Hilbert tensor). The advantages of such definition are 
1) its naturalness even in presence of spin;
2) that it leads to a longitudinal spin sun rule in which all individual terms are gauge invariant;
and 3) that there is a known method to measure $\mbox{L}_q$ (Eq.~(\ref{Eq. Ji SR})), or to
compute it using Lattice Gauge Theory (see Section~\ref{Ji's LGT method}). 
Its caveat is that the nucleon spin decomposition contains only three terms: $\Delta \Sigma$, 
$\mbox{L}_q$ and a global gluon term, thus  without a clear interpretation of the experimentally measured $\Delta G$.
While $\Delta \Sigma$ in the Ji and Jaffe-Manohar decompositions are identical,
the $\mbox{L}_q$ terms are different. 
That several definitions of $\mbox{L}_q$ are possible comes from gauge invariance. To satisfy it, 
quarks do not suffice; gluons must be included, which allows for choices in the separation of $\mbox{L}_q$ and 
$\mbox{L}_g$~\cite{Leader:2013jra, Liu:2015xha}. 
The general physical meaning of $\mbox{L}_q$ is that it is the
torque acting on a quark during the polarized DIS process~\cite{Abdallah:2016xfk, Burkardt:2012sd}: 
Ji's $\mbox{L}_q$ is the OAM before the probing photon is absorbed by the quark, while the Jaffe-Manohar
 $\mbox{L}_q$ is the OAM after the photon absorption, with the absorbing quark kicked out to infinity. 
 These two definitions of $\mbox{L}_q$ have been investigated with several
 models, {\it e.g.},~\cite{Liu:2015xha,Courtoy:2016des}, whose results are shown in 
 Section~\ref{Individual contributions to the nucleon spin}. 

Other definitions of angular moments and gluon fields have been proposed to eliminate the gauge-dependence
problem~\cite{Chen:2008ag}, leading to a spin decomposition
Eq.~(\ref{eq:spin SR}) with four gauge-invariant terms. The complication is that the 
corresponding operators use non-local fields, \emph{viz} fields depending on several
space-time variables or, more generally, a field $A$ for which $A(x) \neq e^{-ipx} A(0) e^{ipx}$.

Recent reviews on angular momentum definition and  separation are given in Ref.~\cite{Leader:2013jra}.
It remains to be added that in practice, to obtain $\mbox{L}_q$ in a \emph{leading-twist} (twist~2) analysis, 
$\Delta \Sigma/2$ must be subtracted, see Eq.~(\ref{Eq. Ji SR}).
Thus, since $\Delta \Sigma$ is renormalization scheme dependent due to the axial anomaly, $\mbox{ L}_q$ is too
(but not their sum $J_q$). A  \emph{higher-twist} analysis of the nucleon spin sum rule allows to separate quark and gluon
spin contributions (twist~2 PDFs/GPDs) from their OAM 
(twist~3 GPD $G_2$)~\cite{Penttinen:2000dg,  Ji:2012sj, Ji:2012ba, Hatta:2011ku, Kanazawa:2014nha}. 
It is expected that OAM are \emph{twist}-3 quantities since they involve the parton's transverse motions.  
However, the quark OAM, as defined in  Eq.~(\ref{Eq. Ji SR}) can be related to \emph{twist}-2 GPDs. 
Beside GPDs, OAM can also be accessed with
GTMDs~\cite{Gutsche:2016gcd, Lorce:2011kd,  Rajan:2016tlg, Burkardt:2008ua}.
It is now traditional to call the Jaffe-Manohar OAM the \emph{canonical} expression and denote it by $l_z$, the Ji OAM is called
 \emph{kinematical} and denoted by $L_z$.
  We will use this convention for the rest of the review.

In summary, the components of  Eq.~(\ref{eq:spin SR}) are scheme and definition (or gauge) dependent.
Thus, when discussing the origin of the nucleon spin,  schemes and definitions must be specified. 
This is not a setback since, as emphasized in the preamble, the main object of spin physics is not to provide
the pie chart of the nucleon spin but rather to use it to verify  QCD's consistency
and understand complex mechanisms involving it, {\it e.g.}, confinement. 
That can be done consistently in fixed schemes and definitions. 
This leads us to the next section where such complex mechanisms start to arise.

\subsection{The resonance region \label{resonance region}}

At smaller values of  $W$ and $Q^2$, namely below the DIS scaling region, the nucleon
reacts increasingly coherently to the photon until it eventually 
responds fully rigidly.  Before reaching this
elastic reaction on the nucleon ground state, scattering may 
excite nucleon states of higher masses where no specific
quark can unambiguously be said to have been struck, 
thus causing interference and coherence effects. 
One thus leaves the DIS domain to enter the resonance region characterized
by bumps in the scattering cross-section, see Fig.~\ref{fig:gross}.
These higher-spin resonances are OAM and radially excited nucleon states.
They then decay by meson or/and photon emission and can be  classified
into two groups: isospin 1/2 (N$^*$ resonances) and
isospin 3/2 ($\Delta^*$ resonances). 

The resonance domain is important for this review since it covers the transition 
from pQCD to nonperturbative QCD.  It also illustrates how spin information can illuminate QCD phenomena.   
Since the resonances are numerous, overlapping
and differ in origin, spin degrees of freedom are needed to identify and characterize them. 
Modern hadron spectroscopy experiments typically involve polarized beams
and targets. However, inclusive reactions are ill suited to disentangle resonances:
final hadronic states must be partly or fully identified. Thus, we will cover this
extensive subject only superficially. 

The nomenclature classifying nucleon resonances originates from  $\pi N$ scattering.  
Resonances are labelled by L$_{2I~2J}$, where L is the OAM \emph{in the $\pi N$ channel} 
(not the hadron wavefunction OAM),  $I$=1/2 or 3/2 is the isospin, and $J$ is the total angular momentum.  
L is labeled by S (for L=0), P (L=1), D (L=2) or F (L=3). 
An important tool to classify resonances and predict their
masses is  the \emph{constituent quark} model, which is discussed next.  Lattice gauge theory
(Section~\ref{LGT}) is now the main technique to predict  and characterize resonances, with
the advantage of being a first-principle QCD approach. Another successful approach based on QCD's 
basic principles and symmetries is LF Holographic QCD (Section~\ref{sec:LFHQCD}), an effective theory which uses
the gauge/gravity duality on the LF, rather than ordinary spacetime, to captures essential aspects of QCD dynamics
 in its nonperturbative domain.

\subsubsection{Constituent quark models \label{CQM}} 

The basic classification of the hadron mass spectra was motivated by the development of
\emph{constituent quark} models obeying an $SU(6) \supset SU(3)_{flavor} \otimes SU(2)_{spin}$ 
 internal symmetry~\cite{Isgur:1978xj, Eichmann:2016yit}.    Baryons are modeled as composites of three 
\emph{constituent quarks} of mass $M/3$ (modulo binding energy corrections which
depend on the specific model) which provides the $J^{PC}$ quantum numbers.  The 
\emph{constituent quark} model predates QCD but is now interpreted and developed in its framework. 
\emph{Constituent quarks} differ from \emph{valence quarks}
-- which also determine the correct quantum numbers of hadrons -- in that they are
not physical (their mass is larger) and are understood as \emph{valence quarks} dressed by virtual partons. 
The large \emph{constituent quark} masses explicitly break both the  \emph{conformal} and chiral symmetries
that are nearly exact for QCD at the classical level; see Sections~\ref{sub:Chipt}.
\emph{Constituent quarks} are assumed to be bound at LO by phenomenological potentials such as the
Cornell potential~\cite{the:Eichten. Cornell pot.}, an approach which was interpreted after the advent of QCD
as gluonic flux tubes acting between quarks. 
The LO spin-independent potential is supplemented by a spin-dependent potential,
{\it e.g.}, by adding exchange of mesons~\cite{Matsuyama:2006rp}, instantons or by 
including the interaction of a spin-1 gluon exchanged between the quarks 
(``hyperfine correction"~\cite{Close:1979bt, Godfrey:1985xj}). 
``Constituent gluons" have also been used to characterize mesons that may exhibit 
explicit gluonic degrees of freedom (``hybrid mesons").    The constituent quark models, which have been built 
to explain hadron mass spectroscopy, can reproduce it well. In particular, they historically lead to the
discovery of color charge. 
Of particular interest to this review, such an approach can also account for 
baryon magnetic moments which can be distinguished from the \emph{constituent quark} pointlike ({\it i.e.}, Dirac) magnetic 
moments. Another feature of these models relevant to this review is that the
physical mechanisms that account for hyperfine corrections are also needed to explain  polarized 
PDFs at large-$x_{Bj}$, see Section~\ref{pqcd high-x}.  Hyperfine corrections
can effectively transfer some of  the quark spin contribution to quark OAM~\cite{Myhrer:1988ap}, consistent with 
the need for non-zero quark OAM in order to describe the PDFs within pQCD~\cite{Avakian:2007xa}. 

In non-relativistic \emph{constituent quark} models, the quark OAM is zero and there are no gluons: the nucleon spin 
comes from the quark spins. SU(6) symmetry and requiring that the non-color
part of the proton wavefunction is symmetric yield~\cite{Close:1979bt, Close:1973xw}:
\vspace{-0.3cm}
\begin{eqnarray}
\vspace{-0.3cm}
\label{SU(6) p wavefunction}
\left|p\uparrow\right\rangle = 
\frac{1}{\sqrt{2}}\left|u\uparrow(ud)_{\mathbf s=0,s=0}\right\rangle +
\frac{1}{\sqrt{18}}\left|u\uparrow(ud)_{\mathbf s=1,s=0}\right\rangle - \\ \nonumber
\frac{1}{3}\left(\left|u\downarrow(ud)_{\mathbf s=1,s=1}\right\rangle -
\left|d\uparrow(uu)_{\mathbf s=1,s=0}\right\rangle +
\sqrt{2}\left|d\downarrow(uu)_{\mathbf s=1,s=1}\right\rangle \right),
\end{eqnarray}
where the arrows indicate the projection of the 1/2 spins along the quantization axis, while the subscripts $\mathbf s$ 
and $s$ denote the total and projected spins of the diquark system, respectively. 
The neutron wavefunction is obtained from the proton wavefunction {\it via}  isospin $u \leftrightarrow d$ interchange.
The spectroscopy of the excited states varies between models, depending in detail on the choice of the quark potential. 

As mentioned in Section~\ref{spin crisis}, the disagreement between the EMC experimental results~\cite{Ashman:1987hv},
and the naive $\Delta \Sigma =1$ expectation from the 
simplest \emph{constituent quark} models has led to the ``spin crisis". 
Myhrer, Bass, and Thomas have interpreted the ``spin crisis" in the \emph{constituent quark} model framework
as a pion cloud effect~\cite{Myhrer:2007cf, Bass:2009ed},  which
together with relativistic corrections and one-gluon exchange,  can transfer part of $\Delta \Sigma$ to 
the quark OAM (mostly to $l^u_q$)~\cite{Tsushima:1988xv}. 
Once these corrections have been  applied, the \emph{constituent quark}
picture  -- which has had success in describing other aspects of the strong force --  
also becomes consistent with the spin structure data.   Relativistic effects, one-gluon exchange and the pion cloud 
reduce the naive $\Delta \Sigma=1$ expectation by 35\%, 25\% and 20\%,
respectively.   The quark spin contribution is transferred to quark OAM, 
resulting in $\Delta \Sigma /2 \approx 0.2$ 
 and $l_q \approx +0.3$.
These predictions apply at the low momentum scale where \emph{DGLAP} evolution starts, estimated to be 
$Q_0^2 \approx 0.16$~GeV$^2$~\cite{Thomas:2008ga}, which could be relevant to the \emph{constituent quark} 
degrees of freedom.  Evolving these numbers from $Q_0^2$ 
to the typical DIS scale of 4~GeV$^2$ using Eqs.~(\ref{eq:LO evol. of spin SR})
decreases  the quark OAM to 0 ($l_q^d \approx - l_q^u \approx 0.1$), transferring it to $\Delta G +\mbox{L}_g$. 
Thus, the Myhrer-Bass-Thomas model yields
$\Delta \Sigma/2 \approx 0.18$, $l_q \approx 0$ and  $\Delta G +\mbox{L}_g \approx 0.32$, with
strange and heavier quarks not directly contributing to $J$.
 
This result is not supported by those of Refs.~\cite{Altenbuchinger:2010sz, Wakamatsu:2009gx} which assessed the value of
$L_q$ at low scales by evolving down large scale LGT estimates of the spin sum rule components. 
A cause of the disagreement might be that Refs.~\cite{Altenbuchinger:2010sz, Wakamatsu:2009gx} use LGT
input, {\it i.e.}, with the quark OAM kinematical definition, while it is unclear which definition applies to the quark OAM 
in constituent quark models, such as that used in Refs.~\cite{Thomas:2008ga}. 
Furthermore, the high scale $L_q$ input of Refs.~\cite{Altenbuchinger:2010sz, Wakamatsu:2009gx} stems
from early LGT calculations which do not include disconnected diagrams. Those are now known to contribute 
importantly to the quark OAM, which makes the $L_q$ input of Refs.~\cite{Altenbuchinger:2010sz, Wakamatsu:2009gx}
questionnable. Finally, the scale evolutions are preformed in~\cite{Altenbuchinger:2010sz, Thomas:2008ga,  Wakamatsu:2009gx}
at leading \emph{twist}, which is known to be insufficient for scales below $Q_0\approx 1$ GeV~\cite{Deur:2014qfa, Deur:2016cxb}. 
(We remark that some \emph{higher-twists} are effectively included when a non-perturbative $\alpha_s$  is employed).
The limitation of these evolutions in the very low scale region characterizing bag models (0.1-0.3 GeV$^2$) 
is in particular studied  in Ref.~\cite{Altenbuchinger:2010sz}. 
The authors improved the cloudy bag model calculation of Ref.~\cite{Thomas:2008ga}
by using the gauge-invariant (kinematical) definition of the spin contributions. 
It yields $Q_0^2 \approx 0.2$~GeV$^2$, $\Delta \Sigma/2=0.23\pm 0.01$,
$L_q=0.53\pm 0.09$ and $\Delta G+\mbox{L}_g=-0.26\pm 0.10$.
The importance of the pion cloud  to $J$ has  also been discussed in Refs.~\cite{Nocera:2016zyg, Speth:1996pz}.

\subsubsection{The resonance spectrum of nucleons}

The first nucleon excited state is the P$_{33}$, also called  the $\Delta(1232)~3/2^+$ ($M_{\Delta}$=1232 MeV) 
in which the three \emph{constituent quark} spins are aligned while in an S-wave. 
Thus, the $\Delta(1232)~3/2^+$ has spin $J= 3/2, $  and its isospin is 3/2.
The $\Delta(1232)~3/2^+$ resonance is the
only one clearly identifiable in an inclusive reaction spectrum. 
It has the largest cross-section and thus contributes dominantly to 
 \emph{sum rules} (Section~\ref{sum rules}) and moments of spin structure functions at moderate $Q^2$.
The nucleon-to-$\Delta$ transition is thus, in this SU(6)-based view, a spin (and isospin) flip;  
{\it i.e.}, a magnetic dipole transition quantified by the $M_{1+}$ multipole amplitude.
Experiments have shown that there is also a small electric quadrupole  component 
$E_{1+}$ ($E_{1+}/M_{1+} < 0.01$ at Q$^2=0$)  which violates SU(6) isospin-spin symmetry.  This effect
can be interpreted as the deformation of the $\Delta(1232)~3/2^+$ charge and current 
distributions in comparison to a spherical distribution.
The nomenclature for multipole longitudinal (also called scalar) amplitudes
$S_{l\pm}$, as well as the transverse $E_{l\pm}$ and $M_{l\pm}$ amplitudes
is given in Ref.~\cite{Burkert:2004sk}.  The small $E_{1+}$ and $S_{1+}$
components are predicted by \emph{constituent quark} models improved with a $M_{1 }$ dipole-type one-gluon exchange 
(see Section~\ref{sec: high-x}).

Due to their similar masses and short lifetimes ({\it i.e.,} large widths in excitation energy $W$), the higher mass
resonances overlap, and thus cannot be readily isolated as distinct contributions to inclusive  cross-sections.
Their contributions can be grouped into four regions whose shapes and mean-$W$ vary with
$Q^2$,  due to the different $Q^2$-behavior of the amplitudes of the individual resonances. 
The second resonance region (the first is the $\Delta(1232)~3/2^+$) is located 
at $W\approx1.5$~GeV and contains the N(1440)~$1/2^+$ P$_{11}$ (Roper resonance),  the 
N(1520)~$3/2^-$ D$_{13}$ and the N(1535)~$1/2^-$ S$_{11}$ which usually dominates over
the first two.  The third region, at $W\approx1.7$~GeV,
includes the $\Delta$(1600)~$3/2^+$ P$_{33}$, N(1680)~$5/2^+ $ F$_{15}$, N(1710)~$1/2^-$ P$_{11}$,
N(1720)~$3/2^+$ P$_{13}$, $\Delta$(1620)~$1/2^-$ S$_{31}$, N(1675)~$5/2^-$ D$_{15}$, $\Delta$(1700)~$3/2^-$ D$_{33}$,
and N(1650)~$1/2^-$ S$_{11}$. The fourth region is located
around $W\approx1.9$~GeV and contains the $\Delta$(1905)~$5/2^+$ F$_{35}$,
$\Delta$(1920)~$3/2^+$ P$_{33}$, $\Delta$(1910)~$1/2^+$ P$_{31}$, $\Delta$(1930)~$5/2^+$ D$_{35}$ and
$\Delta$(1950)~$7/2^+$ F$_{37}$.  Other resonances have been identified
beyond $W=2$~GeV~\cite{Olive:2016xmw}, but their structure cannot be distinguished in an
inclusive experiment not only because of the overlap of their widths, but also because of the dominance
of the ``non-resonant background" -- incoherent scattering similar to DIS at higher $Q^2$. Its
presence is necessary to satisfy the unitarity of the $S$ matrix in the resonance region.

The DIS cross-section formulae remain valid in the resonance domain. 
Although the  intepretation of structure functions as PDFs cannot be applied, the DIS cross-sections can nevertheless be related to
overlaps of LFWFs, as shall be discussed below.

\subsubsection{A link between DIS and resonances: hadron-parton duality}
Bloom and Gilman observed~\cite{Bloom:1970xb} that the unpolarized structure function 
$F_2(x_{Bj},Q^2)$ measured in DIS matches $F_2(x_{Bj},Q^2)$ measured in the resonance domain
if the resonance peaks are suitably smoothed and if the 
$Q^2$-dependence of $F_2$ -- due to pQCD radiations and the non-zero nucleon 
mass  -- is corrected for. This correspondence is known as {\it hadron-parton}  duality. 
It implies that Bjorken scaling, corrected for  
\emph{DGLAP} evolution and non-zero mass terms (kinematic \emph{twists}, see Section~\ref{OPE}),  is effectively valid in the
resonance region if the resonant structures can be averaged over. 
This indicates that the effect of the third source of
$Q^2$-dependence, the parton correlations (dynamical \emph{twists}, see Section~\ref{OPE}),
can be neglected.  Thus the resonance region can be described 
in dual languages -- either hadronic or partonic~\cite{Melnitchouk:2005zr}.   The understanding of
hadron-parton duality  for spin structure functions has also progressed and is discussed in Section~\ref{sec:duality}.

\subsection{Elastic and quasi-elastic scatterings\label{elastic scatt}}
When a leptonic scattering reaction occurs at  low energy transfer 
$\nu = {p \cdot q/ M} $ and/or low photon virtuality $Q^2$, 
nucleon excited states cannot form. Coherent elastic scattering occurs, leaving the target in its
ground state. The transferred momentum is shared by
the target's constituents, the target stays intact and its
structure undisrupted. The 4-momentum of the virtual photon  is spent entirely
as target recoil. The energy transferred is $\nu_{el}=Q^2/(2M)$.
For a nuclear target, elastic scattering may occur on the nucleus itself 
or on an individual nucleon.  If the nuclear structure is disrupted, 
the reaction is called quasi-elastic  (not to
be confused with the ``quasi-elastic'' scattering of neutrinos, which
is charge-exchange elastic scattering; {\it i.e.,} involving $W^{+ / -}$ rather that $Z^0$).
%

For elastic scattering, there is no need for 
``polarized form factors'': the unpolarized and polarized parts 
of the cross-section contain the same form factors. This is 
because in elastic scattering, the final hadronic state is known, from current and angular
momentum conservations. Thus, a hadronic current (a vector) can be constructed, which  requires two parameters.
In contrast, in the inclusive inelastic case, such current cannot be constructed since the final state is
by definition undetermined.  Only the hadronic tensor can be constructed, which requires four parameters.
That the same form factors describe both unpolarized and polarized elastic scattering allowes for accurate 
form factor measurements~\cite{Pacetti:2015iqa}, which illustrates  how
spin is used as a complementary tool for exploring nucleon structure.

The elastic reaction is important for doubly-polarized inclusive scattering experiments.
Since the same form factors control the unpolarized and polarized elastic cross-sections,
the elastic asymmetry is calculable from the well-measured unpolarized elastic scattering.
This asymmetry can be used to obtain or check beam and target polarizations.
Likewise, the unpolarized elastic cross-section can be used to set or to verify the 
normalization of the polarized inelastic cross-section. 
Furthermore, some spin sum rules, {\it e.g.}, Burkhardt-Cottingham sum rule (see
Section \ref{BCSR}), include the elastic contribution. Such sum rules are valid 
for nuclei. Therefore, alongside the nucleon, we provide below the formalism 
of doubly-polarized elastic and quasi-elastic scatterings
for the deuteron and $^3$He nuclei, which are commonly used in doubly-polarized inclusive experiments.

\subsubsection{Elastic cross-section \label{unpo cross-section}}

The doubly polarized elastic cross-section is:
\vspace{-0.1cm}
\footnotesize
\begin{equation}
\vspace{-0.1cm}
\frac{d\sigma}{d\Omega}=\frac{\sigma_{Mott} E' Z^2}{E}\bigg[\bigg(\frac{Q^2}{\overrightarrow{q}^2}\bigg)^2R_{L}(Q^2,\nu)+\big(\tan^2(\theta/2)-\frac{1}{2}\frac{Q^2}{\overrightarrow{q}^2}\big)R_T(Q^2,\nu)\pm\Delta(\theta^{*},\phi^{*},E,\theta,Q^2)\bigg] ,
\label{eq:unpocross}
\end{equation}
\normalsize
\noindent 
where $Z$ is the target atomic number and the angles are defined in Fig.~\ref{fig:spinangles}. $R_{L}$ and $R_T$ are 
the longitudinal and transverse response functions associated with the
corresponding polarizations of the virtual photon.   
The cross-section asymmetry $\Delta$, where $\pm$  refers to the beam helicity sign~\cite{Donnelly:1985ry}, is:
\vspace{-0.3cm}
\begin{equation}
\vspace{-0.3cm}
\Delta=-\left(\tan\frac{\theta}{2}\sqrt{\frac{Q^2}{\overrightarrow{q}^2}+\tan^2\frac{\theta}{2}}R_{T'}(Q^2)\cos\theta^{*}-\frac{\sqrt{2}Q^2}{\overrightarrow{q}^2}\tan\frac{\theta}{2}R_{TL'}(Q^2)\sin\theta^{*}\cos\phi^{*}\right)\label{eq:delasy}.
\nonumber
\end{equation}
Cross-sections for the targets used in  nucleon spin structure experiments are given below:

\noindent \textbf{Nucleon case}

The cross-section for scattering on a longitudinally polarized nucleon is: 
\vspace{-0.4cm}
\begin{eqnarray}
\frac{d\sigma}{d\Omega} & = & \sigma_{Mott}\frac{E'}{E}\left(W_2+2W_1\tan^2(\theta/2)\right)\times\label{eq:hadroncross}\\
 &  & \left(1\pm\sqrt{\frac{\tau_r W_1}{(1+\tau_r)W_2-\tau_r W_1}}\frac{\frac{2M}{\nu}+\sqrt{\frac{W_1}{\tau_r\left((1+\tau_r)W_2-\tau_r W_1\right)}}\frac{2\tau_r M}{\nu}+2(1+\tau_r)\tan^2(\theta/2)}{1+\tau_r\frac{W_1}{\tau_r\left((1+\tau_r)W_2-\tau_r W_1\right)}\left(1+2(1+\tau_r)\tan^2(\theta/2)\right)}\right),
 \nonumber 
 \end{eqnarray}
 \noindent with the recoil term $\tau_r\equiv Q^2/(4M^2)$. 
The hadronic current is usually parameterized by the Sachs form factors,
$G_E(Q^2)$ and $G_M(Q^2)$, rather than $W_1$ and $W_2$: 
\vspace{-0.3cm}
\begin{equation}
\vspace{-0.3cm}
W_1(Q^2)=\tau_r G_M(Q^2)^2, \\\
~~~~~W_2(Q^2)=\frac{G_E(Q^2)^2+\tau_r G_M(Q^2)^2}{1+\tau_r}.
\nonumber
\end{equation}
In the nonrelativistic domain the form factors $G_E$ and $G_M$ can be thought of as 
Fourier transforms of the nucleon charge 
and magnetization spatial densities, respectively.  A rigorous interpretation in term of LF
charge densities is given in Refs.~\cite{Miller:2007uy} (nucleon) and 
\cite{Carlson:2008zc} (deuteron, see next section).
The Dirac and Pauli form factors $F_1(Q^2)$
and $F_2(Q^2)$ can also be used (not to be confused with the DIS structure functions in Section~\ref{DIS}): 
\vspace{-0.3cm}
\begin{equation}
\vspace{-0.3cm}
G_E(Q^2)=F_1(Q^2)-\tau_r\kappa_n F_2(Q^2), \\\
~~~~~G_M(Q^2)=F_1(Q^2)+\kappa_n F_2(Q^2),
\nonumber
\end{equation}
where $\kappa_n$ is the nucleon anomalous magnetic moment.
The helicity conserving current matrix element generates $F_1(Q^2)$. 
$F_2(Q^2)$ stems from the helicity-flip matrix element. 

LF  quantization of QCD provides an interpretation of $F_1(Q^2)$ and $F_2(Q^2)$ which can then be modeled 
using the structural forms for arbitrary twist inherent to the LFHQCD formalism~\cite{Sufian:2016hwn}, 
see Section~\ref{sec:LFHQCD}. In LF QCD, form factors are obtained from the Drell-Yan-West 
formula~\cite{Drell:1969km, West:1970av} as the overlap of the hadronic LFWFs solutions 
of LF Hamiltonian $P^-$, Eq.~(\ref{LF Hamiltonian})~\cite{Brodsky:1980zm}. 
In particular, $F_2(Q^2)$ stems from the overlap 
of $L = 0$  and $L = 1$ LFWFs.  For a ground state system, the \emph{leading-twist} of a reaction, 
that is, its power behavior in $Q^2$ (or in the LF impact parameter $\zeta$, see Section~\ref{OPE}), 
reflects the \emph{leading-twist}  $\tau $ of the target wavefunction, 
which is equal to the number of constituents in the LF valence Fock state with zero 
internal orbital angular momentum. 
This result is intuitively clear,  since in order to keep the target intact after elastic scattering, 
a number $\tau-1$ of gluons of virtuality $\propto Q^2$ must be exchanged between the $\tau$ constituents. 
For example, at high-$Q^2$, all nucleon components are resolved and the \emph{twist}  is $\tau=3$.  
Higher Fock states including additional $q \overline{q}$, $q \overline{q} q \overline{q}$,\ldots  components 
generated by gluons are responsible for the  \emph{higher-twists} corrections.
These constraints are inherent to LFHQCD which can be used to model the LFWFs and thus obtain  
predictions for the form factors.
Alternatively, one can parameterize the general form expected from the 
\emph{twist} analysis in terms of weights reflecting the ratio of the higher Fock state 
probabilities  with respect to the leading Fock state wavefunction. 
These weights provide the probabilities of finding the nucleon in a higher Fock state, computed from the square of the
higher Fock state LFWFs. Two parameters suffice to describe the world data for
the four spacelike nucleon form factors~\cite{Sufian:2016hwn}.

\noindent \textbf{Deuteron case}

The deuteron is a spin-1 nucleus. Three elastic form factors are
necessary to describe doubly polarized elastic cross-sections:
\vspace{-0.3cm}
\begin{equation}
\vspace{-0.3cm}
\frac{d\sigma}{d\Omega}  =  \sigma_{M}\frac{E'}{E}\big(A(Q^2)+B(Q^2)\tan^2(\theta/2)\big)\big(1+A^V + A^T \big),
\label{deuteron pol XS}
\end{equation}
where $A^V$ and $A^T$, the asymmetries stemming respectively from the vector 
and tensor polarizations of the deuteron, are
\vspace{-0.3cm}
\begin{equation}
\vspace{-0.3cm}
A^V=\frac{3P_b P_z}{\sqrt2} \bigg( \frac{1}{\sqrt2}\cos \theta^* T_{10} - \sin \theta^*T_{11}\bigg),
\nonumber
\end{equation}
where $P_b$ is the beam polarization and $P_z$ the deuteron vector polarization, $P_z = (n_+ - n_-)/n_{tot}$. The
$n_i$ are the populations for the spin values $i$, and $n_{tot}=n_+ + n_- + n_0$,
\vspace{-0.3cm}
\begin{equation}
\vspace{-0.3cm}
A^T=\frac{P_{zz}}{\sqrt{2}}\bigg(\frac{3\cos^2 \theta^{*} -1}{2}T_{20}-\sqrt{\frac{3}{2}}\sin(2\theta^{*})\cos \phi^{*} T_{21}+\sqrt{\frac{3}{2}}\sin^2 \theta^{*} \cos(2\phi^{*})T_{22}\bigg),
\nonumber
\end{equation}
with the deuteron tensor polarization $P_{zz}=(n_+ + n_- - 2n_0)/n_{tot}$. 

The seven factors in Eq.~(\ref{deuteron pol XS}),  $A$, $B$, $T_{10}$, $T_{11}$, $T_{20}$, $T_{21}$ 
and $T_{22}$, are combinations of  
three form factors (monopole $G_{C}$, quadrupole $G_Q$ and magnetic dipole $G_M$):
\vspace{-0.3cm}
\begin{eqnarray}
\vspace{-0.3cm}
A & = & G_{C}^2+\frac{8}{9}\tau_r^2G_{Q}^2+\frac{2}{3}\tau_r G_M^2, \\\
B & = & \frac{4}{3}\tau_r(1+\tau_r)G_M^2,
\nonumber
\\
T_{10} & = & -\sqrt{\frac{2}{3}}\tau_r(1+\tau_r)\tan(\theta/2)\sqrt{\frac{1}{1+\tau_r}+\tan^2(\theta/2)G_M^2},
\nonumber
\\
T_{11} & = & \frac{2}{3}\sqrt{\tau_r(1+\tau_r)}\tan(\theta/2)G_M\big(G_C+\frac{\tau_r}{3}G_Q\big),
\nonumber
\\
T_{20} & = & -\frac{1}{\sqrt{2}}\left[\frac{8}{3}\tau_r G_{C}G_{Q}+\frac{8}{9}\tau_r^2G_{Q}^2+\frac{1}{3}\tau_r\left[1+2(1+\tau_r)\tan^2(\theta/2)\right]G_M^2\right],
\nonumber
\\
T_{21} & = & \frac{2}{\sqrt{3}\left[A(Q^2)+B(Q^2)\tan^2(\theta/2)\right]\cos(\theta/2)}\tau_r\left[\tau_r+\tau_r^2\sin^2(\theta/2)G_MG_{C}\right],
\nonumber
\\
\vspace{-0.3cm}
T_{22} & = & \frac{1}{\sqrt{3}\left[A(Q^2)+B(Q^2)\tan^2(\theta/2)\right]}\tau_r G_M^2.
\nonumber
\end{eqnarray}

$P_{zz}$ produces additional quantities in other reactions too:
in DIS, it yields the $b_1(x_{Bj},Q^2)$ and $b_2(x_{Bj},Q^2)$ spin structure functions~\cite{Hoodbhoy:1988am}. The first one,
\vspace{-0.3cm}
\begin{equation}
\vspace{-0.3cm}
b_1(x_{Bj},Q^2)= \sum_i \frac{e_i^2}{2}  \big[2q^0_{\uparrow}(x_{Bj},Q^2)-\big(q^1_{\downarrow}(x_{Bj},Q^2)-q^{-1}_{\downarrow}(x_{Bj},Q^2) \big) \big],
\label{eq:b1 SF}
\end{equation}
has been predicted to be small but measured to be significant by the HERMES experiment~\cite{Airapetian:2005cb}.
For the PDFs $q^{-1,0,1}_{\uparrow,\downarrow}$, the superscript $0$ 
or $\pm1$ indicates the deuteron helicity and the arrow
the quark polarization direction, all of them referring to the beam axis.   

The six quarks of the deuteron eigenstate can be projected onto five different color-singlet Fock  states, 
only one of which corresponds to a proton-neutron bound state.
The other five ``hidden color"  Fock states lead to new QCD phenomena at high $Q^2$~\cite{Brodsky:1976mn}.

\noindent \textbf{Helium 3 case}

The doubly polarized cross-section for elastic lepton-$^3$He scattering is
\vspace{-0.3cm}
\begin{eqnarray}
\vspace{-0.3cm}
\frac{d\sigma}{d\Omega}=\sigma_{Mott}\frac{E'}{E}\left(\frac{G_E^2+\tau_r G_M^2}{1+\tau_r}+2\tau_r G_M^2\tan^2(\theta_{}/2)\right) \bigg(1  \pm   \nonumber \\ 
 \frac{1}{\left(\frac{Q^2}{2M\nu+\nu^2}\right)^2(1+\tau_r)G_E^2+\big(\frac{Q^2}{2M\nu+\nu^2}+2\tan^2(\theta/2)\big)\tau_r G_M^2} \times \nonumber \\
\bigg[2\tau_r G_M^2\cos \theta^{*} \tan(\theta/2)\sqrt{\tan^2(\theta/2)+\frac{Q^2}{2M\nu+\nu^2}}+ \nonumber \\
2\sqrt{2\tau_r(1+\tau_r)}G_MG_E\sin \theta^{*} \cos \varphi^{*} \frac{Q^2}{\sqrt{2}\left(2M\nu+\nu^2\right)}\tan(\theta/2)\bigg]\bigg),
\nonumber
\end{eqnarray}
\normalsize
\noindent where the form factors are normalized to the $^3$He electric charge. 
The magnetic and Coulomb form  factors $F_{m}$ and $F_{c}$ are sometimes 
used~\cite{Amroun:1994qj}. They are related to the response functions of a nucleus ($A$, $Z$) by
$F_c = Z G_E$ and $F_m = \mu_A G_M$ where $\mu_A$ is the
nucleus magnetic moment. 

\subsubsection{Quasi-elastic scattering \label{qel}}

If the target is a composite nucleus and the transferred energy $\nu$ is greater than the 
nuclear binding energy, but still small enough to not resolve
the quarks or excite a nucleon, the scattering loses nuclear coherence. 
For example, the lepton may scatter elastically on one of the
nucleons, and the target nucleus breaks. This is quasi-elastic scattering. Its
threshold with respect to the elastic peak equals the nuclear binding 
energy (2.224 MeV for the deuteron, 5.49 MeV for the $^3$He
two-body breakup and 7.72 MeV for its three-body breakup). 
Unlike elastic scattering, the nucleons are not
at rest in the laboratory frame since they are 
restricted to the nuclear volume.
This Fermi motion causes a Doppler-type broadening of the quasi-elastic peak
around the breakup energy plus $Q^2/(2M)$, the energy transfer in elastic scattering off
a free nucleon. The cross-section shape is nearly Gaussian with 
a width of about 115 MeV (deuteron) or 136 MeV ($^3$He)~\cite{Martinez-Consentino:2017ryk}. 
This model where the nucleon is assumed to be virtually free
(Fermi gas model) provides a qualitative description of the the cross-section,  but  it does not predict the
transverse and longitudinal  components of the cross-section, nor the 
distortions of its Gaussian shape. To
account for this, the approximation of free nucleons is abandoned and
a model for the nucleon-nucleon interaction is introduced.  The
simplest implementation is \emph{via} the ``Plane Wave Impulse Approximation"
(PWIA), where the initial and final particles (the lepton and nucleons) 
are described by plane waves in a mean field. In this approach, all nucleons are quasi-free and therefore 
on their mass-shell, including the nucleon absorbing the virtual photon whose momentum 
is not changed by the mean field.  The other nucleons are passive spectators of the reaction.
The nucleon momentum distribution is given by the spectral function $P(k,E)$.
Thus, the PWIA hypothesis enables the nuclear  tensor to be expressed from the hadronic ones. 
%
%
The PWIA model can  be improved by accounting for 
1) Coulomb corrections on the lepton lines which distort the lepton plane waves. 
This corrects for the long distance electromagnetic interactions  between the lepton
and the nucleus whose interaction is no longer approximated by a single hard photon exchange; 
2) Final state interactions between the nucleon absorbing the hard photon 
and the nuclear debris; 
3) Exchange of mesons between the 
nucleons (meson exchange currents) which is dominated by one pion exchange; and 
4) Intermediate excited nucleon configurations such as the Delta-isobar contribution. 

\subsection{Summary}

We have described the phenomenology for spin-dependent inclusive lepton 
scattering off a nucleus. These reactions, by probing the QCD-ruled nucleon structure, help
to understand QCD's nonperturbative aspects. The spin degrees of freedom allow for additional observables 
which can address more complicated effects.
To interpret the observables and understand what they tell us about QCD, 
a more fundamental theoretical framework is needed. 
We now outline the most important theoretical approaches connected to perturbative and
nonperturbative spin structure studies.

\section{Computation methods \label{Computation methods}}
The strong non-linearity inherent  to the QCD Lagrangian makes
traditional perturbation theory inadequate to study the nucleon structure. 
In this Section, four important approaches are presented. 
Other fruitful approaches to strong-QCD exist, such as solving the Dyson-Schwinger equations,
and the functional renormalization group method or the stochastic quantization method. 
Since they have been less used  in the nucleon spin structure context, they  will not be discussed here.
An overview is given in~\cite{Deur:2016tte}, and an example of  Dyson-Schwinger equations
calculation predicting nucleon spin observables can be found in~\cite{Roberts:2011wy}.
Many other models also exist, some will be briefly described when we compare their
predictions to experimental results. 

The approaches discussed here are the Operator
Product Expansion (OPE), Lattice Gauge Theory (LGT), Chiral Perturbation Theory ($\chi$PT)
and LF Holographic QCD (LFHQCD). 
They cover different QCD domains and are thus complementary:

$\bullet$ The OPE covers the pQCD domain (Section~\ref{DIS}), including  nonperturbative 
\emph{twist} corrections  to the parton model plus the \emph{DGLAP} framework. 
The OPE breaks down at low $Q^2$ due to 1) the
magnitude of the nonperturbative corrections; 2) the precision to which
$\alpha_s(Q^2)$ is known; and 3) the poor convergence of the $1/Q^n$ series. 
The technique is thus typically valid for $Q^2\gtrsim1$~GeV$^2$.

$\bullet$ LGT covers both the nonperturbative and perturbative regimes. 
It is limited at high $Q^2$ by the lattice mesh size $a$ (typically 1/$a \sim$ 2 GeV) and at 
low $Q^2$ by 1) the total lattice size; 2) the large value of the pion mass used in LGT simulations (up to 0.5 GeV);
and 3) the difficulty of treating nonlocal operators.  

$\bullet$ $\chi$PT, unlike OPE and LGT, uses effective degrees of freedom.
However, calculations are limited to small $Q^2$ (a few tenths of GeV$^2$) 
because the momenta involved must be smaller than the pion mass (0.14 GeV).

\noindent The forward Compton scattering amplitude is calculable with the above
techniques. It can also be parameterized at any
$Q^2$ using  \emph{sum rules}, see Section~\ref{sum rules}.
This is important for nucleon structure studies since
it allows to connect the different QCD regimes.

$\bullet$ LFHQCD is typically restricted to $Q^2 \lesssim1$~GeV$^2$, a domain characterized by the hadronic mass scale
$\kappa$ and of higher reach compared to  $\chi$PT. The restriction comes from
ignoring short-distance effects and working in the strong-coupling regime.
However, in cases  involving soft observables, LFHQCD 
may extend to quite large $Q^2$~\cite{Sufian:2016hwn}. For example, 
it describes well the nucleon form factors up to $Q^2 \sim 30$ GeV$^2$~\cite{Brodsky:2014yha}. 
 
 \noindent Although forward Compton scattering amplitudes in the nonperturbative regime have not yet been calculated 
 with the LFHQCD approach (they are available in the perturbative regime, see~\cite{Brodsky:2000xy}),
 LFHQCD plays a important role in connecting the low and high momentum regimes of QCD: 
 the QCD effective charge~\cite{Grunberg:1980ja} can be computed in LFHQCD 
 and then be used in pQCD spin  \emph{sum rules}
 to extend it to the strong QCD domain, thereby linking the hadronic and partonic descriptions of QCD 
 (see Section~\ref{sec:perspectives}).
    
 \subsection{The Operator Product Expansion \label{OPE}}
The OPE technique illuminates the features of matrix elements of the product of local operators. It is used to compute the
$Q^2$-dependence of structure functions and other quantities in the
DIS domain, as well as to isolate nonperturbative contributions
that arise at small $Q^2$.   It also allows the derivations of 
relations constraining physical observables, such as the Callan-Gross and Wandzura-Wilczek relations, 
Eqs.~(\ref{eq:Callan-gross}) and (\ref{eq:g2ww}), respectively, as well as \emph{sum rules}
together with their $Q^2$-dependence.  Due to the parity symmetry of the structure functions under 
crossing symmetry, odd-moment sum rules 
are derived from the OPE for $g_1$ and $g_2$, whereas even-moment sum rules are predicted
for $F_1$ and $F_2$~\cite{Manohar:1992tz}.

The OPE was developed as an alternative to the Lagrangian approach of quantum field theory 
in order to carry out nonperturbative calculations~\cite{Wilson:1969zs}.
The OPE  separates the perturbative contributions of a product of local 
operators from its nonperturbative contributions by
focussing on distances ({\it i.e.}, inverse momentum scales) that are much smaller than the confinement scale.  
Although DIS is LC dominated, 
not short-distance dominated (Section~\ref{LC dominance and LF quantization}),  the LC and 
short-distance criteria are effectively equivalent for DIS in the IMF.  However, there are
instances of LC dominated reactions;  { \it e.g.}, inclusive hadron production in $e^+ e^-$ annihilation, 
for which LC dominance and the short-distance limit are not equivalent~\cite{Jaffe:1996zw}. 
In those cases, the OPE does not apply.

\noindent In the small-distance limit, the product of two local operators can be expanded as:
\vspace{-0.3cm}
\begin{equation}
\vspace{-0.3cm}
\lim_{d\to0}\sigma_a(d)\sigma_{b}(0)=\lim_{d\to0} \sum_{k}C_{abk}(d)\sigma_{k}(0)
\label{eq:OPE}.
\end{equation}
The Wilson coefficients $C_{abk}$ are singular functions containing
perturbative information and are therefore perturbatively calculable. The 
$\sigma_{k}$ are regular operators containing the nonperturbative contributions. In DIS 
this formalism is used to relate the 
product of currents -- such as those needed to calculate
Compton scattering amplitudes -- to a basis of local operators. Such a basis
is given,{\it e.g.,} in Ref.~\cite{Manohar:1992tz}.  
An operator $\sigma_{k}$ contributes to the cross-section by a
factor of $x_{Bj}^{-n}(M/Q)^{D-2-n}$ where $n$ is the spin and $D$ is the
energy dimension of the operator. This defines the \emph{twist}
$\tau \equiv D-n$.  Eq.~(\ref{eq:OPE})  provides a $Q^{2-\tau}$ power series in which
the lowest \emph{twist} $C_{abk}$ functions are the most singular and thus are the most dominant at short
distances (large $Q$). Contrary to what Eq.~(\ref{eq:OPE}) might suggest, the $Q^2$-dependence
of a \emph{twist} term coefficient  ({\it i.e.}, from pQCD radiative corrections) comes mainly 
from the renormalization of the operator $\sigma_{k}$ rather than from the Wilson coefficient $C_{abk}$.

The twist of an operator has a simple origin in the \emph{LF-quantization} formalism: it measures the
excursion out of the LC. That is, it is related to the transverse 
vector $x_\bot$, or equivalently to the invariant impact parameter $\zeta = x_\bot \sqrt{x_{Bj}(1-x_{Bj})}$. 
 The \emph{higher-twist} operators correspond to the number of ``bad" spinor  components
(see Section~\ref{LC dominance and LF quantization}) that enters 
the expression of distribution functions and gives the $\zeta^\tau$ power behavior of the LFWFs.
At high-$Q^2$, \emph{twist} $\tau = 2$ dominates: it is at this order that 
the parton model, with its \emph{DGLAP} corrections, is applicable.

When $Q^2$ becomes small (typically a few GeV$^2$) 
the \emph{higher-twist} operators must be accounted for. These nonperturbative corrections
are of two kinds: 


\noindent $\bullet$ {\bf Dynamical twist corrections.} They are typically due to amplitudes involving \emph{hard} gluon exchange
between the struck quark and the rest of the nucleon, effectively a nonperturbative object. 
Since these \emph{twists} characterize the nucleon structure, they are
relevant to this review. Dynamical \emph{twist} contributions reflect the fact that 
the effects of the binding and confinement of the quarks become 
apparent as $Q^2$ decreases. Ultimately, quarks react coherently when one of them is struck by the virtual photon.
The 4-momentum transfer is effectively distributed among the quarks
by the \emph{hard} gluons whose propagators and couplings generate $1/Q$  \emph{ power corrections}. 
This is also the origin of the QCD \emph{counting rules}~\cite{Brodsky:1973kr}; see Section~\ref{unpo cross-section}. 

\noindent $\bullet$ {\bf Kinematical finite-mass corrections.}
The existence of this additional correction to scale invariance can be understood by 
recalling the argument leading to the invariance: At $Q^2\to\infty$, masses 
are negligible compared to $Q$ and no specific distance scale exists since quarks are pointlike. 
At $Q$ values of a few GeV, however, $M/Q$ is no longer negligible, a scale appears, 
and the consequent scaling corrections must be functions of $M/Q$. Formally, these corrections arise
from the requirement that the local operators $\sigma_{k}$ are traceless~\cite{Jaffe:1996zw}.
These kinematical \emph{higher-twists} are systematically calculable~\cite{Blumlein:1998nv}. 

The Wilson coefficients are calculable perturbatively.  
For an observable $A$ expressed as a power series           \label{twist-mix}
$A=\sum_\tau\frac{\mu_\tau}{Q^{\tau-2}}$, the parameters $\mu_\tau$
are themselves sums of kinematical \emph{twists} $\tau' \leq \tau$, each of them being 
a perturbative series in $\alpha_{s}$ due to pQCD radiative corrections.  
Since $\alpha_{s}$ is itself a series of the QCD  \emph{$\beta$-function}~\cite{Deur:2016tte}, the 
approximant of $A$ is a four-fold sum. 

The nonperturbative nature of \emph{twists} implies that they can only be calculated using 
models or nonperturbative approaches such as Lattice Gauge Theory, 
LFHQCD or Sum Rule techniques. They are also obtainable
from experimental data (see Section~\ref{sub:HT Extraction}).
The construction and evaluation of \emph{higher-twist} contributions using LFWFs, 
in particular for the twist~3 $g_2$,  are given in Ref.~\cite{Braun:2011aw}.

\subsection{Lattice gauge theory \label{LGT}}

LGT employs the path integral formalism~\cite{Dirac:1933xn}. It
provides the evolution probability from an initial state $\left|x_i \right \rangle $
to a final state $\left|x_f \right \rangle$ by summing over all
spacetime trajectories linking $x_i$ to $x_f$. In this sum, a path is weighted according to 
its action $S$.  For instance, the propagator of a one-dimensional system is
$\left\langle x_f\right |e^{-iHt}\left|x_i\right\rangle =\int e^{-iS\left[x(t)\right]/\hbar} Dx(t)$
where $\int Dx$ sums over all possible trajectories with $x(t_f)=x_f$ and $x(t_i)=x_i$. 
Here $\hbar$ is explicitly shown so that the relation between path integrals and the 
principle of least action is manifest;  the classical path ($\hbar\to 0$)
corresponds to the smallest  $S$ value. The fact that  $\hbar\neq 0$ allows for deviations
from the classical path due to quantum effects.

Path integrals are difficult to evaluate analytically, or even numerically, because for a 4-dimension space,
an $n$-dimension integration is required, where $n = 4 \times$(number of possible paths). 
The ensemble of possible paths being infinite, it must be restricted to a representative sample on which the integration
can be done. 
The standard numerical integration method for path integrals is the Monte Carlo technique in Euclidean 
space:  a Wick rotation
$it\rightarrow t$~\cite{Wick:1954eu} provides a weighting factor  $e^{-S_E}$, which makes the integration
tractable, contrary to the oscillating factor $e^{-iS}$ which appears in Minkowski space.
Here, $S_E$ is the Euclidean action. Such an approach allows the computation of correlation functions 
$\left\langle A_1\ldots A_n\right\rangle = \int \mbox{ }A_1\ldots A_ne^{-S_E}Dx/\int  e^{-S_E} Dx$,
where $A_i$ is the gauge field value at $x_i$. 
In particular, the two-point correlation function at $\left\langle x_1x_2\right\rangle $
provides the boson propagator.  No analytical method is known  to compute $\left\langle A_1\ldots A_n\right\rangle$ 
when $S_E$ involves interacting fields, except when the interactions are weak. In that case, the 
integral can be evaluated analytically by expanding the exponential involving the interaction term, effectively a
perturbative calculation.  If the interactions are too strong, the integration must be performed numerically. 
In LGT, the space is discretized as a lattice of sites, and paths linking the sites are generated. 
In the numerical integration program, the path generation probability follows 
its $e^{-S_E}$ weight, with $S_E$ calculated for that specific path. 
This is done using the Metropolis sampling method~\cite{metropolis}. 
The computational time is reduced by using the previous path to produce the next one.  
A path of action  $S_1$ is randomly varied to a new path of
action $S_2$.  If $S_2<S_1$ the new $S_2$ path is added in the sample. Otherwise, it is added
or rejected with probability $S_2-S_1$. However, 
intermediate paths must be generated to provide a path sufficiently decorrelated
from the previously used path. 
Correlation functions are then obtained by summing the integrand over all paths.  
The paths are generated with probability $e^{-S_E}$, corresponding to the weighted
sum $\sum_{path} x_1 \ldots x_n e^{-S_E} \approx \int ~x_1 \ldots x_n e^{-S_E} Dx$.
The statistical precision of the procedure is characterized by the square root of the number of generated paths.  

Gauge invariance in lattice gauge theory is enforced by the introduction  
of \emph{gauge links} between the lattice sites~\cite{Wilson:1974sk}.
The \emph{link variable} is $U_{\overrightarrow{\mu}}=\mbox{exp}(-i\int_{x}^{x+a\overrightarrow{\mu}}gA\mbox{ }dy)$, 
where $\overrightarrow{\mu}$ is an elementary vector of the Euclidean
space, $x$ is a lattice site, $a$ is the lattice spacing and $g$ the bare coupling.   
The link $U_{\overrightarrow{\mu}}$ is explicitly
gauge-invariant and is used to construct closed paths (``\emph{Wilson loop}") 
$U_1\ldots U_n$~\cite{Wilson:1974sk}. 
In the continuum limit ($a\to 0$), the simplest
loop, a square of side $a$, dominates.  However for discretized space, $a\neq 0$,  
corrections from larger loops must be included. 
High momenta are eliminated for $p\lesssim1/a$ by the discretization process, but if $a$ can be made
small enough, LGT results can be matched to pQCD results. The domain where
LGT and pQCD are both valid provides the renormalization procedure for LGT.

The case of \emph{pure gauge} field is described above.
It is not simple to include non-static quarks due to their fermionic nature.  
The introduction of quark fields leads to the ``fermion doubling problem" which 
multiplies the number of fermionic degrees of freedom and creates spurious particles. Several methods exist
to avoid this problem, {\it e.g.},  the Ginsparg-Wilson~\cite{Ginsparg:1981bj} method, which breaks chiral 
symmetry, or the ``staggered fermions" method, which preserves chiral symmetry by using nonlocal 
operators~\cite{Kogut:1974ag}. These fixes significantly  increase the computation time.  When the 
quarks  are included, the action becomes $S_E=S_A-\mbox{ln}\left(\mbox{Det}(K)\right)$
with $S_A$ the pure field action and $K$ is related to the
Dirac equation operator.  Simplifying the computation by ignoring dynamical quarks  corresponds to $\mbox{Det}(K)=1$
(\emph{quenched approximation}). In particular, it eliminates 
the effects of quark anti-quark pair creation from the \emph{instant time} vacuum.

LGT has become the leading method for nonperturbative studies, but it still has serious limitations~\cite{Lin:2017snn}:

\noindent 1) ``Critical slowing down"  limits the statistical precision. 
It stems from the need for  $a$ to be smaller than the studied phenomena's
characteristic scales, such that errors from discretization are small. 
The relevant scale is the correlation length $L_{c}$ defined by
$\left\langle x_1x_2\right\rangle \sim e^{-x/L_{c}}$. 
$L_{c}$ is typically small, except near critical points. Thus, calculations
must be done near such points, but long $L_{c}$ makes the technique used to
generate decorrelated paths inefficient.  For QCD the statistical precision
is characterized by $\left(\frac{L_{R}}{a}\right)^{4}\left(\frac{1}{a}\frac{1}{m_{\pi}^2a}\right)$,
where $m_\pi$ is the pion mass and $L_{R}$ is the lattice size~\cite{Lepage:1998dt}.  The first factor comes from the
number of sites and the second factor from the critical slow down.

\noindent 2) Another limitation is the extrapolation to the physical pion mass. LGT calculations are
often performed where  $m_\pi$ is greater 
than its physical value in order to reduce the critical slow down,  but a new uncertainty arises from
the extrapolation of the LGT results to the physical $m_\pi$ value. This uncertainty can be 
minimized by using $\chi$PT Theory~\cite{Bernard:2006gx} to guide the extrapolation.  Some 
LGT calculations can currently be performed at the physical $m_{\pi}$, 
although this possibility depends on the observable.
A recent calculation of the quark and gluon contributions to the proton spin, 
at the physical $m_{\pi}$, is reported in~\cite{Alexandrou:2016tuo}.

\noindent 3) Finite lattice-size systematic uncertainties arise from having  $a$
small enough so that high momenta reach the pQCD domain, but with the 
number of sites sufficiently small for practical calculations. 
This constrains the total lattice
size which must remain large enough to contain the physical system and minimize boundary effects.

\noindent 4) Local operators are convenient for LGT 
calculations  since the selection  or rejection of a given path entails calculating
the difference between the two actions, $S_2-S_1$.  For local actions, $S_2-S_1$ involves 
only one site and its neighbors (since $S$ contains derivatives). In four dimensions this implies only 9
operations whereas a nonlocal action necessitates calculations at all sites. The quark OAM
in the Ji expansion of Eq.~(\ref{eq:spin SR}) involves local operators 
and is thus suitable for lattice calculations. In contrast,
calculations of nonlocal operators, such as those required to compute structure 
functions, are impractical. Furthermore, quantities
such as PDFs are time-dependent in the \emph{instant form} front, and thus cannot be computed directly since
the lattice time is the Euclidean time $ix^0$. (They are, however, pure spatial correlation functions, {\it i.e.}, time-independent, when using the LF form.)
As discussed below, structure functions can still be calculated in LGT by computing their moments, or 
by using a matching procedure that interpolates the high-momentum LGT calculations and LFQCD distributions.

\subsubsection{Calculations of structure functions} 
An example of a non-local structure function is $g_3$,
Eq.~(\ref{eq:g3}). It depends on
the quark field $\psi$ evaluated at the 0 and $\lambda n$  loci.   As discussed, the OPE provides 
a local operator basis. Calculable quantities involve currents such as the 
quark axial current  $\overline{\psi}\gamma_{\mu}\gamma_{5}\psi$\label{axial current}. These currents
correspond to moments of structure functions. In order to obtain those, e.g. $g_1$,
the moments $\Gamma_1^n\equiv\int x^{n-1}g_1dx$
can be calculated and \emph{Mellin-transformed} from moment-space
to $x_{Bj}$-space.  
However, the larger the value of $n$, the higher the degree of the derivatives in the moments  (see e.g.  
Eqs.~(\ref{eq:a2}) and (\ref{eq:a2op})), which increases their non-locality.
Thus, in practice, only moments up to $n=3$ have been calculated in LGT,
which is insufficient to  accurately obtain structure functions (see {\it e.g.},
Refs.~\cite{Gockeler:1995wg, Gockeler:2000ja, Negele:2002vs, Hagler:2003jd} for 
calculations of $\Gamma_{1,2}^n$ and discussions). 
The  \emph{higher-twist} terms discussed in Section~\ref{OPE} have the same problem, with an
additional one coming from the twist mixing discussed on page~\pageref{twist-mix}.
The mixing brings additional $1/a^2$ terms which diverge when $a\to0$. 
This problem can be alleviated in particular cases by using  \emph{sum rules} which relate a moment of a structure function, 
whatever its twist content, to a quantity calculable on the lattice.

\subsubsection{Direct calculation of hadronic PDFs: 
Matching  LFQCD to LGT \label{Ji's LGT method}}
A method to avoid LGT's non-locality difficulty and compute directly $x$-dependencies of parton distributions
has recently been proposed by X. Ji ~\cite{Ji:2013dva}. 
A direct application of LGT in the IMF is impractical because the $P \rightarrow \infty$ limit  using ordinary time implies that  $a \to 0$.
Since LFQCD is boost invariant (see Section~\ref{LC dominance and LF quantization}) calculating
LC observables using \emph{LF quantization} would fix this problem. However, direct LC calculations are not
possible on the lattice since it is based on Euclidean -- rather than real -- instant time and because the LC gauge
$A^+=0$ cannot be implemented on the lattice.

To avoid these problems, an operator $O(P,a)$ related to the
desired nonperturbative PDF is introduced and computed as usual using LGT; it is  then evaluated at  a large 
3-momentum oriented, {\it e.g.}, toward the $x^3$ direction. The momentum-dependent 
result (in the ``instant front form", except that the time is Euclidean: $ix^0$) is called a quasi-distribution, since it is not
the usual PDF as defined on the LC or IMF.  In particular, the range of  $x_{Bj}$  is not
constrained by $0<x_{Bj}<1$.   The quasi-distribution  computed on the lattice is then related to its LC counterpart 
$o(\mu)$ through a matching condition $O(P,a) = Z(\mu/P) o(\mu) + \sum_{2n} C_n/P^n$, where the sum 
represents higher-order power-law contributions. This matching is possible since the operators $O(P,a)$ and 
$o(\mu)$ encompass the same nonperturbative physics.  The matching coefficient $Z(\mu/P)$ can be computed 
perturbatively~\cite{Ji:2014lra, Xiong:2013bka}.  It contains the effects arising from:
1) the particular gauge choice made in the LGT calculation, although it cannot be the LC gauge $A^+=0$; 
and 
2) choosing a different frame and quantization time when computing quantities using \emph{LF quantization} and 
Euclidean \emph{instant time} quantization in the IMF. 

A special lattice with finer spacing $a$ along $ix^0$ and $x^3$ is needed in order to compensate for the Lorentz 
contraction at large $P^3$. Each of the two transverse directions requires discretization enhanced by a factor $\gamma$ 
(the Lorentz factor of the boost), which becomes large for small-$x_{Bj}$ physics. 
The computed PDFs, {\it i.e.}, the leading \emph{twist} structure functions, can be calculated for high and moderate $x_{Bj}$, as well
as the kinematical and dynamical  \emph{higher-twist} contributions.   
How to compute $\Delta G$ and $L_q$ 
with this method is discussed in Refs.~\cite{Ji:2013fga, Hatta:2013gta, Zhao:2015kca}, 
and Ref.~\cite{Lin:2017snn} reviews the method and prospects. Improvements of Ji's method
have been proposed, such as {\it e.g.}, the use of pseudo-distributions~\cite{Orginos:2017kos} 
instead of quasi-distributions.

The quark OAM definition using either the Jaffe-Manohar or Ji decomposition,
see Section~\ref{SSR components}, corresponds to different choices of the \emph{gauge 
links}~\cite{Ji:2012sj, Rajan:2016tlg, Hatta:2011ku, Burkardt:2012sd,  Engelhardt:2017miy}.
Results of calculations related to nucleon spin structure are given in 
Refs.~\cite{Gamberg:2014zwa, Lin:2014zya, Alexandrou:2015rja, Chen:2016utp}.
In particular, Ji's method was applied recently to computing $\Delta G$~\cite{Yang:2016plb}. 
Although the validity of the matching
obtained in this first computation is not certain, these efforts represent
an important new development in the nucleon spin structure studies.
More generally, the PDFs, GPDs, TMDs and Wigner distributions are in principle calculable 
with the innovative approaches  described here, which are designed to circumvent the inherent 
difficulties in the lattice computation of parton distributions.

\subsection{Chiral perturbation theory \label{sub:Chipt}}

$\chi$PT is an effective low-energy field theory consistent with the chiral symmetry of QCD, in which
the quark masses, the pion mass and the particle momenta can be taken small compared to
the nucleon mass.   Since $M_n \approx 1$ GeV, $\chi$PT is  typically restricted to the domain
$Q^2 \lesssim 0.1$~GeV$^2$. The chiral approach is valuable for nucleon spin studies 
since it allows the extension of photoproduction spin  \emph{sum rules} to non-zero $Q^2$, such as
the Gerasimov-Drell-Hearn sum rule~\cite{Gerasimov:1965et} as well as polarization
sum rules~\cite{Drechsel:2002ar, Lensky:2017dlc},
as first done in Ref.~\cite{Bernard:1992nz}.  
Several chiral-based calculations using different approximations are available~\cite{Bernard:2002bs}-\cite{Kao:2002cp}.
For the most recent applications, see Refs.~\cite{Bernard:2012hb}-\cite{Lensky:2016nui}. 

\subsubsection{Chiral symmetry in QCD \label{chiral_conformal_sym}}
The Lagrangian for a free spin 1/2 particle is $\mathcal{L}$= $\overline{\psi}(i\gamma_{\mu}\partial^{\mu}-m)\psi$.
The left-hand Dirac spinor is defined as $P_{l}\psi=\psi_{l}$, with $P_{l}=(1-\gamma_{5})/2$ the 
left-hand helicity state projection operator. Likewise, $\psi_{r}$ is defined with 
$P_{r}=(1 + \gamma_{5})/2$. If $m=0$ then $\mathcal{L}\mathcal{=L}_l + \mathcal{L}_r$ where 
$\psi_{l}$ and $\psi_{r}$ are the eigenvectors of $P_{l}$ and $ P_{r}$, respectively:
the resulting Lagrangian decouples to two independent contributions. Thus,  two classes of
symmetrical particles with right-handed or left-handed helicities can be distinguished. 

Chiral symmetry is assumed to hold approximately for light quarks.
If quarks were exactly massless, then  
$\mathcal{L}_{QCD} = \mathcal{L}^l _{quarks} + \mathcal{L}^r_{quarks} +  \mathcal{L}_{int} + \mathcal{L}_{gluons}$. 
Massless Goldstone bosons can be generated by spontaneous symmetry breaking.
The pion spin-parity and mass, which is much smaller than that of other hadrons, 
allows the identification of the pion with the Goldstone boson. Non-zero quark masses -- which explicitly 
break chiral symmetry -- then lead to the non-zero pion mass.  The $\chi$PT calculations can be 
extended to massive quarks by adding a perturbative term $\overline{\psi}m\psi$ which explicitly 
breaks the chiral symmetry. The much larger masses of other hadrons
are assumed to come from spontaneous symmetry breaking caused by 
quantum effects; {\it i.e.},  dynamical symmetry breaking.  Calculations of observables at small 
$Q^2$ use an ``effective" Lagrangian expressed in terms of hadronic fields, which incorporates chiral symmetry.
The resulting perturbative series is  a function of $m_{\pi}/M_n$ and the momenta of the on-shell particles involved
in the reaction.

\subsubsection{Connection to conformal symmetry \label{conformal_sym}}

Once the quark masses are neglected, the classical QCD Lagrangian 
$\mathcal{L}_{QCD}$ has no apparent mass scale and is  effectively \emph{conformal}. 
Since there are no dimensionful parameters in 
$\mathcal{L}_{QCD}$, QCD is apparently  scaleless. 
This observation allows one to apply the 
AdS/CFT duality~\cite{Maldacena:1997re} to semi-classical QCD, which is 
the basis for LFHQCD discussed next.
The strong force is effectively \emph{conformal} at high-$Q^2$ (Bjorken scaling), and at low $Q^2$, one observes the
\emph{freezing} of $\alpha_s(Q^2)$~\cite{Deur:2016tte}. The observation of
\emph{conformal} symmetry at high-$Q^2$ is a key feature of QCD. 
More recently, studying the \emph{conformal} symmetry of QCD at low $Q^2$ has provided 
new insights into hadron structure, as will be discussed in the next section.
However, these signals for \emph{conformal} scaling fail at moderate $Q^2$ 
because of quantum corrections --  the QCD coupling $  \alpha_s$ varies strongly near 
$\Lambda_s$, the scale arising from quantum effects and the \emph{dimensional transmutation} 
property arising  from renormalization.   The QCD mass scale can also be expressed as 
$\sigma_{str}$ (the string tension appearing in heavy quark phenomenology) 
and as $\kappa$ (LFHQCD's universal scale) which controls the slope of Regge trajectories.
The pion decay constant $f_\pi$, characterizing the dynamical breaking of chiral symmetry, can also be related to
these mass scales~\cite{Kneur:2011vi}.  Other characteristic mass scales exist, see~\cite{Deur:2016tte}.

\subsection{The light-front holographic QCD approach \label{sec:LFHQCD}}
\emph{LF quantization} allows for a rigorous and exact formulation of QCD, in particular in its nonperturbative domain.
Hadrons. {\it i.e.},  bound-states of quarks, are described on the LF by a relativistic Schr\"{o}dinger-like equation, see 
Section~\ref{LC dominance and LF quantization}. 
All components of this equation can in principle be obtained from the QCD Lagrangian; In
practice, the effective confining potential entering the equation has been obtained 
only in (1+1) dimensions~\cite{Hornbostel:1988fb}.  The complexity of such computations grows quickly with 
dimensions and in (3+1) dimensions, the confining potential must be obtained from other than first-principle calculations. 
An important possibility is to use the LFHQCD framework~\cite{Brodsky:2014yha}. 

LFHQCD is based on the isomorphism between the
group of  isometries of a 5-dimensional \emph{anti-de-Sitter space} (AdS$_5$) and the $SO(4,2)$ 
group of  \emph{conformal} transformations in physical spacetime.
The isomorphism generates a correspondence between a strongly interacting
\emph{conformal field theory} (CFT) in $d$--dimensions
and a weakly interacting, classical gravity-type theory in $d+1$-dimensional AdS space~\cite{Maldacena:1997re}.
Since the strong interaction is approximately  \emph{conformal} and 
strongly coupled at low $Q^2$, gravity calculations 
can be mapped onto the 
boundary of AdS space -- representing the physical 
Minkowski spacetime -- to create an approximation for QCD. 
This approach based on the  ``gauge/gravity correspondence", {\it i.e.}, the mapping of
a gravity theory in a  5-dimensional  AdS space onto  its  4-dimensional boundary, explains the nomenclature ``holographic". 
In this approach, the fifth-dimension coordinate  $z$ of AdS$_5$ space corresponds to the LF 
variable $\zeta_\bot =  x_\bot  \sqrt{{x}(1-x)}$,
the invariant transverse separation between the $q \bar q$ constituents of a meson.   
Here $x$ is the LF fraction $k^+\over P^+$. The holographic correspondence~\cite{deTeramond:2005su} 
relating $z$ to $\zeta$  can be deduced from the fact that the formulae for hadronic 
electromagnetic~\cite{Polchinski:2002jw} and gravitational~\cite{Abidin:2008ku} form factors in AdS space  
match~\cite{Brodsky:2006uqa} their corresponding expressions for form factors of 
composite hadrons in the LF~\cite{Drell:1969km, West:1970av}.

LFHQCD also provides a correspondence between  hadron  eigenstates  and  nonperturbative bound-state 
amplitudes in AdS space, form factors and quark distributions: the analytic structure of the amplitudes leads to a  
nontrivial connection with Regge theory and the hadron spectrum~\cite{deTeramond:2018ecg, Sufian:2018cpj}. It was 
shown in Refs.~\cite{deTeramond:2014asa,Dosch:2015nwa,Brodsky:2016yod} how  
implementing superconformal symmetry  completely fixes  the distortion of AdS space, 
therefore fixing the confining potential of the boundary theory. 
The distortion can be expressed in terms of a specific ``dilaton" profile in the AdS action. 
This specific profile is uniquely recovered by the procedure of Ref.~\cite{deAlfaro:1976vlx} which shows 
how a mass scale can be introduced in the Hamiltonian without affecting 
the \emph{conformal invariance} of the action~\cite{Brodsky:2013ar}.
This   uniquely determines the LF bound-state potential for mesons and 
baryons, thereby making LFHQCD a fully determined approximation to QCD. ``Fully determined" 
signifies here that in the chiral limit LFHQCD has a single free parameter, the minimal number that 
dimensionfull theories using conventional (human chosen) units such as GeV, must have, see {\it e.g.}, 
the discussion in Chapter VII.3 of Ref.~\cite{Zee:2003mt}. 
For LFHQCD this parameter is $\kappa$; for perturbative conventional QCD, it is $\Lambda_s$~\cite{Deur:2014qfa}.
In fact, chiral QCD being independent of  
conventional units such as GeV,  a theory or model of the strong force can 
only provide dimensionless ratios, {\it e.g.}, $M_p / \Lambda_s$ or the proton to $\rho$-meson mass ratio $M_p / M_{\rho}$.

The derived confining potential  has the form of a harmonic oscillator 
$\kappa^4 \zeta^2 $ where $\kappa^2 = \lambda$:   It effectively accounts for the 
gluonic string connecting the 
quark and  
antiquark in a meson. It leads to a massless pion bound state  in the chiral limit and 
explains the mass symmetry between mesons and baryons~\cite{Dosch:2015nwa}.  
The LF harmonic oscillator potential transforms to the well-known nonrelativistic 
confining potential $\sigma_{str} r$ of heavy quarkonia in the \emph{instant form} 
of dynamics~\cite{Trawinski:2014msa} (with $r$, the quark separation). 

%
Quantum fluctuations are not included in the semiclassical LFHQCD computations. 
Although heavy quark masses break \emph{conformal} symmetry, the introduction of a 
heavy mass does not necessarily leads to supersymmetry breaking, since it can 
stem from the  underlying dynamics of color confinement~\cite{deTeramond:2016bre}. Indeed, it was shown in 
 Ref.~\cite{Dosch:2015bca} that supersymmetric relations between 
the meson and baryon masses still hold to a good approximation even for heavy-light
({\it i.e.,} charm and bottom) hadrons, leading to remarkable connections between meson, 
baryon and tetraquark states~\cite{Nielsen:2018uyn}.


A prediction of chiral LFHQCD for the nucleon spin is that the  eigensolutions for the LF wave equation 
for spin 1/2 (plus and minus components) associated with  $L_z = 0$ and 
$L_z = 1$ have equal normalization, see Eq. 5.41 of Ref.~\cite{Brodsky:2014yha}.  Since there is no gluon quanta, 
the gluons being sublimated into the effective 
potential~\cite{Brodsky:2014yha}, the nucleon spin comes from quark OAM 
in the effective quark-diquark two-body Hamiltonian approximation. This agrees with the (pre-EMC)
chiral symmetry prediction obtained in a Skyrme approach, namely, that the nucleon spin is carried by quark 
OAM in the nonperturbative domain~\cite{Brodsky:1988ip}.

\subsection{Summary}

We have outlined  the theoretical approaches that are used to interpret spin-dependent observables. 
Simplifications, both for theory and  experiments, arise when inclusive reactions are considered, 
\emph{viz} reactions in which all hadronic final states are summed over. 
Likewise, summing on all reactions; {\it i.e.}, integrating on $W$ or equivalently over $x_{Bj}$,
to form moments of structure functions yields further simplifications. These moments can
be linked to observables characterizing the nucleon by relations called \emph{sum rules}. They offer
unique opportunities for studying QCD because they are often
valid at any $Q^2$. Thus, they allow tests of the various calculation 
methods applicable at low ($\chi$PT, LFHQCD), intermediate (Lattice QCD, LFHQCD), and high $Q^2$ (OPE).
Spin sum rules will now be discussed following the formalism of Refs.~\cite{Drechsel:2002ar, Drechsel:2004ki}.

\section{Sum rules\label{sum rules}}

Nucleon spin sum rules offer an important opportunity to study QCD. In the last 20 years, the Bjorken sum 
rule~\cite{Bjorken:1966jh}, derived at high-$Q^2$, and the 
Gerasimov-Drell-Hearn (GDH) sum rule~\cite{Gerasimov:1965et}, 
derived at $Q^2=0$, have been studied in detail,  both experimentally and theoretically.   
This primary set of sum rules links the moments of structure functions 
(or equivalently of photoabsorption cross-sections) to the static properties of the nucleon.
Another class of sum rules relate the moments of structure functions to Doubly 
Virtual Compton Scattering (VVCS) amplitudes rather than to static properties. This class includes the generalized 
GDH sum rule~\cite{Ji:1999mr,  Drechsel:2004ki, Anselmino:1988hn} and 
spin polarisability sum rules~\cite{Drechsel:2002ar,  Lensky:2016nui, Drechsel:2004ki}.  The VVCS  amplitudes are
calculable at any $Q^2$ using the techniques described in Section~\ref{Computation methods}. They can then
be compared to the measured moments. Thus, these sum rules are particularly well suited for 
exploring the transition between fundamental and effective descriptions of QCD.

\subsection{General formalism}

Sum rules are generally derived by combining dispersion relations with 
the  \emph{Optical Theorem} ~\cite{Pasquini:2018wbl}.  
Many sum rules can also be
derived using the OPE or QCD on the LC.  In fact, the Bjorken and Ellis-Jaffe~\cite{Ellis:1973kp}
sum rules were originally derived using quark LC current algebra. 
Furthermore, a few years after its original derivation \emph{via} dispersion relations, the GDH sum rule was
rederived using LF current algebra~\cite{Dicus:1972vp}.

A convenient formalism for deriving the sum rules
relevant to this review is given in ~\cite{Drechsel:2002ar, Drechsel:2004ki}.  
The central principle is to apply the  \emph{Optical Theorem} to the VVCS amplitude, thereby  linking virtual 
photoabsorption to the inclusive lepton scattering cross-section. Assuming causality, the VVCS amplitudes 
can be analytically continued in the complex plane. The Cauchy relation 
-- together with the assumption that the VVCS amplitude converges faster than 
$1/\nu$ as $\nu \to \infty$ so that it fulfills Jordan's lemma -- 
yields the widely used Kramer-Kr\"{o}nig relation~\cite{KKR}: 
\vspace{-0.2cm}
\begin{equation}
\vspace{-0.2cm}
\Re e\left(A_{VVCS}(\nu,Q^2)\right)=\frac{1}{\pi}{P\int}_{-\infty}^{+\infty}\frac{\Im m\left(A_{VVCS}(\nu',Q^2)\right)}{\nu'-\nu}d\nu'.
\label{Eq:KKR}
\end{equation}
The crossing symmetry of the VVCS amplitude allows one to restrict the integration range from 0 to $\infty$. 
The  \emph{Optical Theorem} then allows $\Im m\left(A_{VVCS}\right)$ to be expressed
in term of its corresponding photoabsorption cross-section. 
Finally,  after subtracting the target particle pole contribution (the elastic reaction), $\Re e\left(A_{VVCS}\right)$ 
is expanded in powers of $\nu$ using a low energy theorem~\cite{Low:1954kd}.
Qualitatively, the integrand at LO represents the electromagnetic current spatial distribution and  at
NLO reflects the deformation of this spatial distribution due to the probing photon (polarizabilities).
The applicability of Jordan's lemma 
has been discussed extensively.
It has been pointed out~\cite{Abarbanel:1967wk} that an amplitude 
may not vanish as $\nu \rightarrow \infty$ due to fixed  $J=0$ or $J=1$ poles of $\Re e\left(A_{VVCS}\right)$,
leading to sum rule modifications. 
Here, we shall assume the validity of Jordan's lemma. 

\vspace{-0.1cm}
\subsection{GDH and forward spin polarizability sum rules \label{sec:Spin polarizabilities}}
The methodology just discussed applied to the spin-flip VVCS amplitude yields the generalized GDH sum 
rule when the first term of the $\nu$ expansion is considered:
\vspace{-0.2cm}
\begin{eqnarray}
\vspace{-0.2cm}
I_{TT}(Q^2) & = & \frac{M_t^2}{4\pi^2\alpha}\int_{\nu_0}^{\infty}\frac{\kappa_{\gamma^*}(\nu,Q^2)}{\nu}\frac{\sigma_{TT}}{\nu}d\nu\nonumber \\
& = & \frac{2M_t^2}{Q^2}\int_0^{x_0}\Bigr[g_1(x,Q^2)-\frac{4M_t^2}{Q^2}x^2g_2(x,Q^2)\Bigl]dx,
\label{eq:gdhsum_def1}
\end{eqnarray}
where Eq.~(\ref{sigmaTT})  was used for the second equality. $I_{TT}(Q^2)$ 
is the spin-flip VVCS amplitude in the low $\nu$ limit. 
The limits $\nu_0$ and $x_0=Q^2/(2M_t\nu_0)$ 
correspond to the inelastic reaction threshold, and $M_t$ is the target  mass.
For $Q^2 \rightarrow 0$, the low energy theorem relates $I_{TT}(0)$ to  the anomalous magnetic moment $\kappa_t$, 
and Eq.~(\ref{eq:gdhsum_def1}) becomes the GDH sum rule:
\vspace{-0.3cm}
\begin{equation}
\vspace{-0.2cm}
I_{TT}(0)=\int_{\nu_0}^{\infty}\frac{\sigma_{T,1/2}(\nu)-\sigma_{T,3/2}(\nu)}{\nu}d\nu=-\frac{2\pi^2 S \alpha\kappa_t^2}{M_t^2}.
\label{eq:gdh}
\end{equation}
Experiments at MAMI, ELSA and LEGS~\cite{Ahrens:2001qt} have
verified the validity of the proton GDH sum rule within an accuracy of about 10\%.
The low $Q^2$ JLab $\overline{I_{TT}^n}(Q^2)$ measurement extrapolated to $Q^2=0$ is compatible with
the GDH expectation for the neutron within the 20\% experimental uncertainty~\cite{Adhikari:2017wox}.
A recent phenomenological assessment of the sum rule also concludes its validity~\cite{Gryniuk:2015eza}.
The original and generalized GDH sum rules apply to any target, including nuclei, leptons, photons or gluons. 
For these latter massless particles, the sum rule predicts $I_{TT}^{\gamma,~g}(0)=0$~\cite{Bass:1998bw}.

The NLO term of the $\nu$ expansion of the  left-hand side of   Eq.~(\ref{Eq:KKR}) yields the forward spin 
polarizability~\cite{GellMann:1954db}:
\vspace{-0.7cm}
\begin{eqnarray}
\vspace{-0.3cm}
\gamma_0(Q^2) & = & \frac{1}{2\pi^2}\int_{\nu_0}^{\infty}\frac{\kappa_{\gamma^*}(\nu,Q^2)}{\nu}\frac{\sigma_{TT}(\nu,Q^2)}{\nu^{3}}d\nu\nonumber \\
& = &  \frac{16\alpha M_t^2}{Q^{6}}\int_0^{x_0}x^2\Bigl[g_1(x,Q^2)-\frac{4M_t^2}{Q^2}x^2g_2(x,Q^2)\Bigr]dx.
\label{eq:gamma_0}
\end{eqnarray}

Alternatively, the polarized covariant  VVCS amplitude $S_1$ can be considered.  It is connected to the spin-flip 
and longitudinal-transverse interference VVCS amplitudes, $g_{TT}$ and $g_{LT}$ respectively, by:
\vspace{-0.3cm}
 \begin{equation}
 \vspace{-0.3cm}
S_1(\nu,Q^2)=\frac{\nu M_t}{\nu^2+Q^2}\Bigr[g_{TT}(\nu,Q^2)+\frac{Q}{\nu}g_{LT}(\nu,Q^2)\Bigl].
\nonumber
\end{equation}
Under the same assumptions, the dispersion relation yields: 
\vspace{-0.3cm}
\begin{equation}
\vspace{-0.3cm}
\Re e[S_1(\nu,Q^2)-S_1^{pole}(\nu,Q^2)]=\frac{4\alpha}{M_t}I_1(Q^2)+\gamma_{g_1}(Q^2)\nu^2+O(\nu^{4}),
\nonumber
\end{equation}
where the LO term yields a generalized GDH sum rule differing from the one in Eq.~(\ref{eq:gdhsum_def1}): 
\vspace{-0.3cm}
\begin{equation}
\vspace{-0.3cm}
I_1(Q^2)=\frac{2M_t^2}{Q^2}\int_0^{x_0}g_1(x,Q^2)dx.
\label{eq:gdhsum_def2}
\end{equation}
The original GDH sum rule is recovered for
$Q^2=0$ where $I_1(0)=-\frac{1}{4}\kappa_t^2$.  The NLO term
defines the generalized polarizability $\gamma_{g_1}$: 
\vspace{-0.5cm}
\begin{eqnarray}
\vspace{-0.5cm}
\gamma_{g_1}(Q^2) & = & \frac{16\pi\alpha M_t}{Q^{6}}\int_0^{x_0}x^2g_1(x,Q^2)dx.
\nonumber
\end{eqnarray}

\subsection{$\delta_{LT}$ sum rule}
Similarly, the longitudinal-transverse interference VVCS amplitude yields a sum  rule for the $I_{LT}$ 
amplitude~\cite{Drechsel:2002ar, Drechsel:2004ki, Drechsel:1998hk} :
\vspace{-0.5cm}
\begin{eqnarray}
\vspace{-0.5cm}
I_{LT}(Q^2) & = & \frac{M_t^2}{4\pi^2\alpha}\int_{\nu_0}^{\infty}\frac{\kappa_{\gamma^*}(\nu,Q^2)}{\nu}\frac{\sigma_{LT}'(\nu,Q^2)}{Q}d\nu\nonumber \\
& = & \frac{2M_t^2}{Q^2}\int_0^{x}\Bigl[g_1(x,Q^2)+g_2(x,Q^2)\Bigr]dx,
\nonumber
\end{eqnarray}
and defines the generalized LT-interference polarizability:
 \vspace{-0.3cm}
 \begin{eqnarray}
 \vspace{-0.3cm}
\delta_{LT}(Q^2) & = & \left(\frac{1}{2\pi^2}\right)\int_{\nu_0}^{\infty}\frac{\kappa_{\gamma^*}(\nu,Q^2)}{\nu}\frac{\sigma_{LT}'(\nu,Q^2)}{Q\nu^2}d\nu\nonumber \\
& = &  \frac{16\alpha M_t^2}{Q^{6}}\int_0^{x_0}x^2\Bigl[g_1(x,Q^2)+g_2(x,Q^2)\Bigr]dx.
\label{eq:delta_LT SR}
\end{eqnarray}
The quantities $\delta_{LT}$, $\gamma_{g_1}$, $I_{TT}$ and $I_1$ are related by:
\vspace{-0.3cm}
 \begin{eqnarray}
\vspace{-0.3cm}
 M_t \delta_{LT} (Q^2) = \gamma_{g_1}(Q^2)- \frac{2\alpha}{M_t Q^2}\Bigr(I_{TT}(Q^2)-I_1(Q^2)\Bigl).
\nonumber
\end{eqnarray}

It was shown recently that the sum rules of Eqs.~(\ref{eq:gdhsum_def2}) and (\ref{eq:delta_LT SR})
are also related to several other generalized polarizabilities, which are experimentally poorly known,  
but can be constrained by these additional relations~\cite{Pascalutsa:2014zna}.

\subsection{The Burkhardt-Cottingham sum rule \label{BCSR}}
We now consider the second VVCS amplitude $S_2$: 
\vspace{-0.3cm}
 \begin{equation}
 \vspace{-0.3cm}
S_2(\nu,Q^2)=-\frac{M_t^2}{\nu^2+Q^2}\Bigr[g_{TT}(\nu,Q^2)-\frac{\nu}{Q}g_{LT}(\nu,Q^2)\Bigl].
\nonumber
\end{equation}
Assuming a Regge behavior $S_2\to\nu^{-\alpha_2}$ as
$\nu\to\infty$, with $\alpha_2 > 1$,  the dispersion relation for $S_2$ and
$\nu S_2$, including the elastic contribution, requires no subtraction. It thus leads to a 
``super-convergence relation" -- the Burkhardt-Cottingham (BC) sum rule~\cite{Burkhardt:1970ti}: 
\vspace{-0.3cm}
\begin{equation}
\vspace{-0.3cm}
\int_0^1g_2(x,Q^2)dx=0.
\label{eq:bc}
\end{equation}
Excluding the elastic reaction, the sum rule becomes: 
\vspace{-0.3cm}
\begin{equation}
\vspace{-0.3cm}
I_2(Q^2)=\frac{2M_t^2}{Q^2}\int_0^{x_0}g_2(x,Q^2)dx=\frac{1}{4}F_2(Q^2)\Bigl(F_1(Q^2)+F_2(Q^2)\Bigr),
\label{eq:bc_noel}
\end{equation}
where $F_1$ and $F_2$ are the Dirac and Pauli form factors, respectively, see  Section~\ref{elastic scatt}.

The low energy expansion of the dispersion relation leads to: 
\vspace{-0.3cm}
\begin{eqnarray}
\vspace{-0.3cm}
\Re e  \bigl[\nu \bigl( S_2(\nu,Q^2) - S_2^{pole}(\nu,Q^2)\bigr)\bigr] = \nonumber \\
\hspace{-1cm} {2\alpha I_2(Q^2)-\frac{2\alpha}{Q^2}\bigl(I_{TT}(Q^2)-I_1(Q^2)\bigr)\nu^2+\frac{M_t^2}{Q^2}\gamma_{g_2}(Q^2)\nu^{4}+O(\nu^{6}),}
\nonumber
\end{eqnarray}
where the term in $\nu^{4}$ provides the generalized polarisability $\gamma_{g_2}$:
\vspace{-0.3cm}
\begin{eqnarray}
\vspace{-0.3cm}
\gamma_{g_2}(Q^2) =  \frac{16\pi\alpha M_t^2}{Q^{6}}\int_0^{x_0}x_{Bj}^2g_2(x,Q^2)dx
= \delta_{LT}(Q^2)-\gamma_0(Q^2)+\frac{2\alpha}{M_t^2Q^2}\Bigl(I_{TT}(Q^2)-I_1(Q^2)\Bigr).
\nonumber
\end{eqnarray}
%


\subsection{Sum rules for deep inelastic scattering  \label{DIS SR}}
At high-$Q^2$, the OPE 
used on the VVCS amplitude leads to the \emph{twist} expansion: 
\vspace{-0.3cm}
\begin{equation}
\vspace{-0.3cm}
\Gamma_1(Q^2)\equiv\int_0^1g_1(x,Q^2)dx=\sum_{\tau=2,4,...}\frac{\mu_{\tau}(Q^2)}{Q{}^{\tau-2}} ,
\label{eq:Gamma1}
\end{equation}
where the  $\mu_{\tau}$ coefficients correspond to the matrix elements of operators of \emph{twist} $\leq \tau$.
The dominant \emph{twist} term (twist~2) $\mu_2$ is given by
the matrix elements of the axial-vector operator  $\overline{\psi}\gamma_{\mu}\gamma_{5}\lambda^i\psi/2$
summed over quark flavors. $\lambda^i$ are the Gell-Mann matrices for $1 \leq i \leq 8$ and $\lambda^0 \equiv 2$. 
Only $i=0,3$ and $i = 8 $ contribute, 
with matrix elements
\vspace{-0.6cm}
\begin{eqnarray}
\vspace{-0.3cm}
\langle P,S|\overline{\psi}\gamma_{\mu}\gamma_{5}\lambda^0\psi|P,S  \rangle =4Ma_0S_{\mu} ,\nonumber \\
\langle P,S|\overline{\psi}\gamma_{\mu}\gamma_{5}\lambda^3\psi|P,S \rangle =2Ma_3S_{\mu} ,\nonumber \\
\langle P,S|\overline{\psi}\gamma_{\mu}\gamma_{5}\lambda^8\psi|P,S \rangle =2Ma_8S_{\mu}, \nonumber 
\end{eqnarray}
defining the triplet ($a_3$), octet ($a_8$) and singlet ($a_0$) axial charges. Then,
\vspace{-0.3cm}
\begin{eqnarray}
\vspace{-0.3cm}
\mu_2(Q^2) & = & \left(\pm\frac{1}{12}a_3\ +\ \frac{1}{36}a_8\right)+\frac{1}{9}a_0\ +O\big(\alpha_{s}(Q^2)\big),
\label{eq:mu2}
\end{eqnarray}
where $+(-)$ is for the proton (neutron) and $O(\alpha_{s})$
reflects the $Q^2$-dependence derived from pQCD radiation. 
The axial charges can be expressed in the parton model as combinations of quark polarizations:
\vspace{-0.3cm}
\begin{eqnarray}
\vspace{-0.3cm}
a_3 & = & (\Delta u + \Delta \overline{u})-(\Delta d + \Delta \overline{d}), \nonumber \\
a_8 & = &(\Delta u + \Delta \overline{u})+(\Delta d + \Delta \overline{d}) -2(\Delta s + \Delta \overline{s}), \nonumber \\
a_0 & = & (\Delta u + \Delta \overline{u})+(\Delta d + \Delta \overline{d}) +(\Delta s + \Delta \overline{s}).
\nonumber
\end{eqnarray} 
The charges $a_3$ and $a_8$ are $Q^2$-independent; the axial charge $a_0$, 
which is identified with the quark spin contribution to $J$, namely
$\Delta \Sigma$, see Eq.~(\ref{eq:spin SR}), is $Q^2$-independent only at LO in $\alpha_s$. 
At NLO, $a_0$ becomes $Q^2$-dependent because the singlet current is not 
renormalization-group invariant and needs to be
renormalized. (That $a_3$ and $a_8$ remain $Q^2$-independent assumes the validity of $SU(3)_f$.) 
In addition $a_0$ may also depend on the gluon spin contribution $\Delta G$ through the gluon axial 
anomaly~\cite{Efremov:1988zh}. Such a contribution depends on the chosen renormalization scheme. 
In the AB~\cite{Adler:1969er}, CI~\cite{Cheng:1996jr} and JET~\cite{Leader:1998qv, Leader:1998nh} schemes, 
$a_0 = \Delta \Sigma - \frac{f}{2 \pi} \alpha_s(Q^2)\Delta G(Q^2)$, where $f$ is the number of active flavors. 
In the  case of  the $\overline{\mbox{MS}}$ scheme, $\alpha_s(Q^2)\Delta G(Q^2)$ is absorbed in the definition of 
$\Delta \Sigma$ and $a_0 = \Delta \Sigma$. 
At first order, $\Delta G$ evolves as $1/\alpha_s$~\cite{Efremov:1988zh} and
$\alpha_s(Q^2)\Delta G(Q^2)$ is constant at high $Q^2$.  Hence, contrary to the usual case where the scheme
dependence of a quantity disappears at large $Q^2$ due to the dominance of the scheme-independent LO, 
$\Delta\Sigma$ remains scheme-dependent at arbitrarily high $Q^2$. 
The $\alpha_s \Delta G$ term stems from the $g_1$ NLO evolution equations, 
Eqs.~(\ref{g_1 LT evol})-(\ref{gluon LO evol}). In the $\overline{\mbox{MS}}$ scheme, the contribution of the 
gluon evolution to the $g_1$ moment cancels at any order in perturbation theory.  In the AB scheme the Wilson coefficient
controlling the gluon contribution is non-zero, $\Delta C_g=- \frac{f}{2 \pi} \alpha_s$. 
This scheme-dependence and the presence of $1/\alpha_s$, which is
not an observable, emphasize that $\Delta \Sigma$ and $\Delta G$ are also not observables but depend on the convention
used for the renormalization procedure;  {\it e.g.},  how high order ultraviolet divergent diagrams are arranged and regularized. 
The origin of the logarithmic increase of $\Delta G$  is due to the fact that overall, the subprocess in which a gluon splits into 
two gluons of helicity  $+1,$  thereby increasing  $\Delta G$, has a larger probability than subprocesses that
decrease the total gluon helicity, where a gluon splits into a quark-antiquark pair  or a gluon splits into two gluons, one of helicity $+1$ and
the other of  helicity $-1$)~\cite{Ji:1995cu}.  The gluon splitting increases with the probe resolution, leading to the 
logarithmic increase of $\Delta G$ with $Q^2.$

Assuming SU(3)$_f$ quark mass symmetry, the axial charges can be related to the weak decay constants $F$ and $D$: 
$a_3=F+D=g_A$ and $a_8=3F-D$, where 
$g_A$ is well measured from neutron $\beta-$decay: $g_A=1.2723(23)$~\cite{Olive:2016xmw}. 
$a_8$ is extracted from the  weak decay of hyperons, assuming SU(3)$_f$: 
$a_8=0.588(33)$~\cite{Close:1993mv}.  The 0.1 GeV strange quark mass is neglected in SU(3)$_f$,
but its violation is expected to affect $a_8$ only at a level of a few \%. However, 
other effects may alter $a_8$: models based on the 
one-gluon exchange hyperfine interaction as well as meson cloud effects
yield {\it e.g.}, a smaller value, $a_8=0.46(5)$~\cite{Bass:2009ed}. 

If one expresses the axial charges in terms of quark polarizations and assumes that the strange and higher mass quarks do
not contribute to $\Delta \Sigma$, Eqs. (\ref{eq:Gamma1}) and (\ref{eq:mu2}) lead, at \emph{leading-twist}, 
to the Ellis-Jaffe sum rule.   For the proton this sum rule is:
\vspace{-0.1cm}
\begin{equation}
\vspace{-0.1cm}
\Gamma_1^p(Q^2)\equiv\int_0^1g_1^p(x,Q^2)dx \xrightarrow[Q^2 \to \infty] {} \frac{1}{2}\left(\frac{4}{9}\Delta u+\frac{1}{9}\Delta d\right).
\label{eq:Ellis-Jaffe p}
\end{equation}
The neutron sum rule is obtained by assuming isospin symmetry, {\it i.e.}, $u \leftrightarrow d$ interchange.
%
%
The expected asymptotic values are $\Gamma_1^p=0.185\pm0.005$ and $\Gamma_1^n=-0.024\pm0.005$. 
After the  order $\alpha_s^3$ evolution to $Q^2=5$~GeV$^2$ they become $\Gamma_1^p=0.163$ and
$\Gamma_1^n=-0.019$.  Measurements at this $Q^2$ disagree with the sum rule. The most precise ones are from 
E154 and E155. E154 measured $\Gamma^n=-0.041 \pm 0.004 \pm 0.006$~\cite{Abe:1997cx} and 
E155 measured $\Gamma^p=0.118 \pm 0.004 \pm 0.007$ and
$\Gamma^n=-0.058 \pm 0.005 \pm 0.008$~\cite{Anthony:1999rm}. 

The proton-neutron difference for Eqs.~(\ref{eq:Gamma1}) and (\ref{eq:mu2}) gives the non-singlet relation:
\vspace{-0.25cm}
\begin{equation}
\vspace{-0.25cm}
\Gamma_1^p(Q^2)-\Gamma_1^n(Q^2) \equiv  \Gamma_1^{p-n}(Q^2) =\frac{1}{6} g_A + O(\alpha_{s})+O(1/Q^2),
\xrightarrow[Q^2 \to \infty] {} \frac{\Delta u - \Delta d}{6} \nonumber
\end{equation}
which is the Bjorken sum rule for $Q^2\to\infty$~\cite{Bjorken:1966jh}.  Charge symmetry corrections to the Ellis-Jaffe and
Bjorken sum rules are at the 1\% level~\cite{Cloet:2012db}.
\emph{DGLAP} corrections yield~\cite{Kataev:1994gd}:
\vspace{-0.3cm}
 \begin{equation}
 \vspace{-0.2cm}
\Gamma_1^{p-n}(Q^2)=\frac{g_A}{6}\bigg[1-\frac{\alpha_{{\rm {s}}}}{\pi}-3.58\left(\frac{\alpha_{{\rm {s}}}}{\pi}\right)^2
-20.21\left(\frac{\alpha_{{\rm {s}}}}{\pi}\right)^{3} -175.7\left(\frac{\alpha_{{\rm {s}}}}{\pi}\right)^{4}+... \bigg]+O(1/Q^2),
\label{eq:genBj}
\end{equation}
where the series coefficients are given for $n_f=3$.

Eq.~(\ref{eq:genBj}) exemplifies the power of sum rules: these relations connect moments 
integrated over high-energy quantities to low-energy, static characteristics of the nucleon itself.
\label{Bjorken and g_a}  It is clear why 
$g_A \equiv g_A(Q^2=0)$  is involved in the $Q^2 \to \infty$ Bjorken sum rule. The spin-dependent
part of the cross-section comes from the  matrix elements of $\bar \psi \gamma^\mu \gamma^5 \psi$, 
the conserved axial-current 
associated with chiral symmetry: $\psi \to e^{i\phi\gamma^5} \psi$, where the nucleon state 
$\psi$ is projected to its right and left components as defined by the chiral projectors 
$(1\pm\gamma^5$), respectively. 
In elastic scattering, $\bar \psi \gamma^\mu \gamma^5 \psi$ generates the axial form factor 
$g_A(Q^2),$ just as the electromagnetic current $\bar \psi \gamma^\mu \psi$ 
generates the electromagnetic form factors.  And just as  $G_E^N$ provides the 
charge spatial distribution, the Fourier transform of $g_A(Q^2)$
maps the spatial distribution of the nucleon spin; {\it i.e.}, how the net parton polarization evolves from 
the center of the nucleon to its boundary. Thus $g_A(Q^2)$  provides the isovector 
component of the spatial parton polarizations: $g_A(Q^2=0)$ is the parton 
polarizations without spatial resolution;  {\it i.e.} its spatial average, which is directly connected to 
the mean momentum-space parton polarization $\int g_1dx$. 

Comparing Eqs.~(\ref{g_1 LT evol})--(\ref{gluon LO evol}) 
with Eq.~(\ref{eq:genBj}) shows that the $Q^2$-evolution is much simpler for
moments ({\it i.e.}, \emph{Mellin-transforms}) than for structure functions. 
Thus it is beneficial to transform to \emph{Mellin-space} ($N,Q^2$), where $N$ is the moment's order, 
to perform the $Q^2$-evolution and then transform back to $(x_{Bj},Q^2)$ space. 

The coefficient $\mu_{\tau}$ 
in Eq.~(\ref{eq:Gamma1}) would only comprise 
a twist~$\tau$ operator, if not for the effect discussed on page~\pageref{twist-mix} which adds operators 
of \emph{twists}  $\varsigma\leq\tau$. Thus, the twist~4 term,
\vspace{-0.3cm}
\begin{equation}
\vspace{-0.2cm}
\mu_4(Q^2)=M^2\left(a_2(Q^2)+4d_2(Q^2)+4f_2(Q^2)\right)/9,
\label{eq:mu4}
\end{equation}
comprises a twist~2 contribution ($a_2$) and a  twist~3 one ($d_2$) in addition to  the genuine 
twist~4 contribution $f_2$~\cite{Shuryak:1981kj,Ji:1993sv,Stein:1995si,Ji:1997gs}. 
The twist~2 matrix element is:
\vspace{-0.3cm}
\begin{eqnarray}
\vspace{-0.3cm}
a_2\ S^{\{\mu}P^{\nu}P^{\lambda\}} & = & \frac{1}{2}\sum_{f}e_{f}^2\
\langle P,S|\overline{\psi}_{f}\ \gamma^{\{\mu}iD^{\nu}iD^{\lambda\}}\psi_{f}|P,S\rangle,
\label{eq:a2op}
\vspace{-0.3cm}
\end{eqnarray}
where $f$ are the quark flavors and $\{\cdots\}$ signals index symmetrization. 
The third moment of $g_1$ at \emph{leading-twist} gives $a_2 $:
\vspace{-0.5cm}
 \begin{equation}
 \vspace{-0.1cm}
a_2(Q^2)=2\int_0^1 x^2\ g_1^{twist~2}(x,Q^2)  dx,
\label{eq:a2}
\end{equation}
which is thus twist~2. The  twist~3 contribution $d_2$ is defined from the matrix element:
\vspace{-0.3cm}
\begin{equation}
\vspace{-0.3cm}
d_2S^{[\mu}P^{\{\nu]}P^{\lambda\}}=\frac{\sqrt{4\pi}}{8}\sum_{q}\langle P,S|\overline{\psi}_{q}\ 
\sqrt{\alpha_{s}} \widetilde{f}^{\{\mu\nu}\gamma^{\lambda\}}\psi_{q}|P,S\rangle ,
\label{eq:d2op}
\end{equation}
where $\widetilde{f}^{\mu\nu}$ is the dual  tensor of the gluon field:
$\widetilde{f}_{\mu\nu}=(1/2)\epsilon_{\mu\nu\alpha\beta}F^{\alpha\beta}$.
The third moments of $g_1$ and $g_2$ at \emph{leading-twist} give $d_2$: 
\vspace{-0.3cm}
\begin{eqnarray}
\vspace{-0.3cm}
d_2(Q^2)  =  \int_0^1 x^2 2g_1(x,Q^2)+3g_2(x,Q^2)  dx 
 =  3\int_0^1 x^2 g_2(x,Q^2)-g_2^{WW}(x,Q^2) dx,
\label{eq:d2mon}
\end{eqnarray}
where $g_2^{WW}$ is the  twist~2 component of $g_2$:
\vspace{-0.3cm}
\begin{equation}
\vspace{-0.3cm}
g_2^{WW}(x_{Bj},Q^2)={-g}_1(x_{Bj},Q^2)+\int_{x_{Bj}}^1\frac{g_1(y,Q^2)}{y}dy.
\label{eq:g2ww}
\end{equation}
This relation is derived from the Wandzura-Wilczek (WW) sum rule~\cite{Wandzura:1977qf}:
\vspace{-0.3cm}
\begin{equation}
\vspace{-0.3cm}
\int_0^1 x^{n-1}\bigg(\frac{n-1}{n}g_1(x,Q^2)+g_2(x,Q^2)\bigg) dx = 0,
\label{eq:g2ww general}
\end{equation}
where $n$ is odd. The Wandzura-Wilczek sum rule  assumes the validity of the BC sum rule and 
neglects  \emph{higher-twist} contributions to $g_1$ and $g_2$. Eq.~(\ref{eq:g2ww}) furthermore assumes that
the sum rule also holds  for even $n$,  as it is discussed further in Section~\ref{sec:g2-g2ww}. 
Eqs.~(\ref{eq:d2mon})-(\ref{eq:g2ww general}) originate from the OPE-derived expressions
valid at twist~3 and for $n$ odd~\cite{Jaffe:1996zw}:
\vspace{-0.3cm}
\begin{equation}
\vspace{-0.3cm}
\int_0^1 x^{n-1}g_1(x,Q^2) dx = \frac{a_{n-1}}{4}, \\\
\int_0^1 x^{n+1}g_2(x,Q^2) dx = \frac{n+1(d_{n+1} - a_{n+1})}{4(n+2)}.
\nonumber
\end{equation}

The twist~4 component of $\mu_4$ is defined by the matrix element:
\vspace{-0.2cm}
\begin{eqnarray}
\vspace{-0.3cm}
f_2\ M^2S^{\mu} =  \frac{1}{2}\sum_{q}e_{q}^2\
\langle N|g\ \overline{\psi}_i\ \widetilde{f}^{\mu\nu}\gamma_{\nu}\ \psi_i|N\rangle ,
\label{eq:f2op}
\end{eqnarray}
and, in terms of moments: 
\vspace{-0.3cm}
\begin{equation}
\vspace{-0.3cm}
f_2(Q^2)=\frac{1}{2}\int_0^1 x^2\Bigl(7g_1(x,Q^2)+12g_2(x,Q^2)-9g_3(x,Q^2)\Bigr) dx,
\label{eq:f2}
\end{equation}
where $g_3$  (not to be confused with a spin structure function also denoted $g_3$ and
appearing in neutrino scattering off a polarized target~\cite{Anselmino:1994gn}) is the twist~4 function:
\vspace{-0.3cm}
\begin{equation}
\vspace{-0.3cm}
g_3(x_{Bj})=\frac{1}{2\pi \Lambda_s^2}\int e^{i\lambda x_{Bj}}\left\langle PS\right|\overline{\psi}(0)\gamma_{5}{\displaystyle {\not}p}\psi(\lambda n)\left|PS\right\rangle  d\lambda
\label{eq:g3}
\end{equation}
with $p=\frac{1}{2}\left(\sqrt{M^2+P^2}+P\right)(1,0,0,1)$ and
$n=\frac{1}{M^2}\left(\sqrt{M^2+P^2}-P\right)(1,0,0,-1)$.
Since only $g_1$ and $g_2$ are measured, $f _{2}$ must  be
extracted  using Eqs. (\ref{eq:Gamma1}) and (\ref{eq:mu4}).
This is discussed in Section~\ref{HT}.

As mentioned in Section~\ref{OPE}, the OPE provides only odd moment sum rules 
for $g_1$ and $g_2$ (and even moment sum rules for $F_1$ and $F_2$) due to their positive parity under 
crossing symmetry. DIS spin sum rules involving even moments do exist for inclusive observables,
such as the Efremov-Leader-Teryaev (ELT) sum rule~\cite{Efremov:1996hd}:
\vspace{-0.4cm}
\begin{equation}
\vspace{-0.2cm}
\int_0^1 x \big(g_1^V(x,Q^2) + 2g_2^V(x,Q^2) \big)  dx = 0,
\nonumber
\end{equation}
where the superscript  $V$ indicates \emph{valence} distributions. 
Like the BC sum rule, the ELT prediction is a superconvergent relation.
The fact that \emph{sea quarks} do not contribute
minimizes complications from the low-$x_{Bj}$ domain that hinders the experimental checks  of sum rules. 
The ELT sum rule is not derived from the OPE, but instead follows from gauge invariance 
or, more generally, from the structure and gauge properties of
hadronic matrix elements involved in $g_1$ and $g_2$. 
It is an exact sum rule, but with the caveat that it
neglects  \emph{higher-twist} contributions as OPE-derived sum rules do 
(although  \emph{higher-twists} can be subsequently added, 
see  {\it e.g.}, the twist~4 contribution to the Bjorken sum rule given by Eq.~(\ref{eq:mu4})). 
Assuming that the \emph{sea} is isospin invariant leads to an isovector DIS sum rule, 
\vspace{-0.3cm}
\begin{equation}
\vspace{-0.3cm}
\int_0^1 x (g_1^p+2g_2^p - g_1^n-2g_2^n )   dx = 0, 
\nonumber
\end{equation}
which agrees with its experimental value at $\langle Q^2\rangle = 5$~GeV$^2$, 0.011(8). It can be re-expressed as: 
\vspace{-0.3cm}
\begin{equation}
\vspace{-0.3cm}
\int_0^1 x  \big(g_2^p(x,Q^2) - g_2^n(x,Q^2) \big) dx = 
\frac{-1}{12}\int_0^1 x \big( \Delta u_V(x,Q^2) -\Delta d_V(x,Q^2) \big) dx,
\label{Eq:ELT SR}
\end{equation}
which can be verified by comparing $g_2$ measurements for the l.h.s to PDF global fits for the r.h.s.
Neglecting twist~3 leads to a sum rule similar to the Wandzura-Wilczek sum rule, 
Eq.~(\ref{eq:g2ww general}), but for $n$ even ($n=2$):
\vspace{-0.3cm}
\begin{equation}
\vspace{-0.3cm}
\int_0^1 x \big(g_1 + 2g_2 \big) dx = 0.
\nonumber
\end{equation}

\subsection{Color polarizabilities}
The twist~3 and 4 operators discussed in the previous section describe the response of the
electric and magnetic-like components of the color field to the nucleon spin. They are therefore akin to polarizabilities, but
for the strong force rather than electromagnetism. Expressing the  twist~3 and 4 matrix elements 
as functions of the components of $\widetilde{f}^{\mu\nu}$
in the nucleon rest frame, $d_2$ and $f_2$ can be related to the electric
and magnetic color polarizabilities defined as~\cite{Shuryak:1981kj,Stein:1995si,Ji:1993sv,Ji:1997gs}:
\vspace{-0.3cm}
\begin{equation}
\vspace{-0.3cm}
\chi_E\ 2M_t^2\vec{J}=\langle N|\ \vec{j}_a\times\vec{E}_a\ |N\rangle\ ,\ \ \
\chi_{B}\ 2M_t^2\vec{J}=\langle N|\ j_a^0\ \vec{B}_a\ |N\rangle\ ,
\nonumber
\end{equation}
where $\vec{J}$ is the nucleon spin, $j_a^{\mu}$ is the
quark current, $\vec{E}_a$ and $\vec{B}_a$ are the color electric 
and magnetic fields, respectively. They relate to $d_2$ and $f_2$ as:
\vspace{-0.3cm}
\begin{equation}
\vspace{-0.3cm}
\chi_E(Q^2)=\frac{2}{3}\left(2d_2(Q^2)\ +\ f_2(Q^2)\right),\ \ \
\chi_{B}(Q^2)=\frac{1}{3}\left(4d_2(Q^2)\ -\ f_2(Q^2)\right).
\label{eq:chi}
\end{equation}

\section{World data and global analyses \label{sec:data}}
\subsection {Experiments and world data \label{sec: world data}} 

As mentioned in Section~\ref{LC dominance and LF quantization}, a hadron non-zero anomalous 
magnetic moment requires a non-zero quark transverse OAM~\cite{Brodsky:1980zm, Burkardt:2005km}
and thus, information on the nucleon's internal angular momenta can be traced back at least as far as the 
1930s with Stern and Frisch's discovery of the proton anomalous magnetic moment~\cite{Stern:1933}. However,  
the first direct experimental information on the internal components making the nucleon spin
came from doubly-polarized DIS experiments. 
They took place at SLAC, CERN, DESY, and are continuing at JLab and CERN.
The development of polarized beams~\cite{Sinclair:2007ez} and targets~\cite{Goertz:2002vv} has enabled this program. 
It started at SLAC in the late 1970s and early 1980s with the pioneering E80 and 
E130 experiments~\cite{Alguard:1976bm, Alguard:1978gf}. It continued in the 1990s with 
E142~\cite{Anthony:1996mw}, E143~\cite{Abe:1994cp} -- which also forayed in the resonance region -- 
E154~\cite{Abe:1997cx,Abe:1997qk}, E155~\cite{Anthony:1999rm} and E155x~\cite{Anthony:1999py}
(an extension of E155 focused on $g_2$ and $A_2$).  
The CERN experiments started in 1984 with EMC~\cite{Ashman:1987hv} 
-- whose results triggered the ``spin crisis" -- continued with
SMC~\cite{Adeva:1993km}, and are ongoing with COMPASS~\cite{Alexakhin:2006oza}.
At the DESY accelerator, the HERMES experiment~\cite{Ackerstaff:1997ws, Airapetian:2006vy} ran from 1995 to 2007.
The inclusive program of these experiments focused on the Bjorken sum rule (Eq.~(\ref{eq:genBj})) and the longitudinal
nucleon spin structure, although $g_2$ or $A_2$, and resonance data were also taken. HERMES and COMPASS
also provided important SIDIS and GPDs data. 
The JLab doubly polarized inclusive program started in 1998 with a first
set of experiments in the resonance region: E94-010~\cite{Amarian:2002ar} and 
EG1a~\cite{Yun:2002td} measured the generalized GDH sum (Eqs.~(\ref{eq:gdhsum_def1}) or (\ref{eq:gdhsum_def2})), 
$g_1$ and $g_2$ and their moments for  $0.1<Q^2< 1$~GeV$^2$.
Then, the RSS experiment~\cite{Wesselmann:2006mw, Slifer:2008xu} 
covered the resonance domain at $\left\langle Q^2\right\rangle = 1.3$~GeV$^2$. 
In early 2000, another set  of experiments was performed: EG1b~\cite{Dharmawardane:2006zd, Prok:2008ev, Bosted:2006gp, Fersch:2017qrq} 
extended EG1a up to $Q^2= 4.2$~GeV$^2$ with improved statistics, E99-117~\cite{Zheng:2003un} 
covered the high-$x_{Bj}$ region at $Q^2=5$~GeV$^2$, 
E97-103~\cite{Kramer:2005qe}  measured $g_2^n$ in the DIS, 
and E01-012~\cite{Solvignon:2013yun, Solvignon:2008hk}  
covered the resonance region at $Q^2>1$~GeV$^2$. 
Furthermore, E97-110~\cite{E97110} and EG4~\cite{Adhikari:2017wox} 
investigated $\Gamma_1$, $\Gamma_2$, $g_1$ and $g_2$ in the $Q^2 \to 0$ limit. 
EG1dvcs~\cite{Prok:2014ltt} extended EG1 
to $Q^2= 5.8$~GeV$^2$ with another large improvement in  statistics, 
and the SANE experiment~\cite{Rondon:2015mya} focused on $g_2$ and 
the twist~3 moment $d_2$ up to $Q^2= 6.5$~GeV$^2$ and $0.3<x_{Bj}<0.85$. 
Finally, E06-014 precisely measured $d_2^n$ 
at $Q^2= 3.2$ and 4.3 GeV$^2$~\cite{Posik:2014usi, Parno:2014xzb}.
These JLab experiments are inclusive, although EG1a~\cite{DeVita:2001ue}, 
EG1b~\cite{Chen:2006na}, EG4~\cite{Zheng:2016ezf} and EG1dvcs~\cite{Seder:2014cdc} 
also provided semi-inclusive, exclusive and  DVCS data.
The JLab polarized $^3$He SIDIS program
comprised  E06-010/E06-011~\cite{Qian:2011py}, while E07-013~\cite{Katich:2013atq} 
used spin degrees of freedom to study
the effect of two \emph{hard} photon exchange in DIS.
(Experiments using polarized beam on unpolarized protons and measuring the proton recoil polarization 
had already revealed the importance of such reaction for the proton electric form factor~\cite{Jones:1999rz}.)
Data at $Q^2=0$ or low $Q^2$ from  MIT-Bates, LEGS, MAMI and TUNL
also exist. 

\begin{figure}
\center
\protect\includegraphics[scale=0.35]{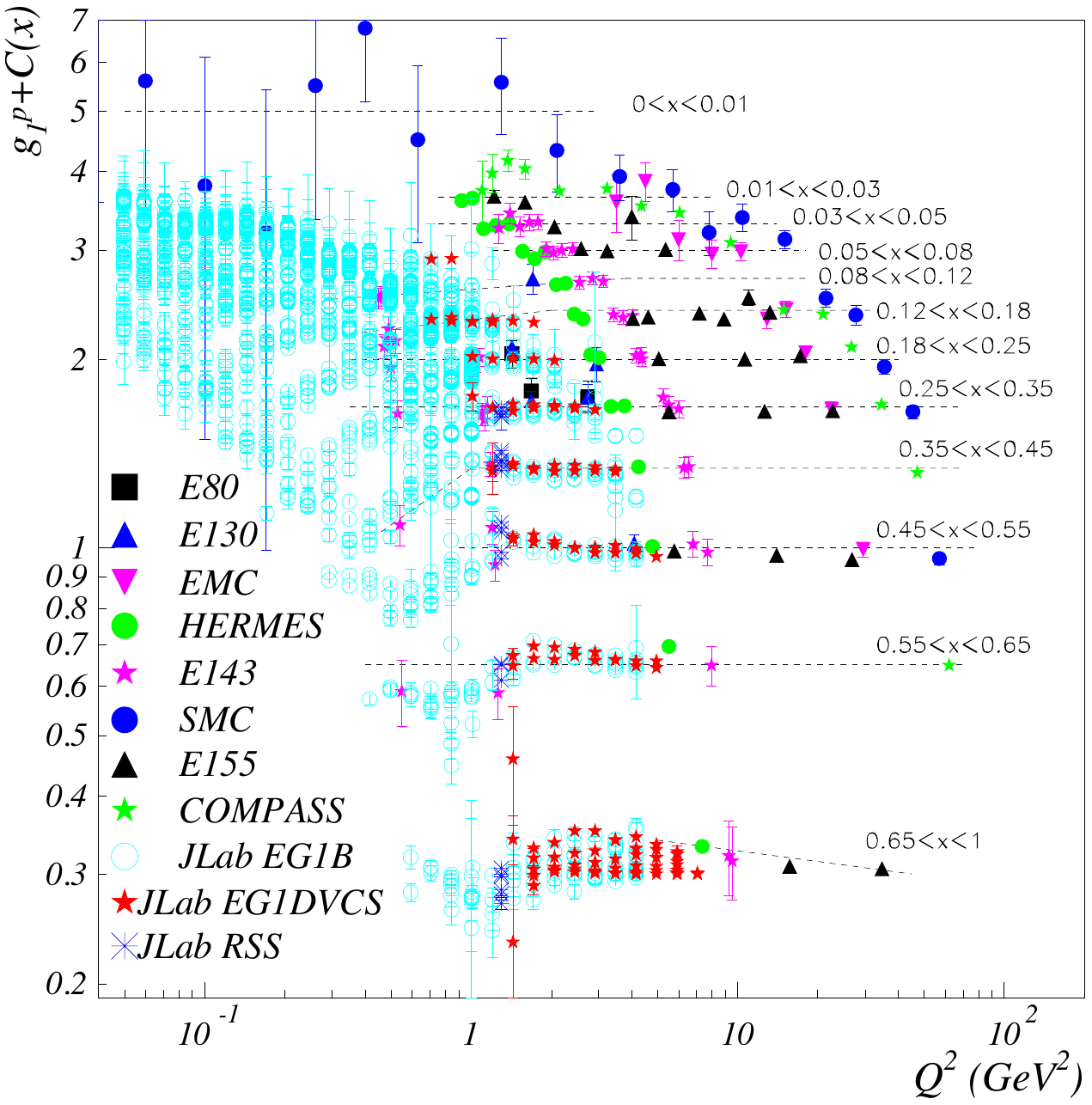}
\protect\includegraphics[scale=0.35]{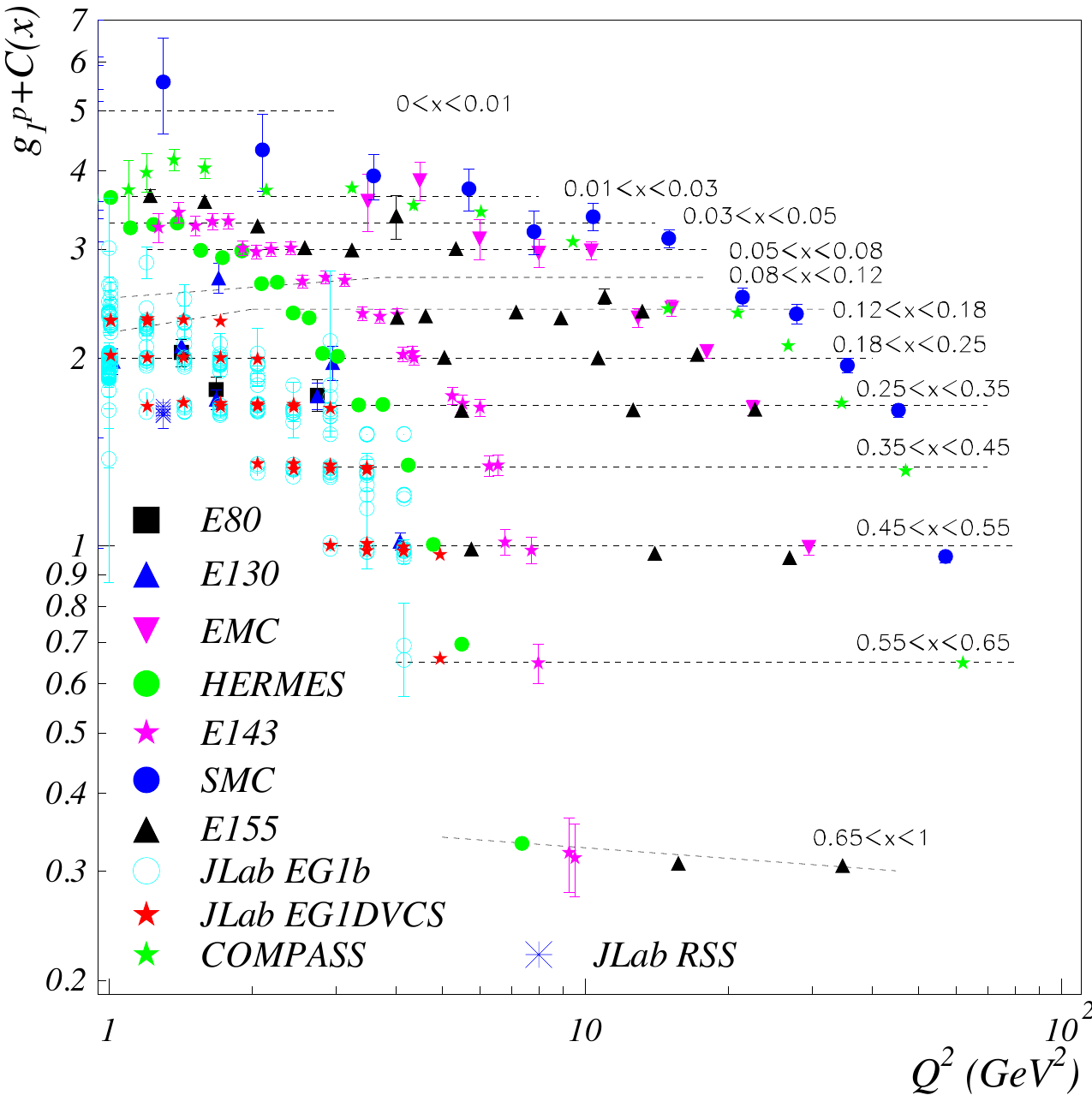}
\vspace{-1.6cm}
\caption{\label{fig:g1p} \small{Left: Available world data on $g_1^p$ as of 2017. An offset $C(x_{Bj})$ is added 
to $g_1^p$ for visual clarity. Only two of the four energies of
experiment EG1b are shown. The dotted lines mark a particular $x_{Bj}$ bin and do not represent
the $Q^2$-evolution. 
Right: Same as left but for DIS data only.
Despite the modest energy, part of JLab's data reaches the DIS and, thanks to 
JLab's high luminosity, they contribute significantly to the global data. 
}}
\end{figure}

These experiments, their observables and kinematics are listed in Table~\ref{table_exp}. 
The world data for $g_1^p$, as of 2017, is shown in Fig.~\ref{fig:g1p}.
Not included in Table~\ref{table_exp} because they are not discussed in this review, are the doubly or singly 
polarized inclusive experiments measuring nucleon form factors~\cite{Pacetti:2015iqa}, including 
the strange ones~\cite{Armstrong:2012bi}, or probing the resonance and DIS~\cite{Armstrong:2012bi}
or the Standard Model~\cite{Wang:2014bba} using parity violation.
\begin{table}
\footnotesize
\caption{\label{table_exp}\small Lepton scattering experiments on the nucleon spin structure and their kinematics. 
The column ``Analysis" indicates wether the analysis was primarily 
conducted in terms of asymmetries ($A_{1,2}$, or single spin asymmetry) or of cross-sections 
($g_{1,2}$), and if transverse data were taken in addition to the longitudinal data.}
\hspace{-0.2cm}
\begin{tabular}{|p{3.2cm}|p{2.8cm}|p{0.9cm}|p{2.1cm}|p{1.6cm}|p{2.8cm}|p{1.8cm}|}
\hline 
Experiment & Ref. & Target & Analysis & $W$ (GeV)& $\hspace{1cm}x_{Bj}$ & $Q^{2}$  (GeV$^{2}$)\tabularnewline
\hline
\hline 
E80 (SLAC) & \cite{Alguard:1976bm} & p & $A_{1}$ & 2.1 to 2.6  & 0.2 to 0.33 & 1.4 to 2.7 \tabularnewline
\hline 
E130 (SLAC) & \cite{Alguard:1978gf} & p & $A_{1}$ & 2.1 to 4.0  & 0.1 to 0.5 & 1.0 to 4.1 \tabularnewline
\hline 
EMC (CERN) & \cite{Ashman:1987hv} & p & $A_{1}$ & 5.9 to 15.2  & $1.5\times10^{-2}\mbox{ to }0.47$ & 3.5 to 29.5 \tabularnewline
\hline 
SMC (CERN) & \cite{Adeva:1993km} & p, d & $A_{1}$ & 7.7 to 16.1  & $10^{-4}\mbox{ to }0.482$ & 0.02 to 57 \tabularnewline
\hline 
E142 (SLAC) & \cite{Anthony:1996mw} & $^{3}$He & $A_{1}$, $A_{2}$ & 2.7 to 5.5  & $3.6\times10^{-2}\mbox{ to }0.47$ & 1.1 to 5.5 \tabularnewline
\hline 
E143 (SLAC) & \cite{Abe:1994cp} & p, d & $A_{1}$, $A_{2}$ & 1.1 to 6.4  & $3.1\times10^{-2}\mbox{ to }0.75$ & 0.45 to 9.5 \tabularnewline
\hline 
E154 (SLAC) & \cite{Abe:1997cx,Abe:1997qk} & $^{3}$He & $A_{1}$, $A_{2}$ & 3.5 to 8.4  & $1.7\times10^{-2}\mbox{ to }0.57$ & 1.2 to 15.0 \tabularnewline
\hline 
E155/E155x (SLAC) & \cite{Anthony:1999rm,Anthony:1999py} & p, d & $A_{1}$, $A_{2}$ & 3.5 to 9.0  & $1.5\times10^{-2}\mbox{ to }0.75$ & 1.2 to 34.7 \tabularnewline
\hline 
{\scriptsize HERMES (DESY)} & \cite{Ackerstaff:1997ws, Airapetian:2006vy} & p, $^{3}$He & $A_{1}$ & 2.1 to 6.2  & $2.1\times10^{-2}\mbox{ to }0.85$ & 0.8 to 20 \tabularnewline
\hline 
E94010 (JLab)& \cite{Amarian:2002ar}  & $^{3}$He & $g_{1}$, $g_{2}$ & 1.0 to 2.4  & $1.9\times10^{-2}\mbox{ to }1.0$ & 0.019 to 1.2 \tabularnewline
\hline 
EG1a (JLab) & \cite{Yun:2002td} & p, d & $A_{1}$ & 1.0 to 2.1  & $5.9\times10^{-2}\mbox{ to }1.0$ & 0.15 to 1.8 \tabularnewline
\hline 
RSS (JLab) & \cite{Wesselmann:2006mw, Slifer:2008xu}  & p, d & $A_{1}$, $A_{2}$ & 1.0 to 1.9  & $0.3\mbox{ to }1.0$ & 0.8 to 1.4 \tabularnewline
\hline 
{\scriptsize COMPASS (CERN) DIS} & \cite{Alexakhin:2006oza} & p, d & $A_{1}$ & 7.0 to 15.5  & $4.6\times10^{-3}\mbox{ to }0.6$ & 1.1 to 62.1 \tabularnewline
\hline 
{\scriptsize COMPASS low-$Q^2$}& \cite{Nunes:2016otf} & p, d & $A_{1}$ & 5.2 to 19.1  & $4\times10^{-5}\mbox{ to }4\times10^{-2}$ & 0.001 to 1. \tabularnewline
\hline 
EG1b (JLab) &~\cite{Dharmawardane:2006zd, Prok:2008ev, Bosted:2006gp, Fersch:2017qrq}  & p, d & $A_{1}$ & 1.0 to 3.1  & $2.5\times10^{-2}\mbox{ to }1.0$ & 0.05 to 4.2 \tabularnewline
\hline 
E99-117  (JLab) & \cite{Zheng:2003un} & $^{3}$He & $A_{1}$, $A_{2}$ & 2.0 to 2.5  & $0.33\mbox{ to }0.60$ & 2.7 to 4.8 \tabularnewline
\hline 
E97-103  (JLab) & \cite{Kramer:2005qe} & $^{3}$He & $g_{1}$, $g_{2}$ & 2.0 to 2.5  & $0.16\mbox{ to }0.20$ & 0.57 to 1.34 \tabularnewline
\hline 
E01-012  (JLab) & \cite{Solvignon:2013yun, Solvignon:2008hk} & $^{3}$He & $g_{1}$, $g_{2}$ & 1.0 to 1.8  & $0.33\mbox{ to }1.0$ & 1.2 to 3.3 \tabularnewline
\hline 
E97-110  (JLab) & \cite{E97110}  & $^{3}$He & $g_{1}$, $g_{2}$ & 1.0 to 2.6  & $2.8\times10^{-3}\mbox{ to }1.0$ & 0.006 to 0.3 \tabularnewline
\hline 
EG4  (JLab) & \cite{Adhikari:2017wox} & p, n & $g_{1}$ & 1.0 to 2.4  & $7.0\times10^{-3}\mbox{ to }1.0$ & 0.003 to 0.84 \tabularnewline
\hline 
SANE  (JLab) & \cite{Rondon:2015mya} & p & $A_{1}$, $A_{2}$ & 1.4 to 2.8  & $0.3\mbox{ to }0.85$ & 2.5 to 6.5 \tabularnewline
\hline 
EG1dvcs (JLab) & \cite{Prok:2014ltt} & p & $A_{1}$ & 1.0 to 3.1  & $6.9\times10^{-2}\mbox{ to }0.63$ & 0.61 to 5.8 \tabularnewline
\hline 
E06-014  (JLab) & \cite{Posik:2014usi, Parno:2014xzb} & $^{3}$He & $g_{1}$, $g_{2}$ & 1.0 to 2.9  & $0.25\mbox{ to }1.0$ & 1.9 to 6.9 \tabularnewline
\hline 
E06-010/011  (JLab) & \cite{Qian:2011py} & $^{3}$He & single spin asy. & 2.4 to 2.9  & $0.16\mbox{ to }0.35$ & 1.4 to 2.7 \tabularnewline
\hline 
E07-013  (JLab)  & \cite{Katich:2013atq} & $^{3}$He & single spin asy. & 1.7 to 2.9  & $0.16\mbox{ to }0.65$ & 1.1 to 4.0 \tabularnewline
\hline 
E08-027  (JLab)  & \cite{g2p} & p & $g_{1}$, $g_{2}$ & 1. to 2.1  & $3.0\times10^{-3}\mbox{ to }1.0$ &  0.02 to 0.4 \tabularnewline
\hline
\end{tabular}
\end{table}
\normalsize

Global DIS data analyses~\cite{Ball:1995td}-\cite{Shahri:2016uzl} are discussed next. 
Their primary goal  is to provide the polarized
PDFs $\Delta q(x_{Bj})$ and $\Delta g(x_{Bj})$, as well as their integrals $\Delta \Sigma$ and $\Delta G$, which
enter the spin sum rule, Eq.~(\ref{eq:spin SR}). 
Then, we present the specialized DIS experiments focusing on large $x_{Bj}$.
Next, we review the information on the nucleon spin structure emerging 
from experiments with kinematics below the DIS. 
%
Afterward, we review the parton correlations (\emph{higher-twists}) information obtained with
these low energy data together with the DIS ones and the closely related phenomenon of hadron-parton duality.
Finally, we conclude this section with our present knowledge on the nucleon spin at high energy,
in particular the components of the spin sum rule, Eq.~(\ref{eq:spin SR}), and discuss the origin of their values.
We conclude on the consistency of the data and remaining questions.

\subsection{Global analyses }

DIS experiments are analyzed in the pQCD framework. 
Their initial goal was to test QCD using the Bjorken sum rule, Eq.~(\ref{eq:mu4}). 
After 25 years of studies, it is now checked to almost 
5\% level~\cite{Alekseev:2009ac, Alekseev:2010ub, Adolph:2015saz, Alexakhin:2005iw}. 
Meanwhile, the nucleon spin structure started to be uncovered. Among the main results
of these efforts is the determination of the small contribution of the quark spins $\Delta\Sigma$, Eq.~(\ref{eq:mu2}),
which implies that the quark OAM or/and of the gluon contribution $\Delta G+\mbox{L}_g$ are important.
Global analyses, which now include not only DIS but SIDIS, p-p and $e^+$-$e^-$
collisions provide fits of PDFs and are the main avenue of interpreting the 
data~\cite{Ball:1995td, Leader:2010rb, deFlorian:2009vb, Jimenez-Delgado:2013boa, Nocera:2014gqa}. 
These analysis are typically at NLO in $\alpha_s$, although NNLO has become available recently~\cite{Moch:2014sna}. 
Several groups have carried out such analyses. Beside data, the analyses are constrained by general principles, 
including \emph{positivity constraints} (see Section~\ref{parton model}) and often other constraints such as 
SU(2)$_f$ and SU(3)$_f$ symmetries (see Section~\ref{DIS SR}),
\emph{counting rules}~\cite{Brodsky:1973kr} and integrability ({\it i.e.}, the matrix elements of the axial current 
are always finite).
A crucial difference between the various analyses
is the choice of initial PDF ansatz, particularly for $\Delta g(x_{Bj})$, and of methods
to minimize the bias stemming from such choice, which is the leading contribution
to the systematic uncertainty. Two methods are used to optimize the PDFs starting from
the original ansatz. One is to start from polynomial PDFs and optimize them with respect to the data and general
constraints using Lagrange multipliers or Hessian techniques. 
The other approach determines the best PDFs using neural networks. 
Other differences between analyses are the choice of renormalization schemes 
(recent analyses typically use $\overline{MS}$), of factorization schemes and of
\emph{factorization scale}. Observables are in principle independent of these arbitrary choices but not
in practice because of the necessary truncation of the pQCD series: calculating perturbative 
coefficients at high orders quickly becomes overbearing. Furthermore, pQCD series are \emph{Poincar\'{e} series} 
that diverge beyond an order approximately given by $\pi/\alpha_{s}$. Thus, they
must be truncated at or before this order. However, at the typical scale $\mu^2=5$~GeV$^2$, 
$\pi/\alpha_{s} \approx 11$ so this is currently not a limitation. The truncations make the perturbative approximant of an
observable to retain a dependence on the arbitrary choices made by the DIS analysts. In principle, this dependence
decreases with $Q^2$: at high enough $Q^2$ where the observable is close to the LO value of its perturbative 
approximant, unphysical dependencies should disappear since LO is renormalization 
scheme independent (with some exceptions
however, such as non-zero renormalons~\cite{Deur:2016tte}. Another noticeable example is $\Delta\Sigma$'s perturbative 
approximant which contains a non-vanishing contribution at $Q^2 \to \infty$ from the gluon anomaly, see
Section~\ref{DIS SR}). 
Evidently, at finite $Q^2$, observables also depend on the $\alpha_s$ order at which the analysis is carried out. 
DIS analysis accuracy is limited by these unphysical dependencies.
Optimization methods exist to minimize them. For instance, the  \emph{factorization scale} $\mu$  
can be determined by comparing nonperturbative calculations to their corresponding perturbative 
approximant, see { \it e.g.}, Refs.~\cite{Deur:2014qfa, Deur:2016cxb}. 
That  $\mu$ depends on the renormalization scheme (and of the pQCD order) 
illustrates the discussion: at
N$^3$LO $\mu = 0.87 \pm 0.04$~GeV in the $\overline{MS}$ scheme, $\mu = 1.15 \pm 0.06$~GeV 
in the $MOM$ scheme and $\mu = 1.00 \pm 0.05$~GeV in the $V$ scheme.
Another example of optimization procedure is implementing the renormalization group criterium
that an observable cannot depend on conventions such as the 
renormalization scheme choice. Optimizing a pQCD series is then  achieved by minimizing the 
renormalization scheme dependence. One such approach is the \emph{BLM} procedure~\cite{the:BLM}.
The \emph{Principle of Maximum Conformality} (PMC)~\cite{the:PMC} generalizes it and 
sets unambiguously order-by-order in pQCD the \emph{renormalization scale}, 
{\it i.e.}, the scale at which the renormalization procedure subtracts the ultraviolet divergences 
(often also denoted $\mu$ but not to be confused with the  \emph{factorization scale} just discussed).
By fulfilling renormalization group invariance the PMC provides approximants
independent of the choice of renormalization scheme.

While polarized  DIS directly probes $\Delta q(x_{Bj},Q^2)$, 
$\Delta g(x_{Bj},Q^2)$ is also accessed  through the pQCD evolution
equations, Eq.~(\ref{quark LO evol}). However, the present precision and kinematics coverage of the data do not 
constrain it well. It will be significantly improved by the 
12 GeV spin program at JLab that will cover the largely unconstrained $x_{Bj} >0.6$ region, and
by the polarized EIC (electron-ion collider) that will cover the low-$x_{Bj}$ domain~\cite{Accardi:2012qut}.
(The EIC may also constrain the gluon OAM~\cite{Ji:2016jgn}). But $\Delta g(x_{Bj},Q^2)$ is best accessed \emph{via}
semi-exclusive DIS involving photon-gluon fusion, $\gamma^{*}g \to  q \overline{q}$. 
This was evaluated by the SMC, HERMES and COMPASS experiments. 
Polarized p-p (RHIC-spin) provides other channels that 
efficiently access $\Delta g(x_{Bj},Q^2)$, see Section~\ref{n-n scattering}. 

Global analysis results are discussed in Section~\ref{nucleon spin structure at high energy}
which gives the current picture of the nucleon spin structure at high energy. 
They are listed in Tables~\ref{table Delta Sigma 1}-\ref{table Delta G 1} in  the Appendix.

%

%
\vspace{-0.3cm}
\subsection{PQCD in the high-$x_{Bj}$ domain\label{sec: high-x}} 

The high-$x_{Bj}$ region should be relatively simple: as $x_{Bj}$ grows, the \emph{valence quark}
distribution starts prevailing over the ones of gluons and of 
$q$-$\overline{q}$ pairs materializing from gluons, see Fig.~\ref{fig:pdf_dist}. 
This prevalence allows the use of \emph{constituent quark} models (see page \pageref{CQM})~\cite{Isgur:1978xj}. 
Thus the high-$x_{Bj}$  region is particularly interesting. 
It has been studied with precision by the JLab collaborations E99-117, EG1b, E06-014 and EG1dvcs, 
and by the CERN's COMPASS collaboration. 

This region has been precisely studied only recently since there, unpolarized PDFs (Fig.~\ref{fig:pdf_dist})  are small, which 
entails small cross-sections that, furthermore, have kinematic factors
varying at first order as $1/x_{Bj}$. Thus, early data high-$x_{Bj}$ lacked the 
precision necessary to extract polarized PDF. The high polarized luminosity 
of JLab has allowed to explore this region more precisely.

\subsubsection{$A_1$ in the DIS at high-$x_{Bj}$}\label{pqcd high-x}
Assuming that quarks are in a $S$ state, {\it i.e.}, they have no OAM, a quark carrying all the nucleon momentum
($x_{Bj}\to 1$) must carry  the nucleon helicity~\cite{Farrar:1975yb}. 
This implies $A_1 \xrightarrow[x_{Bj} \to 1] {}1$. This is a rare example of absolute prediction from QCD: generally 
pQCD predicts only  the $Q^2$-dependance of observables, see 
Sections~\ref{sub:MecaDIS} and~\ref{sub:pQCD}. 
(Other examples are the processes involving the chiral anomaly, such as $\pi^0 \to \gamma \gamma$.)
Furthermore, the \emph{valence quarks}
dominance makes known the nucleon wavefunction, see Eq.~(\ref{SU(6) p wavefunction}). 
The  BBS~\cite{Brodsky:1994kg} and LSS~\cite{Leader:2001kh}) global fits 
include these two constraints. The $x_{Bj}$-range where the  $S$-state dominates
is the only significant assumption of these fits which have been improved to include 
the $\left|\mbox{L}_z(x_{Bj})\right|=1$ wavefunction components~\cite{Avakian:2007xa}.
The phenomenological predictions~\cite{Brodsky:1994kg, Avakian:2007xa, Leader:2001kh} for 
$A_1 (x_{Bj} \to 1)$ are thus based on solid premises. Model predictions also exist and
are discussed next. 

\subsubsection{Quark models and other predictions of $A_1$ for high-$x_{Bj}$ DIS}
%
\begin{wrapfigure}{r}{0.44\textwidth} 
\vspace{-0.7cm}
\includegraphics[width=0.44\columnwidth]{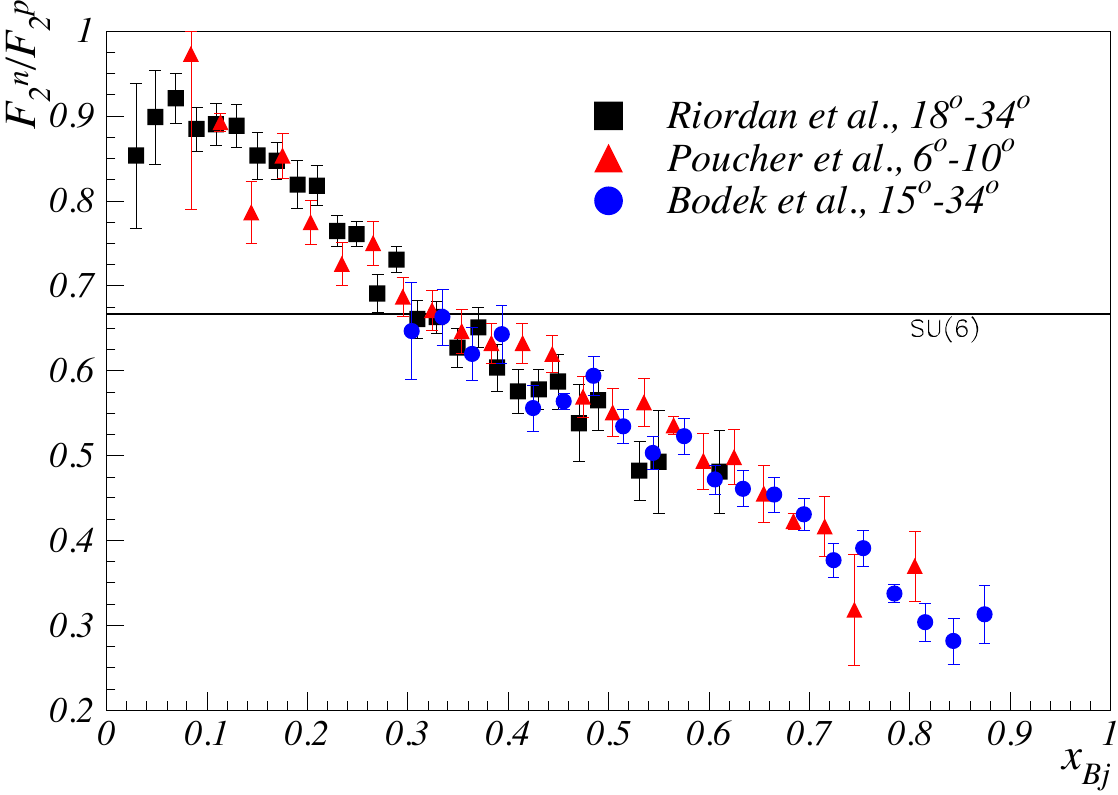}
\vspace{-1.cm}
\caption{\label{fig:f2n/f2p}\small{$F_2^n/F_2^p$ SLAC data~\cite{Bodek:1973dy}.  SU(6) predicts
$F_2^n/F_2^p=2/3$.}
\vspace{-0.6cm}
}
\end{wrapfigure}
Modeling the nucleon as made of three \emph{constituent quarks} is justified in the high-$x_{Bj}$ 
DIS domain since there, \emph{valence quarks} dominate. This finite number of partons and 
the SU(6) flavor-spin symmetry allow one to construct  a simple nucleon wavefunction, 
see Eq.~(\ref{SU(6) p wavefunction}), leading to $A_1^p=5/9$ and $A_1^n=0$. 
However SU(6) is broken, as clearly indicated  {\it e.g.}, by
the nucleon-$\Delta$ mass difference of 0.3 GeV or the failure of the
SU(6) prediction that $F_2^n/F_2^p=2/3$, see Fig.~\ref{fig:f2n/f2p}.
The one-gluon exchange  (pQCD ``hyperfine interaction'', see page \pageref{CQM}) 
breaks SU(6) and can account for the nucleon-$\Delta$ mass difference.
It predicts the same  $x_{Bj}\to1$ limits as for pQCD:
$A_1^p= A_1^n=1$. 
A  prediction
of the \emph{constituent quark} model improved with the hyperfine interaction~\cite{Isgur:1998yb}
is shown in Fig.~\ref{fig:A1}.

Another approach to refine the \emph{constituent quark} model is using 
chiral \emph{constituent quark} models~\cite{Manohar:1983md}.
Such models assume a $\approx 1$~GeV scale for chiral symmetry breaking,  
significantly higher than $\Lambda_s$ ($0.33$~GeV in the $\overline{MS}$ scheme) 
and use an effective Lagrangian~\cite{Weinberg:1978kz}  with
\emph{valence quarks} interacting \emph{via} Goldstone bosons as 
effective degrees of freedom. The models include \emph{sea quarks}. 
$x_{Bj}$-dependence is  included phenomenologically in recent models, {\it e.g.}, in the 
prediction~\cite{Dahiya:2016wjf}.

Augmenting quark models with meson clouds provides another possible SU(6) breaking 
mechanism~\cite{Myhrer:2007cf, Myhrer:1988ap}.
Ref.~\cite{Signal:2017cds} compares $A_1$ predictions with this approach
and that of the ``hyperfine" mechanism.

Other predictions for $A_1$  at high-$x_{Bj}$ exist and are shown in Fig.~\ref{fig:A1}. They are:

\noindent ~$\bullet$ The statistical model of  Ref.~\cite{Bourrely:2001du}. It describes the nucleon as fermionic and bosonic gases 
in equilibrium at an empirically determined temperature;

\noindent ~$\bullet$ The hadron-parton duality (Section~\ref{sec:duality}). It relates 
well-measured baryons form factors (elastic or $\Delta(1232)~3/2^+$ reactions, all at high-$x_{Bj}$) to DIS
structure functions at the same $x_{Bj}$~\cite{Close:2003wz}. Predictions depend on the mechanism
chosen to break SU(6), with two examples shown in Fig.~\ref{fig:A1};

\noindent ~$\bullet$ Dyson-Schwinger Equations with contact or realistic interaction. 
They predict $A_1(1)$ values significantly smaller than pQCD~\cite{Roberts:2011wy};

\noindent ~$\bullet$ The bag model of Boros and Thomas, in which three free quarks are confined in a sphere
of nucleon diameter. Confinement is provided by the boundary conditions
requiring that the quark vector current cancels on the sphere surface~\cite{Boros:1999tb}.

\noindent ~$\bullet$ The quark model of Kochelev~\cite{Kochelev:1997ux} in which the quark polarization
is affected by instantons representing non-perturbative fluctuations of gluons. 

\noindent ~$\bullet$ The chiral soliton models of Wakamatsu~\cite{Wakamatsu:2003wg} and 
Weigel \emph{et al.}~\cite{Weigel:1996kw} in which the quark degrees of freedom explicitly generate
the hadronic chiral soliton properties of the Skyrme nucleon model.

\noindent ~$\bullet$ The quark-diquark model of Cloet \emph{et al.}~\cite{Cloet:2005pp}.

\subsubsection{ $A_1$ results } 
\begin{figure}
\center
\includegraphics[scale=0.4]{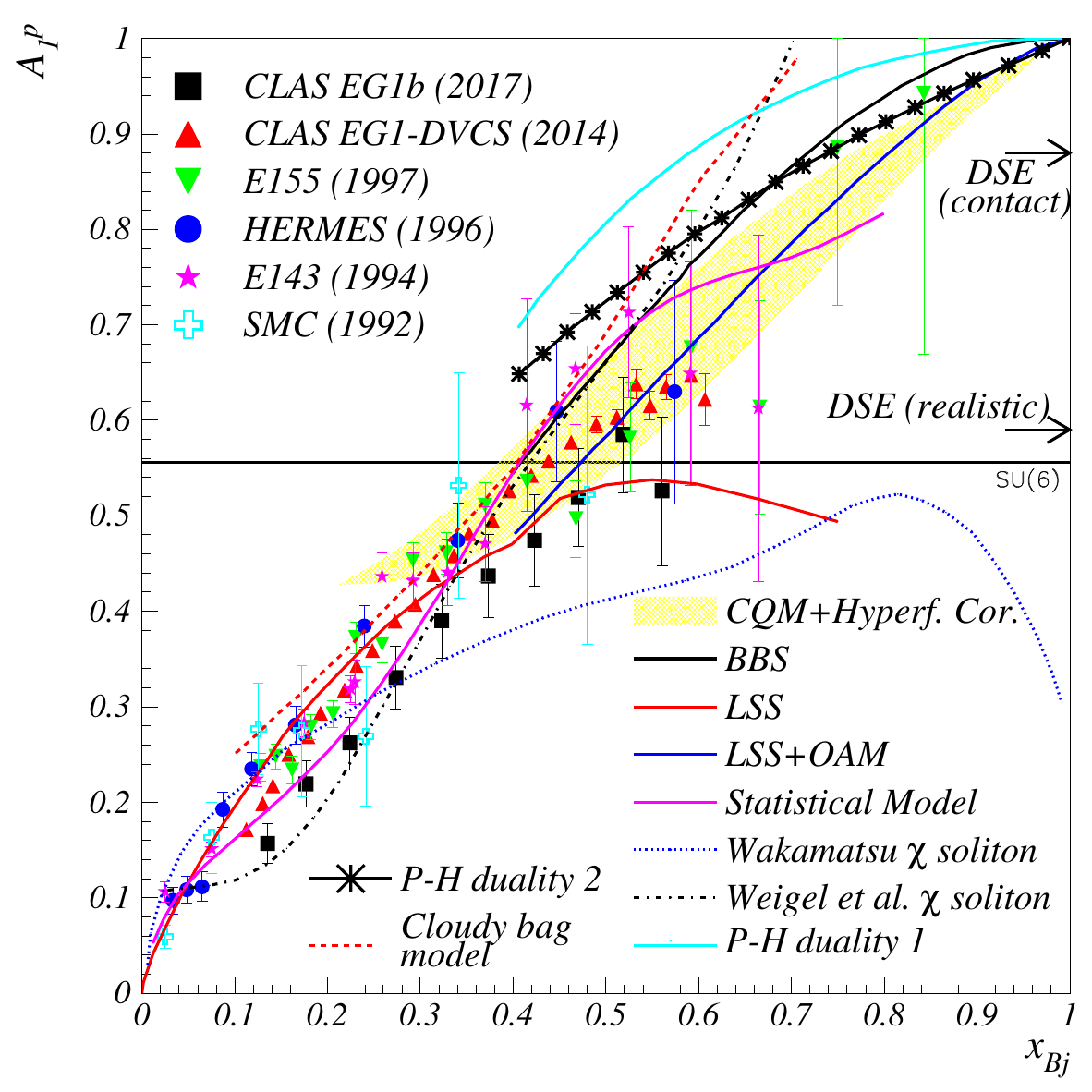}\includegraphics[scale=0.4]{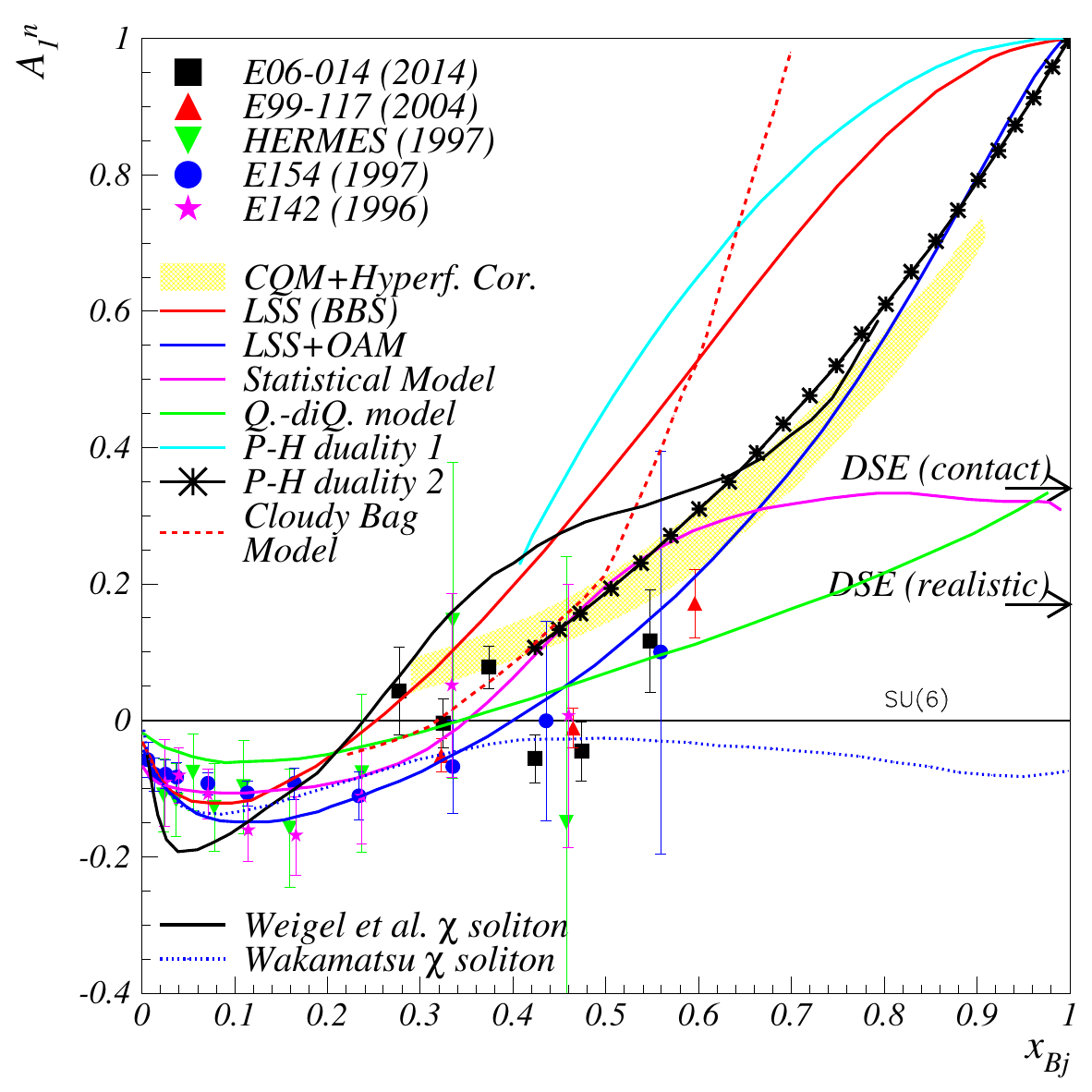}
\vspace{-0.6cm}
\caption{\label{fig:A1}\small{$A_1$ DIS data on the proton (left) and neutron (right).
The $Q^2$ values of the various results are not necessarily the same, but
$A_1$'s $Q^2$-dependence is weak.}}
\vspace{-0.6cm}
\end{figure}
Experimental results on $A_1$~\cite{Anthony:1999rm, Anthony:1996mw,  Dharmawardane:2006zd, 
Prok:2008ev, Fersch:2017qrq, Zheng:2003un,  Prok:2014ltt, Parno:2014xzb} 
are shown in Fig.~\ref{fig:A1}. They  confirm that SU(6), whose prediction is 
shown by the flat lines in Fig.~\ref{fig:A1}, is broken. 
The  $x_{Bj}$-dependence of $A_1$ is well reproduced by the \emph{constituent quark} model
with ``hyperfine" corrections. The systematic shift  for $A_1^n$ at
$x_{Bj}<0.4$ may be a \emph{sea quark} effect.
The BBS/LSS fits to pre-JLab data disagrees with these data. The fits are constrained by pQCD but 
assume no quark OAM. Fits including it~\cite{Avakian:2007xa} agree with the
data, which suggests the importance of the quark OAM. 
However, the relation between the effect of states $\left|\mbox{L}_z(x_{Bj})\right|=1$ 
at high $x_{Bj}$ and $\Delta L$ in Eq.~(\ref{eq:spin SR}) remains to be elucidated. 
To solve this issue, the nucleon wavefunction at low $x_{Bj}$ must be known. 
While the data have excluded some of the models (bag model~\cite{Boros:1999tb}, or specific SU(6) breaking mechanisms 
in the duality approach), high-precision data at higher $x_{Bj}$ are needed to test the remaining predictions.
Such data will be taken at JLab in 2019~\cite{large-x A_1 12 GeV exps}.

\subsection{Results on  the polarized partial cross-sections $\sigma_{TT}$ and $\sigma_{LT}'$\label{sec:g1, g2, stt, slt}} 

The pairs of observables ($g_1$, $g_2$), ($A_1$, $A_2$), or ($\sigma_{TT}$, $\sigma_{LT}'$) 
all contain identical spin information. 
$A_1$  at high-$x_{Bj}$ was discussed in the previous section.
The $g_1$ DIS data at smaller $x_{Bj}$ are discussed in the Section~\ref{nucleon spin structure at high energy}, 
and the  $g_2$  data are discussed in Section~\ref{sec:g2-g2ww}. 
Here, $\sigma_{TT}$ and $\sigma_{LT}'$, Eq.~(\ref{sigmaTT}), are discussed. 

Data on $\sigma_{TT}$ and ${\sigma}_{LT}'$ on $^3$He are available 
in the strong-coupling QCD region~\cite{Amarian:2002ar, E97110}
for $0.04<Q^2<0.90$~GeV$^2$ and $0.9<W<2$~GeV.  Neutron data are unavailable
since for $x_{Bj}$-dependent quantities such as $g_1$ or $\sigma_{TT}$, 
there is no known accurate method to extract the neutron from $^3$He. 
 Yet, since in $^3$He, protons contribute little to polarized observables, the results of 
Refs.~\cite{Amarian:2002ar, E97110} suggest how neutron data may look like. Neutron information
can be extracted for moments, see Sections~\ref{GDHsum} and \ref{sec:Spin polarizabilities}.

A large trough is displayed at the $\Delta(1232)~3/2^+$ resonance by $\sigma_{TT}$. 
It is also present for other resonances, but not as marked. 
The $\Delta(1232)~3/2^+$ dominates because it is the lightest resonance (see Eq.~(\ref{sigmaTT})) 
and because its spin 3/2 makes the nucleon-$\Delta$ transition largely 
transverse. \label{sub:sigmaTT} Since $\sigma_{TT} = (\sigma_{T,1/2} - \sigma_{T,3/2}) /2$,
where {\scriptsize 1/2} and  {\scriptsize 3/2} refer to the spin of the intermediate state, here the  $\Delta(1232)~3/2^+$, 
$\sigma _{TT}$ is maximum and negative. At large $Q^2$, chiral symmetry is restored, which forbids
spin-flips and makes $\sigma _{T,1/2}$ dominant. 
This shrinkage of the $\Delta(1232)~3/2^+$ trough is seen in the 
$1 \leq Q^2 \leq 3.5$~GeV$^2$ data used to study duality, Section~\ref{sec:duality}.
All this implies that at low $Q^2$ 
the $\Delta(1232)~3/2^+$ contribution dominates the generalized GDH integral ($\propto\int\sigma_{TT}/\nu\, d\nu$),
a dominance further amplified by the $1 / \nu$ factor in the integral.
This latest effect is magnified in higher moments, such as those of 
generalized polarizabilities, Eqs.~(\ref{eq:gamma_0}) and (\ref{eq:delta_LT SR}).

${\sigma}_{LT}'$ is rather featureless 
compared to $\sigma_{TT}$ and in particular shows no structure at the
$\Delta(1232)~3/2^+$ location. It confirms that the nucleon-to-$\Delta$ transition occurs mostly via spin-flip
(magnetic dipole transition) induced by
transversely polarized  photons. The longitudinal photons contributing little,  the 
longitudinal-transverse interference cross-section ${\sigma}_{LT}'$ is almost zero.
At higher $W$, ${\sigma}_{LT}'$ becomes distinctly positive.

\subsection{Results on the generalized GDH integral}

\label{GDHsum} The generalized GDH integral $I_{TT}(Q^2)$, Eq.~(\ref{eq:gdhsum_def1}),
was measured for the neutron and proton at DESY (HERMES)~\cite{Airapetian:2000yk}  and 
JLab~\cite{Adhikari:2017wox, Amarian:2002ar, E97110}. 
The measurements cover the energy range from the pion production threshold up to typically $ W \approx 2.0$~GeV. 
The higher-$W$ contribution is estimated with parameterizations, {\it e.g.}, that of Ref.~\cite{Bianchi:1999qs}. 
%
%
%
At low $ Q^2$, $I_{TT}$ can be computed using $\chi$PT~\cite{Bernard:2002bs, 
Bernard:2002pw, Ji:1999pd, Ji:1999mr, Kao:2002cp, Bernard:2012hb, Lensky:2014dda}. 
The Ji-Kao-Lensky \emph{et al.} calculations~\cite{Ji:1999pd, Kao:2002cp, Lensky:2014dda}
and data agree, up to about $Q^2=0.2$~GeV$^2 $. After this, the calculation uncertainties become too large for a 
relevant comparison. The Bernard \emph{et al.} calculations and 
data~\cite{Bernard:2002pw, Bernard:2002bs, Bernard:2012hb} also agree, although marginally. 
The MAID model underestimates the data~\cite{Drechsel:1998hk}.
($I_{TT}(Q^2)$ constructed with MAID is integrated only up to $W\le2$~GeV
and thus must be compared to data without large-$W$ extrapolation. 
The extrapolation of the p+n data~\cite{Adhikari:2017wox} together with the proton GDH sum rule 
world data~\cite{Ahrens:2001qt} yield $I_{TT}^{n}(0)= -0.955\pm 0.040\,(stat) \pm 0.113 \,(syst)$, which agrees 
with the sum rule expectation.

\subsection{Moments of $g_1$ and $g_2$ \label{sec:Gamma1s} } 

 \subsubsection{Extractions of the $g_1$ first moments}
 
\begin{figure}[ht!]
 \center
\includegraphics[scale=0.4]{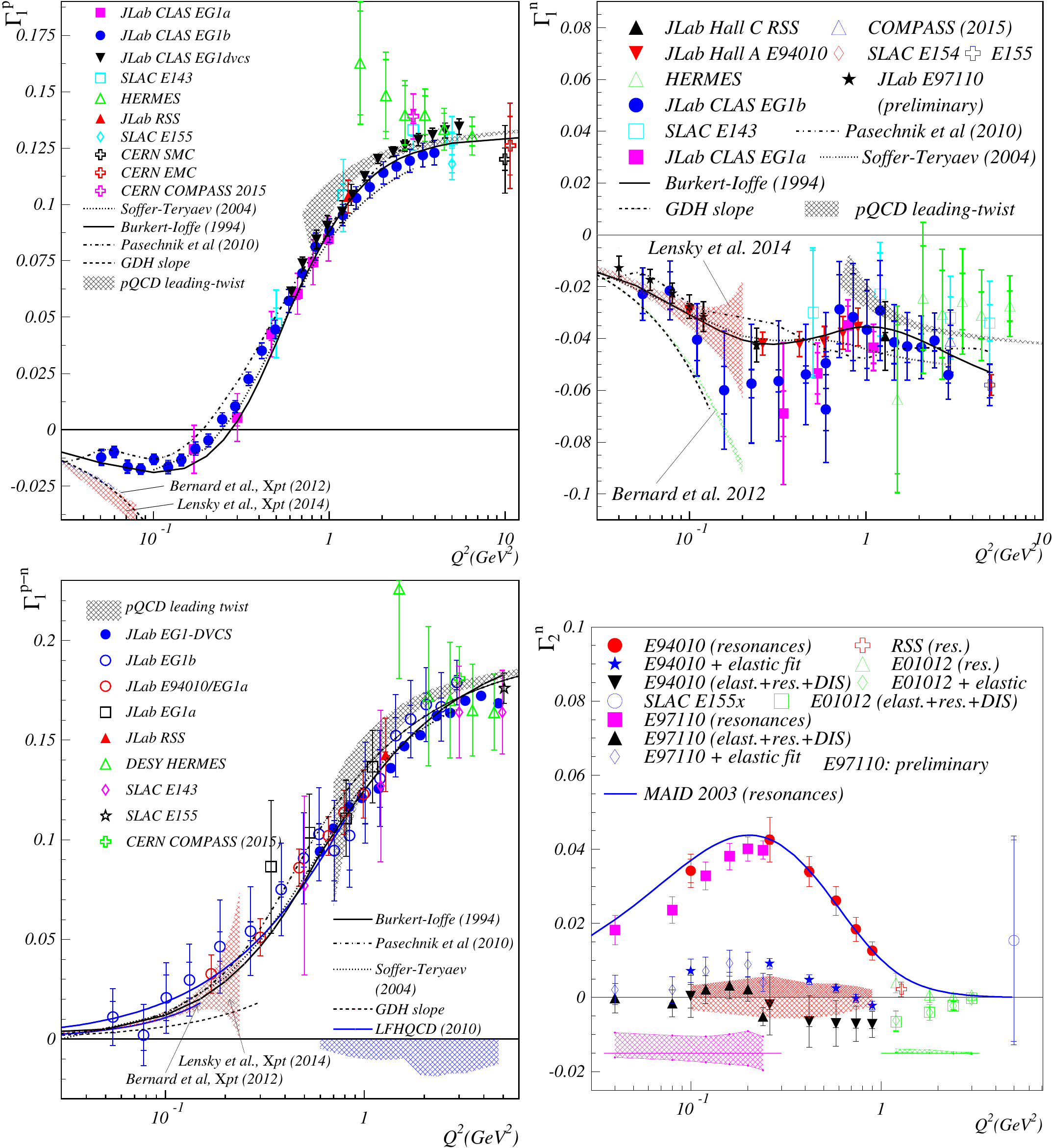}
\vspace{-0.3cm}
\caption{\small The moments {$\Gamma_1^p$ (top left), {$\Gamma_1^n$
(top right) and the Bjorken integral (bottom left), all without elastic contribution. The 
derivatives at $ Q^2 = 0$ are predicted by the GDH sum rule.
In the DIS, the \emph{leading-twist} pQCD evolution is shown by the gray band. 
Continuous lines and bands at low $Q^2$ are $\chi$PT predictions. 
$\Gamma_2^n$, with and without elastic contribution, is shown on the 
lower right panel wherein the upper bands are experimental 
systematic uncertainties. The lower bands in the figure are the systematic 
uncertainties from the unmeasured part below $x_{Bj,min}$.
($ \Gamma_2^p$ is not shown since only two points, from E155x and RSS, are presently available.)
The Soffer-Teryaev~\cite{Soffer:2004ip}, Burkert-Ioffe~\cite{Burkert:1992tg}, Pasechnik 
\emph{et al.}~\cite{Pasechnik:2010fg} and MAID~\cite{Drechsel:1998hk} 
models are phenomenological parameterizations.
}}}
\label{fig:gamma1pn} 
\end{figure}

\noindent \textbf{$\Gamma_1^p$ and $\Gamma_1^n$  moments:}
The measured $ \Gamma_1(Q^2)$ is constructed by integrating $g_1$
from $x_{Bj,min}$ up to the pion production threshold. $x_{Bj,min}$, the minimum $x_{Bj}$ reached,
depends on the beam energy and minimum detected scattering angle for a given $Q^2$ point.
Table~\ref{table_exp} on page \pageref{table_exp} provides these limits. 
When needed, contributions below $x_{Bj,~min}$ are estimated using low-$x_{Bj}$ 
models~\cite{Bianchi:1999qs, Bass:1997fh}.
For the lowest $Q^2$, typically below the GeV$^2$ scale, the large-$x_{Bj}$ 
contribution (excluding elastic) is also added when it is not measured. 
The data for $ \Gamma_1$, shown in Fig.~\ref{fig:gamma1pn},
are from SLAC~\cite{Anthony:1996mw}-\cite{Anthony:1999py},
CERN~\cite{Ashman:1987hv, Adeva:1993km}-\cite{Ageev:2007du, Alekseev:2010hc, 
Alexakhin:2005iw, Alekseev:2009ac}-\cite{Adolph:2015saz},
DESY~\cite{Airapetian:2000yk} and
JLab~\cite{Amarian:2002ar}-\cite{Prok:2008ev, Fersch:2017qrq, 
Solvignon:2013yun,  Solvignon:2008hk}-\cite{Prok:2014ltt, Deur:2004ti, Deur:2008ej}.

\noindent \textbf{Bjorken sum $\Gamma_1^{p-n}$:}
The proton and neutron (or deuteron) data can be combined to form the isovector moment  $ \Gamma_1^ {p -n}$.
The Bjorken sum rule predicts that 
{\vspace{-0.1cm}{$ \Gamma_1^ {p -n} \xrightarrow[Q^2 \to \infty] {}  g_A/6$}}~\cite{Bjorken:1966jh}. 
The prediction is generalized to finite $Q^2$ using OPE, resulting in a relatively 
simple leading--twist $Q^2$-evolution in which only non-singlet coefficients remain, see Eq.~(\ref{eq:genBj}).
The sum rule has been experimentally validated,  most precisely by E155~\cite{Anthony:1999rm}:
$\Gamma_1^{p-n}=0.176 \pm 0.003 \pm 0.007$ at $Q^2$ = 5 GeV$^2$, while the sum rule prediction at the same
$Q^2$ is $\Gamma_1^{p-n} = 0.183 \pm  0.002$. $\Gamma_1^{p-n}$ was first measured by 
SMC~\cite{Adeva:1993km} and then E143~\cite{Abe:1994cp},  E154~\cite{Abe:1997cx}, 
E155~\cite{Anthony:1999rm} and HERMES~\cite{Airapetian:2000yk}. 
Its $Q^2$-evolution was  mapped at JLab~\cite{Fersch:2017qrq, Deur:2004ti, Deur:2008ej}.
The latest measurement (COMPASS) yields 
$\Gamma_1^{p-n}= 0.192 \pm 0.007$ (stat) $\pm 0.015$ 
(syst)~\cite{Alekseev:2009ac, Alekseev:2010ub, Adolph:2015saz, Alexakhin:2005iw, Alekseev:2010hc}.
\\
As an isovector quantity,  $ \Gamma_1^ {p -n}$ has no $\Delta(1232)~3/2^+$ resonance contribution. 
This simplifies $\chi$PT calculations, which may remain valid to higher $Q^2$ than typical for 
$\chi$PT~\cite{Burkert:2000qm}. 
In addition, a non-singlet moment is simpler to calculate with LGT since the CPU-expensive 
disconnected diagrams (quark loops) do not contribute. (Yet, the axial charge $g_A$ and the axial form factor $g_A(Q^2)$
remain a challenge for LGT~\cite{Capitani:2012gj} because
of their strong dependence to the lattice volume. Although the calculations are
improving~\cite{Liang:2016fgy}, the LGT situation for $g_A$ is still unsatisfactory.) 
Thus, $ \Gamma_1^ {p -n}$ is especially convenient to test the techniques discussed in Section~\ref{Computation methods}.
As for all moments, a limitation is the impossibility to measure the $x_{Bj} \to 0$ contribution, which would require
infinite beam energy. The Regge behavior $g_1^{p-n}(x_{Bj}) = (x_0/x_{Bj})^{0.22}$ may provide 
an adequate low-$x_{Bj}$ extrapolation~\cite{Bass:1997fh} (see also~\cite{Kirschner:1983di,Bartels:1995iu,Kovchegov:2015pbl}).

\subsubsection{Data and theory comparisons}

At $Q^2 = 0$, the GDH sum rule, Eq.~(\ref{eq:gdh}), predicts $d \Gamma_1/d Q^2 $ 
(see Fig.~\ref{fig:gamma1pn}). At small $Q^2$, $ \Gamma_1(Q^2) $ can be computed using 
$\chi$PT. The comparison between data and $\chi$PT results on moments is given in Table~\ref{xpt-comp}
in which one sees that in most instances, tensions exist between data and claculations of $\Gamma_1$.

\begin{table}
{\small
\caption{Comparison between $\chi$PT results and data for moments. The bold symbols denote moments for which
$\chi$PT was expected to provide robust predictions. ``{\color{blue}{\bf{A}}}" means that data and calculations agree up to at least
$Q^2=0.1$ GeV$^2$, ``{\color{red}{\bf{X}}}" that they disagree and ``-" that no calculation is available.
The  $p + n$  superscript indicates either deuteron data without deuteron break-up channel, 
or proton+neutron moments added together with neutron information either from D or $^3$He.
\label{xpt-comp}}
\vspace{0.30cm}
\begin{tabular}{|c|c|c|c|c|c|c|c|c|c|c|}
\hline 
Ref. & $\Gamma_1^p$ & $\Gamma_1^n$ & $\pmb{\Gamma_1^{p-n}}$ & $\Gamma_1^{p+n}$ &  $\gamma_0^p$ & $\gamma_0^n$ & $\pmb{\gamma_0^{p-n}}$ & $\gamma_0^{p+n}$ & $\pmb{\delta_{LT}^n}$ & $d_2^n$  \tabularnewline
\hline
\hline 
Ji 1999  \cite{Ji:1999pd,Ji:1999mr}  & {\color{red}\bf{X}} & {\color{red}\bf{X}} & {\color{blue}\bf{A}} & {\color{red}\bf{X}} & - & - & - & - & - & - \tabularnewline
\hline 
Bernard 2002 \cite{Bernard:2002bs,Bernard:2002pw} & {\color{red}\bf{X}} & {\color{red}\bf{X}} & {\color{blue}\bf{A}} & {\color{red}\bf{X}} & {\color{red}\bf{X}} & {\color{blue}\bf{A}} & {\color{red}\bf{X}} & {\color{red}\bf{X}} & {\color{red}\bf{X}} & {\color{red}\bf{X}}\tabularnewline
\hline 
Kao 2002 \cite{Kao:2002cp}  & - & - & - & - & {\color{red}\bf{X}} & {\color{blue}\bf{A}} & {\color{red}\bf{X}} & {\color{red}\bf{X}} & {\color{red}\bf{X}} &  {\color{red}\bf{X}}\tabularnewline
\hline 
Bernard 2012 \cite{Bernard:2012hb}  & {\color{red}\bf{X}} & {\color{red}\bf{X}} & {\color{blue}\bf{A}} & {\color{red}\bf{X}} & {\color{red}\bf{X}} & {\color{blue}\bf{A}} & {\color{red}\bf{X}} & {\color{red}\bf{X}} & {\color{red}\bf{X}} & -\tabularnewline
\hline 
Lensky 2014 \cite{Lensky:2014dda} & {\color{red}\bf{X}} & {\color{blue}\bf{A}} &  {\color{blue}\bf{A}} & {\color{blue}\bf{A}} &  {\color{blue}\bf{A}} & {\color{red}\bf{X}} & {\color{red}\bf{X}} & {\color{red}\bf{X}} &  {\color{blue}\bf{$\sim$ A}} &  {\color{blue}\bf{A}}\tabularnewline
\hline
\end{tabular}
}
\vspace{-0.6cm}
\end{table}

%
The models of Soffer-Teryaev~\cite{Soffer:2004ip}, Burkert-Ioffe~\cite{Burkert:1992tg} and Pasechnik  
\emph{et al.}~\cite{Pasechnik:2010fg} agree well with the data, as does the 
LFHQCD calculation~\cite{Brodsky:2010ur}. 
The Soffer-Teryaev model uses
 the weak $Q^2$-dependence of $\Gamma_T=\Gamma_1+\Gamma_2$
to robustly interpolate $\Gamma_T$ between its zero value and known derivative at $Q^2 = 0$
 and its known values  at large $Q^2$. $\Gamma_1$ is 
 obtained from $\Gamma_T$ using the BC sum rule, 
 Eq.~(\ref{eq:bc_noel}), where PQCD radiative corrections and  \emph{higher-twists} are accounted for.
Pasechnik ~\emph{et al.}  improved this model by 
using for the pQCD and \emph{higher-twist} corrections  a strong coupling $\alpha_s$ analytically continued
at low-$Q^2$, which removes the unphysical Landau-pole divergence at $Q^2=\Lambda_s^2$, 
and minimizes {\emph{higher-twist} effects}~\cite{Deur:2016tte}.
This extends pQCD calculations to lower $Q^2$ than typical. 
The improved  $\Gamma_1$ is continued to $Q^2=0$ by
using $\Gamma_1(0)=0$ and $d\Gamma_1(0)/dQ^2$ from the GDH sum rule.
The Burkert-Ioffe model is based on a parameterization of the resonant 
and non-resonant amplitudes~\cite{Burkert:1992yk}, 
complemented with a DIS parameterization~\cite{Anselmino:1988hn} based on vector dominance. 
In LFHQCD, the effective charge $\alpha_{g_1}$  (\emph{viz} the coupling $\alpha_s$ that includes
the pQCD gluon radiations and  \emph{higher-twist} effects of $\Gamma_1^{p-n}$~\cite{Deur:2016tte}) 
is computed and used in the leading order expression of the Bjorken sum to obtain $ \Gamma_1^{p-n}$. 

The \emph{leading-twist} $Q^2$-evolution is shown in Fig.~\ref{fig:gamma1pn} (gray bands). 
The values $ a_8 $ = 0.579, $g_A $ = 1.267 and $ \Delta \Sigma^p $ = 0.15 ($ \Delta \Sigma^n $ = 0.35)  
are used to anchor the $\Gamma_1^{p (n)} $ evolutions, 
see Eq.~(\ref{eq:mu2}). For $ \Gamma_1^{p-n} $, $g_A$ suffices to fix the absolute scale. In all 
cases,  \emph{leading-twist} pQCD follows the data down to surprisingly low $Q^2$, 
exhibiting hadron-parton global duality {\it i.e.}, an overall suppression of  \emph{higher-twists},
see Sections~\ref{sub:HT Extraction} and~\ref{sec:duality}.

\subsubsection{Results on $\Gamma_2$ and on the BC and ELT sum rules}
\noindent \textbf{Neutron results:}
$ \Gamma_2^n(Q^2)$ from E155x~\cite{Anthony:1999py},  E94-010~\cite{Amarian:2002ar}, 
E01-012~\cite{Solvignon:2013yun}, RSS~\cite{Slifer:2008xu} and
E97-110~\cite{E97110} is shown
in Fig.~\ref{fig:gamma1pn}. Except for  
E155x  for which the resonance 
contribution is negligible, measurements comprise essentially
the whole resonance region. This region contributes positively and significantly yielding 
$ \Gamma_2^{n,res.} \approx - \Gamma_1^{n,res.}$,
as expected since there, $g_2 \approx -g_1$ (see Section~\ref{sec:g2-g2ww}). 
The MAID parameterization (continuous line) agrees well with these data.
The elastic contribution, estimated from the parameterization in Ref.~\cite{Mergell:1995bf}, 
is of opposite sign and nearly cancels the resonance contribution, 
as expected from the BC sum rule $\Gamma_2(Q^2)=0$. 
The unmeasured part below $x_{Bj,min}$ is estimated assuming 
$ g_2 = g_2^{WW}$, see Eq.~(\ref{eq:g2ww}). (While at \emph{leading-twist} $g_2^{WW}$
satisfies the BC sum rule, $ \int g_2^{WW} dx = 0 $, the low-$x_{Bj}$ 
contribution is the non-zero partial integral 
$ \int_0^{x_{Bj,min}} g_2(Q^2,y) dy = x_{Bj,min}\big[g_2^{WW}(Q^2,x_{Bj,min}) + 
g_1(Q^2,x_{Bj,min})\big]$.)
The resulting $ \Gamma_2^n $ fulfills the BC sum rule.
The  interesting fact that the elastic contribution nearly cancels that of the resonances  
accounts for the sum rule validity at low and moderate $Q^2$.

\noindent \textbf{Proton results:}
The E155x proton result  ($Q^2 = 5$~GeV$^2$)~\cite{Anthony:1999py} agrees with the BC sum rule:
$ \Gamma_2^p = -0.022 \pm 0.022$ where, as  for the JLab data, a 100\% uncertainty is assumed 
on the unmeasured low-$x_{Bj}$ contribution estimated to be 0.020 using Eq.~(\ref{eq:g2ww}). 
Neglecting  \emph{higher-twists}  for the low-$x_{Bj}$ extrapolation, RSS yields, 
$\Gamma_2^p = \big(-6 \pm8 $(stat)$\pm20$(syst)\big)$\times10^{-4}$ at  
$Q^2 = 1.28$~GeV$^2$~\cite{Slifer:2008xu}, which agrees with the BC sum rule.
Finally $g_2^p$ has been measured at very low $Q^2$~\cite{g2p},
from which $\Gamma_2^p$ should be available soon.

\noindent \textbf{Conclusion:}
Two conditions for the BC sum rule validity are that
1) $ g_2 $ is well-behaved, so that  $ \Gamma_2 $ is finite, and 
2) $ g_2 $ is not singular at $x_{Bj}= 0 $.
The sum rule validation implies that the conditions
are  satisfied. Moreover, since $ g_2^{WW}$ fulfills the sum rule at  large $Q^2$, 
these conclusions can be applied to twist~3 contribution describing the quark-gluon correlations. Finally, since
the sum rule seems verified from $Q^2\sim0$ to 5 GeV$^2$
and since the contributions of \emph{twist}-$\tau$ are $Q^{2-\tau}$-suppressed, 
the conclusion ensuring that the $ g_2 $ function is regular should be true for all
the terms of the twist series that represents $g_2$.

\noindent \textbf{The Efremov-Leader-Teryaev sum rule:} 
The ELT sum rule, Eq.~(\ref{Eq:ELT SR}), is compatible with the current world data. However, 
the recent global PDF fit KTA17~\cite{Shahri:2016uzl} indicates that the sum rule for $n=2$ 
and twist~2 contribution only is violated at $Q^2=5$~GeV$^2$, 
finding $\int_0^1 x \big(g_1 + 2g_2 \big) dx = 0.0063(3) $ rather
than the expected null sum. If this is true, it would suggest a contribution of  \emph{higher-twists} even at $Q^2=5$~GeV$^2$.

\vspace{-0.1cm}
\subsection{Generalized spin polarizabilities $\gamma_0$, $\delta_{LT}$} 
\vspace{-0.1cm}
Generalized spin polarizabilities offer another test of strong QCD calculations. 
Contrary to $\Gamma_1$ or $\Gamma_2$, the kernels of the polarizability integrals, 
Eqs.~(\ref{eq:gamma_0}) and (\ref{eq:delta_LT SR}), have a $1 / \nu^2$ factor that
suppresses the low-$x_{Bj}$ contribution. 
Hence, polarizability integrals converge faster and have smaller low-$x_{Bj}$ uncertainties.
At low $Q^2$, generalized polarizabilities have been calculated
using  $\chi$PT, see Table~\ref{xpt-comp}.
It is difficult to include in these calculations the resonances, in particular $\Delta(1232)~3/2^+$. 
It was however noticed that this excitation is suppressed in $\delta_{LT}$, making it ideal
to test $\chi$PT calculations for which the $\Delta(1232)~3/2^+$ is not included, or included 
phenomenologically~\cite{Kao:2002cp, Bernard:2002bs}.
Measurements of $\gamma_0$ and $\delta_{LT}$ are available for the neutron (E94-010 and E97-110) 
for $0.04<Q^2<0.9$~GeV$^2$~\cite{Amarian:2002ar, E97110}. 
JLab CLAS results are also available for $\gamma_0$ for the proton, neutron and 
deuteron~\cite{Prok:2008ev, Dharmawardane:2006zd, Fersch:2017qrq, Adhikari:2017wox} 
for approximately $0.02<Q^2<3$~GeV$^2$.  
\vspace{-0.1cm}

\subsubsection{Results on $\gamma_0$}  
\vspace{-0.1cm}
The $ \gamma_0^n$ extracted either from $^3$He~\cite{Amarian:2002ar} 
or D~\cite{Dharmawardane:2006zd} agree well with each other. 
The  MAID phenomenological model~\cite{Drechsel:1998hk} agrees with the 
$\gamma_0^n$ data, and so do the $\chi$PT results (Table~\ref{xpt-comp}), except
the recent  Lensky \emph{et al.} calculation~\cite{Lensky:2014dda}.
For $ \gamma_0^p$, the situation is reversed: only Ref.~\cite{Lensky:2014dda} agrees well
with the data, but not the others (including MAID).
This problem motivated an isospin analysis of $ \gamma_0$~\cite{Deur:2008ej} since, e.g. 
axial-vector meson exchanges in the $t-$channel  (short-range interaction) that are not included 
in computations could be important for only one of the isospin components of  $\gamma_0$. 
$\chi$PT calculations disagree with $ \gamma_0^{p +n} $ but  MAID  agrees. 
Alhough the $\Delta(1232)~3/2^+$ is  suppressed in $ \gamma_0^{p-n}$, $\chi$PT disagrees with the  data. 
Thus, the disagreement on $ \gamma_0^p $ and $ \gamma_0^n $ cannot be assigned to the $\Delta(1232)~3/2^+$. 
MAID also disagrees with $ \gamma_0^{p-n} $.   
\vspace{-0.1cm}

\subsubsection{The $\delta_{LT}$ puzzle} 
\vspace{-0.1cm}
Since the $\Delta(1232)~3/2^+$ is suppressed in $\delta_{LT}$, it was
expected that its $\chi$PT calculation would be robust.
However, the $\delta_{LT}^n$ data~\cite{Amarian:2002ar} disagreed 
with the then available $\chi$PT results. 
This discrepancy is known as the ``$\delta_{LT}$ puzzle". 
Like $\gamma_0$, an isospin analysis of $ \delta_{LT}$ may help with this puzzle.
The needed $ \delta_{LT}^p $ data are becoming available~\cite{g2p}.
The second generation of $\chi$PT calculations on 
$ \delta_{LT}^n$~\cite{Bernard:2012hb, Lensky:2014dda} agrees better with the data.
At larger $Q^2$ (5~GeV$^2 $), the E155x data~\cite{Anthony:1999py} agree with a quenched LGT 
calculation~\cite{Gockeler:1995wg, Gockeler:2000ja}. 
At large $Q^2$, generalized spin polarisabilities are expected to scale as
$1/ Q^{6}$, with the usual additional softer dependence from pQCD radiative 
corrections~\cite{Drechsel:2002ar, Drechsel:2004ki}. 
Furthermore, the Wandzura-Wilczek relation, Eq.~(\ref{eq:g2ww}), relates $\delta_{LT}$ to $\gamma_0$:
\vspace{-0.3cm}
\begin{equation}
\vspace{-0.3cm}
\delta_{LT}(Q^2)\to{\frac{1}{3}}\gamma_0(Q^2)\ \ \ {\rm if~} g_2 \approx g_2^{WW} 
\label{eq:relation gam0 deltaLT}
\end{equation}
The available data being mostly at 
$Q^2 <1 $~GeV$^2$, this relation and the scaling law have not been tested yet. 
Furthermore, the signs of the $\gamma_0 $ and $\delta_{LT}$ data disagree with 
Eq.~(\ref{eq:relation gam0 deltaLT}). These facts are not worrisome: 
for $ \Gamma_1 $ and $ \Gamma_2 $,
scaling is observed for $Q^2 \gtrsim1 $~GeV$^2$, when the overall effect of \emph{higher-twists} 
decreases. For higher moments, resonances contribute more, so scaling should begin at larger $Q^2$. 
The violation of Eq.~(\ref{eq:relation gam0 deltaLT}) is consistent with the fact that 
$ g_2 \neq g_2^{WW}$ in the resonance domain, see Section~\ref{sub:g2 in res}.

\subsection{$d_2$ results}
Another combination of second moments, $ d_2$ (Eqs.~(\ref{eq:d2op}) and (\ref{eq:d2mon})), 
is particularly interesting because it is interpreted as part of the transverse confining force acting on
quarks~\cite{Burkardt:2008ps, Abdallah:2016xfk}, see Section~\ref{color pol.}. 
Furthermore, $d_2$ offers another possibility to study the nucleon spin structure at 
large  $Q^2$ since it  has been calculated by 
LGT~\cite{Gockeler:1995wg, Gockeler:2000ja, Gockeler:2005vw} and modeled with
LC wave functions~\cite{Braun:2011aw}.
$d_2$ can also be used 
to study the transition between large and small $Q^2$. $ \overline{d_2}(Q^2)$ is shown in Fig.~\ref{fig:d2} 
(the bar over $ d_2 $ indicates that the elastic contribution is excluded). 
The experimental results are from JLAB  (neutron from 
$^3$He~\cite{Amarian:2002ar, Zheng:2003un, Solvignon:2013yun, Posik:2014usi}
and from D~\cite{Slifer:2008xu}, and proton~\cite{Wesselmann:2006mw}), from
SLAC (neutron from D and proton)~\cite{Anthony:1999py}, and from global analyses
(JAM~\cite{Jimenez-Delgado:2013boa, Sato:2016tuz}, KTA17~\cite{Shahri:2016uzl}), which
contain only DIS contributions.

\subsubsection{Results on the neutron}
At moderate $Q^2$, $ \overline{d_2}^n$ is positive and reaches a maximum at 
$Q^2 \gtrsim0.4 $~GeV$^2$. Its sign is uncertain at large $Q^2$.
At low $Q^2$ the comparison with $\chi$PT is summarized in Table~\ref{xpt-comp}.
MAID agrees with the data. 
That MAID and the RSS datum (both covering only the resonance region) match 
the DIS-only global fits and E155x datum suggests
that hadron-parton duality is valid for $d_2^n$, albeit uncertainties are large. 
The LGT~\cite{Gockeler:1995wg, Gockeler:2000ja, Gockeler:2005vw}, Sum Rule 
approach~\cite{Balitsky:1989jb}, Center-of-Mass bag model~\cite{Song:1996ea} 
and Chiral Soliton model~\cite{Weigel:1996jh}  all yield a small $d_2^n$ at $Q^2 > 1$~GeV$^2$, which agrees
with data. At these large $Q^2$, the data precision is still insufficient to discriminate between these predictions.
The negative $d_2^n$  predicted with a  LC model~\cite{Braun:2011aw}  disagrees with the data.
\subsubsection{Results on the Proton} 
\begin{figure}[ht!]
\center
\includegraphics[scale=0.36]{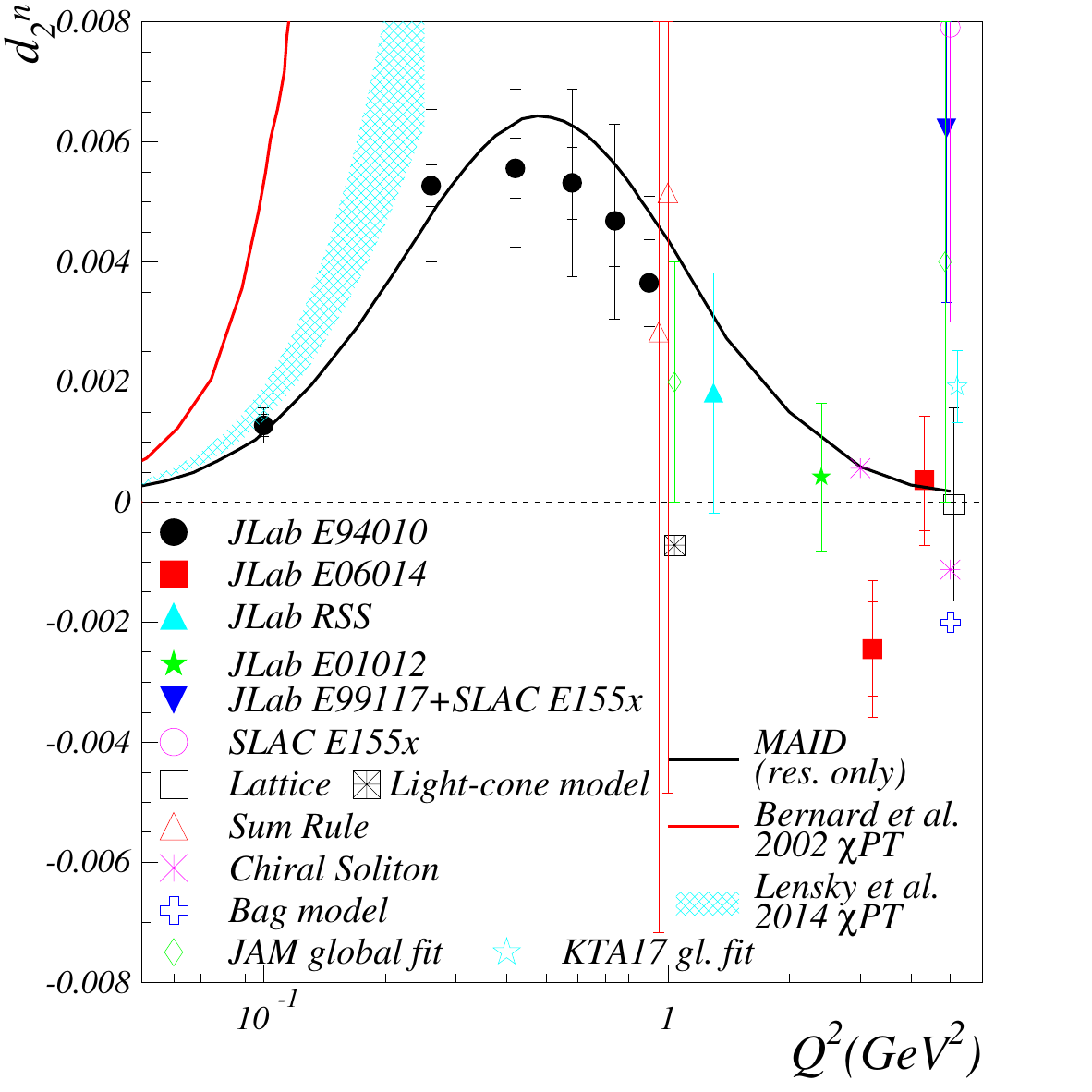}
\hspace{0.5cm}
\includegraphics[scale=0.36]{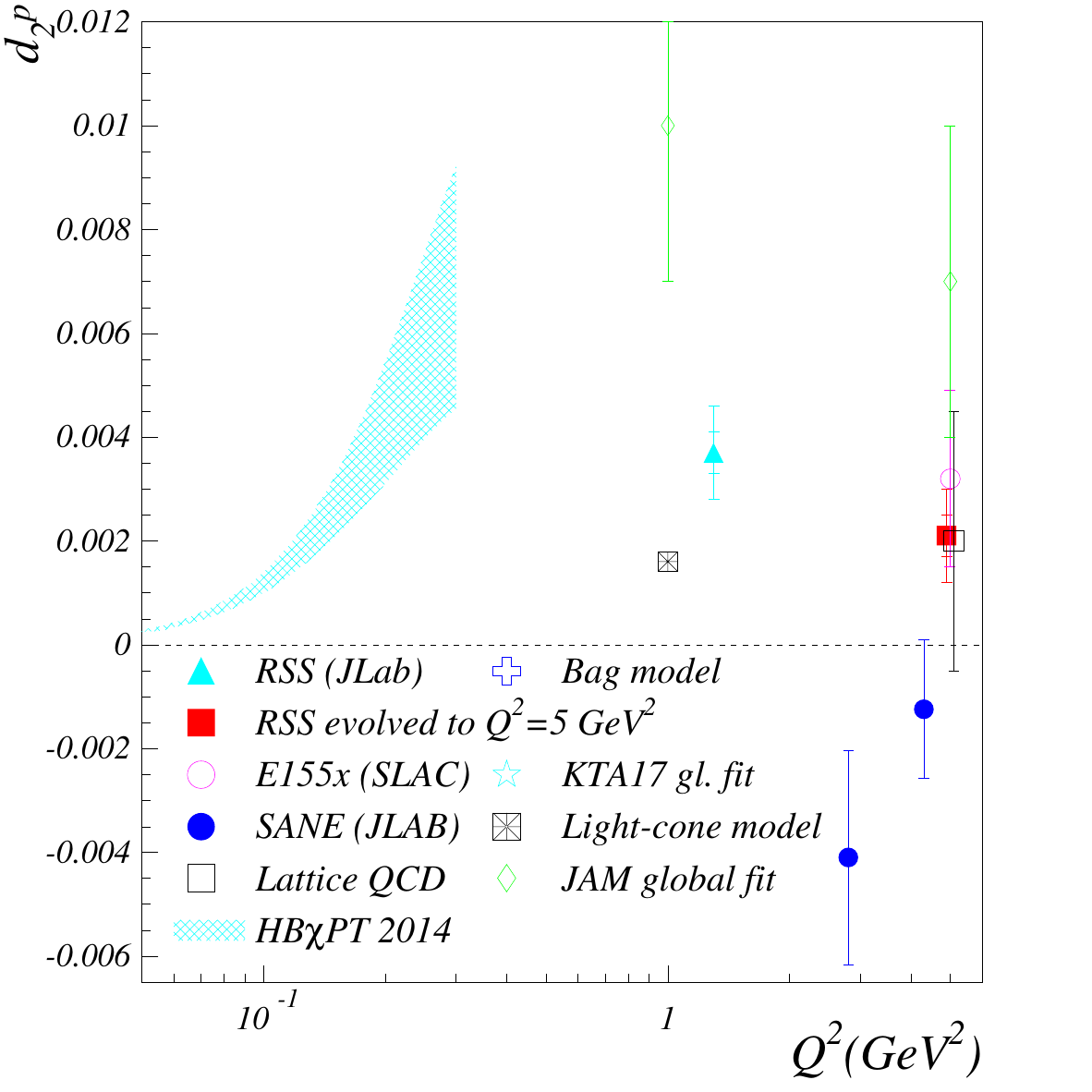}
\vspace{-0.5cm}
\caption{\small{$\overline{d_2}$ data from SLAC, JLab and PDF global analyses, compared to 
LGT~\cite{Gockeler:1995wg, Gockeler:2000ja, Gockeler:2005vw}, 
$\chi$PT~\cite{Bernard:2002pw, Lensky:2014dda}
and models~\cite{Braun:2011aw, Balitsky:1989jb,Song:1996ea,Weigel:1996jh}.
Left panel: neutron data (the inner error bars are statistical. The outer ones are for 
the systematic and statistic uncertainties added in quadrature). Right: proton data. }}
\label{fig:d2} 
\end{figure}

Proton data are scarce, with a datum from RSS~\cite{Wesselmann:2006mw} and 
one from E155x~\cite{Anthony:1999py}. In Fig.~\ref{fig:d2}, the RSS point was evolved to the E155x $Q^2$ 
assuming the $ 1/Q $-dependence expected for a twist~3 dominated quantity (neglecting
the weak log dependence from pQCD radiation). The E155x and RSS results  agree although
RSS measured only the resonance contribution. As for $d_2^n$, this suggests that hadron-parton duality 
is valid for $d_2^p$. However, this conclusion is at odds with the mismatch between the (DIS-only) JAM global PDF 
fit~\cite{Sato:2016tuz} and the (resonance-only) result from RSS.

\subsubsection{Discussion}
Overall, $ \overline{d_2}$ is small compared to the twist~2 term ($| \Gamma_1|  \approx 0.1$ 
typically at $Q^2 = 1 $~GeV$^2$, see Fig.~\ref{fig:gamma1pn}) or to the twist~4 term ($f_2 \approx 0.1 $, see Fig.~\ref{fig:f2}). 
This smallness was predicted by several models.
The high-precision JLab experiments measured a clearly non-zero $ \overline{d_2} $.
More data for $ \overline{d_2^p}$ are needed and will be provided shortly at low $Q^2$~\cite{g2p}
and in the DIS~\cite{Rondon:2015mya}, see Table~\ref{table_exp}. 
Then, the 12 GeV upgrade of JLAB will provide $ \overline{d_2}$ in the DIS 
with refined precision, in particular  with the SoLID detector~\cite{SoLID}.

\subsection{ \emph{Higher-twist} contributions to $\Gamma_1$, $g_1$ and $g_2$ \label{sub:HT Extraction}} 
Knowledge of \emph{higher-twists} is important 
since for inclusive lepton scattering, they are the
next nonperturbative distributions beyond the PDFs, correlating them.
\emph{higher-twists} thus underlie the parton-hadron transition, {\it i.e.}, the process of 
strengthening the quark binding as the probed distance increases. In fact, some 
\emph{higher-twists} are interpreted as confinement forces~\cite{Burkardt:2008ps, Abdallah:2016xfk}. 
Furthermore, knowing \emph{higher-twists} permits one to set the limit of applicability to pQCD and
extend it to lower $Q^2$, see {\it e.g.},  Massive Perturbation Theory~\cite{Pasechnik:2010fg, Natale:2010zz}. 
Despite their phenomenological importance, \emph{higher-twists} have 
been hard to measure accurately because they are often surprisingly small. 
 %
 \subsubsection{Leading and higher-twist analysis of $\Gamma_1$ \label{HT} }
 \begin{wrapfigure}{r}{0.5\textwidth} 
\center
\vspace{-1.65cm}
\includegraphics[scale=0.4]{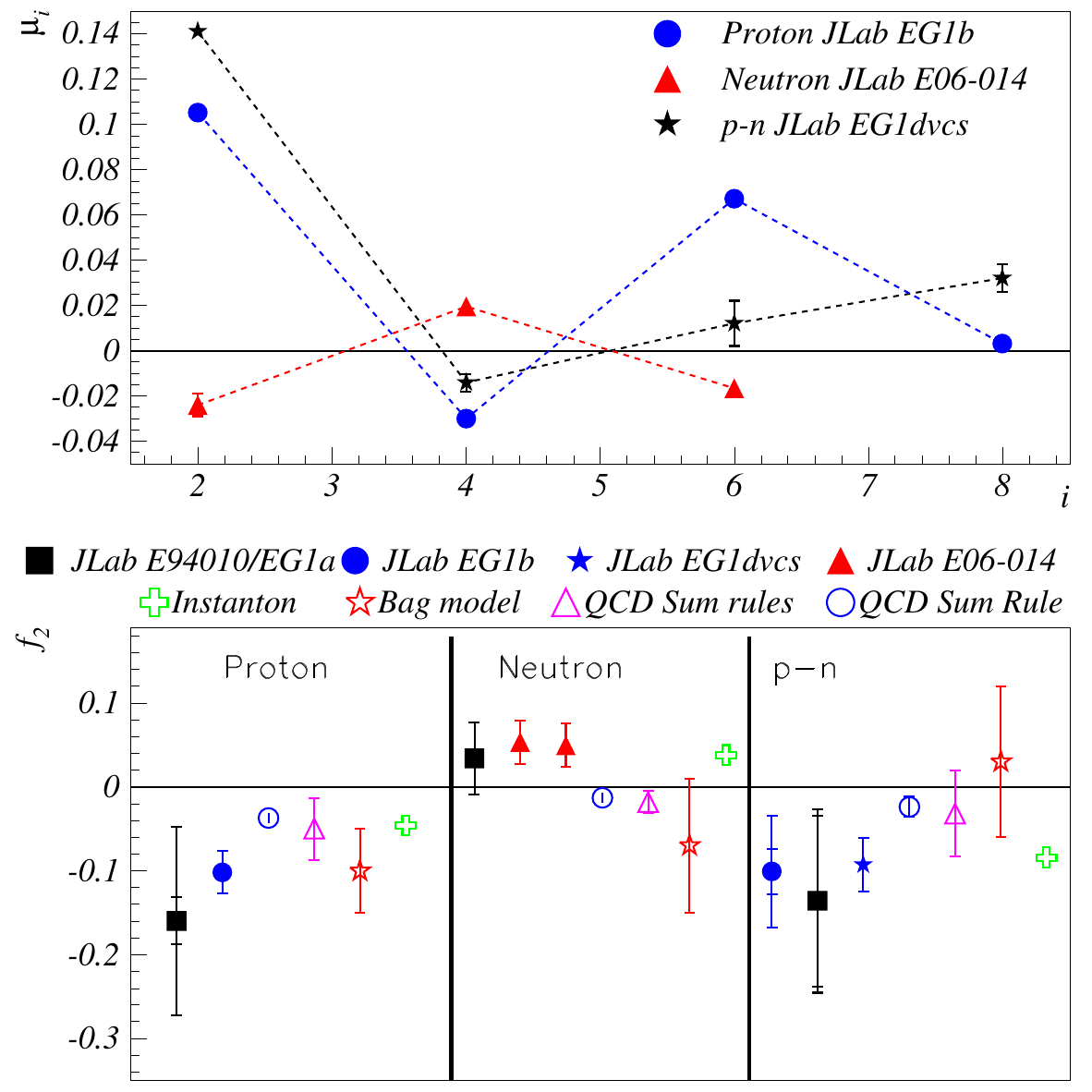}
\vspace{-0.45cm}
\caption{\label{fig:f2}\small{Top: twist coefficients $\mu_i$ {\it vs}. $i$. 
The lines linking the points show the oscillatory behavior.  
Bottom: twist~4 $f_2$. Newer results ({\it e.g.}, EG1dvcs) include the older data ({\it e.g.}, EG1a).}}
\vspace{-0.5cm}
\end{wrapfigure}
The \emph{higher-twist} contribution to $\Gamma_1$ can be obtained by fitting its data with
a function conforming to Eqs.~(\ref{eq:Gamma1})-(\ref{eq:mu2}) and (\ref{eq:genBj})-(\ref{eq:mu4}).
The perturbative series is truncated to an order relevant to the data accuracy. 
Once $\mu_4 $ is extracted, the pure twist~4 matrix element $ f_2 $ is obtained by subtracting $ a_2 $  (twist~2)
and $d_2 $ (twist~3) from Eq.~(\ref{eq:mu4}). 
For $ \Gamma_1^{p,n} $, $\mu_2^{p,n} $ is set by fitting high-$Q^2$ data,  {\it e.g.}, $Q^2 \ge 5 $~GeV$^2$,
and assuming that \emph{higher-twists}  are negligible there. For
$ \Gamma_1^{p-n} $, $\mu_2^{p-n} $ is set by $g_A =1.2723(23)$~\cite{Olive:2016xmw}. 
The resulting $\mu_2^{p,n} $, together with $ a_8 $ from the hyperons $\beta$-decay, yield
$ \Delta \Sigma = 0.169 \pm 0.084 $ for the proton and $\Delta \Sigma = 0.35 \pm 0.08 $ 
for the neutron~\cite{Fersch:2017qrq, Meziani:2004ne, Deur:2005jt}. The discrepancy 
may come from the low-$x_{Bj }$ part of
$ \Gamma_1^n $, which is still poorly constrained, as the COMPASS deuteron data~\cite{Alexakhin:2006oza} 
suggest. Specifically, it may be the  low-$x_{Bj }$ contribution to the isoscalar quantity $ \Gamma_1^{n+p}$,
since $ \Gamma_1^{p-n} $ agrees well with the Bjorken sum rule. 
Another possibility is a SU(3)$_f$ violation.
The $ \Delta \Sigma$ obtained from global analyses (see Section~\ref{nucleon spin structure at high energy})
mix the proton and neutron data and agree with the averaged value of $ \Delta \Sigma^p $ and $ \Delta \Sigma^n $. 

Fit results~\cite{Fersch:2017qrq,  Posik:2014usi,  Deur:2004ti, Deur:2008ej, Meziani:2004ne, Deur:2005jt} 
are shown and compared to available 
calculations~\cite{ Ji:1997gs, Balitsky:1989jb, Ioffe:1981kw, Lee:2001ug, Sidorov:2006vu} 
in Fig.~\ref{fig:f2}. There are no predictions yet for \emph{twists} higher than $ f_2 $.  
We note the sign alternation between  $ \mu_2 $, $ \mu_4 $ and $ \mu_6 $. 
All higher power corrections are folded in
$\mu_8$, which is thus not a clean term and does not follow the alternation. 
This one decreases the  \emph{higher-twist} effects and could explain the global 
quark-hadron spin duality (see Section~\ref{sec:duality}). The sign alternation is opposite for proton and neutron,
as expected from isospin symmetry, see Eq.~(\ref{eq:mu2}) in which the non-singlet $ g_A/ 12 \approx 0.1 $ dominates the 
singlet terms $ \Delta \Sigma / 9 \approx 0. 03 $ and $ a_8 / 36 \approx 0.008 $.
The discrepancy between $ \Delta \Sigma^p$ and $\Delta \Sigma^n $ explains why  the value of $ f_2 $ 
extracted from $ \Gamma_1^{p-n}$ differs from the $f_2$ values extracted individually. 
Indeed, $ \Delta \Sigma $ vanishes in the Bjorken sum rule whose derivation does not assume SU(3)$_f$ symmetry.

Although the overall effect of  \emph{higher-twists} is small at $Q^2 > 1 $~GeV$^2$, $ f_2 $  itself is large:
$|f_2^p| \approx 0.1$, to compare to $\mu_2^p=0.105(5)$;
$f_2^n \approx 0.05$ for $|\mu_2^n|=0.023(5)$; 
$|f_2^{p-n}| \approx 0.1$ for $\mu_2^{p-n} = 0.141(11)$.
These large values conform to the intuition that nonperturbative effects should be important at moderate $Q^2$. 
The smallness of the total \emph{higher-twist} effect is due to the factor $M^2/9 \approx 0.1$ in Eq.~(\ref{eq:mu4}),
and to the $\mu_i$  alternating signs. Such oscillation can be understood with vector  meson
dominance~\cite{Weiss_pc}. 

\subsubsection{Color polarizabilities and confinement force \label{color pol.}}

Electric and magnetic color polarizabilities can be determined using Eq.~(\ref{eq:chi}).
For the proton,
$\chi_E^p = -0.045(44)$ and $\chi_B^p = 0.031(22)$~\cite{Fersch:2017qrq}. 
For the neutron, $\chi_E^n =0.030(17) $, $\chi_B^n =-0.023(9) $. The Bjorken sum data yield
$\chi_E^{p-n} = 0.072(78)$, $\chi_B^{p-n} = -0.020(49)$.
These values are small and the proton and neutron have opposite signs. Since $ f_2 \gg d_2 $, this
reflects the dominance of the non-singlet term $g_A$. 
The electric and magnetic Lorentz transverse confinement forces are  proportional 
to the color polarizabilities~\cite{Burkardt:2008ps, Abdallah:2016xfk}:
\vspace{-0.3cm}
\begin{equation}
\vspace{-0.3cm}
F^y_E=-\frac{M^2}{4}\chi_E , ~~~F_B=-\frac{M^2}{2}\chi_B.
\label{eq:color forces}
\end{equation}
Their size of a few $10^{-2}$~GeV$^2$ can be compared to the string tension
$\sigma_{str} = 0.18$~GeV$^2$ obtained from heavy quarkonia. Several coherent processes
prominent for the proton and neutron, {\it e.g.},  the $\Delta(1232)~3/2^+$, are nearly inexistent for
$ \Gamma_1^{p-n}$~\cite{Burkert:2000qm}. This may explain why $ \Gamma_1^{p-n}$ is suited to extract
$\alpha_s$ at low $Q^2$~\cite{Deur:2016tte, Deur:2005cf}.

\subsubsection{\emph{Higher-twist} studies for $g_1$, $A_1$, $g_2$ and $A_2$\label{sec:g2-g2ww}}
\begin{wrapfigure}{r}{0.52\textwidth}
\vspace{-2.1cm}
\center
\includegraphics[scale=0.39]{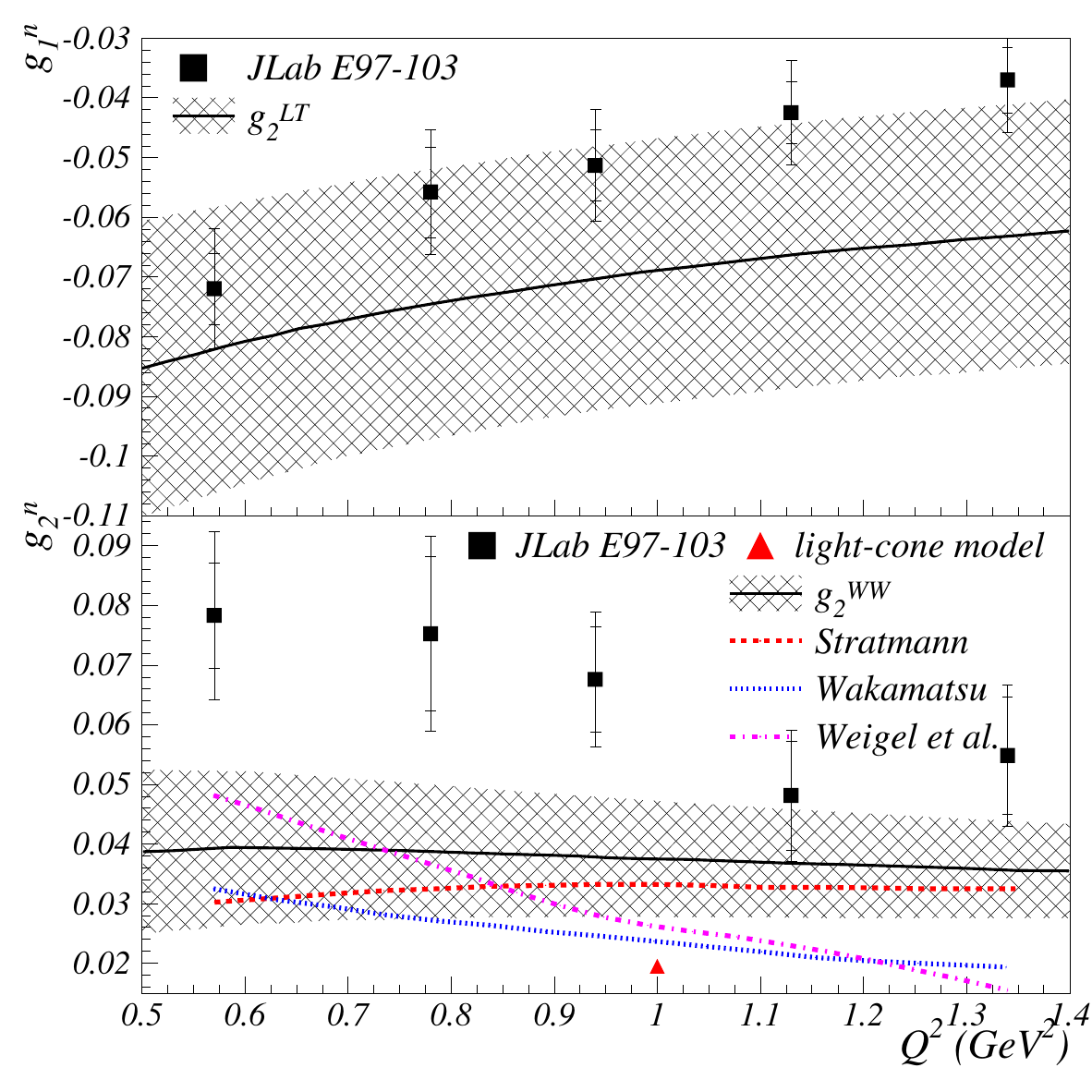}
\vspace{-0.6cm}
\caption{\label{fig:E97-103 g2}\footnotesize{Top: $g_1^n(Q^2)$ from E97103 
(symbols). The inner error bars give the 
statistical uncertainty while outer bars are the systematic and statistical uncertainties added in quadrature.
The continuous line is a global fit of the world data on $g_1^n$~\cite{Bluemlein:2002be}, 
with its uncertainty given by the hatched band. 
Bottom: Corresponding $g_2^n$ data with various models and  $g_2^{WW}$ computed from the global fit on $g_1^n$.
The data are at $x_{Bj} \approx 0.2$.
}}
\vspace{-1.8cm}
\end{wrapfigure}

\vspace{-0.2cm}
\emph{Higher-twists} and their $x_{Bj}$-dependence have been extracted from spin structure 
data~\cite{Anthony:1999py, Zheng:2003un,Kramer:2005qe},
in particular by global fits~\cite{Leader:2006xc, Blumlein:2010rn, Leader:2002ni}. 
More \emph{higher-twists} data are expected soon~\cite{Rondon:2015mya}.

\noindent \textbf{Study of $g_2$ in the DIS}
We consider first $g_2$ data in the DIS. Lower  $W$ or $Q^2$ data are discussed afterwards.

The Wandzura-Wilczek term $g_2^{WW}$, Eq.~(\ref{eq:g2ww}), is the twist~2 part of $ g_2$. 
Nevertheless, due to the asymmetric part of the axial matrix element entering the 
OPE~\cite{Manohar:1992tz, Jaffe:1990qh},
it contributes alongside the twist~3 part of $ g_2$, similarly to {\it e.g.}, the twist-2 term $a_2$ and 
twist-3  term $d_2$  contributing alongside the twist-4 term $f_2$ in Eq.~(\ref{eq:mu4}). 
Indeed, in Eq.~(\ref{eq:sigmapar}), $g_2$ is suppressed as $Q/(2E) = 2Mx_{Bj}/Q$ compared to $g_1$.
%
%
Just like there is no reason in $\mu_4$ that 
$ a_2 \gg d_2 \gg f_2 $ (which is indeed \\
not the case), there is no obvious reason for \\
having $g_2^{WW}\gg g_2^{twist~3}$ and thus $g_2\approx g_2^{WW}$. \\
This is, however, the empirical observation: all the  $g_2^{p,n}$ 
DIS data  (SMC~\cite{Adeva:1993km}, E143~\cite{Abe:1995dc}, 
E154~\cite{Abe:1997qk} and E155x~\cite{Anthony:1999py}, 
E99-117~\cite{Zheng:2003un}, E97-103~\cite{Kramer:2005qe}, E06-104 \cite{Posik:2014usi} and 
HERMES~\cite{Airapetian:2011wu}) are compatible with $g_2^{WW}$. 
Below $Q^2=1$~GeV$^2$, E97-103~\cite{Kramer:2005qe}  did observe that 
$g_2^n>g_2^{WW,n}$, see Fig.~\ref{fig:E97-103 g2}. 
Its data cover $0.55<Q^2<1.35$~GeV$^2$, at a fixed $x_{Bj} \approx 0.2$ to isolate 
the $Q^2$-dependence. The deviation seems 
to decrease with $ Q^2 $ as expected for  \emph{higher-twists}. 
Models~\cite{Braun:2011aw, Weigel:1996jh, Wakamatsu:2000ex, Stratmann:1993aw} predict a 
negative contribution from  \emph{higher-twists} 
while the data indicate none, or a positive one for $Q^2 \lesssim 1$~GeV$^2$.

The \emph{leading-twist} part of $g_1$, \emph{viz} $g_1^{LT}$, is needed to form $ g_2^{WW}$. To check the
PDF~\cite{Bluemlein:2002be} used to compute $g_1^{LT}$, $ g_1$ was measured by 
E97-103, see Fig.~\ref{fig:E97-103 g2}. No  \emph{higher-twist} is 
seen: $g_1 \approx g_1^{LT}$. However,
at such $x_{Bj}$ and $Q^2$, the LSS global fit~\cite{Leader:2006xc}
saw a twist-4 contribution $h/Q^2=0.047(29)$ at $Q^2=1$~GeV$^2$, which E97-103 should have seen.
While the large uncertainties preclude firm conclusions, this implies 
either a tension between  LSS and  E97-103, or that kinematical and dynamical 
 \emph{higher-twists} compensate each other.

The BC sum rule, Eq.~(\ref{eq:bc}) implies a zero-crossing of $g_2(x_{Bj})$.
The E99-117~\cite{Zheng:2003un} and E06-104~\cite{Posik:2014usi} DIS 
data suggest it is near $x_{Bj} \approx 0.6$ for the neutron.
E143~\cite{Abe:1995dc}, E155x~\cite{Anthony:1999py} and 
HERMES~\cite{Airapetian:2011wu}  indicate it is between  $0.07<x_{Bj}<0.2$ for the proton.

\vspace{-0.5cm}
\paragraph{Study of $g_2$ in the resonance domain} \label{sub:g2 in res}
\begin{wrapfigure}{r}{0.5\textwidth} 
\vspace{-1.cm}
\center
\includegraphics[scale=0.43]{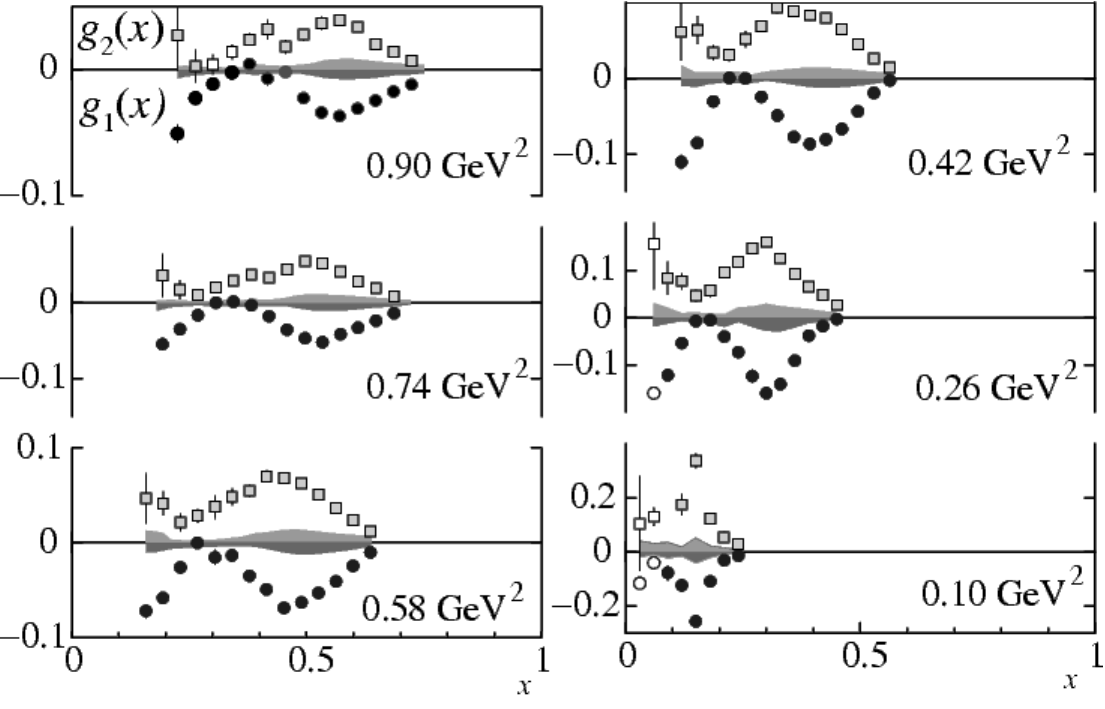}
\vspace{-0.6cm}
\caption{\label{fig:g1g2 E94010}\small{The symmetry between $g_1$ and $g_2$. 
(JLab $^3$He data from E94-010~\cite{Amarian:2002ar}.)
}}
\end{wrapfigure}
So far, $g_2$ DIS data have been discussed. Many data at $W<2$~GeV and 
$6 \times 10^{-3} <Q^2< 3.3$~GeV$^2$ also exist.
Being derived using OPE, the Wandzura-Wilczek relation, Eq.~(\ref{eq:g2ww}), should 
not apply there.  Yet, it is
instructive to compare $ g_2 $ and $ g_2^{WW}$ in this region. (In fact, it was done 
when $d_2(Q^2)$ was discussed, since $d_2=\int x^2[g_2-g_2^{WW}]dx$.)
Proton and deuteron $ g_2$ data are available from the RSS experiment at
$\left\langle Q^2\right\rangle =1.3$~GeV$^2$ and for $0.3 \leq x_{Bj} \leq 0.8$~\cite{Wesselmann:2006mw, Slifer:2008xu}. 
The $x_{Bj}$-dependences of $g_2$ and $g_2^{WW*}$ (the $^*$ means it is formed using $g_1$ 
measured by RSS and thus is not leading \emph{twist})  are similar  except that generally 
$ |g_2^p| < |g_2^{WW*,p}|$, while $ |g_2^n| > |g_2^{WW*,n}|$. 
The inequality indicates either  \emph{higher-twist} effects or coherent resonance 
contributions. The ranks and types of the  \emph{higher-twists} are unclear since $ g_2^{WW*}$ 
itself contains  \emph{higher-twists} whereas its OPE expression is twist~2. 
A  similar study  on $ g_2^{^3He}$ from E97-110~\cite{E97110} was done
for $6 \times 10^{-3} <Q^2< 0.3$~GeV$^2$.
Again, $ g_2^{^3He}$ 
is close to $ g_2^{WW*,^3He}$. 
Their difference may come from  \emph{higher-twists} or 
coherence effects, but now also possibly from nuclear effects.
Resonance data on $g_2^{^3He}$ are also available from E01-012~\cite{Solvignon:2013yun} and were
compared to $g_2^{WW,^3He}$ computed at \emph{leading-twist}. 
It results that $g_2^{WW,^3He}$ provides an accurate approximation of 
$ g_2^{^3He}$, maybe facilitated by the smearing of resonances in nuclei. Such analysis amounts 
to assessing the size of twist-3 and higher in $g_2$, neglecting structures due to resonances. It also tests hadron-parton 
spin-duality for  $^3$He, see Section~\ref{sec:duality}.

A feature of the $g_1$ and $g_2$ resonance data is the symmetry around 0 of their 
$x_{Bj}$-behavior, see Fig.~\ref{fig:g1g2 E94010}.
It is observed for the proton~\cite{Wesselmann:2006mw} and for 
$^3$He~\cite{Amarian:2002ar, Kramer:2005qe, Solvignon:2013yun, E97110}. 
DIS data do not display the symmetry. It arises from the smallness of ${\sigma}_{LT}'$:
since ${\sigma}_{LT}'\propto(g_1+g_2)$, then $g_1 \approx -g_2$. In particular,
for the $\Delta(1232)~3/2^+$, ${\sigma}_{LT}'\approx0$ because the dipole component $M_{1+}$ 
dominates the nucleon-$\Delta$ transition.  
This holds at low $Q^2$ where $M_{1+} \gg E_{1+}$ and  $S_{1+}$. At larger $Q^2$, 
another reason arises:  resonances being at high $x_{Bj} $, 
$\int_{x_{Bj}}^1(g_1/y )dy$
in Eq.~(\ref{eq:g2ww})  is negligible and since $ g_2^{WW} \approx g_2$, then $g_2 \approx -g_1$.

\subsection{Study of the hadron-parton spin duality \label{sec:duality}}%
Hadron-parton duality is the observation that a structure function in the DIS appears as a precise 
average of its  measurement in the resonance domain. 
This coincidence can be understood as a dearth of dynamical  \emph{higher-twists}. 
Duality is thus related to the study of parton correlations. 
In the last two decades precise data were gathered to test duality on $g_1$. 
%
Duality on $ g_1^p $, $ g_1^n $, $ g_1^d $ and $ g_1^{^3He}$ has been studied using the SLAC and JLab data from
E143~\cite{Abe:1994cp}, 
E154~\cite{Abe:1997cx,Abe:1997qk}, 
E155~\cite{Anthony:1999rm,Anthony:1999py},
E94010~\cite{Amarian:2002ar},
E97-103~\cite{Kramer:2005qe},
E99-117~\cite{Zheng:2003un},
E01-012~\cite{Solvignon:2013yun, Solvignon:2008hk} (which was dedicated to studying spin-duality),
EG1b~\cite{Dharmawardane:2006zd, Prok:2008ev, Bosted:2006gp, Fersch:2017qrq} and
RSS~ \cite{Wesselmann:2006mw, Slifer:2008xu}.
The $x_{Bj} $-value at which duality appears depends on $Q^2$. At low $Q^2$, duality is 
violated around the $\Delta(1232)~3/2^+$. This is expected since there, $ g_1 < 0$ due to 
the $M_{1+}$ transition dominance; see discussions about $\sigma_{TT}$ on page~\pageref{sub:sigmaTT}
and about the $g_1$ and $g_2$ symmetry page~\pageref{fig:g1g2 E94010}. 
(The discussion applies to $ g_1 $ because 
$\sigma_{TT}\propto(g_1-\gamma^2g_2)\approx g_1(1+\gamma^2)\propto g_1$
 at the $\Delta(1232)~3/2^+$.) Above $Q^2 = 1.2$~GeV$^2$ duality seems to be valid at all $x_{Bj}$. 
 Duality's onset for $ g_1^d $ and $ g_1^{^3He}$ appears at smaller $Q^2$  
 than for $ g_1^p $ as expected, since duality 
 is aided by the nucleon Fermi motion inside a composite nucleus.

Spin-duality was also studied using the $ A_1 $ and $ A_2 $ asymmetries using the SLAC, HERMES~
\cite{Airapetian:2002rw} and JLab data. Duality in $ A_1 $ arises for $Q^2 \gtrsim2.6 $~GeV$^2$. 
At lower $Q^2$, it is  invalidated by the $\Delta(1232)~3/2^+$.  
The $A_1^{^3 He}$ $Q^2$-dependence is weak for both DIS 
and resonances (except near the $\Delta(1232)~3/2^+$). 
This is expected in DIS since $ A_1 \approx g_1 / F_1 $ and $ g_1 $ and $ F_1 $ 
have the same $Q^2$-dependence at LO of \emph{DGLAP} and \emph{leading-twist}. 
The weak $Q^2$-dependence in the resonances signals duality. 
Duality in $ A_1 $ seems to arise at greater $Q^2$ 
than for $ g_1 $. Duality in $ A_2 $ arises at lower $Q^2$ than  
for $ A_1 $ because the  $\Delta(1232)~3/2^+$ is suppressed in $ A_2 $, since $ A_2 \propto \sigma_{LT}' $.

The similar $Q^2$- and $x_{Bj}$-dependences of the DIS and 
resonance structure functions discussed so far is called ``local duality".  ``Global duality"
considers the moments. It is tested by forming the partial moments 
$\widetilde{\Gamma}^{res}$ integrated only over the resonances. 
They are compared to $\widetilde{\Gamma}^{DIS}$ moments covering the 
same $x_{Bj}$ interval and formed using \emph{leading-twist} structure functions. 
$\widetilde{\Gamma}^{DIS}$ is corrected for pQCD radiation and kinematical \emph{twists}. 
Global duality has been tested  on $\widetilde{\Gamma}_1^p$ 
and $\widetilde{\Gamma}_1^d$~\cite{Bosted:2006gp}, and 
on $\widetilde{\Gamma}_1^n$ and $\widetilde{\Gamma}_1^{^{3}\textnormal{He}}$~\cite{Solvignon:2013yun}.
For the proton, duality arises for $Q^2 \gtrsim1.8 $~GeV$^2$ or $Q^2 \gtrsim1.0 $~GeV$^2$ if the elastic reaction
is included. For the deuteron, $^3$He and neutron (extracted from the previous nuclei), duality arises earlier,
as expected from Fermi motion.

\subsection{Nucleon spin structure at high energy \label{nucleon spin structure at high energy}}

In this section we will discuss the picture of the nucleon spin structure painted by both high-energy experiments and theory. 
The PDFs quoted here are for the proton. The neutron PDFs should be nearly identical after SU(2) isospin symmetry rotation.

\subsubsection{General conclusions}

The polarized inclusive DIS experiments from SLAC, CERN and DESY laid the foundation 
for our understanding of the nucleon spin structure and showed  that:

\noindent$\bullet$ The strong force is well described by pQCD, 
even when spin degrees of freedom are accounted for.  Since QCD is the accepted paradigm,
the contribution of inclusive, doubly polarized DIS  experiments to nucleon spin studies provided an  important
test of the theory.  For example, the verification of the Bjorken sum rule, Eq.~(\ref{eq:genBj}), has played a central role. 
To emphasize this, one can recall the oft-quoted statement of Bjorken~\cite{Bjorken:1996dc}:
``Polarization data has often been the graveyard of fashionable theories. If theorists had their way,
they might well ban such measurements altogether out of self-protection." 

\noindent$\bullet$ QCD's fundamental quanta,  the quarks and gluons,  
and their OAM should  generate the nucleon spin,  
see Eq.~(\ref{eq:spin SR}):
\vspace{-0.4cm}
\begin{equation}
\vspace{-0.15cm}
J=\frac{1}{2}=\frac{1}{2}\Delta\Sigma+\mbox{L}_q+\Delta G+\mbox{L}_g
\nonumber
\end{equation}
Estimates for each of the components are discussed in the next Section.  Recent determinations suggest
$\Delta \Sigma \approx 0.30(5)$, 
$\mbox{L}_q \approx 0.2(1)$, and 
$\Delta G+\mbox{L}_g \approx 0.15(10)$  at $Q^2=4$~GeV$^2$.  Thus the nucleon spin is shared
between the three components, with the quark OAM possibly the largest contribution.   
This result includes the PDF evolution effects from the low $Q^2$ nonperturbative domain  to the 
experimental resolution at $Q^2=4$~GeV$^2$.  

\noindent$\bullet$ The PDFs extracted from diverse DIS data and evolved to the 
same $Q^2$ are generally consistent. Global analyses
show that the up quark polarization in the proton is large and positive, 
$ \Delta \Sigma_u  \approx 0.85$, whereas the down quark one is smaller and negative, 
$\Delta \Sigma_d \approx -0.43$. The $x_{Bj}$-dependences of $\Delta u+\Delta\overline{u}$ and $\Delta d+\Delta\overline{d}$  
are well determined in the kinematical domains of the experiment.

\noindent$\bullet$  The gluon axial anomaly~\cite{Efremov:1988zh} is small and cannot 
explain the ``spin crisis".

\noindent$\bullet$ The  contribution of the gluon spin,
which  is only indirectly accessible  in inclusive experiments, seems to be moderate.

\noindent$\bullet$ Quark OAM, which is required in the baryon LFWF to have 
nonzero Pauli form factor and anomalous magnetic moment~\cite{Brodsky:1980zm}, 
is the most difficult  component to measure from DIS;
however,  an analysis of DIS data at high-$x_{Bj}$, GPD data, as well as LGT suggest it is a major contribution to $J$. 

\noindent$\bullet$ The Ellis-Jaffe sum rule, Eq.~(\ref{eq:Ellis-Jaffe p}), is violated for both nucleons. 
This either implies a large $\Delta s$, large SU(3)$_f$ breaking effects, 
or an inaccurate value of $a_8$~\cite{Bass:2009ed}. Global fits indicate 
$ \Delta s\approx-0.05(5)$, which is too small to fully explain the violation.  

\noindent$\bullet$ \emph{Higher-twist} power-suppressed contributions are small at  $Q^2 > 1$~GeV$^2$.

\subsubsection{Individual contributions to the nucleon spin \label{Individual contributions to the nucleon spin}} 

\paragraph{Total quark spin contribution}
The most precise determinations of $\Delta \Sigma$ are from global fits, see
Table~\ref{table Delta Sigma 1} in the Appendix. In average, $\Delta \Sigma \approx 0.30(5)$.
A selection of LGT results is shown in Table~\ref{table Delta Sigma 2}. 
The early calculations typically did not include the disconnected diagrams 
that are responsible for the \emph{sea quark} contribution. They account for the 
larger uncertainty in some recent LGT analyses and reduce  the predicted 
$\Delta\Sigma$ by about 30\%~\cite{Alexandrou:2016tuo, Alexandrou:2018xnp}.
(An earlier result indicated only a 5\% reduction, but this was evaluated with $m_{\pi} = 0.47$~GeV~\cite{Chambers:2015bka}). 
The determination of $\Delta \Sigma$ from SIDIS at COMPASS~\cite{Alexakhin:2005iw} 
agrees with the  inclusive data~\cite{Alexakhin:2006oza}.  The two analyses have  similar statistical precision.

\paragraph{Individual quark spin contributions}
Inclusive DIS data on proton, neutron, and deuteron  targets can be used to
separate the contributions from different quark polarizations  assuming SU(3)$_f$ validity. 
SIDIS, which tags the struck 
quark allows the identification of individual quark spin contributions without this
assumption.  However, the domain where PDFs can be safely 
extracted assuming factorizations demands a larger momentum scale than untagged DIS. It is presently
unclear whether the kinematical range of available data has reached this domain. 

Tables~\ref{table Delta q 1} and~\ref{table Delta q 2}  list $\Delta q$ from experiments, 
models and results from LGT. Overall, $ \Delta \Sigma_u  \approx 0.85$ and $\Delta \Sigma_d \approx -0.43$.
$\Delta s$ is of special interest since it could explain 
the violation of the Ellis-Jaffe sum rule (Eq.~(\ref{eq:Ellis-Jaffe p})), and also underlines 
the limitations of \emph{constituent quark} models.  Ref.~\cite{Chang:2014jba} reviewed recently
the nucleon \emph{sea}, including $\Delta s$. The current favored value for $\Delta s, $ approximately $-0.05(5)$, is
barely enough to reconcile the Ellis-Jaffe sum rule, which predicts $\Delta \Sigma^{EJ} = 0.58(12)$ without $\Delta s,$
with the measured $\Delta \Sigma \approx 0.30(5)$.  Recent LGT data yield an even smaller $\Delta s$
value, about $-0.03(1)$. (Early quenched LGT data yielded a larger $\Delta s=0.2(1)$, agreeing with the EMC
initial determination.) Thus, this suggests that SU(3)$_f$ breaking also contributes to 
the Ellis-Jaffe sum rule violation~\cite{Bass:2009ed}. 
This conclusion is supported by recent global analyses from
DSSV~\cite{deFlorian:2009vb}, NNPDF14~\cite{Nocera:2014gqa} and in particular JAM~\cite{Ethier:2017zbq}.
Nevertheless, this question remains open since 
for example, LGT investigations of hyperon axial couplings show no evidence
of SU(3)$_f$ violation~\cite{Lin:2007ap}. 

There is also tension  between the values for $\Delta s$ derived from DIS and from kaon SIDIS data.
Those suggest that the $x_{Bj}$-dependence of $\Delta s + \Delta \overline{s}$ 
flips sign and thus contributes less to $J$ than indicated by DIS data. 
For example, COMPASS obtains $\Delta s + \Delta \overline{s}= -0.01 \pm 0.01$(stat) $\pm 0.01$(syst) from SIDIS 
whereas a PDF fit of inclusive asymmetries  yields
$\Delta s + \Delta \overline{s}= -0.08 \pm 0.01$(stat) $\pm 0.02$(syst), in clear disagreement. This 
suggests that even at the large CERN energies, we may not yet be in the factorization domain for  SIDIS.
Furthermore, a LSS analysis showed that the SIDIS $\Delta s$  is very sensitive to the 
parameterization of the fragmentation functions and that the lack of their precise 
knowledge may  cause the tension~\cite{Leader:2011tm}.
However, the JAM analysis recently suggested~\cite{Ethier:2017zbq} that the  tension 
comes from imposing SU(3)$_f$, which is consistent
with the likely explanation of the Ellis-Jaffe sum rule violation ~\cite{Bass:2009ed, Leader:2000dw}. 
The JAM analysis, done at NLO and in the $\overline{MS}$ scheme, was
aimed at determining $\Delta s + \Delta \overline{s} (x_{Bj})$ with minimal bias.  It used
DIS, SIDIS and $e^+ e^-$ annihilation data without imposing SU(3)$_f$, and allowed for \emph{higher-twist} contributions.
It finds $\Delta s + \Delta \overline{s} = -0.03 \pm 0.10$ at $Q^2=5$~GeV$^2$. 
Fragmentation function data from LHC, COMPASS, HERMES, BELLE and BaBar may clarify the
situation. Measurements of $\overrightarrow{p} p \to \overrightarrow{\Lambda}X$  may also  help since the $\Lambda$
polarization depends on $\Delta s$. 
Reactions utilizing parity violation are also useful: proton 
strange form factor data, together with neutrino scattering data yield 
$g_A^s=\Delta s + \Delta \overline{s}= -0.30 \pm 0.42$~\cite{Pate:2008va}. 
New parity violation data on $g_A^s$ should be available soon~\cite{Woodruff:2017kex}  and can be complemented  
with measurements using the future SoLID detector at JLab ~\cite{SoLID}.  
A polarized $^3$He target  and unpolarized electron beam can
provide $g_1^{\gamma Z,n}$ and $g_5^{\gamma Z,n}$ from $Z^0$--$\gamma$ parity-violating interference.
These measurements, combined with the existing $g_1^p$ and $g_1^n$ data, can determine $\Delta s$ 
without assuming SU(3)$_f$~\cite{LOI12-16-007}.

\begin{figure}
\center
\vspace{-0.5cm}
\includegraphics[scale=0.39]{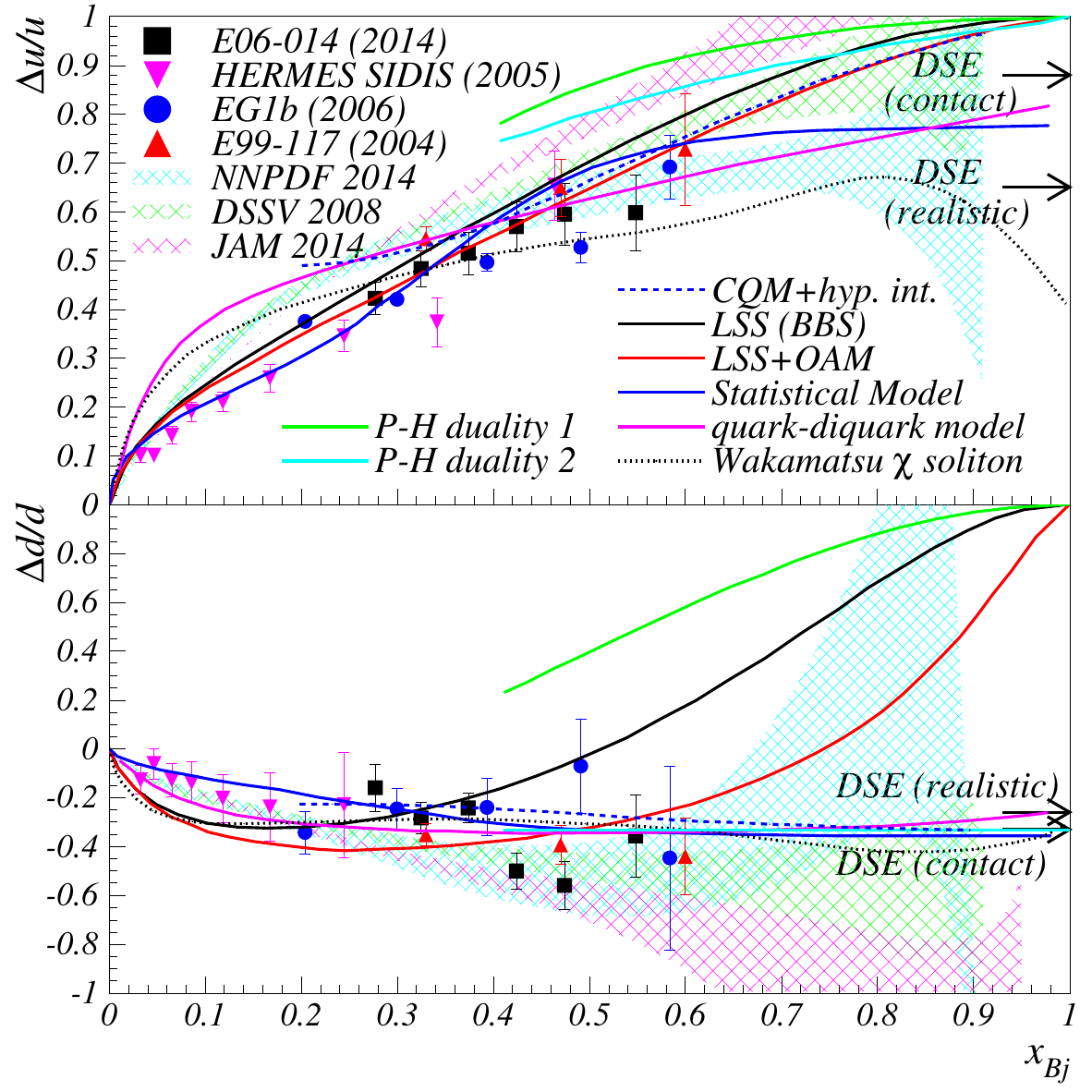}\includegraphics[scale=0.4]{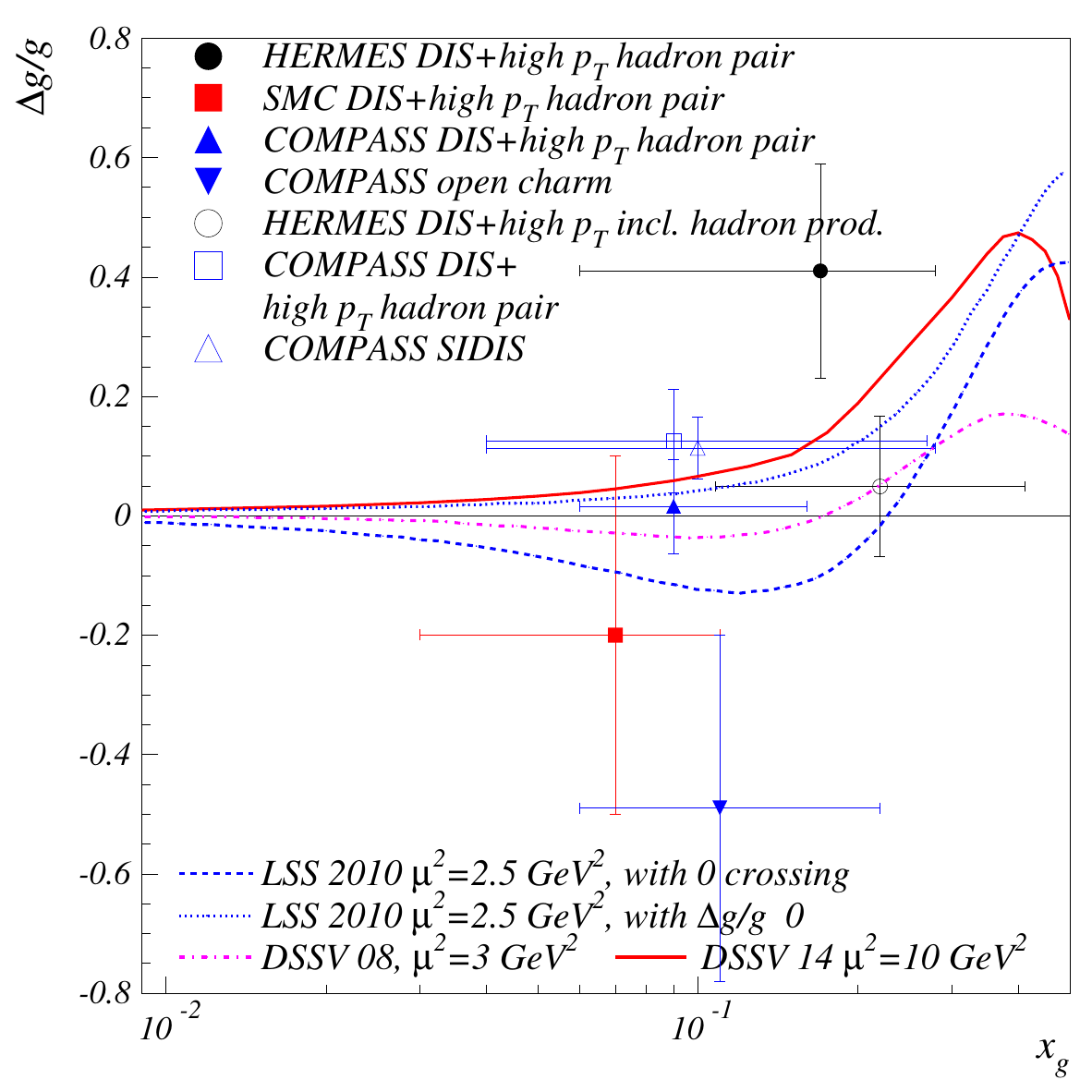}
\vspace{-0.6cm}
\caption{\label{fig:partons_polar_vs_x} \small{Data and global fits for 
$\Delta q/q$ vs quark momentum fraction $x_{Bj}$ (left), and for
$\Delta g/g$ vs the gluon momentum fraction $x_g$ (right). 
}}
\vspace{-0.5cm}
\end{figure}

The $x_{Bj}$-dependence of $\Delta u$ and $\Delta d$ can be obtained 
from $A_1\approx g_1/F_1$ at high $x_{Bj}$  (see Section~\ref{pqcd high-x})
and from SIDIS at lower $x_{Bj}$. At high $x_{Bj}$, \emph{sea quarks} contribute little so
$F_1$ and $g_1$ mostly depend on $u^+$, $u^-$, $d^+$ and $d^-$ 
(see Eqs.~(\ref{eq:eqf1parton}) and (\ref{eq:eqg1parton})). They can thus be extracted from 
$F_1^p$, $F_1^n$, $g_1^p$ and $g_1^n$ assuming  isospin symmetry. 
The results for $\Delta u/u$ and $\Delta d/d$ extracted 
from  $A_1$~\cite{Dharmawardane:2006zd, Zheng:2003un, Parno:2014xzb, Airapetian:2004zf} 
are shown in Fig.~\ref{fig:partons_polar_vs_x}. 
For clarity, only the most precise data are plotted. 
Smaller $x_{Bj}$ points are from SIDIS data~\cite{Ackerstaff:1999ey}. 
Global fits are also shown~\cite{ Leader:2001kh, Jimenez-Delgado:2014xza, deFlorian:2009vb, Nocera:2014gqa}. 
The latter Ref. used the high-$x_{Bj}$ pQCD constraints 
discussed in Section~\ref{pqcd high-x} and assumed no quark OAM. 
OAM is included in the results from Refs.~\cite{Avakian:2007xa, Jimenez-Delgado:2014xza}.

The $\Delta d/d$ data are negative, agreeing with most models but not with pQCD evolution
which predicts that $\Delta d/d>0$ for $x_{Bj} \gtrsim 0.5$ without quark OAM. 
Including OAM pushes the zero crossing to $x_{Bj} \approx 0.75$,
which agrees with the data. PQCD's validity being established, this suggests that quark OAM is important. 
Integrating $\Delta u(x_{Bj})$ and $\Delta d(x_{Bj})$ over $x_{Bj}$ yield a large positive $\Delta u$ and a
moderate negative $\Delta d$. 

\noindent 
First results on $\Delta u - \Delta d$ from LGT are becoming available~\cite{Chen:2016utp, Alexandrou:2018pbm}.

\paragraph{The $ \Delta \overline{u}  - \Delta \overline{d}$ difference}
Global fits and LGT calculations indicate a nonzero total polarized \emph{sea} difference 
$ \Delta \overline{u}  - \Delta \overline{d}$. 
(We use the term ``sea difference" rather than the conventional  ``\emph{sea} asymmetry"  
in order to avoid confusion with spin asymmetry, a central object of this review.)
Ref.~\cite{Chang:2014jba} recently reviewed 
the nucleon \emph{sea} content, including its polarization. An unpolarized 
non-zero \emph{sea} difference $\overline{u} -  \overline{d} \approx -0.12$
has been known since the early 1990s~\cite{Baldit:1994jk, Amaudruz:1991at}. 
Such phenomenon must be nonperturbative  since the perturbative process $g \to q \bar{q}$ 
generating \emph{sea quarks} is nearly symmetric, and Pauli blocking for 
$g \to u \bar{u}$ in the proton ($g \to d \bar{d}$ in the neutron) is expected to be very small.  
Many of the nonperturbative processes proposed for $ \overline{u} - \overline{d} \neq 0$ also predict  
$ \Delta \overline{u} - \Delta \overline{d} \neq 0$.   
As mentioned,  $\overline{u} -  \overline{d}$ may be related to the total OAM, see Eq.~(\ref{Eq. L propto sea}).
Table~\ref{table sea asy} provides data and predictions for $ \Delta \overline{u}  - \Delta \overline{d}$. 
Other predictions are provided in Refs.~\cite{Kumano:2001cu}.

 \paragraph{Spin from intrinsic heavy-quarks}
 
 More generally, the nonperturbative contribution  to the nucleon  spin
arising from its ``intrinsic" heavy quark Fock states -- intrinsic strangeness, charm, and 
bottom~\cite{Brodsky:1984nx} -- is an interesting question. 
Such contributions arise from $Q \bar Q$ pairs which are 
multiply connected to the valence quarks. One can show from the OPE that the probability of heavy 
quark Fock states such as $|uud Q\bar Q \rangle$ in the proton scales as 
$1/M^2_Q$ ~\cite{Brodsky:1984nx, Franz:2000ee}.   
In the case of Abelian theory, a Fock state such as $ |e^+ e^- L \bar L \rangle$ in positronium atoms 
arises from the heavy lepton loop light-by-light insertion in the self-energy of positronium.  
In the Abelian case the probability scales as $1/M^4_L$.
The proton spin $J^z$ can receive contributions from the spin $S^z$ of the heavy quarks 
in the $|uud Q \bar Q\rangle$ Fock state. For example, the least off-shell hadronic contribution
to the $|uud s \bar s\rangle$  Fock state has a dual representation as  a 
$|K^+(u\bar s)  \Lambda(uds) \rangle$ fluctuation where the polarization of the $\Lambda$  hyperon is 
opposite to the proton spin $J^z$~\cite{Brodsky:1996hc}. Since the spin of the $s$ quark is 
aligned with $\Lambda$ spin, the $s$ quark will have spin $S^z_s$ opposite to the proton $J^z$. 
The $\bar s$ in the $K^+$ is unaligned. Similarly, the spin $S^z_c$ of the intrinsic charm quark 
from the $|D^+(u\bar c)  \Lambda_c(udc) \rangle$ fluctuation of the proton will also be anti-aligned to 
the proton spin.  The magnitude of the spin correlation of the intrinsic Q quark with the proton is thus 
bounded by the $|uudQ\bar Q\rangle$ Fock state probability. The net spin correlation of 
the intrinsic heavy quarks can be bounded using the OPE~\cite{Polyakov:1998rb}.
It is also of interest to consider the intrinsic heavy quark distributions of nuclei. 
For example, as shown in Ref.~\cite{Brodsky:2018zdh}, the gluon and intrinsic heavy quark 
content of the deuteron will be enhanced due its ``hidden-color" degrees of 
freedom~\cite{Brodsky:1983vf, Bashkanov:2013cla}, such as $ |(uud)_{8_C}  (ddu)_{8_C} \rangle$.

\paragraph{The gluon contribution to the proton spin}
$\Delta g/g (x_{Bj})$ and $\Delta g (x_{Bj})$ 
have been determined from either global fits to $g_1$ data
\emph{via} the sensitivity introduced by the \emph{DGLAP} equations, or from more direct semi-exclusive processes. 
Tables~\ref{table Delta G 1} and~\ref{table Delta G 2} summarize the current information on $\Delta G$ and 
$\Delta G+\mbox{L}_g$.  Results on $\Delta g/g$ are shown in Fig.~\ref{fig:partons_polar_vs_x}.
The averaged value is  $\Delta g/g = 0.113 \pm 0.038$(stat)$\pm0.035$(syst).

\paragraph{Orbital angular momenta \label{OAM}}
Of all the nucleon spin  components, the OAMs are  the hardest to measure.  
Quark OAM can be extracted \emph{via} the GPDs $E$ and $H$, see Eq.~(\ref{Eq. Ji SR}), 
the two-parton twist~3 GPD $G_2$, see Eq.~(\ref{Eq. quark OAM from twist-3}),
or GTMDs. They can also be assessed using TMDs
with nucleon structure models~\cite{Gutsche:2016gcd}. 
While GPDs yield the kinematical OAM, 
GTMDs provide the canonical definition, see Section~\ref{SSR components}.
GPD and GTMD measurements are difficult and, in order to obtain the OAM, 
must be extensive since sum rule analyses are required. 
The present dearth of data can be alleviated by models if 
the data are sufficiently constraining  so that the model dependence
is minimal. See Refs.~\cite{Bacchetta:2011gx, Courtoy:2016des} for examples of such work.  
In Ref.~\cite{Bacchetta:2011gx}, 
a model is used to connect $E$ and the Sivers TMD. 
The fit to the single-spin transverse asymmetries allows to extract the TMD, to which $E$ is connected and then used to extracted $J_q$. 
Thus $L_q=J_q-\Delta q/2 $  can be obtained.
In Ref.~\cite{Courtoy:2016des}, the quark OAM is computed within a
bag model using Eq.~(\ref{Eq. quark OAM from twist-3}).
A LF analysis of the deuteron single-spin transverse asymmetry~\cite{Alexakhin:2005iw} also constrains OAM and 
suggests a small  value for $\mbox{L}_g$~\cite{Brodsky:2006ha}. Similar conclusions are reached 
using measurements of the $pp^\uparrow \to \pi^0X$ single spin transverse asymmetry \cite{Anselmino:2006yq}. 
LGT can predict $L_q$ by calculating $J_q$ and
subtracting the computed or experimentally known $\Delta q/2 $.  $\mbox{L}_g$ is obtained
likewise. 
Alternatively, a first direct LGT calculation of quark OAMs obtained from the cross-product of 
position and momentum is outlined in~\cite{Engelhardt:2017miy}. Quark OAMs are obtained 
from GTMDs~\cite{Lorce:2011kd, Hatta:2011ku, Rajan:2016tlg} and can be set to
follow the canonical $l_q$ or kinematical $L_q$ OAM definition, or any definition in between
by varying the shape of the Wilson link chosen for the calculation. $l_q$ and $L_q$ can be 
compared, as well as how they transform into each other, and it is found that $l_q > L_q$.

Early LGT calculations, which indicated small $L_q$ values, did not include the contributions of disconnected diagrams.
More recent calculations including the disconnected diagrams yield
larger values for the quark OAM, in agreement with several observations: 
A) The predictions from  LF  at first order and the Skyrme model  that in the nonperturbative domain, 
the spin of the nucleon comes entirely from the quark OAM, see Section~\ref{sec:LFHQCD}.
B) The relativistic quark model which predicts $ l_q\approx 0.2$~\cite{Jaffe:1989jz, Brodsky:1994fz}; 
C) The $\Delta d/d$ high-$x_{Bj}$ data that is understood within pQCD only 
if the quark OAM is sizeable~\cite{Avakian:2007xa}, see Section~\ref{pqcd high-x};
and 
D) A non-zero nucleon anomalous magnetic moment implies a non-zero quark OAM~\cite{Brodsky:1980zm,Burkardt:2005km}. 
Although $L_q$ is dominated by disconnected diagrams in LGT, they are absent in the LF and quark models, 
and highly suppressed for the large-$x_{Bj}$ data. 
Thus, although the various approaches agree that the quark OAM is important, the underlying mechanisms are evidently different.

Tables~\ref{table OAM 1} and~\ref{table OAM 2}  provide the LGT 
results and the indirect phenomenological determinations from single spin asymmetries.
If only the quark OAM or quark total angular momenta are provided in a reference, we have computed the other one assuming
$\Delta u/2 = 0.41(2)$,  $\Delta d/2 = -0.22(2)$ and $\Delta s/2 = -0.05(5)$. One notices on the tables
that the strange quark OAM seems to be of opposite sign to $\Delta s$, effectively suppressing the total strange quark contribution to the nucleon
spin.

\subsubsection{High-energy picture of the nucleon spin structure}
\begin{figure}
\center
\includegraphics[scale=0.52]{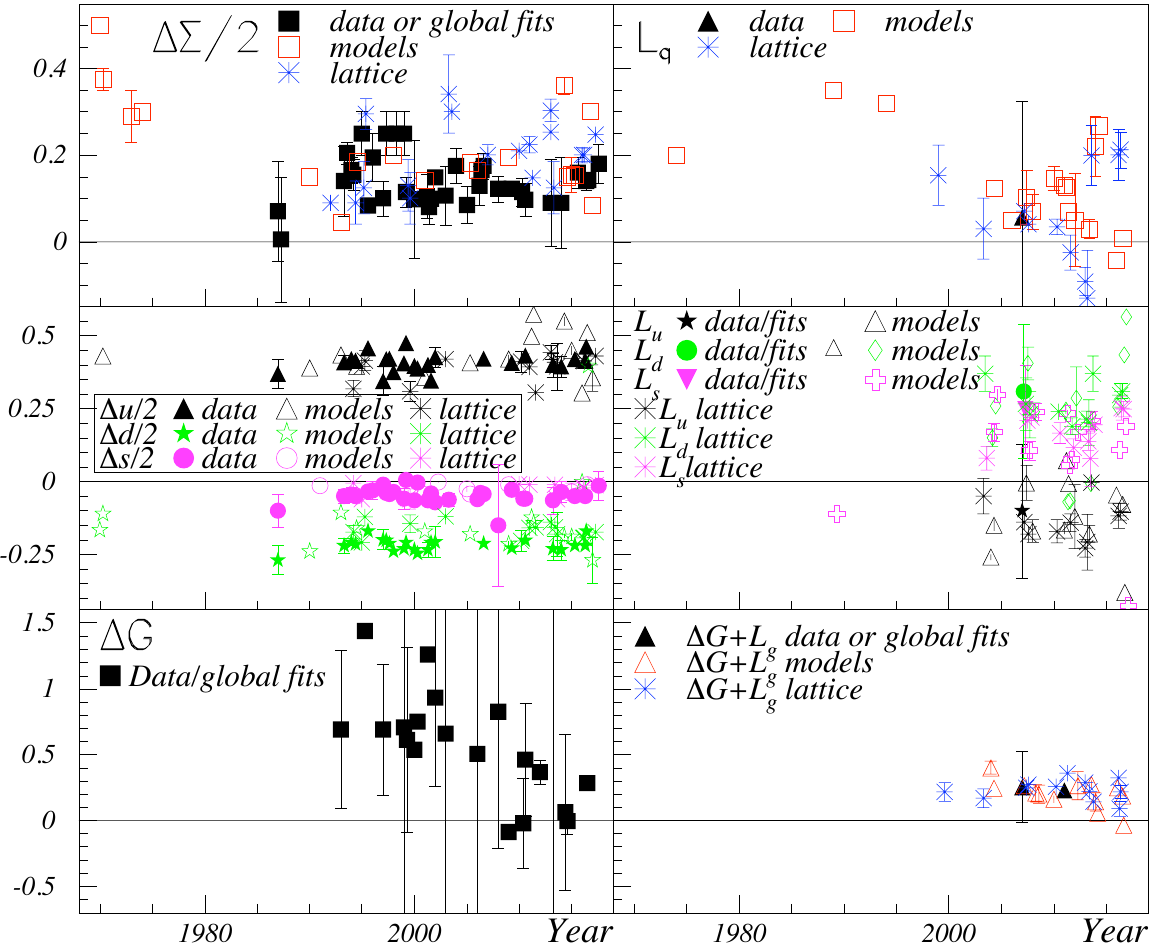}
\vspace{-0.45cm}
\caption{\label{spin history}\small{History of the measurements, models and LGT results on 
$\Delta \Sigma/2$ (top left panel);
$\mbox{L}$ (top right panel);
$(\Delta q+ \Delta\bar{q})/2$ (middle left panel);
quark OAM for light flavors  (middle right panel);
$\Delta G$ (bottom left panel);
and $\Delta G+\mbox{L}_g$ (bottom right panel).
The results shown, from Tables~\ref{table Delta Sigma 1}-\ref{table OAM 2}, are not comprehensive. 
The determinations of $\mbox{L}$ use
different definitions, and may thus not be directly comparable, see Section.~\ref{SSR components}.
The data points are significantly correlated since 
they use the same data set and/or related assumptions and/or similar approximations, {\it e.g.}, quench approximation
or neglecting disconnected diagrams for the earlier LGT results. Values were LO-evolved to $Q^2=4$ GeV$^2$. 
The uncertainties, when available, were not evolved.}}
\end{figure}

The contributions to $J$  listed in Tables~\ref{table Delta Sigma 1}-\ref{table OAM 2} 
are shown on Fig.~\ref{spin history}. It allows for a visualization of the evolution of our knowledge. 
While the measured $\Delta u+\Delta \overline{u}$ agrees with the relativistic quark model, 
its prediction for $\Delta d+\Delta \overline{d}$ is 50\% smaller than the data.  Thus  the 
failure of the relativistic quark model stems in part from neglecting the \emph{sea quarks}, 
chiefly $\Delta \overline{d}$ and $\Delta s$  to a lesser extent. 
The situation for the quark OAM is still unclear due the data scarcity. 
%
The indication that $\mbox{L}_s$ and $\Delta s$ have opposite signs reduces the overall strange 
quark contribution to $J$ to a second-order effect. 
Finally, $\Delta G+\mbox{L}_g$ appears to be of moderate size and thus not as important as initially thought.

The picture of the nucleon spin structure arising from these high-energy results is as follows: 
The nucleon appears as a mixture of quasi-free quarks and bremsstrahlung-created gluons, 
which in turn generate \emph{sea quarks}. At $Q^2 \sim 4$~GeV$^2$,
the \emph{valence quarks} carry between 30\% to 40\% of $J$. 
The \emph{sea} quarks contribute a smaller value and have opposite 
sign -- about $-10\%$; it is dominated by $\Delta \bar{d}$. 
The gluons carry about 20\% to 40\% of $J$. 
The remainder, up to ~50\%, comes from the quark OAM. 
This agrees with the asymptotic  prediction  $L_q \to \Delta\Sigma(Q_0) + \frac{3n_f}{32+6n_f}$,
assuming  $Q_0 \approx$~ 1 GeV for the \emph{DGLAP} evolution starting scale. 
This, together with the  LFHQCD first order prediction that the spin of the nucleon comes entirely from the quark OAM, and hence
$\Delta\Sigma(Q_0)=0$ yields $L_q \xrightarrow[Q^2 \to \infty]~0.52 \,J$ at LO.
Part of this physics can be understood as a relativistic effect, the consequence of the Dirac equation 
for light quarks in a confining potential. In the \emph{constituent quark} model, this effect  
is about $0.3 \,J$. 

Finally, DIS experiments indicate small \emph{higher-twist} contributions, 
{\it i.e.}, power-law suppressed contributions from parton correlations such as quark-quark interactions, even though
the lower $Q$ values of the SLAC or HERMES experiments are of the GeV order, close to the 
$\kappa \approx 0.5 $~GeV confinement scale~\cite{Brodsky:2016yod}. 
This is surprising since such correlations are related to quark confinement. 
(We refer to $\kappa$ rather than $\Lambda_s$ which is renormalization scheme dependent and hence ambiguous. 
Typically $0.3 < \Lambda_s <1$~GeV~\cite{Deur:2016tte}.) 


\subsubsection{Pending Questions}

\noindent The polarized DIS experiments leave several important questions open: 

\noindent$\bullet$ Why is scale invariance precocious ({\it i.e.}, why are \emph{higher-twist} effects small)?

\noindent$\bullet$ What are precisely the values of  $\Delta G$,   $\mbox{L}_q$  and $\mbox{L}_g$?

\noindent$\bullet$ What are the values and roles of parton correlations (\emph{higher-twists}), and their connection
to strong-QCD phenomena such as confinement and hadronic degrees of freedom?

\noindent$\bullet$ Is the nucleon simpler to understand at high $x_{Bj} $? 

\noindent$\bullet$ How does the transverse momentum influence  the nucleon spin structure?

\noindent$\bullet$ What is the behavior of the polarized PDFs at small $x_{Bj}$? 

Except for the two last points, recent inclusive data at lower energy have partially addressed these questions, as will be discussed below.
Experiments which measure  GPDs and GTMDs are relevant to all of these questions, 
except for the last point 
which can be addressed  by future polarized EIC experiments.

\subsubsection{Contributions from lower energy data}

The information gained from low energy experiments includes parton correlations, the high-$x_{Bj}$ 
domain of structure functions, the various contributions to the nucleon spin, the transition between the hadronic  
and partonic degrees of freedom, and tests of nucleon structure models.

\noindent$\bullet$ Parton correlations:
Overall \emph{higher-twist}  leads only to small deviations from Bjorken scaling even at $Q^2 \approx 1$~GeV$^2$. 
In fact, the low-$Q^2$ data allow us to  quantify the characteristic scale $Q_0$ at which 
\emph{leading-twist} pQCD fails, see Section~\ref{sec:perspectives}. 
In the $\overline{MS}$ scheme and N$^4$LO, $ Q_0\approx0.75$~GeV.
Individual  \emph{higher-twist} contributions, however, can be significant. 
For example,  for $\Gamma_1(Q^2 = 1$~GeV$^2)$
$f_2$ (twist~4) has similar strength as  $\Gamma_1^{\mbox{\scriptsize{twist~2}}}$. 
The overall  smallness of  the total \emph{higher-twist} effect comes from the sign 
alternation of the $Q^{2-twist}$ series
and the similar magnitude of its coefficients near $Q^2 = 1$~GeV$^2$.

\noindent$\bullet$ The $x_{Bj}$-dependence of the effect of parton correlations has 
been determined for $g_1$;   
the dynamical \emph{higher-twist} contribution was found to be significant at moderate $x_{Bj}$ but becomes 
less important at high and low $x_{Bj} $.  Since $ g_1 $ is itself small at high $x_{Bj} $, 
 \emph{higher-twists} remain important there. 
This conclusion can agree with the absence of large \emph{higher-twist} contribution in $g_1^n $ 
for $Q^2 \sim 1$~GeV$^2$ (Fig.~\ref{fig:E97-103 g2}), if 
kinematical  \emph{higher-twist} contribution cancels the dynamical contribution.

\noindent$\bullet$ The verification of the Burkhardt-Cottingham sum rule, 
Eq.~(\ref{eq:bc_noel}), implies that $ g_2$ is not singular. 
This should apply to each term of the $ g_2 $ \emph{twist} series.

\noindent$\bullet$ At $Q^2 <1 $~GeV$^2$,  \emph{higher-twist} effects become noticeable:
For example, at $Q^2 = 0.6 $~GeV$^2$, their contribution to $g_2^n $ appears to be similar  to the \emph{twist}-2 term contributing to 
 $ g_2^{WW}$ (Fig.~\ref{fig:E97-103 g2}), although uncertainties remain important. 
%
%
%
 
The indications that the overall  \emph{higher-twist}  contributions are under control 
allow one to extend  the database used to extract the polarized PDFs
~\cite{Leader:2006xc, Jimenez-Delgado:2013boa, Shahri:2016uzl}.

\noindent \textbf{High-$x_{Bj}$ data}

Measurements from  JLab experiments have provided the first significant  constraints on polarized PDFs at 
high $x_{Bj}$. 
\emph{Valence quark} dominance is confirmed. 

\noindent \textbf{Information on the nucleon spin components}

The data at  high $x_{Bj}$ have constrained  
$ \Delta \Sigma $, the quark OAM and $ \Delta G $. For example, in the global analysis  of 
Ref.~\cite{Leader:2006xc}, the uncertainty on $ \Delta G $ 
has decreased by a factor of 2 at $x_{Bj} = 0.3 $ and by a factor of 4 at $x_{Bj } =  0.5$. 
Furthermore, these data have revealed the importance of the quark OAM. 
However, to reliably obtain its value, the 
quark wave functions of the nucleon have to be known for all $x_{Bj}$, rather 
than only at high $x_{Bj} $.  

Fits of the $ \Gamma_1$ data at $Q^2> 1$~GeV$^2$ indicate $ \Delta \Sigma^p = 0.15 \pm 0.07 $  
and $ \Delta \Sigma^n = 0.35 \pm 0.08 $. 
This difference suggests an  insufficient knowledge $ g_1 $ at low $x_{Bj}$, rather than a breaking of isospin
symmetry. 
 
\noindent \textbf{The transition between partonic and hadronic descriptions}

At large $Q^2$, data and pQCD predictions agree well without  the  need to account for parton correlations;  
this  is at first surprising, but it can be understood in terms of  \emph{higher-twist} contributions of alternating signs.
At intermediate $Q^2$, the transition between descriptions based on partonic 
versus hadronic descriptions of the strong force such as the $\chi$PT approach,  is 
characterized by a marked $Q^2$-evolution for most moments. However, the evolution 
is smooth  {\it e.g.}, without indication of phase transition, an important fact in the context 
of Section~\ref{sec:perspectives}. 
At lower $Q^2$, $\chi$PT predictions  initially disagreed with most of 
the data for structure function moments. 
Recent  calculations agree better, but some challenges remain for $\chi$PT. 
New LGT methods are being  developed 
which should  allow tractable, reliable first principle calculations of PDFs. 

\noindent \textbf{Neutron information}

Constraints on neutron structure extracted from experiments using deuteron and 
$^3$He targets appear to be consistent; 
this validates the use of light nuclei  as effective polarized neutron targets in the $Q^2$  range of the data. 
These results provide 
complementary checks on nuclear effects: such effects are small ($\approx10\%$) for $^3$He due 
to the near cancelation between proton spins,  
but nuclear corrections are difficult to compute since the $^3$He nucleus
is tightly bound. Conversely, the corrections are large ($\approx50\%$) for the deuteron 
but more computationally tractable because the deuteron is a weakly bound $n-p$ object.

\section{Perspectives:  Unexpected connections \label{sec:perspectives}}

Studying nucleon structure is fundamental since nucleons represent most of the known matter. 
It provides primary information on the strong force and the confinement of quarks and gluons.
We provide here an example of what has been learned from doubly polarized inclusive 
experiments at moderate $Q^2$ from JLab. 
These experiments determined the $Q^2$-dependence of spin observables and 
thus constrained the connections between partonic and hadronic degrees of freedom. 
A goal of these  experiments was  to motivate new 
nonperturbative theoretical approaches and insights into understanding nonperturbative QCD. 
We  discuss here how this goal was achieved. 

As discussed at the end of the previous Section, the data at the transition 
between the perturbative and  nonperturbative-QCD domains evolve smoothly.
A dramatic behavior could have been expected from the pole structure of the perturbative running coupling;
 $\alpha_s  \xrightarrow[Q \to \Lambda_s]{}\infty$. However, this  \emph{Landau pole}
is unphysical and only signals the breakdown of pQCD~\cite{Deur:2016tte} rather than the
actual behavior of $\alpha_s$.   In contrast, a smooth behavior is observed   {\it e.g.} for the Bjorken 
sum $\Gamma_1^{p-n}$, see Fig.~\ref{fig:gamma1pn}.
At low $Q^2$, $\Gamma_1^{p-n}$ is effectively $Q^2$-independent, {\it i.e.}, QCD's 
approximate \emph{conformal} behavior seen at large $Q^2$ (Bjorken scaling) is recovered
at low $Q^2$  (see Section~\ref{conformal_sym}). 
This permits us to use the AdS/CFT correspondence~\cite{Maldacena:1997re}  an incarnation of which is
the LFHQCD framework~\cite{Brodsky:2014yha}, see Section~\ref{sec:LFHQCD}, which predicts that
%
$\Gamma_1^{p-n} (Q^2) = \big(1- e^{-\frac{Q^2}{4\kappa^2}} \big)/6 $~\cite{Brodsky:2010ur}.
%
Data~\cite{Deur:2005cf} and LFHQCD prediction agree well;  see 
Fig.~\ref{fig:gamma1pn}. Remarkably, the prediction 
has no adjustable parameters since $\kappa$ is fixed by hadron masses (in  Fig.~\ref{fig:gamma1pn},
$\kappa=M_{\rho}/\sqrt{2}$).

The LFHQCD prediction is valid up to $Q^2 \approx 1$~GeV$^2$. 
At higher $Q^2$, gluonic corrections not included in LFHQCD become important. However, there
pQCD's Eq.~(\ref{eq:mu4}) may be applied. The validity domains of LFHQCD and pQCD  overlap 
around $Q^2 \approx 1$~GeV$^2$;  matching the magnitude and the first derivative of their predictions 
allows one to relate the pQCD parameter $\Lambda_s$ to the LFHQCD
parameter $\kappa$ or equivalently to hadronic masses~\cite{Deur:2016cxb}. 
For example, in ${\overline{MS}}$ scheme at LO,
$\Lambda_{\overline{MS}}=M_\rho e^{-a}/\sqrt{a},$
 \label{eq: Lambda LO analytical relation}
where $a=4\big(\sqrt{\ln(2)^{2}+1+\beta_0/4}-\ln(2)\big)/\beta_0$. For $n_f = 3$ quark flavors, $a\approx 0.55$. 

\begin{figure}
\centerline{\includegraphics[width=.4\textwidth]{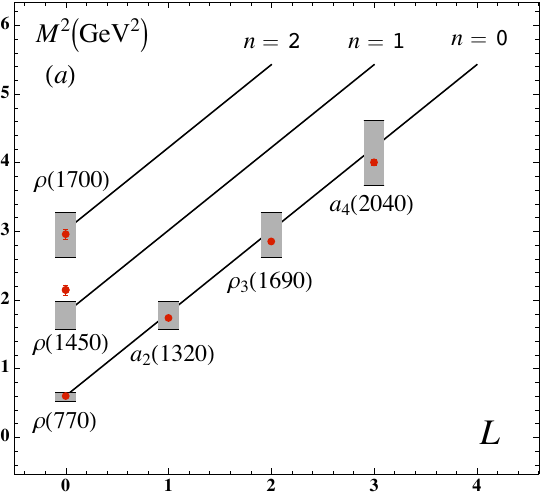} 
\includegraphics[width=.4\textwidth]{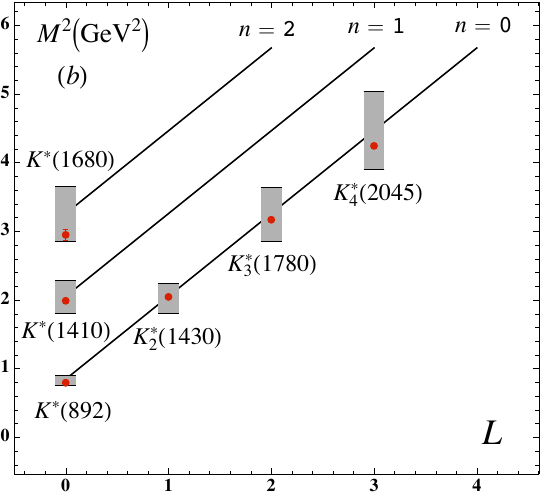}}
\vspace{-0.45cm}
\caption{\label{Fig:masses}\small The mass spectrum for unflavored (a) and 
strange light vector mesons (b) predicted by LFHQCD
using only $\Lambda_s$ as input~\cite{Deur:2014qfa} . 
The gray bands provide the uncertainty. The points indicate the experimental values.}
\vspace{-0.7cm}
\end{figure}

The $\rho$ meson is the ground-state solution of the  quark-antiquark LFHQCD Schr\"{o}dinger
equation including the spin-spin interaction~\cite{Brodsky:2016yod, deTeramond:2008ht}, {\it i.e.}, the solution with radial 
excitation $n=0$ and internal OAM $L=0$ and $S = 1$. Higher mass mesons are 
described with $n > 0$ or/and $L > 0$. They are shown in Fig.~\ref{Fig:masses}. 
The baryon spectrum can be obtained similarly or \emph{via} the mass symmetry between baryons and mesons 
using superconformal algebra~\cite{Dosch:2015nwa}.
%
Computing the hadron spectrum from $\Lambda_s$, such as shown in Fig.~\ref{Fig:masses}, 
has been a long-thought goal of the strong force studies. 
LFHQCD is not QCD but it represents a semiclassical approximation that successfully incorporates basic aspects of QCD's 
nonperturbative dynamics that are not explicit from its Lagrangian.   
Those include confinement and the emergence of a related mass scale, 
universal Regge trajectories, and a massless pion in the chiral limit~\cite{deTeramond:2016htp}.  
The confinement potential is determined by implementing QCD's \emph{conformal 
symmetry}, following de Alfaro, Fubini and Furlan who showed how a 
mass scale can be introduced in the  Hamiltonian without affecting the action 
\emph{conformal} invariance~\cite{deAlfaro:1976vlx, Brodsky:2013ar}.  
The potential is also related by LFHQCD to a dilaton-modified representation  
of the \emph{conformal} group in AdS$_5$ space,
Thus, the connection of the hadron mass spectrum~\cite{Deur:2014qfa} to key 
results derived from the QCD Bjorken sum rule represents an exciting progress 
toward long-sought goals of physics, and it provides an example of how spin studies 
foster progress in our understanding of fundamental physics.  Another profound connection 
relates the holographic structure of form factors (and unpolarized quark distributions), which 
depends on the number of components of a bound state, to the properties of the Regge 
trajectory of the vector meson that couples to the quark current in a given 
hadron~\cite{deTeramond:2018ecg, Sufian:2018cpj}. This procedure  
incorporates axial currents and the axial-vector meson spectrum to describe 
axial form factors and the structure of polarized quark distributions in the LFHQCD 
approach~\cite{HLFHS:2019}.

\section{Outlook\label{cha:Futur-results}}

We reviewed in Section~\ref{sec:data} the constraints on the composition of nucleon spin 
which have been obtained from existing doubly polarized inclusive data.
In Section~\ref{sec:perspectives}, we gave an example of the exciting 
advances obtained from these data.  In this section we will discuss constraints which can be 
obtained from presently scheduled future spin experiments. 
Most of these experiments are dedicated to measurements of 
GPDs and TMDs, which now provide the main avenue for spin structure studies.

JLab's upcoming experimental studies will utilize the upgrade of the electron beam energy from 6 to 12 
GeV.\footnote{Halls A, B and C, the halls involved in nucleon spin structure studies, are 
limited to 11 GeV, the 12 GeV beam being deliverable only to Hall D.}
The upgraded JLab retains its high polarized luminosity (several $10^{36}$ cm$^{-2}$s$^{-1}$)
which will allow larger kinematic coverage of the DIS region. 
In particular, higher values of $x_{Bj}$ will be reached, allowing for
$\Delta u / u $ and $\Delta d / d$ measurements up to $x_{Bj} \approx 0.8$ for $W> 2$~GeV. 
The quark OAM analysis discussed in Section~\ref{OAM} will thus be improved. 
Three such experiments have been approved for running: 
one on neutron utilizing  a $^3$He target  in JLab Hall A, one in Hall B on 
proton and neutron (Deuteron) targets, and  
the third one, planned in Hall C with a neutron ($^3$He) target~\cite{large-x A_1 12 GeV exps},
is scheduled to run very soon (2019).

The large solid angle detector CLAS12~\cite{Burkert:2008rj} in Hall B is well suited to measure $\Gamma_1$
up to $Q^2 = $ 6 GeV$^2$ and to minimize the low-$x_{Bj}$ uncertainties
at the values of $Q^2$ reached at 6 GeV.
These data will also refine the determination of \emph{higher twists}. 
In addition, inclusive data  from CLAS12 will  significantly constrain the 
polarized PDFs of the nucleons~\cite{Leader:2006xc}:
the precision on  $\Delta G$ extracted from lepton DIS \emph{via} \emph{DGLAP} analysis 
is expected to improve by  a factor of 3 at moderate and low $x_{Bj}$. It will complement the  $\Delta G$
measurements from p-p reactions at RHIC. The precision on $\Delta u$ 
and $\Delta d$ will improve by a factor of 2.
Knowledge of $\Delta s$ will be less improved since the inclusive data only give weak constraints.
Constrains on $\Delta s$  can be obtained in Hall A using the  SoLID~\cite{SoLID} 
experiment without assuming SU(3)$_f$  symmetry~\cite{LOI12-16-007}.
Measurements of $\Delta G$ at RHIC are expected to continue for another decade 
using the upgraded STAR and sPHENIX  detectors~\cite{Aschenauer:2016our},
until the advent of the electron-ion collider (EIC)~\cite{Accardi:2012qut}.

The GPDs are among the most important quantities to be measured at the upgraded 
JLab~\cite{12-06-114, Hall B DVCS 12 GeV prog.}. A first experiment
has already taken most of its data~\cite{12-06-114}. Since at $Q^2$ of a few GeV$^2$, 
$L_q$ appears to be the largest contribution to the nucleon spin,
the JLab GPD program is clearly crucial.  Information on the quark OAM will also be provided by measurements of the nucleon GTMDs 
on polarized H, D and $^3$He targets~\cite{12 GeV prog. SIDIS program} utilizing 
the Hall A and B SIDIS experimental programs.

The ongoing SIDIS and Drell-Yan measurements which access TMDs 
are expected to continue at CERN using the COMPASS phase-III upgrade.  
TMDs can also be measured with the upgraded STAR and sPHENIX detectors at 
RHIC~\cite{Aschenauer:2016our}. Spin experiments are also possible  at the LHC 
with polarized nucleon and nuclear targets using the proposed fixed-target facility AFTER@LHC~\cite{Brodsky:2012vg}.

Precise DIS data are lacking at $x_{Bj} \lesssim 0.05$  (see {\it e.g.}, the DSSV14 global 
fit~\cite{deFlorian:2014yva}).  The proposed EIC can access this domain 
with a luminosity of up to $10^{34}$ cm$^{-2}$s$^{-1}$. It will 
allow for  traditional polarized DIS, DDIS, SIDIS, exclusive
and charged current ($W^{+/-}$) DIS measurements.  
Precise inclusive data over a much extended $x_{Bj}$ range will yield
$\Delta G$ with increased precision from \emph{DGLAP} global fits. 
The discrepancy between $ \Delta \Sigma^p $ and $ \Delta \Sigma^n $ (Section~\ref{HT}),
which is most likely due to the paucity of low-$x_{Bj}$ data, should thus be clarified. Furthermore, the
tension between the $\Delta s$ from DIS and SIDIS can be solved by DIS charged current 
charm production with a high-luminosity collider such as the EIC. Charged current DIS will 
allow for flavor separation at high $Q^2$ and a first glance at
the  $g_5^{\gamma Z,n}$ structure function~\cite{Anselmino:1994gn}.

Other future facilities for nucleon spin structure studies are NICA (Nuclotron-based Ion Collider Facilities) 
at JINR in Dubna~\cite{Savin:2014sva}, and possibly an EIC in China (EIC@HIAF).
The NICA collider at Dubna was  approved in 2008; it  will provide  
polarized proton and deuteron beams up to $\sqrt  s =27$~GeV.
These beams will allow polarized Drell-Yan studies of TMDs and direct photon 
production which can access $\Delta G$.
China's HIAF (High Intensity Heavy Ion Accelerator Facility) was approved in 2015. EIC@HIAF, the 
facility relevant to nucleon spin studies, is not yet approved as of 2018.  The EIC@HIAF collider would provide a 3 GeV 
polarized electron beam colliding with 15 GeV polarized protons. It would  measure $\Delta s$, 
$\overline{\Delta u}-\overline{\Delta d}$, GPDs and TMDs over 
$0.01\leq x_{Bj}\leq0.2$ with a luminosity of about $5\times10^{32}$ cm$^{-2}$s$^{-1}$.
Improvements of the polarized sources, beams, and targets 
are proceeding at these facilities.

The success of the \emph{constituent quark} model in the early days of QCD suggested a simple picture for the origin 
of the nucleon spin: it was expected to come from the quark spins, $\Delta \Sigma = 1$. However, 
the first nucleon spin structure experiments, in particular EMC, showed that the nucleon spin 
composition is far from being trivial. This complexity means that spin degrees of  freedom reveal interesting 
information on the nucleon structure and on the strong force nonperturbative mechanisms.
The next  experimental step was the verification of the Bjorken sum rule, thereby 
verifying that QCD is valid even when spin degrees of  freedom are involved. 
The inclusive programs of SLAC, CERN and DESY also provided a mapping of the $x_{Bj}$ and $Q^2$ 
dependences of the $g_1$ structure function, yielding knowledge on the quark polarized distributions $\Delta q (x_{Bj})$ and some constraints 
on the gluon spin distribution $\Delta G$ and on \emph{higher twists}. 
The main goal of the subsequent JLab program was to study how partonic degrees 
of freedom merge to produce hadronic systems. These data have led to  
advances that permit an analytic computation of the
hadron mass spectrum with $\Lambda_s$ as the sole input. 
Such a calculation represents exciting progress toward reaching the long-sought and primary goals of strong force studies. 
The measurements and theoretical understanding discussed in this review, which has been focused 
on  doubly-polarized inclusive observables, have provided
testimony on the importance and dynamism of studies of the spin structure of the nucleon. The 
future prospects discussed here show that this research remains as dynamic as it was in the 
aftermath of the historical EMC measurement.

\paragraph{Acknowledgments} The authors thank S. \v{S}irca and R. S. Sufian for their useful comments on the manuscript. 
This review is based in part upon work supported by the U.S. Department of Energy, the Office of Science, 
and the Office of Nuclear Physics under contract DE-AC05-06OR23177. This work is also supported by the 
Department of Energy contract DE--AC02--76SF00515. 

\noindent
\begin{table}
\Large{\bf{Appendix.}}\normalsize~Tables of the contributions to the nucleon spin

~

\scriptsize{
\center
\begin{tabular}{|c|c|c|c|}
\hline 
Ref. & $Q^{2}$ (GeV$^{2}$) & $\Delta\Sigma$ & Remarks\tabularnewline
\hline
    - & - & 1  & naive quark model \tabularnewline
 \hline
 \cite{Jaffe:1989jz} & - & 0.75$\pm$0.05 & relativistic quark model \tabularnewline
\hline
 \cite{Ellis:1973kp} & - & 0.58$\pm0.12$ & Ellis-Jaffe SR \tabularnewline
\hline
 \cite{Sehgal:1974rz}  & - & 0.60  & quark parton model \tabularnewline
 \hline
\textbf{\cite{Ashman:1987hv}}  & \textbf{10.7} & \textbf{0.14$\pm$0.23} & \textbf{EMC} \tabularnewline
 \hline
\textbf{\cite{Jaffe:1989jz}}  & \textbf{10.7} & \textbf{0.01$\pm$0.29} & \textbf{EMC (Jaffe-manohar analysis)} \tabularnewline
\hline
\cite{Li:1990xr}  & - & 0.30 & Skyrme model \tabularnewline
\hline
\cite{Dorokhov:1993fc} &  - & $0.09 $ & Instanton model \tabularnewline
\hline
\textbf{\cite{Adeva:1993km}}  & \textbf{10}  & \textbf{0.28$\pm$0.16} & \textbf{SMC} \tabularnewline
 \hline
 \textbf{\cite{Close:1993mv}}  &  \textbf{-} &  \textbf{0.41$\pm0.05$} &  \textbf{global analysis} \tabularnewline
\hline
\textbf{\cite{Abe:1994cp}}  & \textbf{3}  & \textbf{0.33$\pm$0.06} & \textbf{E143} \tabularnewline
\hline
\textbf{\cite{Brodsky:1994kg}}  & \textbf{10} & \textbf{0.31$\pm$0.07} & \textbf{BBS} \tabularnewline
\hline
\cite{Cheng:1994zn}&  -  & 0.37 & $\chi$ quark model \tabularnewline
\hline
\textbf{\cite{Ball:1995td}}  & \textbf{1} & \textbf{0.5$\pm$0.1} & \textbf{global fit} \tabularnewline
\hline
\textbf{\cite{Gluck:1995yr}}&  \textbf{4}  & \textbf{0.168} & \textbf{GRSV 1995} \tabularnewline
\hline
\textbf{\cite{Anthony:1996mw}}  & \textbf{2} & \textbf{0.39$\pm$0.11} & \textbf{E142} \tabularnewline
\hline
\textbf{\cite{Abe:1997cx}}  & \textbf{5} & \textbf{0.20$\pm$0.08} & \textbf{E154} \tabularnewline
\hline
\textbf{\cite{Leader:1997kw}}  & \textbf{4} & \textbf{0.342} & \textbf{  LSS 1997} \tabularnewline
\hline
\cite{Qing:1998at}  & - & $0.4$ & relativistic quark model \tabularnewline
\hline
\textbf{\cite{Altarelli:1998nb}}  & \textbf{1} & \textbf{0.45$\pm$0.10} & \textbf{ABFR 1998} \tabularnewline
\hline
\textbf{\cite{Goto:1999by}}  & \textbf{5} & \textbf{0.26$\pm$0.02} & \textbf{AAC 2000} \tabularnewline
\hline
\textbf{\cite{Anthony:1999rm}}  & \textbf{5} & \textbf{0.23$\pm$0.07} & \textbf{E155} \tabularnewline
\hline
\textbf{\cite{Gluck:2000dy}}  & \textbf{5} &  \begin{tabular}{@{}c@{}} \textbf{0.197} \\ \textbf{0.273} \end{tabular} & \begin{tabular}{@{}c@{}} \textbf{Standard GRSV 2000}  \\ \textbf{SU(3)$_f$ breaking}\end{tabular}  \tabularnewline
\hline
\cite{Bourrely:2001du} & 4 & 0.282 & Stat. model \tabularnewline
\hline
\textbf{\cite{Leader:2001kh}}  & \textbf{1} & \textbf{0.21$\pm$0.10} & \textbf{LSS 2001} \tabularnewline
\hline
\textbf{\cite{Forte:2001ph}}  & \textbf{4} & \textbf{0.198} & \textbf{ABFR 2001} \tabularnewline
\hline
\textbf{\cite{Filippone:2001ux}} & \textbf{5} & \textbf{0.16$\pm$0.08} & \textbf{Global analysis} \tabularnewline
\hline
\textbf{\cite{Bluemlein:2002be}}  & \textbf{4} & \textbf{0.298} & \textbf{BB 2002} \tabularnewline
\hline
\textbf{\cite{Hirai:2003pm}}  & \textbf{5} & \textbf{0.213$\pm$0.138} & \textbf{AAC 2003} \tabularnewline
\hline
\textbf{\cite{Meziani:2004ne}} & \textbf{5} &  \textbf{0.35$\pm$0.08} & \textbf{Neutron ($^3$He) data (Section~\ref{HT})} \tabularnewline
\hline
\textbf{\cite{Fersch:2017qrq}} & \textbf{5} &  \textbf{0.169$\pm$0.084} & \textbf{Proton data (Section~\ref{HT})} \tabularnewline
 \hline
\cite{Silva:2005fa} & - &  0.366 &  $\chi$ quark soliton model \tabularnewline
\hline
\cite{Wakamatsu:2006dy, Wakamatsu:2007ar} & $\infty$ & $0.33 $&chiral quark soliton model.  $n_f=6$ \tabularnewline
\hline
\textbf{\cite{Hirai:2006sr}}  & \textbf{5} & \textbf{0.26$\pm$0.09} & \textbf{AAC 2006} \tabularnewline
\hline
\textbf{\cite{Airapetian:2006vy}} & \textbf{5} & \textbf{0.330$\pm$0.039}& \textbf{HERMES Glob. fit} \tabularnewline
\hline
\textbf{\cite{Alexakhin:2006oza}}& \textbf{10} & \textbf{0.35$\pm$0.06} & \textbf{COMPASS} \tabularnewline
\hline
\textbf{\cite{Hirai:2008aj}}  & \textbf{5} & \textbf{0.245$\pm$0.06} & \textbf{AAC 2008} \tabularnewline
\hline
\cite{Bass:2009ed} & $\approx0.2$ & 0.39 & cloudy bag model w/ SU(3)$_f$ breaking\tabularnewline
\hline 
\textbf{\cite{deFlorian:2009vb}} & \textbf{4} & \textbf{0.245} & \textbf{DSSV08} \tabularnewline
\hline
\textbf{\cite{Leader:2010rb}}  & \textbf{4} & \textbf{0.231$\pm$65} & \textbf{LSS 2010} \tabularnewline
\hline
\textbf{\cite{Blumlein:2010rn}} & \textbf{4}  & \textbf{0.193$\pm$75} &  \textbf{BB 2010} \tabularnewline
\hline
\cite{Altenbuchinger:2010sz} & $\approx0.2$  & $0.23\pm0.01$ & Gauge-invariant cloudy bag model \tabularnewline
\hline
\textbf{\cite{Ball:2013lla}}  & \textbf{4} & \textbf{0.18$\pm$0.20} & \textbf{NNPDF 2013} \tabularnewline
\hline 
\textbf{\cite{Nocera:2014gqa}}& \textbf{10} & \textbf{0.18$\pm$0.21} & \textbf{NNPDF 2014} \tabularnewline
\hline
\cite{Bijker:2014ila} & - & 0.72$\pm$0.04 & unquenched quark mod. \tabularnewline
\hline
\cite{Brodsky:2014yha}   &  5 &  0.30 & LFHQCD  \tabularnewline 
\hline
\cite{Li:2015exr} & 3 & $0.31\pm0.08$ & $\chi$ effective $\mathcal{L}$ model\tabularnewline
\hline
\cite{Liu:2015jna} &  - & 0.308 & LFHQCD  \tabularnewline
\hline
\textbf{\cite{Adolph:2015saz}}  & \textbf{3} & \textbf{0.32$\pm$0.07} & \textbf{COMPASS 2017 deuteron data} \tabularnewline
\hline
\textbf{\cite{Sato:2016tuz} }  & \textbf{5} & \textbf{0.28$\pm$0.04} & \textbf{JAM 2016}\tabularnewline
\hline
\cite{Dahiya:2016wjf} & $\approx 1$ & 0.602 & chiral quark model \tabularnewline
\hline
\textbf{\cite{Shahri:2016uzl}}& \textbf{5} & \textbf{0.285} &  \textbf{KTA17 global fit}  \tabularnewline
\hline
\cite{Maji:2017ill}    & 1 & 0.17 & AdS/QCD q-qq model  \tabularnewline
\hline 
\textbf{\cite{Ethier:2017zbq}} & \textbf{5} & \textbf{0.36$\pm$0.09} &  \textbf{JAM 2017} \tabularnewline\hline
\end{tabular}
\caption{\label{table Delta Sigma 1} \small Determinations of $\Delta\Sigma$ from  
experiments and models. Experimental results, including global fits,  are in bold 
and  are given in the $\overline{MS}$ scheme. The model list is indicative rather than comprehensive.
}
}
\end{table}
\noindent
\begin{table}
\scriptsize{
\center
\begin{tabular}{|c|c|c|c|}
\hline 
Ref. & $Q^{2}$ (GeV$^{2}$) & $\Delta\Sigma$ & Remarks\tabularnewline
\cite{Altmeyer:1992nt} & - & $0.18\pm0.02$& Altmeyer, Gockeler \emph{et al.} Quenched calc.   \tabularnewline
\hline
\cite{Fukugita:1994fh} & - & $0.18 \pm0.10$ & Fukugita \emph{et al.} Quenched calc. w/ $\chi$ extrap. \tabularnewline
\hline
\cite{Dong:1995rx} & - & $0.25 \pm0.12$ & U. Kentucky group. Quenched calc. w/ $\chi$ extrap.\tabularnewline
\hline
\cite{Gockeler:1995wg} & 2 & $0.59 \pm0.07$ & Gockeler \emph{et al.} u,d only. Quenched calc. w/ $\chi$ extrap.\tabularnewline
\hline
\cite{Mathur:1999uf}& 3  & $0.26 \pm 0.12$ &  U. Kentucky group. Quenched calc. w/ $\chi$ extrap.\tabularnewline
\hline
\cite{Gusken:1999xy} & 5 & $0.20 \pm 0.12$ & SESAM 1999. $\chi$ extrap. Unspecified RS \tabularnewline
\hline
 \cite{Hagler:2003jd} & 4  & $0.682 \pm 0.18$ &  LHPC 2003. u, d only. $\chi$ extrap.  \tabularnewline
\hline
\cite{Gockeler:2003jfa}  & 4 & $0.60 \pm 0.02$ &  QCDSF 2003. u, d only. Quenched calc. w/ $\chi$ extrap. \tabularnewline
\hline
\cite{Brommel:2007sb} & 4  & $0.402 \pm 0.048$ & QCDSF-UKQCD 2007. u, d only. $\chi$ extrap. \tabularnewline
\hline
\cite{Bratt:2010jn} & 5 & $0.42 \pm 0.02$ & LHPC. u, d only. $\chi$ extrap. \tabularnewline
 \hline
\cite{QCDSF:2011aa} & 7.4 & $0.448 \pm 0.037$  & QCDSF 2011.  $m_{\pi}$=285 MeV. Partly quenched calc.  \tabularnewline
\hline
 \cite{Alexandrou:2011nr} & 4  & $0.296 \pm 0.010$ & Twisted-Mass 2011 u, d only.   W/ $\chi$ extrap. \tabularnewline
\hline
 \cite{Alexandrou:2013joa} & 4  & 0.606$\pm$0.052 & Twisted-Mass 2013 u, d only. $m_{\pi}$=213 MeV \tabularnewline
\hline 
\cite{Abdel-Rehim:2013wlz}  &  4& 0.507$\pm$0.008 & Twisted-Mass 2013. Phys. q masses  \tabularnewline
\hline
\cite{Deka:2013zha} & 4 & $0.25 \pm 0.12$ & $\chi$QCD 2013. Quenched calc. w/ $\chi$ extrap. \tabularnewline
\hline
\cite{Alexandrou:2016mni} &  4 & $0.400 \pm 0.035$ & Twisted-Mass 2016. Phys. $\pi$ mass  \tabularnewline
\hline
\cite{Alexandrou:2016tuo} & 4 & 0.398$\pm0.031$ & Twisted-Mass 2017. Phys. $\pi$ mass   \tabularnewline
\hline
\cite{Green:2017keo} & 4 & $0.494 \pm 0.019$ & Partly quenched calc. $m_\pi = 317$ MeV \tabularnewline
\hline
\end{tabular}
\caption{\label{table Delta Sigma 2} \small Continuation of Table~\ref{table Delta Sigma 1}, for LGT results. They are given in the $\overline{MS}$ scheme unless stated otherwise. The list  is not comprehensive.
}
}
\end{table}
\noindent
\begin{table}
\scriptsize{
\center
\begin{tabular}{|c|c|c|c|c|c|}
\hline 
Ref. &  \begin{tabular}{@{}c@{}} $Q^{2}$ \\ (GeV$^{2}$) \end{tabular} & $\Delta u + \Delta \overline{u}$ & $\Delta d + \Delta \overline{d}$ & $\Delta s + \Delta \overline{s}$ & Remarks\tabularnewline
\hline
-  & - & 4/3 & -1/3 & 0 & quark model \tabularnewline
\hline
\cite{Jaffe:1989jz}  & - & 0.86 & -0.22 & 0 & relat. q. mod. \tabularnewline
\hline
\textbf{\cite{Ashman:1987hv}}  & \textbf{10} & \textbf{0.74(10)} & \textbf{-0.54(10)} & \textbf{-0.20(11)} & \textbf{EMC} \tabularnewline
\hline
\cite{Li:1990xr}  & - & 0.78 & -0.48 & 0 & Skyrme model \tabularnewline
\hline
\cite{Park:1991fb}& - & - &  -& -0.03 & $g_a^s$ SU(3) skyrme model \tabularnewline
\hline
\cite{Dorokhov:1993fc} &  - & 0.867 & -0.216 & - & Instanton model \tabularnewline
\hline
\textbf{\cite{Adeva:1993km}}  & \textbf{10}  & \textbf{0.82(5)} &\textbf{-0.44(5)} & \textbf{-0.10(5)} & \textbf{SMC} \tabularnewline
\hline
\textbf{\cite{Abe:1994cp}}  & \textbf{3} & \textbf{0.84(2)} & \textbf{-0.42(2)} & \textbf{-0.09(5)} & \textbf{E143} \tabularnewline
\hline
\textbf{\cite{Brodsky:1994kg}}  & \textbf{10} &\textbf{0.83(3)} &  \textbf{-0.43(3)} & \textbf{-0.10(3)} & \textbf{BBS} \tabularnewline
\hline
\cite{Cheng:1994zn}&  -  & 0.79 & -0.32 & -0.10 & $\chi$ quark model \tabularnewline
\hline
\textbf{\cite{Gluck:1995yr}}&  \textbf{4}  & \textbf{0.914} & \textbf{-0.338} & \textbf{-0.068} & \textbf{GRSV 1995} \tabularnewline
\hline
\textbf{\cite{Anthony:1996mw}}  & \textbf{2} & - & -& \textbf{-0.06(6)} & \textbf{E142} \tabularnewline
\hline
\textbf{\cite{Abe:1997cx}}  & \textbf{5} & \textbf{0.69$^{(15)}_{(5)}$}  & \textbf{-0.40$^{(8)}_{(5)}$} & \textbf{ -0.02$^{(1)}_{(4)}$} & \textbf{E154} \tabularnewline
\hline
\textbf{\cite{Leader:1997kw}}  & \textbf{4} & \textbf{0.839} & \textbf{-0.405} & \textbf{-0.079} & \textbf{LSS 1997} \tabularnewline
\hline
\textbf{\cite{Ackerstaff:1997ws}} & \textbf{5}& \textbf{0.842(13)} &  \textbf{-0.427(13)} &   \textbf{-0.085(18)} & \textbf{HERMES (1997)}  \tabularnewline
\hline
\cite{Qing:1998at}  & - & 0.75 & -0.48 & -0.07 & relat. quark model \tabularnewline
\hline
\textbf{\cite{Goto:1999by}}  & \textbf{5} & \textbf{0.812} & \textbf{-0.462} & \textbf{-0.118(74)} & \textbf{AAC 2000 global fit} \tabularnewline
\hline
\textbf{\cite{Anthony:1999rm}}  & \textbf{5} & \textbf{0.95} & \textbf{-0.42} & \textbf{0.01} & \textbf{E155} \tabularnewline
\hline
\textbf{\cite{Gluck:2000dy}}  & 5 & 
\textbf{ \begin{tabular}{@{}c@{}} 0.795 \\ 0.774  \end{tabular}} & 
\textbf{ \begin{tabular}{@{}c@{}} -0.470 \\ -0.493 \end{tabular}} & 
\textbf{ \begin{tabular}{@{}c@{}} -0.128 \\ -0.006 \end{tabular}} & 
\textbf{ \begin{tabular}{@{}c@{}} \textbf{Standard GRSV 2000}  \\ \textbf{SU(3)$_f$ breaking}\end{tabular}}  \tabularnewline
 \hline
\cite{Bourrely:2001du} & 4 & 0.714 & -0.344 & -0.088 &  Stat. model \tabularnewline
\hline
\textbf{\cite{Leader:2001kh}}  & \textbf{1} & \textbf{0.80(3)} & \textbf{-0.47(5)} & \textbf{-0.13(4)} & \textbf{LSS 2001} \tabularnewline
\hline
\textbf{\cite{Forte:2001ph}}  & \textbf{4} & \textbf{0.692} & \textbf{-0.418} & \textbf{-0.081} & \textbf{ABFR 2001} \tabularnewline
\hline
\textbf{\cite{Bluemlein:2002be}}  & \textbf{4} & \textbf{0.854(66)} & \textbf{-0.413(104)} & \textbf{-0.143(34)} & \textbf{BB 2002} \tabularnewline
\hline
\cite{Lyubovitskij:2002ng}& - & - &  -& -0.0052(15) & $g_a^s$ chiral quark model \tabularnewline
\hline
\textbf{\cite{Hirai:2003pm}}  & \textbf{5} & \textbf{-} & \textbf{-} & \textbf{-0.124(46)} & \textbf{AAC 2003} \tabularnewline
\hline
\cite{An:2005cj}& - & - &  -& - 0.05(2) & $g_a^s$ pentaquark model \tabularnewline
\hline
\cite{Silva:2005fa} & - & 0.814 & -0.362 & -0.086 &  $\chi$ quark soliton model \tabularnewline
\hline
\textbf{\cite{Hirai:2006sr}}  & \textbf{5} & \textbf{-} & \textbf{-} & \textbf{-0.12(4)} & \textbf{AAC 2006} \tabularnewline
\hline
\textbf{\cite{Alexakhin:2006oza}}& \textbf{10} & \textbf{-} & \textbf{-}& \textbf{-0.08(3)} & \textbf{COMPASS} \tabularnewline
\hline
\textbf{\cite{Airapetian:2006vy}} & \textbf{5} & \textbf{0.842(13)} & \textbf{-0.427(13)} & \textbf{-0.085(18)} & \textbf{HERMES Glob. fit} \tabularnewline
\hline
\textbf{\cite{Pate:2008va}}  & \textbf{-} &  \textbf{-} & \textbf{-}  & \textbf{-0.30(42)} & \textbf{PV + $\nu$ data}  \tabularnewline
 \hline
\cite{Bass:2009ed} & - & 0.84(2) & -0.43(2) & -0.02(2) & cloudy bag model w/ SU(3)$_f$ breaking\tabularnewline
\hline
\textbf{\cite{deFlorian:2009vb}} & \textbf{4} & \textbf{0.814} & \textbf{-0.456} & \textbf{-0.056} & \textbf{DSSV08} \tabularnewline 
\hline
\textbf{\cite{Leader:2010rb}}  & \textbf{4} & \textbf{-} & \textbf{-} & \textbf{-0.118(20)} & \textbf{LSS 2010} \tabularnewline
\hline
\textbf{\cite{Blumlein:2010rn}}  & \textbf{4} & \textbf{0.866(0)} & \textbf{-0.404(0)} & \textbf{-0.118(20)} & \textbf{BB 2010} \tabularnewline
\hline
\cite{Lorce:2011kd} &- &  0.996 & -0.248 & -  & LC const. quark mod. \tabularnewline
\hline
\cite{Lorce:2011kd} &- & 1.148  & -0.286 & -  & LC $\chi$ qu. solit. mod. \tabularnewline
\hline
\cite{Altenbuchinger:2010sz} & $\approx0.2$ &  $0.38\pm0.01$ & $-0.15\pm0.01$ & -  &  Gauge-invariant cloudy bag model \tabularnewline
\hline
\textbf{\cite{Ball:2013lla}}  & \textbf{1} & \textbf{0.80(8)} & \textbf{-0.46(8)} & \textbf{-0.13(9)} & \textbf{NNPDF 2013 } \tabularnewline
\hline
\textbf{\cite{Nocera:2014gqa}}& \textbf{10} & \textbf{0.79(7)}  & \textbf{-0.47(7)}   & \textbf{-0.07(7)} &  \textbf{NNPDF (2014)} \tabularnewline
\hline
\cite{Bijker:2014ila} & - & 1.10(3) & -0.38(1) & 0 & unquenched quark mod. \tabularnewline
\hline
\cite{Li:2015exr} & $\approx0.5$ & 0.90(3) & -0.38(3) & -0.07($^4_7$) & $\chi$ effective $\mathcal{L}$ model\tabularnewline
\hline
\textbf{\cite{Adolph:2015saz}} & \textbf{3} & \textbf{0.84(2)} & \textbf{-0.44(2)} & \textbf{-0.10(2)} & \textbf{D COMPASS} \tabularnewline
\hline 
\cite{Gutsche:2016gcd}& 1 & 0.606 & -0.002 & - & LF  quark mod. \tabularnewline
\hline
\textbf{\cite{Sato:2016tuz}} & \textbf{1} & \textbf{0.83(1)} & \textbf{-0.44(1)} & \textbf{-0.10(1)} & \textbf{JAM16} \tabularnewline
\hline
\textbf{\cite{Shahri:2016uzl}}& \textbf{5} & \textbf{0.926} & \textbf{-0.341} & \textbf{-} & \textbf{KTA16 global fit}  \tabularnewline
\hline
\cite{Dahiya:2016wjf} & $\approx 1$ & 1.024 & -0.398 & -0.023 & chiral quark model \tabularnewline
\hline
\cite{Chakrabarti:2016yuw}    & - & 1.892 & 0.792 & - & AdS/QCD q-qq model  \tabularnewline
\hline
\cite{Maji:2017ill}    & 1 & 0.71(9) & -0.54$^{19}_{13}$ & - & AdS/QCD q-qq model  \tabularnewline
\hline
\textbf{\cite{Ethier:2017zbq}} & \textbf{5} & \textbf{-} & \textbf{-} & \textbf{-0.03(10)} & \textbf{JAM17} \tabularnewline
\hline
\end{tabular}
\caption{\label{table Delta q 1}\small Same as Table~\ref{table Delta Sigma 1} but for  $\Delta q$. Results are ordered chronologically. The list for models is indicative rather than comprehensive.
}
}
\end{table}
\noindent
\begin{table}
\scriptsize{
\center
\begin{tabular}{|c|c|c|c|c|c|}
\hline 
Ref. &  \begin{tabular}{@{}c@{}} $Q^{2}$ \\ (GeV$^{2}$) \end{tabular} & $\Delta u + \Delta \overline{u}$ & $\Delta d + \Delta \overline{d}$ & $\Delta s + \Delta \overline{s}$ & Remarks\tabularnewline
\hline
\hline
\cite{Fukugita:1994fh} & - & 0.638(54) & -0.347(46) & -0.0109(30) & Fukugita \emph{et al.} Quenched calc. w/ $\chi$ extrap.\tabularnewline
\hline
\cite{Dong:1995rx} & - & 0.79(11) & -0.42(11) & -0.12(1) & U. Kentucky group. Quenched calc. w/ $\chi$ extrap. \tabularnewline
\hline
\cite{Gockeler:1995wg} & 2 & 0.830(70) & -0.244(22) & -& Gockeler \emph{et al.} u,d only. Quenched calc. w/ $\chi$ extrap.\tabularnewline
\hline
\cite{Mathur:1999uf} & 3  & - & - & -0.116(12) &  U. Kentucky group. Quenched calc.  w/ $\chi$ extrap. \tabularnewline
\hline
\cite{Gusken:1999xy} & 5  & 0.62(7)& -0.29(6)& -0.12(7) &  SESAM 1999. $\chi$ extrap. Unspecified RS \tabularnewline
\hline
\cite{Gockeler:2003jfa}  & 4  & 0.84(2)& -0.24(2)& -  &  QCDSF 2003. u, d only.  Quenched calc. w/ $\chi$ extrap.\tabularnewline 
\hline
\cite{Babich:2010at} & -  & - & - & -0.019(11) & Unrenormalized result.    W/ $\chi$ extrap.  \tabularnewline
\hline 
\cite{Bratt:2010jn} & 5   & 0.822(72) & -0.406(70) & - & LHPC 2010. u, d only. $\chi$ extrap.  \tabularnewline
\hline 
\cite{QCDSF:2011aa} &7.4   & 0.787(18) & -0.319(15) & -0.020(10) &   QCDSF 2011. $m_{\pi}$=285 MeV. Partly quenched calc. \tabularnewline
\hline
 \cite{Alexandrou:2011nr} & 4  & 0.610(14) & -0.314(10) & - & Twisted-Mass 2011 u, d only.   W/ $\chi$ extrap. \tabularnewline
\hline 
\cite{Abdel-Rehim:2013wlz}  &  4& 0.820(11) &-0.313(11) & -0.023(34) & Twisted-Mass 2013. Phys. q masses  \tabularnewline
\hline
 \cite{Alexandrou:2013joa} & 4  & 0.886(48) & -0.280(32) & - & Twisted-Mass 2013 u, d only. $m_{\pi}$=213 MeV \tabularnewline
\hline 
\cite{Deka:2013zha} & 4  & 0.79(16) & -0.36(15)  & -0.12(1) &  $\chi$QCD 2013. Quenched calc. w/ $\chi$ extrap. \tabularnewline
\hline 
\cite{Alexandrou:2016mni} &  4 & 0.828(32) & -0.387(20) & -0.042(10) &  Twisted-Mass 2016. Phys. $\pi$ mass \tabularnewline
\hline 
\cite{Alexandrou:2016tuo}& 4  & 0.826(26) & -0.386(14) & -0.042(10) & Twisted-Mass 2017. Phys. $\pi$ mass  \tabularnewline
\hline
\cite{Green:2017keo} & 4 & 0.863(17) & -0.345(11) & -0.0240(24) & Partly quenched calc. $m_\pi = 317$ MeV  \tabularnewline
\hline
\end{tabular}
\caption{\label{table Delta q 2}\small Continuation of Table~\ref{table Delta q 1}, for LGT results. They are given in the $\overline{MS}$ scheme unless stated otherwise. The list  is not comprehensive.
}
}
\end{table}
\noindent
\begin{table}
\center
\scriptsize{
\begin{tabular}{|c|c|c|c|c|c|} 
\hline 
Ref. & \begin{tabular}{@{}c@{}}  $Q^{2}$ \\ (GeV$^{2}$) \end{tabular}& $ \Delta \overline{u}  - \Delta \overline{d}$ & $ \Delta \overline{u}$  & $\Delta \overline{d}$ & Remarks\tabularnewline
\hline 
\hline
\cite{Thomas:1983fh} & - & 0  & 0 & 0 & $\pi$-cloud model \tabularnewline 
\hline
\cite{Dorokhov:1993fc} &  4 & 0.215  & - & - & Instanton model \tabularnewline 
\hline
\cite{Fries:1998at} & 2 & 0.014(13)   & - & - & $\rho$-cloud model \tabularnewline 
\hline
\textbf{\cite{Adeva:1997qz}} & \textbf{10} &  \textbf{0.00(19)}  & \textbf{0.01(6)}& \textbf{0.01(18)} & \textbf{SMC} \tabularnewline 
\hline
\cite{Boreskov:1998hp} & 4 & 0.76(1)  & -& -& cloud model, $\rho$-$\pi$ interf.  \tabularnewline 
\hline
\cite{Dressler:1998zi} & - & 0.31 & - & - & $\chi$ soliton model \tabularnewline
\hline
\textbf{\cite{Ackerstaff:1999ey}}  & \textbf{2.5} & \textbf{0.01(6)} & \textbf{-0.01(4)} &  \textbf{-0.02(5)} &  \textbf{HERMES} \tabularnewline
\hline
\textbf{\cite{Gluck:2000dy}}  & \textbf{5} & 
\begin{tabular}{@{}c@{}} \textbf{0} \\  \textbf{0.32} \end{tabular} & 
\begin{tabular}{@{}c@{}} \textbf{-0.064} \\ \textbf{0.085}  \end{tabular} & 
\begin{tabular}{@{}c@{}} \textbf{-0.064} \\  \textbf{-0.235} \end{tabular} & 
\begin{tabular}{@{}c@{}} \textbf{Standard GRSV 2000}  \\ \textbf{SU(3)$_f$ breaking}\end{tabular}  \tabularnewline
\hline
\cite{Cao:2001nu} & 4 & 0.023(31) & - & - & meson cloud bag model \tabularnewline
\hline
\cite{Dorokhov:2001pz} & - & 0.2 & - &-  &  Instanton model \tabularnewline
\hline
\cite{Bourrely:2001du} & 4 & 0.12 & 0.046 & -0.087 & Stat. model \tabularnewline
\hline
\cite{Steffens:2002zn}& - & 0.2 &  -& - & sea model with Pauli-blocking\tabularnewline
\hline
\cite{Fries:2002um}& 1 & 0.12 & - & - &  cloud model $\sigma$-$\pi$ interf.  \tabularnewline
\hline
\textbf{\cite{Airapetian:2004zf}}  & \textbf{2.5} & \textbf{0.048(64)} & \textbf{-0.002(23)} & \textbf{-0.054(35)} & \textbf{HERMES} \tabularnewline
\hline
\textbf{\cite{Alekseev:2007vi}}  & \textbf{10} & \textbf{0.00(5)} & - & - & \textbf{COMPASS} \tabularnewline
\hline
\textbf{\cite{deFlorian:2009vb}} & \textbf{5} & \textbf{0.15 }& \textbf{0.036} & \textbf{-0.114} &\textbf{DSSV08} \tabularnewline
\hline
\textbf{\cite{Alekseev:2009ac}}  & \textbf{3} & \textbf{-0.04(3)}  & - & - & \textbf{COMPASS} \tabularnewline
\hline
\textbf{\cite{Alekseev:2010ub}}  & \textbf{3} & \textbf{0.06(5)} &\textbf{ 0.02(2)} & \textbf{-0.05(4)} & \textbf{COMPASS} \tabularnewline
\hline
\textbf{\cite{Nocera:2014gqa}} & \textbf{10} & \textbf{0.17(8)} & \textbf{0.06(6)}  & \textbf{-0.11(6)} &  \textbf{NNPDF (2014)} \tabularnewline
\hline
 \begin{tabular}{@{}c@{}} \textbf{\cite{Adamczyk:2014xyw}} \\ \textbf{\cite{Adare:2015gsd}} \end{tabular}  & - &
- & 
\textbf{$>0$} &  - &  \begin{tabular}{@{}c@{}} \textbf{0.05$<x_{Bj}<$0.2. STAR}  \\  \textbf{and PHENIX $W^{\pm}$, $Z$ prod.} \end{tabular}  \tabularnewline
\hline
\textbf{\cite{Ethier:2017zbq}} & \textbf{5} & \textbf{0.05(8)} & - & - & \textbf{global fit  (JAM 2017)} \tabularnewline
\hline
\textbf{\cite{Lin:2014zya}} &  \textbf{4} & \textbf{0.24(6)}  & - & - & \textbf{$m_{\pi}$=310 MeV} \tabularnewline
\hline
\end{tabular}
\caption{\label{table sea asy} \small Phenomenological (top) and LGT (bottom) results on the sea asymmetry $ \Delta \overline{u}  - \Delta \overline{d}$. Results are in the $\overline{MS}$ scheme. 
The lists for models and LGT are ordered chronologically and are not comprehensive.
}}
\end{table}
\noindent
\begin{table}
\center
\scriptsize{
\vspace{-1.5cm}
\begin{tabular}{|c|c|c|c|}
\hline 
Ref. & $Q^{2}$ (GeV$^{2}$) & Contribution & Remarks\tabularnewline
\hline
\hline
\textbf{\cite{Adeva:1993km}}  & \textbf{5} & \textbf{$\Delta G$=0.9(6)} & \textbf{SMC incl. \emph{DGLAP}}\tabularnewline
\hline
\textbf{\cite{Brodsky:1994kg}}  & \textbf{1} & \textbf{$\Delta G$=0.5} & \textbf{BBS global fit} \tabularnewline
\hline
\textbf{\cite{Ball:1995td}}  & \textbf{1} & \textbf{$\Delta G$=1.5(8)} & \textbf{Ball \emph{et al.} global fit} \tabularnewline
\hline
\textbf{\cite{Gluck:1995yr}}&  \textbf{4}  &  \textbf{$\Delta G$=1.44} & \textbf{GRSV 1995} \tabularnewline
\hline
\textbf{\cite{Abe:1997cx}}  & \textbf{5} & \textbf{$\Delta G$=0.9(5)} & \textbf{E154 incl. \emph{DGLAP}}\tabularnewline
\hline
\textbf{\cite{Altarelli:1998nb}}  & \textbf{1} & \textbf{$\Delta G$=1.5(9)} & \textbf{ABFR 1998} \tabularnewline
\hline
\textbf{\cite{Goto:1999by}}  & \textbf{5} & \textbf{$\Delta G$=0.920(2334)} & \textbf{AAC 2000} \tabularnewline
\hline
\textbf{\cite{Airapetian:1999ib}} &  \textbf{2} &  \begin{tabular}{@{}c@{}}  \textbf{$\Delta g /  g$=0.41(18)} \\ at \textbf{$\left\langle x_{g} \right\rangle$= 0.17} \end{tabular}  &  \begin{tabular}{@{}c@{}} \textbf{HERMES DIS+high-$p_T$} \\ \textbf{hadron pairs} \end{tabular} \tabularnewline
\hline
\textbf{\cite{Anthony:1999rm}}  & \textbf{5} & \textbf{$\Delta G$=0.8(7)} & \textbf{E155 incl. \emph{DGLAP}} \tabularnewline
\hline
\textbf{\cite{Gluck:2000dy}}  & \textbf{5} &  \begin{tabular}{@{}c@{}} \textbf{$\Delta G$=0.708} \\ \textbf{$\Delta G=0.974$} \end{tabular} & \begin{tabular}{@{}c@{}} \textbf{Standard GRSV 2000}  \\ \textbf{SU(3)$_f$ breaking}\end{tabular}  \tabularnewline
\hline
\textbf{\cite{Leader:2001kh}}  & \textbf{1} & \textbf{$\Delta G$=0.68(32)} & \textbf{LSS 2001} \tabularnewline
\hline
\textbf{\cite{Forte:2001ph}}  & \textbf{4} & \textbf{$\Delta G=1.262$} & \textbf{ABFR 2001} \tabularnewline
\hline
\textbf{\cite{Bluemlein:2002be}}  & \textbf{4} & \textbf{$\Delta G=0.931(669)$} & \textbf{BB2002} \tabularnewline
\hline
\textbf{\cite{Hirai:2003pm}}  & \textbf{5} & \textbf{$\Delta G=0.861(2185)$} & \textbf{AAC 2003} \tabularnewline
\hline
\textbf{\cite{Adeva:2004dh}} &  \textbf{13} &  \begin{tabular}{@{}c@{}}   \textbf{$\Delta g /  g$=-0.20(30)} \\ \textbf{at $\left\langle x_{g} \right\rangle = 0.07$} \end{tabular} &  \begin{tabular}{@{}c@{}} \textbf{SMC DIS+high-$p_T$} \\ \textbf{hadron pairs} \end{tabular} \tabularnewline
\hline 
\cite{Diehl:2004cx} & 4  & $\Delta G+\mbox{L}_g=0.40(5) $   &Valence only. GPD constrained w/  nucl. form factors \tabularnewline
\hline 
\cite{Guidal:2004nd} & 2  & $\Delta G+\mbox{L}_g= 0.22$  & GPD model \tabularnewline
\hline
\textbf{\cite{Procureur:2006sg}}  &  \textbf{3} &  \begin{tabular}{@{}c@{}}  \textbf{$\Delta g /  g$=0.016(79)} \\ \textbf{at $\left\langle x_{g} \right\rangle = 0.09$} \end{tabular}  &  \begin{tabular}{@{}c@{}} \textbf{COMPASS quasi-real high-$p_T$} \\ \textbf{hadron pairs prod.}\end{tabular} \tabularnewline
\hline
\textbf{\cite{Hirai:2006sr}}  & \textbf{5} & \textbf{$\Delta G=0.67(186)$} & \textbf{AAC 2006} \tabularnewline
\hline
\begin{tabular}{@{}c@{}}  \textbf{\cite{Ellinghaus:2005uc}} \\  \textbf{\cite{Mazouz:2007aa}}  \end{tabular} & \textbf{1.9}  & \textbf{$\Delta G+\mbox{L}_g$=0.23(27)}  & \begin{tabular}{@{}c@{}} \textbf{JLab and HERMES} \\  \textbf{DVCS data} \end{tabular}   \tabularnewline
\hline
\begin{tabular}{@{}c@{}}  \cite{Wakamatsu:2006dy} \\ \cite{Wakamatsu:2007ar}  \end{tabular} & 
$\infty$ & $\Delta G+\mbox{L}_g=0.264 $    & 
\begin{tabular}{@{}c@{}}  $\chi$ quark solit. \\ mod.  $n_f=6$  \end{tabular}\tabularnewline
 \hline
\textbf{\cite{Hirai:2008aj}}  & \textbf{5} & \textbf{$\Delta G=1.07(104)$} & \textbf{AAC 2008} \tabularnewline
\hline
\begin{tabular}{@{}c@{}}\cite{Myhrer:2007cf}, \\ \cite{Thomas:2008ga} \end{tabular}&  4 & $\Delta G+\mbox{L}_g=0.208(63) $   & \begin{tabular}{@{}c@{}} quark model \\  w/ pion cloud \end{tabular} \tabularnewline 
\hline
\cite{Goloskokov:2008ib} & 4  &  $\Delta G + \mbox{L}_g =  0.20(7)$ & GPD model \tabularnewline
\hline
\textbf{\cite{Alekseev:2009ad}} & \textbf{13} &  \begin{tabular}{@{}c@{}}  \textbf{$\Delta g /  g$=-0.49(29)} \\ \textbf{at $\left\langle x_{g} \right\rangle$ = 0.11} \end{tabular} & \begin{tabular}{@{}c@{}} \textbf{COMPASS  Open} \\ \textbf{Charm} \end{tabular} \tabularnewline
\hline
\textbf{\cite{deFlorian:2009vb}} & \textbf{5} & \textbf{$\Delta G$=-0.073} &\textbf{DSSV08} \tabularnewline
 \hline
\textbf{\cite{Airapetian:2010ac}} & \textbf{1.35} &  \begin{tabular}{@{}c@{}}  $\Delta g /  g =0.049(35)(^{126}_{~99})$ \\ \textbf{at $\left\langle x_{g} \right\rangle$ = 0.22} \end{tabular} & \begin{tabular}{@{}c@{}} \textbf{HERMES DIS +} \\ \textbf{high-$p_T$ incl. hadron production} \end{tabular} \tabularnewline
\hline
\cite{Garvey:2010fi} & -  & $\Delta G+\mbox{L}_g=0.163(28) $ & \begin{tabular}{@{}c@{}} quark model+unpol. sea \\ asym. (Garvey relation) \end{tabular} \tabularnewline
\hline
\textbf{\cite{Leader:2010rb}}  & \textbf{4} & \textbf{$\Delta G=-0.02(34)$} & \textbf{LSS 2010} \tabularnewline
\hline
\textbf{\cite{Blumlein:2010rn}}  & \textbf{4} & \textbf{$\Delta G=0.462(430)$} & \textbf{BB 2010} \tabularnewline
\hline
\cite{Altenbuchinger:2010sz} & $\approx0.2$ &  $\Delta G+\mbox{L}_g= -0.26(10)$  &  Gauge-invariant cloudy bag model \tabularnewline
\hline
\cite{Bacchetta:2011gx} & 4 &  $\Delta G+\mbox{L}_g= 0.23(3)$ & single spin trans. asy. \tabularnewline
\hline
\cite{Bass:2011zn} & 5 & $\Delta G \lesssim 0.4$ &  c-quark axial-charge constraint \tabularnewline
\hline
\textbf{\cite{Adolph:2012ca}} &  \textbf{13} & $ \begin{tabular}{@{}c@{}}  $\Delta g /  g$\textbf{=-0.13(21)} \\ \textbf{at $\left\langle x_{g} \right\rangle$= 0.2} \end{tabular}$ & \textbf{COMPASS open charm} \tabularnewline
\hline
\textbf{\cite{Adolph:2012ca}} &  \textbf{3} & \textbf{$\Delta G$=0.24(9)} & \textbf{Global fit+COMPASS open charm} \tabularnewline
\hline
\cite{GonzalezHernandez:2012jv} &  4 & $\Delta G+\mbox{L}_g=0.263(107) $   &   GPD constrained w/  nucl. form factors \tabularnewline
\hline
\textbf{\cite{Adolph:2012vj}} & \textbf{3} & \begin{tabular}{@{}c@{}}  \textbf{$\Delta g /  g$=0.125(87)} \\ \textbf{at $\left\langle x_{g} \right\rangle$=0.09} \end{tabular} & \begin{tabular}{@{}c@{}} \textbf{COMPASS  DIS +}  \\\textbf{high-$p_T$ hadron pairs}  \end{tabular} \tabularnewline
\hline
\textbf{\cite{Ball:2013lla}}  & \textbf{4} & \textbf{$\Delta G=-0.9(39)$} & \textbf{NNPDF 2013} \tabularnewline
\hline
\cite{Diehl:2013xca} & 4  & $\Delta G+\mbox{L}_g=0.274(29) $   &  \begin{tabular}{@{}c@{}}  GPD constrained w/ \\ nucl. form factors  \end{tabular}  \tabularnewline
\hline
\cite{Bijker:2014ila} & -  &  $\Delta G+\mbox{L}_g=0.14(7) $  & \begin{tabular}{@{}c@{}} unquenched  \\ quark model \end{tabular} \tabularnewline
\hline
\cite{Brodsky:2014yha}   &  5 &   $\Delta G + \mbox{L}_g =0.09$   &  LFHQCD \tabularnewline
\hline
\textbf{\cite{deFlorian:2014yva}} & \textbf{10} & \textbf{$\int_{0.001}^1 \Delta g dx$=0.37(59)} & \textbf{DSSV14} \tabularnewline
\hline
\textbf{\cite{Adamczyk:2014ozi}} & \textbf{10}  & \textbf{$\Delta G$=0.21(10)} & \textbf{NNPDF~\cite{Nocera:2014gqa}  including STAR data}  \tabularnewline
\hline
\textbf{\cite{Adolph:2015cvj}}& \textbf{3} &  \begin{tabular}{@{}c@{}}  \textbf{$\Delta g /  g$=0.113(52)} \\ \textbf{at $\left\langle x_{g} \right\rangle$= 0.1} \end{tabular} &  \begin{tabular}{@{}c@{}} \textbf{COMPASS SIDIS} \\ \textbf{deuteron data}  \end{tabular}  \tabularnewline
\hline
\cite{Gutsche:2016gcd} & 1 & $\Delta G+\mbox{L}_g=0.152$  & LF quark model \tabularnewline
\hline
\textbf{\cite{Shahri:2016uzl}}& \textbf{5} & \textbf{$\Delta G$=0.391} &  \textbf{KTA17 global fit}  \tabularnewline
\hline
\cite{Dahiya:2016wjf} & $\approx 1$ & $ \Delta G +\mbox{L}_g= 0$ & chiral quark model \tabularnewline
\hline
\cite{Chakrabarti:2016yuw}    & - & $\Delta G+\mbox{L}_g=-0.035$ & AdS/QCD scalar quark-diquark model \tabularnewline
 \hline
\end{tabular}
\caption{\label{table Delta G 1}\small Same as Table~\ref{table Delta Sigma 1} but for gluon contributions. $x_g$ is the gluon momentum fraction. Results are in the $\overline{MS}$ scheme. The lists for models and LGT are ordered chronologically and are not comprehensive.
}
}
\end{table}
\noindent
\begin{table}
\vspace{-1.5cm}
\center
\scriptsize{
\begin{tabular}{|c|c|c|c|}
\hline 
Ref. & $Q^{2}$ (GeV$^{2}$) &  $\Delta G+\mbox{L}_g$ & Remarks\tabularnewline
\hline
 \hline
 \cite{Mathur:1999uf}& 3 & 0.20(7) & U. Kentucky group. Quenched calc.  w/ $\chi$ extrap. \tabularnewline
\hline
\cite{Gockeler:2003jfa}  & 4 & 0.17(7) &  QCDSF 2003. u, d only. Quenched calc. w/ $\chi$ extrap. \tabularnewline
\hline
\cite{Dorati:2007bk} & 4  & 0.249(12) &  CC$\chi$PT. u, d only.  W/ $\chi$ extrap. \tabularnewline
\hline
\cite{Brommel:2007sb} & 4  & 0.274(11)  & QCDSF-UKQCD. u, d only. $\chi$ extrap. \tabularnewline 
\hline
\cite{Bratt:2010jn} & 5 & 0.262(18)  & LHPC 2010. u, d only. $\chi$ extrap. \tabularnewline
\hline
 \cite{Alexandrou:2011nr} & 4  & 0.358(40)   &  Twisted-Mass 2011 u, d only.   W/ $\chi$ extrap. \tabularnewline
\hline
 \cite{Alexandrou:2013joa} & 4  & 0.289(32)    &  Twisted-Mass 2013 u, d only.  $m_{\pi}$=0.213 GeV\tabularnewline
\hline
\cite{Abdel-Rehim:2013wlz} & 4 & 0.220(110)  &  Twisted-Mass 2013. Phys. q masses \tabularnewline
 \hline
\cite{Deka:2013zha} & 4 & 0.14(4) &  $\chi$QCD col. w/ $\chi$ extrap. \tabularnewline
\hline
\cite{Alexandrou:2016mni} & 4 & 0.325(25)   &  Twisted-Mass 2016. Phys. $\pi$ mass \tabularnewline
\hline
\cite{Yang:2016plb} & 10 &  0.251(47) &  $\chi$QCD 2017. Phys. $\pi$ mass  \tabularnewline
\hline
\cite{Alexandrou:2016tuo} & 4 &  0.09(6) &  Twisted-Mass 2017. Phys. $\pi$ mass  \tabularnewline
\hline
\end{tabular}
\vspace{-0.3cm}
\caption{\label{table Delta G 2} \small Continuation of Table~\ref{table Delta G 1}, for LGT results. They are given in the $\overline{MS}$ scheme.
}
}
\end{table}
\noindent
\begin{table}
\center
\scriptsize{
\begin{tabular}{|c|c|c|c|c|c|c|}
\hline 
Ref. & \begin{tabular}{@{}c@{}}  $Q^{2}$ \\ (GeV$^{2}$) \end{tabular}&  
\begin{tabular}{@{}c@{}} $\mbox{L}_{u}$ \\ $J_{u}$ \end{tabular}  & 
\begin{tabular}{@{}c@{}} $\mbox{L}_{d}$ \\ $J_{d}$ \end{tabular} & 
\begin{tabular}{@{}c@{}} $\mbox{L}_{s}$ \\ $J_{s}$ \end{tabular} & 
\begin{tabular}{@{}c@{}} disc. \\  diag.? \end{tabular} & Remarks\tabularnewline
\hline
\hline
 \cite{Sehgal:1974rz}  & - & \multicolumn{3}{|l|}{$  \hspace{3cm} \mbox{L}_q=0.20$} & N/A & quark parton model \tabularnewline
\hline
\begin{tabular}{@{}c@{}}  \cite{Jaffe:1989jz} \\ \cite{Thomas:2008ga} \end{tabular}& -  &   \begin{tabular}{@{}c@{}} 0.46 \\   0.89  \end{tabular} &
\begin{tabular}{@{}c@{}}  -0.11  \\ -0.22  \end{tabular} &
\begin{tabular}{@{}c@{}} 0 \\ 0 \end{tabular}   &
N/A &
\begin{tabular}{@{}c@{}}  relat. quark model \\ Canonical def. \end{tabular}   \tabularnewline
\hline
\cite{Cheng:1994zn}&  -  & \multicolumn{3}{|l|}{$  \hspace{3cm} \mbox{L}_q=0.32$} & N/A & $\chi$ quark model \tabularnewline
\hline
\cite{Gluck:2000dy}  & 5 & \multicolumn{4}{|l|}{\begin{tabular}{@{}c@{}} \hspace{3cm}$\mbox{L}_{q+g}=0.18$ \\ \hspace{3cm}$\mbox{L}_{q+g}=0.08$  \end{tabular} }& \begin{tabular}{@{}c@{}} Standard GRSV 2000  \\ SU(3)$_f$ breaking\end{tabular}  \tabularnewline
\hline 
\cite{Guidal:2004nd} & 2  & 
\begin{tabular}{@{}c@{}} -0.12(2)  \\  0.29  \end{tabular} &
\begin{tabular}{@{}c@{}}  0.20(2)  \\ -0.03  \end{tabular} &
\begin{tabular}{@{}c@{}} 0.07(5) \\ 0.02 \end{tabular} &
N/A & GPD model \tabularnewline
\hline 
\cite{Diehl:2004cx} & 4  & 
\begin{tabular}{@{}c@{}} -0.26(1)  \\  0.15(3)  \end{tabular} &
\begin{tabular}{@{}c@{}}  0.17(3)  \\  -0.05(4) \end{tabular} &
\begin{tabular}{@{}c@{}} - \\ - \end{tabular} &
\begin{tabular}{@{}c@{}} Valence  \\ contr. only \end{tabular}  &
 \begin{tabular}{@{}c@{}}  GPD constrained w/ \\ nucl. form factors  \end{tabular} \tabularnewline
\hline
\begin{tabular}{@{}c@{}}  \cite{Wakamatsu:2006dy} \\ \cite{Wakamatsu:2007ar}  \end{tabular} & 
$\infty$ & \multicolumn{3}{|l|}{\begin{tabular}{@{}c@{}}  $L_{u+d}=0.050$ \\  $J_{u+d}=0.236$  \end{tabular}  }&
\begin{tabular}{@{}c@{}} Valence  \\ contr. only \end{tabular}  & 
\begin{tabular}{@{}c@{}}  $\chi$ quark solit. \\ mod.  $n_f=6$  \end{tabular}\tabularnewline
\hline
\begin{tabular}{@{}c@{}}\cite{Myhrer:2007cf}, \\ \cite{Thomas:2008ga} \end{tabular}&  4 & 
\begin{tabular}{@{}c@{}} -0.005(60)  \\  0.405(57)  \end{tabular} &
\begin{tabular}{@{}c@{}}  0.107(33)  \\  -0.113(26) \end{tabular} &
\begin{tabular}{@{}c@{}} - \\ - \end{tabular} &
 N/A & \begin{tabular}{@{}c@{}} quark model \\  w/ pion cloud \end{tabular} \tabularnewline 
\hline
\begin{tabular}{@{}c@{}} \textbf{\cite{Ellinghaus:2005uc}} \\  \textbf{\cite{Mazouz:2007aa}}  \end{tabular} & \textbf{1.9}  & 
\begin{tabular}{@{}c@{}}  \textbf{-0.03(23)} \\  \textbf{0.38(23)}  \end{tabular} &
\begin{tabular}{@{}c@{}}  \textbf{0.11(15)}  \\  \textbf{-0.11(15)} \end{tabular} &
\begin{tabular}{@{}c@{}} \textbf{-} \\ \textbf{-} \end{tabular} &
\textbf{N/A} & \begin{tabular}{@{}c@{}} \textbf{JLab and HERMES} \\  \textbf{DVCS data} \end{tabular}   \tabularnewline
\hline 
\cite{Goloskokov:2008ib} & 4  & 
\begin{tabular}{@{}c@{}} -0.17(4)  \\  0.24(3)  \end{tabular} &
\begin{tabular}{@{}c@{}}   0.24(3) \\  0.02(3)  \end{tabular} &
\begin{tabular}{@{}c@{}} 0.07(6) \\ 0.02(3) \end{tabular} &
N/A & GPD model \tabularnewline
\hline
\cite{Garvey:2010fi} & -  & \multicolumn{3}{|l|}{\begin{tabular}{@{}c@{}} $l_{u+d+s}=0.147(27)$ \\  $J_{u+d+s}=0.337(28)$ \end{tabular}}   &
N/A & \begin{tabular}{@{}c@{}} quark model+unpol. sea \\ asym. (Garvey relation) \end{tabular} \tabularnewline
\hline
\begin{tabular}{@{}c@{}}  \cite{Altenbuchinger:2010sz}  \end{tabular}& $\approx0.2$  &   \begin{tabular}{@{}c@{}} 0.34(13) \\   0.72(14)  \end{tabular} &
\begin{tabular}{@{}c@{}}  0.19(13)  \\ 0.04(14)  \end{tabular} &
\begin{tabular}{@{}c@{}} - \\ - \end{tabular}   &
N/A & 
\begin{tabular}{@{}c@{}}Gauge-invariant \\ cloudy bag model \end{tabular} \tabularnewline
\hline
\cite{Bacchetta:2011gx} & 4 & 
\begin{tabular}{@{}c@{}} -0.166(15) \\ 0.244(11) \end{tabular} &
\begin{tabular}{@{}c@{}} 0.235(12) \\  0.015(6)$(^{20}_{~5})$ \end{tabular} &
\begin{tabular}{@{}c@{}} 0.062($^5_9$) \\ 0.012 $(^{2}_{8})$ \end{tabular} &
N/A & \begin{tabular}{@{}c@{}}  single spin \\ trans. asy. \end{tabular} \tabularnewline
\hline
\cite{Lorce:2011kd} & - & 
\begin{tabular}{@{}c@{}} 0.071 \\ 0.569 \end{tabular} &
\begin{tabular}{@{}c@{}} 0.055 \\ -0.069 \end{tabular} &
\begin{tabular}{@{}c@{}} - \\ - \end{tabular} &
N/A & \begin{tabular}{@{}c@{}} LC constituent \\ quark model \end{tabular}   \tabularnewline
\hline
\cite{Lorce:2011kd} & - & 
\begin{tabular}{@{}c@{}} -0.008 \\ 0.566 \end{tabular} &
\begin{tabular}{@{}c@{}}  0.077 \\ -0.066 \end{tabular} &
\begin{tabular}{@{}c@{}} - \\ - \end{tabular} &
N/A & \begin{tabular}{@{}c@{}} $\chi$ quark \\  soliton model \end{tabular}   \tabularnewline
\hline
\cite{GonzalezHernandez:2012jv} &  4 & 
\begin{tabular}{@{}c@{}}  -0.12(11) \\  0.286(107)  \end{tabular} &
\begin{tabular}{@{}c@{}}  0.17(2)  \\ -0.049(7)  \end{tabular} &
\begin{tabular}{@{}c@{}} - \\ - \end{tabular} &
N/A &  \begin{tabular}{@{}c@{}}  GPD constrained w/ \\ nucl. form factors  \end{tabular}  \tabularnewline
\hline
\cite{Diehl:2013xca} & 4  & 
\begin{tabular}{@{}c@{}}  -0.18(3) \\  $0.230(^9_{24})$  \end{tabular} &
\begin{tabular}{@{}c@{}}   0.21(3) \\  $-0.004(^{10}_{16})$ \end{tabular} &
\begin{tabular}{@{}c@{}} - \\ - \end{tabular} &
N/A &  \begin{tabular}{@{}c@{}}  GPD constrained w/ \\ nucl. form factors  \end{tabular}  \tabularnewline
\hline
\cite{Brodsky:2014yha}  &  5  &  \multicolumn{3}{|l|}{
 \begin{tabular}{@{}c@{}}  $\mbox{L}_{u+d+s}=0.25$ \\ $J_{u+d+s}=0.31$\end{tabular}} &
N/A & LFHQCD.  \tabularnewline
\hline
\cite{Bijker:2014ila} & -  & \multicolumn{3}{|l|}{$l_{u+d+s}=0.221(41)$,  $J_{u+d+s}=0.36(7)$}    &
N/A & \begin{tabular}{@{}c@{}} unquenched  \\ quark model \end{tabular} \tabularnewline
\hline
\cite{Gutsche:2016gcd} & 1 & 
\begin{tabular}{@{}c@{}} 0.055 \\ 0.358 \end{tabular} &
\begin{tabular}{@{}c@{}} -0.001 \\ -0.010 \end{tabular} &
\begin{tabular}{@{}c@{}} - \\ - \end{tabular} &
N/A & LF quark model \tabularnewline
\hline
\cite{Dahiya:2016wjf} & $\approx 1$ & 
\begin{tabular}{@{}c@{}} 0.265 \\ 0.777 \end{tabular} &
\begin{tabular}{@{}c@{}} -0.066 \\ -0.265 \end{tabular} &
\begin{tabular}{@{}c@{}} 0 \\ -0.012 \end{tabular} &
N/A & chiral quark model \tabularnewline
\hline
\cite{Chakrabarti:2016yuw}    & - & 
\begin{tabular}{@{}c@{}} -0.3812 \\ 0.565 \end{tabular} &
\begin{tabular}{@{}c@{}} -0.4258 \\ -0.030 \end{tabular} &
\begin{tabular}{@{}c@{}}  \\  \end{tabular} &
 & \begin{tabular}{@{}c@{}} AdS/QCD scalar \\ quark-diquark model \end{tabular} \tabularnewline
\hline
\end{tabular}
\vspace{-0.3cm}
\caption{\label{table OAM 1}\small Phenomenological results on quark 
$\mbox{L}_q=\mbox{L}_u+\mbox{L}_d+\mbox{L}_s$ and total angular momenta 
$J_q=\mbox{L}_q+\Delta \Sigma_q/2$. Results are in the $\overline{MS}$ scheme.  They use
different definitions of $\mbox{L}_q$, and may thus not be directly comparable, see Section~\ref{SSR components}. 
The list is ordered chronologically and is not comprehensive.
}}
\end{table}
\noindent
\begin{table}
\center
\scriptsize{
\begin{tabular}{|c|c|c|c|c|c|c|}
\hline 
Ref. & \begin{tabular}{@{}c@{}}  $Q^{2}$ \\ (GeV$^{2}$) \end{tabular}&  
\begin{tabular}{@{}c@{}} $L_{u}$ \\ $J_{u}$ \end{tabular}  & 
\begin{tabular}{@{}c@{}} $L_{d}$ \\ $J_{d}$ \end{tabular} & 
\begin{tabular}{@{}c@{}} $L_{s}$ \\ $J_{s}$ \end{tabular} & 
\begin{tabular}{@{}c@{}} disc. \\  diag.? \end{tabular} & Remarks\tabularnewline
\hline

\hline
\cite{Mathur:1999uf} & 3  & \multicolumn{3}{|l|}{$L_{u+d+s}=0.17(6)$,  $J_{u+d+s}=0.30(7)$}    &
yes & \begin{tabular}{@{}c@{}}  U. Kentucky group. Quenched \\  calc.  w/ $\chi$ extrap.  \end{tabular} \tabularnewline
\hline
 \cite{Hagler:2003jd} & 4  & \multicolumn{3}{|l|}{$  \hspace{3cm} J_{q}=0.338(4)$} & No  &  LHPC 2003. u, d only.  $\chi$ extrap. \tabularnewline
\hline
\cite{Gockeler:2003jfa} &  4 & 
\begin{tabular}{@{}c@{}} -0.05(6) \\ 0.37(6)  \end{tabular} &
\begin{tabular}{@{}c@{}}  0.08(4) \\ -0.04(4) \end{tabular} &
\begin{tabular}{@{}c@{}} - \\ - \end{tabular} &
no &  \begin{tabular}{@{}c@{}} QCDSF. u, d only. Quenched \\  calc. w/ $\chi$ extrap. \end{tabular} \tabularnewline
\hline
\cite{Dorati:2007bk} & 4  &  
\begin{tabular}{@{}c@{}}  -0.14(2) \\   0.266(9) \end{tabular} &
\begin{tabular}{@{}c@{}}   0.21(2) \\  -0.015(8) \end{tabular} &
\begin{tabular}{@{}c@{}} - \\ - \end{tabular} &
no &  \begin{tabular}{@{}c@{}} CC$\chi$PT. u, d only.  \\  W/ $\chi$ extrap.\end{tabular} \tabularnewline
\hline
\cite{Brommel:2007sb} & 4  & 
\begin{tabular}{@{}c@{}}  -0.18(2) \\   0.230(8) \end{tabular} &
\begin{tabular}{@{}c@{}}   0.22(2)  \\  -0.004(8) \end{tabular} &
\begin{tabular}{@{}c@{}} - \\ - \end{tabular} &
 no & QCDSF-UKQCD. u, d only. $\chi$ extrap. \tabularnewline
\hline
\cite{Bratt:2010jn} & 5 & 
\begin{tabular}{@{}c@{}} -0.175(40) \\ 0.236(18) \end{tabular} &
\begin{tabular}{@{}c@{}} 0.205(35) \\  0.002(4) \end{tabular} &
\begin{tabular}{@{}c@{}} - \\ - \end{tabular} &
no & LHPC 2010. u, d only. $\chi$ extrap. \tabularnewline
\hline
 \cite{Alexandrou:2011nr} & 4  & 
\begin{tabular}{@{}c@{}}  -0.141(30) \\   0.189(29) \end{tabular} &
\begin{tabular}{@{}c@{}}   0.116(30) \\  -0.047(28) \end{tabular} &
\begin{tabular}{@{}c@{}} - \\ - \end{tabular} &
no &  \begin{tabular}{@{}c@{}} Twisted-Mass 2011 u, d only.  \\  W/ $\chi$ extrap.\end{tabular} \tabularnewline
\hline
 \cite{Alexandrou:2013joa} & 4  & 
\begin{tabular}{@{}c@{}}  -0.229(30) \\   0.214(27) \end{tabular} &
\begin{tabular}{@{}c@{}}   0.137(30) \\  -0.003(17) \end{tabular} &
\begin{tabular}{@{}c@{}} - \\ - \end{tabular} &
no &  \begin{tabular}{@{}c@{}} Twisted-Mass 2013 u, d only.  \\  $m_{\pi}$=0.213 GeV\end{tabular} \tabularnewline
\hline
 \cite{Deka:2013zha} & 4  & 
\begin{tabular}{@{}c@{}}  -0.003(8) \\  0.37(6)  \end{tabular} &
\begin{tabular}{@{}c@{}}  0.195(8)  \\ -0.02(4)  \end{tabular} &
\begin{tabular}{@{}c@{}}  0.07(1) \\ 0.012(4) \end{tabular} &
yes & $\chi$QCD col. w/ $\chi$ extrap. \tabularnewline
\hline
\cite{Abdel-Rehim:2013wlz} & 4 & 
\begin{tabular}{@{}c@{}} -0.208(95) \\ 0.202(78) \end{tabular} &
\begin{tabular}{@{}c@{}}  0.078 (95)\\  0.078(78) \end{tabular} &
\begin{tabular}{@{}c@{}}  - \\ - \end{tabular} &
yes &  Twisted-Mass 2013. Phys. q masses \tabularnewline
\hline
 \cite{Alexandrou:2016mni} & 4 & 
\begin{tabular}{@{}c@{}} -0.118(43) \\ 0.296(40) \end{tabular} &
\begin{tabular}{@{}c@{}} 0.252(41) \\ 0.058(40) \end{tabular} &
\begin{tabular}{@{}c@{}} 0.067(21) \\  0.046(20)\end{tabular} &
yes & Twisted-Mass 2016. Phys. $\pi$ mass  \tabularnewline
\hline
\cite{Alexandrou:2016tuo} & 4 & 
\begin{tabular}{@{}c@{}} -0.104(29) \\ 0.310(26) \end{tabular} &
\begin{tabular}{@{}c@{}} 0.249(27) \\ 0.056(26) \end{tabular} &
\begin{tabular}{@{}c@{}} 0.067(21) \\ 0.046(21) \end{tabular} &
yes & Twisted-Mass 2017. Phys. $\pi$ mass \tabularnewline
\hline
\end{tabular}
\caption{\label{table OAM 2} \small Same as Table~\ref{table OAM 1} but for  LGT results.
}}
\end{table}
\FloatBarrier

\pagebreak

\huge{\bf{Lexicon and acronyms}}\normalsize

\vspace{10pt}

To make this review more accessible to non-specialists,
we provide here specific terms associated with  the nucleon structure, with short explanations 
and links to where they are first discussed in the review. For convenience, we also provide 
the definitions of the acronyms used in this review.

\begin{itemize}

\item  AdS/CFT: \emph{anti-de-Sitter/conformal field theory}.
\item  AdS/QCD: \emph{anti-de-Sitter/quantum chromodynamics}.
\item anti-de-Sitter (AdS) space: a maximal symmetric space endowed with a constant negative curvature.
\item Asymptotic freedom: QCD's property that its strength decreases at short distances.
\item Asymptotic series: see Poincar\'{e} series.
\item $\beta$-function: the logarithmic derivative of $\alpha_{s}$: $\beta\left(\mu^{2}\right)=\frac{d\alpha_{s}\left(\mu\right)}{d\mbox{\footnotesize{ln}}(\mu)}$
where $\mu$ is the \emph{subtraction point}. In the perturbative
domain, $\beta$ can be expressed as a perturbative series \textbf{$\beta=-\frac{1}{4\pi}\sum_{n=0}\left(\frac{\alpha_{s}}{4\pi}\right)^{n}\beta_{n}$}.
\item Balitsky-Fadin-Kuraev-Lipatov (BFKL) evolution equations: the equations controlling the low-$x_{Bj}$ behavior of structure functions.
\item  BBS: Brodsky-Burkardt-Schmidt.
\item  BC: Burkhardt-Cottingham.
\item BLM: Brodsky-Lepage-Mackenzie. See Principle of Maximal Conformality (PMC).
\item CERN: Conseil Europ\'een pour la Recherche Nucl\'eaire.
\item  $\chi$PT: chiral perturbation theory.
\item CEBAF: continuous electron beam accelerator facility.
\item CLAS: CEBAF large acceptance spectrometer.
\item COMPASS: common muon and proton apparatus for structure and spectroscopy.
%
%
\item Condensate (or Vacuum Expectation Value, VEV): the vacuum expectation value of a given local operator.
  Condensates allow one to parameterize the  nonperturbative \emph{OPE}'s power
corrections.    Condensates and vacuum loop diagrams do not appear in the frame-independent 
light-front Hamiltonian since all lines have $k^+ = k^0 + k^3 \ge 0$ 
and the sum of $+$ momenta is conserved at every vertex.
In the light-front formalism condensates are associated with physics of the hadron 
wavefunction and are called ``in-hadron"  condensates, which  refers to physics 
possibly contained in the higher LF  Fock states of the hadrons \cite{Casher:1974xd}. In the case of the Higgs theory,
the usual Higgs VEV of the \emph{instant form} Hamiltonian is replaced by a ``zero mode", a background field with $k^+=0$ \cite{Brodsky:2012ku}.
\item Conformal behavior/theory: the behavior of a quantity or a theory
that is scale invariant. In a conformal theory the \emph{$\beta$-function} vanishes.  
More rigorously, a conformal  theory  is invariant under  both  
dilatation and the special conformal transformations which involve coordinate inversion. 
\item Cornwall-Norton moment: the moment $\int_0^1 x^N g(x,Q^2) dx$ 
of a structure function $g(x_{Bj},Q^2)$. See Mellin-transform.
\item Constituent quarks: unphysical particles of approximately a third of the nucleon mass and ingredients
of \emph{constituent quark} models. They provide the $J^{PC}$ quantum numbers describing the hadron.  
Constituent quarks can be viewed as \emph{valence quarks} dressed by virtual pairs of partons. 
\item  DDIS: diffractive deep inelastic scattering.
\item DESY: Deutsches Elektronen-Synchrotron.
\item Dimensional transmutation: the emergence of a mass or momentum scale in a quantum theory with 
a classical Lagrangian devoid of  explicit mass or energy parameters \cite{Coleman:1973sx}.
\item  DIS: deep inelastic scattering.
\item Distribution amplitudes: universal quantities describing the \emph{valence quark} structure of hadrons and nuclei. 
\item Dokshitzer-Gribov-Lipatov-Altarelli-Parisi (DGLAP) evolution equations: 
the equations controlling the $Q^2$ behavior of structure functions, except at extreme $x_{Bj}$ (low- and large-$x_{Bj}$).  
 The DGLAP equations are used in global determinations of parton distributions by evolving the distribution functions from an initial  to a final scale.
\item DVCS: deeply virtual Compton scattering.
\item Effective charge: an effective coupling defined from a perturbatively calculable observable. It  includes
all perturbative and relevant nonperturbative effects~\cite{Grunberg:1980ja}.
\item Effective coupling: the renormalized (running) coupling, in contrast
with the constant unphysical bare coupling.
\item EFL: Efremov-Leader-Teryaev.
\item Efremov-Radyushkin-Brodsky-Lepage (ERBL) evolution equations: the equations controlling the evolution of the  \emph{Distribution amplitudes} in $\ln (Q^2)$.
\item EIC: electron-ion collider.
\item EMC: european muon collaboration.
%
%
\item Factorization scale: the scale at which nonperturbative effects become negligible.
\item Factorization theorem: the ability to separate at short distance the perturbative coupling of the probe to the nucleon, from
the nonperturbative nucleon structure~\cite{factorization theorem}.
\item Freezing: the loss of scale dependence of finite $\alpha_{s}$ in the infrared.  See
also conformal behavior. 
%
%
\item Gauge link  or link variable: in Lattice QCD, the segment(s) linking two lattice sites to which a unitary matrix is 
associated to implement gauge invariance.  While quarks reside at the lattice sites, gauge links effectively represent the gluon field.
Closed links are  \emph{Wilson loops} used to construct the LGT Lagrangian.
\item  GDH: Gerasimov-Drell-Hearn.
\item Ghosts: ghosts referred to unphysical fields. For example in certain gauges 
in QED and QCD, such as the Feynman gauge, there are four vector-boson fields: 
two transversely polarized bosons (photons and  gluons, respectively), 
a longitudinally polarized one, and 
a scalar one with a negative metric. This later is referred to as a ghost photon/gluon and is unphysical since it does not
represent  an independent  degree of freedom: While vector-bosons have in principle 4-spin degrees of freedom, 
only three are independent due to the additional constraint from gauge invariance.
In Yang-Mills theories, Faddeev-Popov ghosts are fictitious particles of spin zero but that obey
the Fermi--Dirac statistics (negative-metric particles). These characteristics are chosen so that the
ghost propagator complements the non-transverse term in the gluon propagator to 
make it transverse, and thus insure current conservation. 
In radiation or Coulomb gauge, the scalar and longitudinally polarized vector-bosons 
are replaced by the Coulomb interaction. 
Axial gauges where vector-bosons are always transverse, in particular the LC gauge $A^+$, can alternatively
be used to avoid introducing ghosts.
\item GPD: generalized parton distributions.
\item GTMD: generalized transverse momentum distributions
\item Hard reactions or hard scattering: high-energy processes, in particular in which the quarks are resolved. 
\item HIAF: high intensity heavy ion accelerator facility.
\item Higher-twist: See \emph{Twist}
\item  HLFHS: holographic light-front hadron structure collaboration.
\item IMF: infinite momentum frame.
\item Instant form, or instant time quantization: the traditional second quantization of a field theory, 
done at instant time $t$;  one of the forms of relativistic dynamics introduced by Dirac. 
See \emph{Light-front quantization} and Sec.~\ref{LC dominance and LF quantization}.
%
%
\item JAM: JLab angular momentum collaboration
\item JINR: Joint Institute for Nuclear Research.
\item JLab: Jefferson Laboratory.
\item Landau pole, Landau singularity or Landau ghost:  the point where a perturbative coupling
diverges. At first order (1-loop) in pQCD, this occurs at the \emph{scale parameter}
$\Lambda_s$. The value can depend on the choice of renormalization scheme, the order $\beta_i$
at which the coupling series is estimated, the number of flavors $n_f$
and the approximation chosen to solve the QCD $\beta$ equation. The Landau pole is unphysical.
\item  LC: light cone.
\item LEGS: laser electron gamma source.
\item  LF: light-front.
\item  LFHQCD: light-front holographic QCD.
\item Light-front quantization: second   quantizaton of a field theory done at fixed LF-time $\tau$, 
rather than at \emph{instant time} $t$; one of the relativistic forms introduced by Dirac.
The equal LF-time condition defines a plane, rather than a cone, tangent to the light-cone. 
Thus the name "Light-Front". 
See \emph{Instant form} and Sec.~\ref{LC dominance and LF quantization}.
\item  LFWF: light-front wave function.
\item  LGT: lattice gauge theory.
\item LO: leading order.
\item LSS: Leader-Sidorov-Stamenov.
\item LT: longitudinal-transverse.
\item  MAMI: Mainz Microtron.
\item Mellin transform: the moment $\int _0^1 x^N g(x,Q^2) dx$, typically of a structure function $g(x_{Bj},Q^2)$. 
It transforms $g(x_{Bj},Q^2)$ to Mellin space ($N,Q^2$), with $N$ the moment's order. 
Advantages are 1) that  the $Q^2$-evolution of moments are simpler than that 
of structure function $Q^2$-evolution, since the nonperturbative $x_{Bj}$-dependence is integrated over. 
Furthermore, convolutions of PDFs partition functions (see 
Eqs.~(\ref{g_1 LT evol})--(\ref{gluon LO evol})) become simple products in \emph{Mellin-space}.
The structure functions are then recovered by inverse transforming back to the $x_{Bj},Q^2$ space; and 2)
low-$N$ moments are computable on the lattice with smaller noise than (non-local) structure functions. 
Structure functions can be obtain by inverse transform the 1- to $N$-moments, if $N$ is large enough.
\item NICA: nuclotron-based ion collider facilities.
\item NLO: next-to-leading order.
\item NNLO: next-to-next-to-leading order.
\item OAM: orbital angular momentum.
%
%
\item Operator Product Expansion (OPE). See also higher-twist:  
the \emph{OPE} uses the \emph{twist} of effective operators to predict the power-law fall-off of an amplitude.
It thus can be used to distinguish 
logarithmic leading \emph{twist} perturbative corrections from the $1/Q^{n}$ \emph{power corrections}. The \emph{OPE}
typically does not provide values for the nonperturbative \emph{power correction} coefficients.
\item Optical Theorem: the relation between a cross-section and its corresponding photo-absorption amplitude.
Generally speaking, the dispersion of a beam is related to the transition amplitude. 
This results from the \emph{unitarity} of a reaction. The theorem expresses
the fact that the dispersive part of a process (the cross-section) is proportional 
to the imaginary part of the transition amplitude. The
case is similar to classical optics, where complex refraction indices
are introduced to express the dispersion of a beam of light in a medium
imperfectly transparent. This explains the name of the theorem.
\item PDF: parton distribution functions
%
%
%
\item Poincar\'{e} series (also Asymptotic series). See also ``renormalons". 
A series that converges up to an order 
$k$ and then diverges. The series reaches its best convergence at
order $N_{b}$ and then diverges for orders $N\gtrsim N_{b}+\sqrt{N_{b}}$. 
Quantum Field Theory series typically are asymptotic and converge 
up to an order $N_b \simeq 1/a$, with $a$ the expansion coefficient. 
IR \emph{renormalons}  generate an $n!\beta^{n}$ factorial growth
of the $n$th coefficients in \emph{nonconformal} ($\beta \neq 0$) theories. Perturbative calculation
to high order ($\alpha_{s}^{20}$) has been performed on the lattice 
\cite{the:Renormalon growth} to check 
the asymptotic behavior of QCD series. Factorial growth is seen up
to the 20th order of the calculated series. 
\item Positivity constraint: the requirement on PDF functions that scattering cross-sections must be positive.
\item Power corrections. See ``Higher-twist" and ``Renormalons".
\item pQCD: perturbative quantum chromodynamics.
\item Principle of Maximal Conformality (PMC):  a method used to set the \emph{renormalization scale}, 
order-by-order in perturbation theory, by shifting all $\beta$ terms  in the pQCD series  into the 
\emph{renormalization scale} of the running QCD coupling at each order.  The resulting coefficients of the 
series then match the coefficients of the corresponding \emph{conformal} theory with $\beta=0$.   The PMC 
generalizes the Brodsky Lepage Mackenzie BLM method to all orders.  In the Abelian $N_C\to 0$ limit, the 
PMC reduces to the standard Gell-Mann--Low method used for scale setting in QED \cite{Brodsky:1997jk}.
\item   Pure gauge sector, pure Yang Mills or pure field. Non Abelian field theory
without fermions. See also \emph{quenched} approximation.
\item PV: parity violating.
\item PWIA: plane wave impulse approximation.
\item QCD: quantum chromodynamics.
\item QCD counting rules:  the asymptotic constraints imposed on  form factors and transition amplitudes by the minimum number of partons involved in the elastic scattering. 
\item   QCD scale parameter $\Lambda_s$: the UV scale ruling the energy-dependence
of $\alpha_{s}$. It also provides the
scale at which $\alpha_{s}$ is expected to be large, and  nonperturbative
treatment of QCD is required~\cite{Deur:2016tte}.
\item QED: quantum electrodynamics.
\item Quenched approximation: calculations where the fermion loops are neglected.
It differs from the \emph{pure gauge}, pure Yang Mills case in that heavy (static)
quarks are present. 
\item Renormalization scale: the argument of the running coupling. See also ``Subtraction point".
\item Renormalon: the residual between the physical value of an observable
and the \emph{Asymptotic series} of the observable at its best convergence
order $n \simeq 1/\alpha_{s}$.  The terms of a pQCD calculation which involve the \emph{$\beta$-function} typically diverge as $n!$:  {\it i.e.}, as a renormalon.
Borel summation techniques indicate that IR  renormalons  can often be interpreted as \emph{power
corrections}. Thus, IR  renormalons  should be related to the  \emph{higher
twist} corrections of the \emph{OPE} formalism \cite{the:Renormalons}.
The existence of IR  renormalons  in \emph{pure gauge} QCD is supported by lattice 
QCD \cite{the:Renormalon growth}.  See also ``Asymptotic series".
\item RHIC: relativistic heavy ion collider (RHIC).
\item RSS: resonance spin structure.
\item Sea quarks: quarks stemming from gluon splitting $g \to q \bar{q}$ and from QCD's vacuum fluctuations.
This second contribution is frame dependent and avoided in the light-front formalism.  
Evidence for \emph{sea quarks} making up the nucleon structure in addition to the \emph{valence quarks} 
came from DIS data yielding PDFs that strongly rise at low-$x_{Bj}$.  
%
%
%
\item  SIDIS: semi-inclusive deep inelastic scattering.
\item  SLAC: Stanford Linear Accelerator Center.
\item SMC: spin muon collaboration.
\item SoLID: solenoidal large intensity device.
\item  SSA: single-spin asymmetry.
\item Subtraction point $\mu$: the scale at which the renormalization
procedure subtracts the UV divergences.   
\item Sum rules: a relation between the moment of a structure function, a form factor or a photoabsorption cross-section, and static properties of the nucleon. A more general definition includes
relations of moments to double deeply virtual Compton scattering amplitudes rather than to a static property.  
\item Tadpole corrections: in the context of lattice QCD, tadpole terms
are unphysical contributions to the lattice action which arise from
the discretization of space-time. They contribute at NLO of the bare
coupling $g^{bare}=\sqrt{4\pi\alpha_{s}^{bare}}$ to the expression
of the \emph{gauge link} variable $U_{\overrightarrow{\mu}}$. (The LO corresponds
to the continuum limit.) To suppress these contributions, one can redefine
the lattice action by adding larger  \emph{Wilson loops} or by rescaling the
\emph{link variable}. 
\item TMD: transverse momentum distributions.
\item TT: transverse-transverse.
\item TUNL: Triangle Universities Nuclear Laboratory.
\item Twist:  the twist $\tau$ of an elementary operator is given by its dimension minus its spin. 
For example, the  quark operator $\psi$  has dimension $3$, spin $1/2$ and thus $\tau=1$.   
For elastic scattering at high $Q^2$, LF QCD gives $ \tau=n-1$ with $n$ is the number of effective constituents of a hadron. 
For DIS, structure functions are dominated by $ \tau=2$, the \emph{leading-twist}. 
 \emph{Higher-twist} are $Q^{2-\tau}$ \emph{power corrections} to those, typically 
derived from the \emph{OPE} analysis of  the nonperturbative effects of multiparton interactions.
 \emph{Higher-twist} is sometimes interpretable as kinematical phenomena, {\it e.g.} the mass $M$ of a 
nucleon introduces a \emph{power correction} beyond the pQCD scaling violations, or as dynamical 
phenomena, {\it e.g.},  the intermediate distance transverse forces that confine 
quarks \cite{Burkardt:2008ps, Abdallah:2016xfk}.
\item Unitarity: conservation of the probability:
the sum of probabilities that a scattering occurs with any reaction, or does not occur, must be 1.
\item Unquenched QCD: see  \emph{pure gauge} sector and \emph{quenched} approximation.
\item Valence quarks: the nucleon quark content once all quark-antiquark pairs (\emph{sea quarks}) are excluded.
Valence quarks determine the correct quantum numbers of hadrons.
\item VEV: vacuum expectation value.
\item VVCS: doubly virtual Compton scattering.
\item Wilson line: a Wilson line represents all of the final-state interactions between the 
struck quark in DIS and the target spectators. It generates both leading and higher \emph{twists} effects:
for example the exchange of a gluon between the struck quark and the proton's spectators after the 
quark has been struck yields the Sivers effect~\cite{Sivers:1989cc}. It also contributes to DDIS at leading twist.
\item Wilson Loops: closed paths linking various sites in a lattice \cite{Wilson:1974sk}. 
They are used to define the lattice action and \emph{Tadpole corrections}. 
(See Section \ref{LGT}.)

\end{itemize}

\end{document}